\documentclass[universe,review,accept,pdftex,moreauthors]{Definitions/mdpi} 
\def\hb{H$\beta$\/}
\def\oiii{[{O}{\sc iii}]$\lambda\-\lambda$\-4959,\-5007\/}
\def\oiiionly{[{O}{\sc iii}]\/}

\def\civ{{C}{\sc iv}$\lambda$1549\/}
\def\mgii{{Mg}{\sc ii}$\lambda$2800\/}

\def\siiv{{Si}{\sc iv}$\lambda$1397\/}
\def\oiv{{O}{\sc iv}]$\lambda$1402\/}

\def\aliii{{Al}{\sc iii}$\lambda$1860\/}
\def\heiiuv{{He}{\sc ii}$\lambda$1640\/}
\def\feii{{Fe}{\sc ii}\/}
\def\kms{km\,s$^{-1}$\/}

\def\ciii{{C}{\sc iii}]$\lambda$1909\/}
\def\siiii{{Si}{\sc iii}]$\lambda$1892\/}
\def\rfe{$R_\mathrm{{Fe}{\textsc{ii}}}$\/}
\def\mbh{$M_\mathrm{\rm BH}$\/}
\def\lledd{$L/L_\mathrm{\rm Edd}$\/}

\def\c14{$c$(1/4)\/}
\def\ergs{erg s$^{-1}$\/}
\def\apjs{ApJS}
\def\aap{A\&Ap}
\def\mnras{MNRAS\/}
\def\apjl{ApJL\/}
\def\apj{ApJ\/}
\def\pasj{PASJ\/}
\def\pasp{PASP\/}
\def\aj{AJ\/}
\def\nat{Nature\/}
\def\apss{ApSS\/}
\def\feq{Fe{\sc ii}$\lambda$4570\/}
\def\araa{ARA\&Ap}
\firstpage{1} 
\pubvolume{1}
\issuenum{1}
\articlenumber{0}
\pubyear{2025}
\copyrightyear{2025}
\externaleditor{Firstname Lastname}
\datereceived{7 January 2025} 
\daterevised{10 February 2025}
\dateaccepted{13 February 2025} 
\datepublished{} 
\hreflink{https://doi.org/} 

\Title{Super-Eddington Accretion in Quasars}

\TitleCitation{Super-Eddington Accretion in Quasars}

\Author{{Paola} 
 Marziani $^{1,}$*\orcidA{}, 
Karla Garnica Luna $^{2}$, 
Alberto Floris $^{1,3,4}$, 
Ascensi\'on del Olmo $^5$, 
 {Alice Deconto--Machado} 
 $^{5,6}$, 
l{Tania M. Buendia-Rios} $^2$, 
{C. Alenka} 
 {  Negrete} $^2$ 
and 
Deborah Dultzin $^2$ } 

\AuthorNames{Paola Marziani, Karla Garnica Luna, Alberto Floris, Ascensi\'on del Olmo, Alice Deconto--Machado, 
Tania M. Buendia-Rios, C. Alenka Negrete and Deborah Dultzin}

\AuthorCitation{Marziani, P.; Garnica Luna, K.; Floris, A.; del Olmo, A.; Deconto--Machado, A.; Buendia-Rios, T.; Negrete, C.A.; Dultzin, D.}

\address{%
$^{1}$ \quad {Astronomical Observatory of Padova, National Institute for Astrophysics (INAF), }
  IT-35122 Padova, Italy; {afloris@physics.uoc.gr} 
\\
$^{2}$ \quad {Instituto de Astronom\'{\i}a, Universidad Nacional Aut\'onoma de M\'exico}
, AP 70-264, {Mexico City} 
 04510,  {Mexico}
; {kgarnica@astro.unam.mx (K.G.L.); tbuendia@astro.unam.mx (T.B.-R.); alenka@astro.unam.mx (C.A.N.); deborah@astro.unam.mx (D.D.)}\\
$^{3}$ \quad Department of Physics{,} 
 University of Crete, Voutes University Campus, 70013 Heraklion, Greece\\
$^{4}$ \quad Institute of Astrophysics, Foundation for Research and Technology Hellas, N. Plastira 100, Vassilika Vouton, 70013 Heraklion, Greece\\
$^{5}$ \quad Instituto de Astrof\'\i sica de Andaluc\'\i a (IAA), Consejo Superior de Investigaciones Cient\'{\i}ficas {(CSIC)}
, Glorieta de Astronom\'\i a s/n, ES18008 Granada, Spain; {chony@iaa.es (A.d.O.); alice.deconto@inaf.it (A.D.-M.)}\\
$^{6}$ \quad {Istituto di Astrofisica Spaziale e Fisica Cosmica, National Institute for Astrophysics (INAF), }
  \mbox{Via Alfonso Corti 12}, I-20133 Milano, Italy}

\corres{Correspondence: paola.marziani@inaf.it}

\abstract{This review provides an observational perspective on the fundamental properties of super-Eddington accretion onto supermassive black holes in quasars. It begins by outlining the selection criteria, particularly focusing on optical and UV broad-line intensity ratios, used to identify a population of unobscured super-Eddington candidates. Several defining features place these candidates at the extreme end of the Population A in main sequence of quasars: among them are the highest observed singly-ionized iron emission, extreme outflow velocities in UV resonance lines, and unusually high metal abundances. These key properties reflect the coexistence of a virialized sub-system within the broad-line region alongside powerful outflows, with the observed gas enrichment likely driven by nuclear or circumnuclear star formation. The most compelling evidence for the occurrence of super-Eddington accretion onto supermassive black holes comes from recent observations of massive black holes at early cosmic epochs. These black holes require rapid growth rates that are only achievable through radiatively inefficient super-Eddington accretion. Furthermore, extreme Eddington ratios, close to or slightly exceeding unity, are consistent with the saturation of radiative output per unit mass predicted by accretion disk theory for super-Eddington accretion rates. The extreme properties of super-Eddington candidates suggest that these quasars could make them stable and well-defined cosmological distance indicators, leveraging the correlation between broad-line width and luminosity expected in virialized systems. Finally, several analogies with accretion processes around stellar-mass black holes, particularly in the high/soft state, are explored to provide additional insight into the mechanisms driving super-Eddington accretion.
 }


\begin{document}




\section{Introduction}

 The astonishing luminosity of quasars (which may reach $\sim$$10^{48}$ \ergs, 10,000 the luminosity of a giant galaxy) suggests a highly efficient energy release mechanism, much beyond the efficiency yielded by nuclear reactions in stars. Such high efficiency can only be achieved through infall into the deep gravitational well of a compact object. Any lingering doubt regarding the fundamental tenet of what is now known as the standard model of active galactic nuclei (AGN) (e.g.,~\cite{peterson97,elvis00,franketal02,donofrioetal12,netzer13} and references therein), namely accretion onto a supermassive black hole, has definitively passed into history \citep{eht19}, {{while} 
 certain aspects of unification schemes based on orientation are still subject to debate}~\cite{urrypadovani95,dultzin-hacyanetal99,bianchietal08,villarroelkorn14,shangguanho19,zouetal19,lopez-navasetal23}. 

As quasar samples from recent surveys grow in size, now reaching the scale of \mbox{$10^5$--$10^6$ objects}~\cite{storey-fisheretal24}, new and diverse phenomenologies are beginning to emerge. Super-Eddington accretion { involves} a scenario that borders the extraordinary: the infalling material onto the black hole is so large that it surpasses the theoretical Eddington limit, the maximum luminosity a black hole can sustain balancing gravitational pull and radiation pressure. The  evidence supporting the existence of super-Eddington accretion on stellar-mass black holes has been robust, with~examples dating back several decades, particularly in X-ray binaries such as SS433 and GRS 1915+105 \citep{margon1984,greiner2001}. In contrast, evidence for quasar activity exceeding the Eddington limit has been more limited and sporadic. For~a long time, the concept of super-Eddington accretion in quasars remained primarily theoretical, with~no specific ability to identify super-Eddington accretors (e.g.,~\cite{teerikorpi05}). 

Identifying black holes accreting at super-Eddington (SE) rates is important for several reasons. First, while accretion disks are expected to form in basically any realistic scenario {\cite{paczynskywiita80,paczynskiabramowicz82}}, as matter usually has too much angular momentum to accrete directly, super-Eddington accretion requires consideration of physical processes that are needed to understand the transport of gas within the accretion disk towards the black hole~\cite{bh91,bh98}, which go much beyond the local viscosity parameter considered in the model of steady accretion disks at sub-Eddington rates~\cite{shakurasunyaev73,franketal02}.
Second, accurately evaluating the true accretion and mass outflow rates of SE black holes is essential for determining their growth rates, which has significant implications for understanding quasar evolution across all cosmic epochs, especially during the cosmic dawn.
Third, SE processes could be { among} the prime movers of galactic evolution, if~ accretion is occurring on SMBH powering very luminous quasars \citep{dimatteoetal05,king05,kingmuldrew16}. Finally, they hold promising potential as cosmological distance indicators because SE accretion leads to some extreme, stable properties \citep{marzianisulentic14,dultzinetal20,czernyetal21,marzianietal21}.
 
This review aims to delineate the observational properties of SE accretion in quasars\endnote{The term quasar is used here as an umbrella term for type-1 AGN, {  independent of} their luminosity, { unless otherwise noted}.} and { to} assess some elementary implications for both astrophysical and cosmological contexts. We will first provide a definition of SE regime (Section~\ref{define}), then select criteria for their identification from optical and UV spectroscopy (Section~\ref{find}). Our ability to identify at least an (admittedly biased) sample of SE candidates rests on the Eigenvector 1 main sequence (MS \cite{sulenticetal00a,marzianisulentic14,duetal16a}). Several multifrequency properties, from~the radio to the X-ray domain (Section~\ref{ms}), are also potential markers of SE accretion. We sketch a structural and dynamical model of the emitting regions involving a virialized and wind system that has been around since the late 1980s, but~which still needs to be developed in many of its physical and dynamical aspects (Section~\ref{virial}). 
A major source of uncertainty concerns the accretion disk winds, which appear to be a natural consequence of SE accretion modes (Section~\ref{galev} \cite{paczynski90,lupietal24,yangyuan24}). While evidence of outflows is widespread in AGN, some parameters (for example outflow speed and metallicity) reach extreme values in SE candidates. At~the same time, a~virialized sub-system offers the potential to use the line widths of prominent emission lines in SE quasars for estimating cosmological distances, thereby contributing to constraints on the Universe expansion history (\cite{wangetal13,marzianisulentic14}, Section~\ref{cosmo}). Finally, we emphasize the intriguing parallels between stellar-mass and supermassive black holes (Section~\ref{stars}), as well as  several unresolved issues that remain open for future investigations (Section~\ref{conclusion}).



\section{Defining Super-Eddington Accretion}
\label{define}

Super-Eddington accretion refers to a situation where the mass accretion rate onto a compact object—such as a black hole, neutron star, or~white dwarf—is so high that the associated luminosity exceeds the Eddington limit, at~which radiation pressure would normally counteract the inward pull of gravity. It can be expressed as the \mbox{Eddington luminosity}
\begin{equation}
 L_\mathrm{\text{Edd}} = {4 \pi G M c}/{\kappa}, 
 \end{equation}
where \(G\) is the gravitational constant, \(c\) is the speed of light, and \(\kappa\) is the opacity of the accreted material. \(\kappa\) typically represents a measure of how transparent a material is to radiation, and is the amount of absorption and scattering that occurs as light passes through a material. In ionized gases where the number of free electrons plays a role in determining the pressure and other gas properties, the opacity is related to the mean molecular mass per free electron (\(\mu_\mathrm{e}\)) which represents the average mass of particles in a gas divided by the number of free electrons. The  $\mu_\mathrm{e}$ can be expressed as \(
\mu_\mathrm{e} = {1}/{\sum_i X_i {Z_i}/{A_i}} \approx 1.2
\), 
where \(X_i\) is the mass fraction of element \(i\), \(Z_i\) is the number of protons (and thus, free electrons) in element \(i\), and
 \(A_i\) is the atomic mass of element \(i\) (c.f.~\cite{netzer13}). In a fully ionized hydrogen gas, where electron scattering dominates the opacity, the specific opacity \(\kappa_\mathrm{es}\) is given by \( \kappa_\mathrm{{es}} = {\sigma_T}/{(m_\mathrm{p} \mu_\mathrm{e})}.\) In this context, \(\kappa\) is inversely proportional to \(\mu_\mathrm{e}\). If~\(\mu_\mathrm{e}\) is higher (meaning there are fewer free electrons per unit mass), the opacity due to electron scattering will be lower. Conversely, a~lower \(\mu_\mathrm{e}\) (more free electrons per unit mass) results in higher opacity. The Eddington luminosity can then be written as \( L_\mathrm{\text{Edd}} = 4 \pi G M c m_\mathrm{p} \mu_\mathrm{e}/{\sigma_\mathrm{T}} \approx 1.26 \cdot 10^{46} \left({M}/{10^8 M_\mathrm{\odot}}\right) \mu_\mathrm{e} \approx 1.5 \cdot 10^{46} \left({M}/{10^8 M_{\odot}}\right) \). Here, $\mu_\mathrm{e}$ is assumed $ \approx 1.2$ and is obtained by a gas mixture with about 64.7\% hydrogen, 33.3\% helium, and a small amount of metals (about 2\%). Note that for a fully ionized gas $\mu_\mathrm{e} \approx 1.156$. If~a fraction of gas is neutral, the opacity can be much larger and the $L_\mathrm{Edd}$ correspondingly~lower. 




The dimensionless accretion rate $\dot{m}$\ is a way to express the accretion rate in terms of a unit-less quantity, making it easier to compare across different systems without dependence on specific units. This rate is typically defined by normalizing the actual accretion rate to a characteristic rate related to the Eddington limit:
\begin{equation}
 \dot{m} = {\dot{M}}/{\dot{M}_\mathrm{\text{Edd}}},
 \end{equation}
where 
 \(\dot{M}\) is the mass accretion rate in units of M$_\mathrm{\odot}$ yr$^{{-1}}$\ or g s$^{{-1}}$ and \(\dot{M}_\mathrm{\text{Edd}}\) is the Eddington accretion rate. The  \(\dot{M}_\mathrm{\text{Edd}}\) is the rate at which the luminosity of the accreting object reaches the Eddington luminosity \(L_\mathrm{\text{Edd}}\). It can be defined as follows: \( \dot{M}_\mathrm{\text{Edd}} = {L_\mathrm{\text{Edd}}}/{\eta c^2} \), 
where \(L_\mathrm{\text{Edd}}\) is the Eddington luminosity, \(\eta\) is the efficiency of converting accreted mass into radiation, and \(c\) is the speed of light. Note that there is no consistency in the $\eta$ used in the literature because $\eta$\ is a function of the accretion rate itself. We assume $\eta =1$ in the definition of \mbox{$\ \dot{M}_\mathrm{\text{Edd}} = L_\mathrm{\text{Edd}}/{c^2} $\ \citep{duetal16a}}, so that the Eddington ratio can be written as follows:
\begin{equation}
 {L}/{L_\mathrm{Edd}} = \eta \dot{m},
 \end{equation} 
where $L$ is the bolometric luminosity. The Eddington ratio is a basic observable, as it represents the ratio between the bolometric luminosity and the black hole mass \mbh\ (\lledd\ $\propto L/$\mbh), i.e.,~between radiation and gravitation forces. Its importance is comparable to the $M/L$\ ratio of galaxies~\cite{donofrioetal16a,donofrioetal17,donofriomarziani18}. An estimate of the sub-Eddington $\dot{m}$\ can be achieved following the geometrically thin, optically thick accretion disk models, once the luminosity at 5100 \AA\ and the black hole mass are known~\cite{shakurasunyaev73,davislaor11,duetal16a,pandaetal18}, but~this is intrinsically more difficult in the case of a SE disk~\cite{kubotadone19,jiangetal19}.







When the accretion rate surpasses the Eddington limit, the excess radiation pressure should theoretically blow away the infalling material, preventing further accretion. In practice, black holes may accrete at SE rates due to two main factors: first, accretion can occur in a non-spherical manner, such as through a disk, which allows material to funnel onto the object while avoiding direct opposition by radiation pressure~\cite{paczynskiabramowicz82,inayoshietal16,toyouchietal21}; second, photons can be trapped and advected inward with the gas, reducing the outward radiation pressure that counteracts accretion. Advection-dominated accretion flows (ADAFs) arise in regions where the majority of the energy released by accreting gas is retained and transported inward by the flow, rather than being efficiently radiated away. A ``slim disk'' is expected to occur at moderately SE accretion rates~\cite{abramowiczetal88}, and  have higher radiation pressure leading to a vertical puffing up, making the disk geometrically thick and optically thick. Even when the disk becomes geometrically thick due to increased radiation pressure, the gas remains radiatively inefficient, despite being hot and dense. The  local viscosity prescription used in the thin-disk configuration~\cite{shakurasunyaev73} no longer applies, and global considerations, such as angular momentum transport via global torques or outflows, become more important~\mbox{\cite{sirkogoodman03,thompson05}}. Optically thick, stationary slim disks are non-local solutions~\cite{thompson05}, obtained by numerical integration of the two-dimensional Navier–Stokes equations with a transsonic point~\cite{sadowski09,sadowskietal14}. 

At the highest $\dot{m}$, the flow remains advection dominated, optically thick, and supported by radiation pressure, similar to slim disks, but~with even greater radiative inefficiency \citep{abramowicz05}. The thick disk is expected to take on the form of thick torus with two narrow funnels along the rotation axis (a ``Polish doughnut’’), which collimates radiation into beams with highly SE luminosities. The thickness of a slim disk is considerable compared to the geometrically thin standard disks~\cite{abramowiczstaub14}, with~a height to radius $H/R$ ratio in the range of 0.1 to 0.3, but~is not as extreme as that of a Polish doughnut ($H/R \gtrsim 1$). 

A related prediction of geometrically thick, optically thick accretion disk models is anisotropy in the emission (e.g.,~\cite{madau88,calvanietal89,wangzhou99,jinetal17}), and the existence of two regions: one shielded from the hottest continuum \citep{wangetal14a}, and one conical region exposed to the radiation of the funnel’s walls (which in the slim disks at moderate accretion may be also of large aperture, see Figure~3 of \citet{wangetal14a}). 

A second observational property predicted by SE disk models, if~the accretion flow becomes advection dominated, is the radiative efficiency decrease with $\dot{m}$, for \mbox{$\dot{m}\gg 1$~\cite{abramowiczetal88,mineshigeetal00,sadowski09,sadowskietal14}}. A convenient parameterization for non-rotating black holes \( a = 0 \) \mbox{is as follows \mbox{(\cite{madauetal14}, c.f.~\cite{inayoshietal20})}:}
\begin{equation}
{L}/{L_\mathrm{Edd}} \approx 2 \cdot [1 + \ln {(\dot{m}}/{50})].
 \end{equation}

The implication is that the Eddington ratio \lledd\ tends to ``saturate'' towards limiting values of order unity; correspondingly, the radiative efficiency $\eta$ rapidly decreases { roughly} with $1/\sqrt{\dot{m}}$. Slightly larger values can be obtained if the spin parameter is $a > 0$.
 
Hydrodynamical simulations confirm this basic framework. They illustrate that with increasing accretion rates a flat disk thickens, forming regions where gas velocities exceed local escape velocities \citep{ohsugaetal05,jiangetal19,ogawaetal17}. Initially, the accretion flow remains flat, but~with higher accretion rates, the disk thickens and develops regions where gas velocities surpass the local escape velocity. This leads to the generation of powerful winds, which are a hallmark of SE accretion. These winds are intimately connected to the accretion process and play a significant role in the overall dynamics of the system \citep{yangyuan24}. 

Three major properties are, therefore, expected to be associated with SE accretion: (1) the development of an inner thick structure ({ although see~\cite{pognanetal20}}), optically thick, with~implication for the illumination of the line emitting gas; (2) the extreme value of \lledd{,}\ that is, a~radiative output that is growing little with the accretion rate; and (3) presence of high-ionization winds. 


\section{Identifying Super-Eddington Candidates}
\label{find}

\subsection{The Quasar Main Sequence: A Synopsis}

\begin{figure}[H]
\includegraphics[width=13.5 cm]{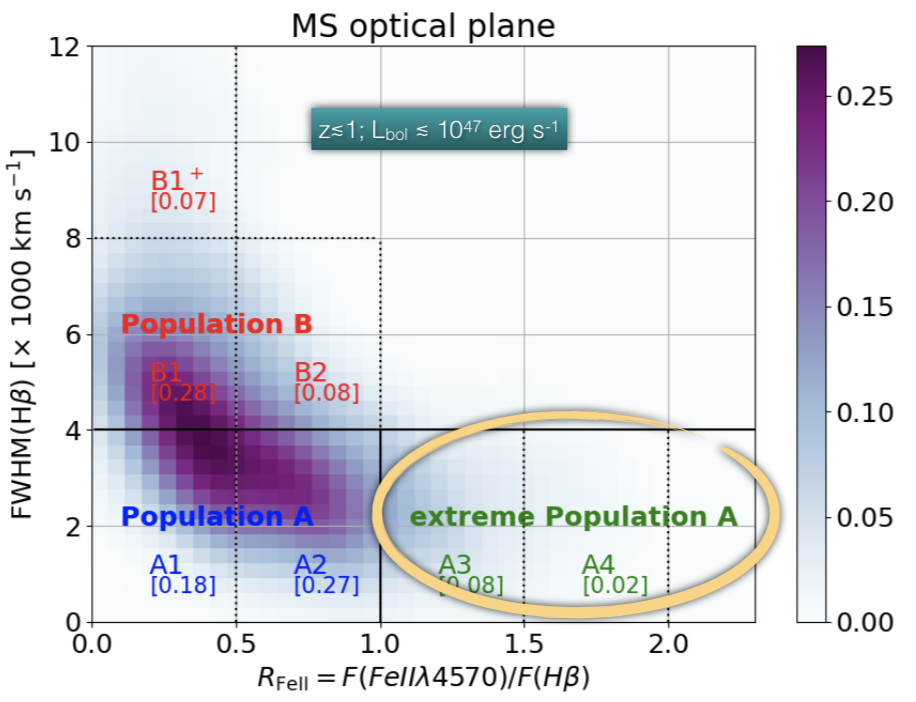}
\caption{{The} 
 optical plane of the quasar main sequence, full width half maximum (FWHM) \hb\ vs. \rfe\ (see text for its definition). Spectral types are as identified by {\citet{sulenticetal02},} 
 in intervals of \rfe\ and FWHM \hb. The numbers in square brackets are the prevalence of each spectral type in a large optically selected sample of low--$z$\ ($\lesssim$1) type 1 AGN. The cartouche emphasizes the area of extreme Pop. A sources with \rfe\ $\ge 1$. 
 \label{fig:mslabeled}}
\end{figure}

\begin{figure}[H]
\includegraphics[width=4.5 cm]{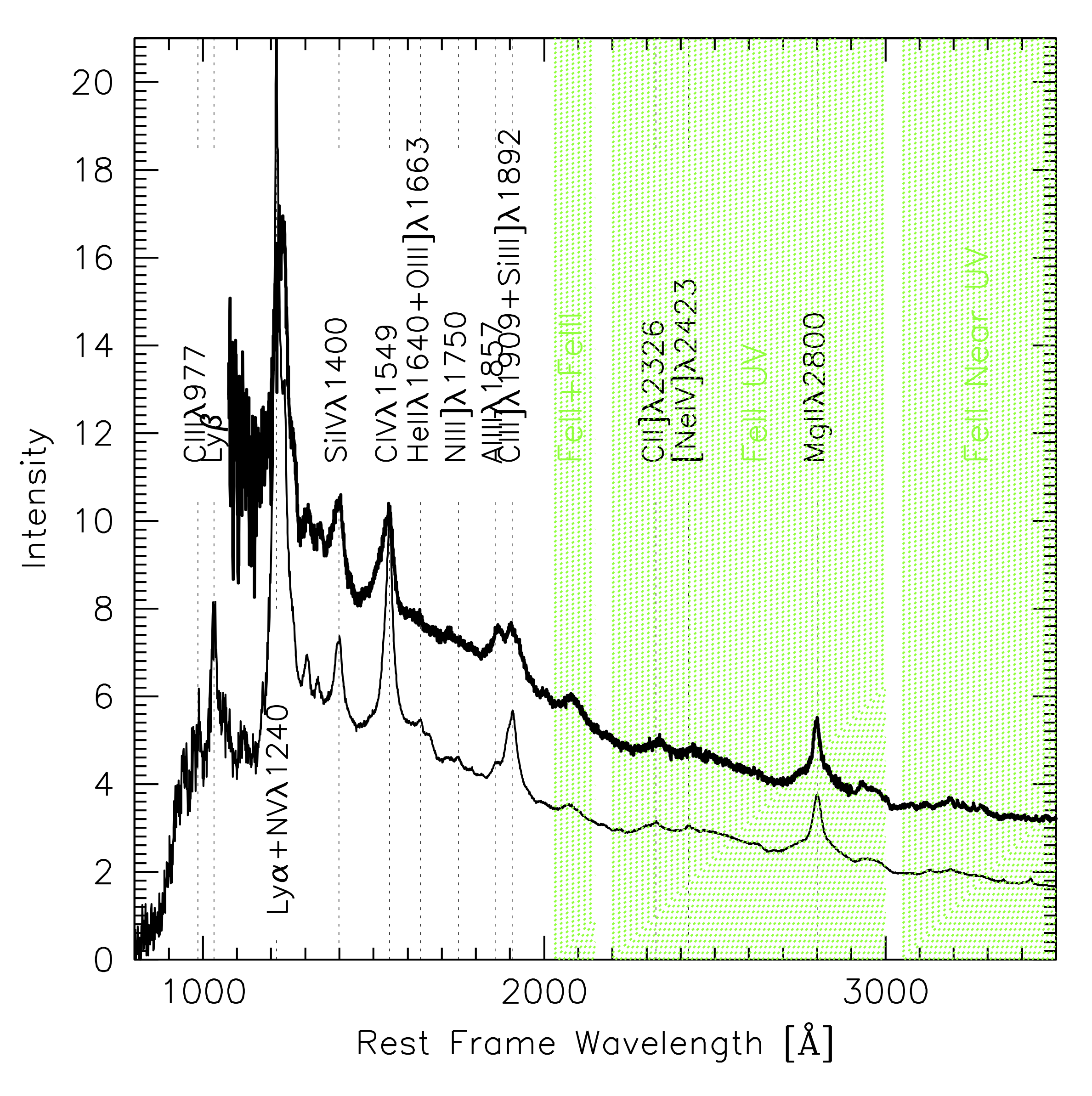}\hspace{-0.2cm}
\includegraphics[width=4.7 cm]{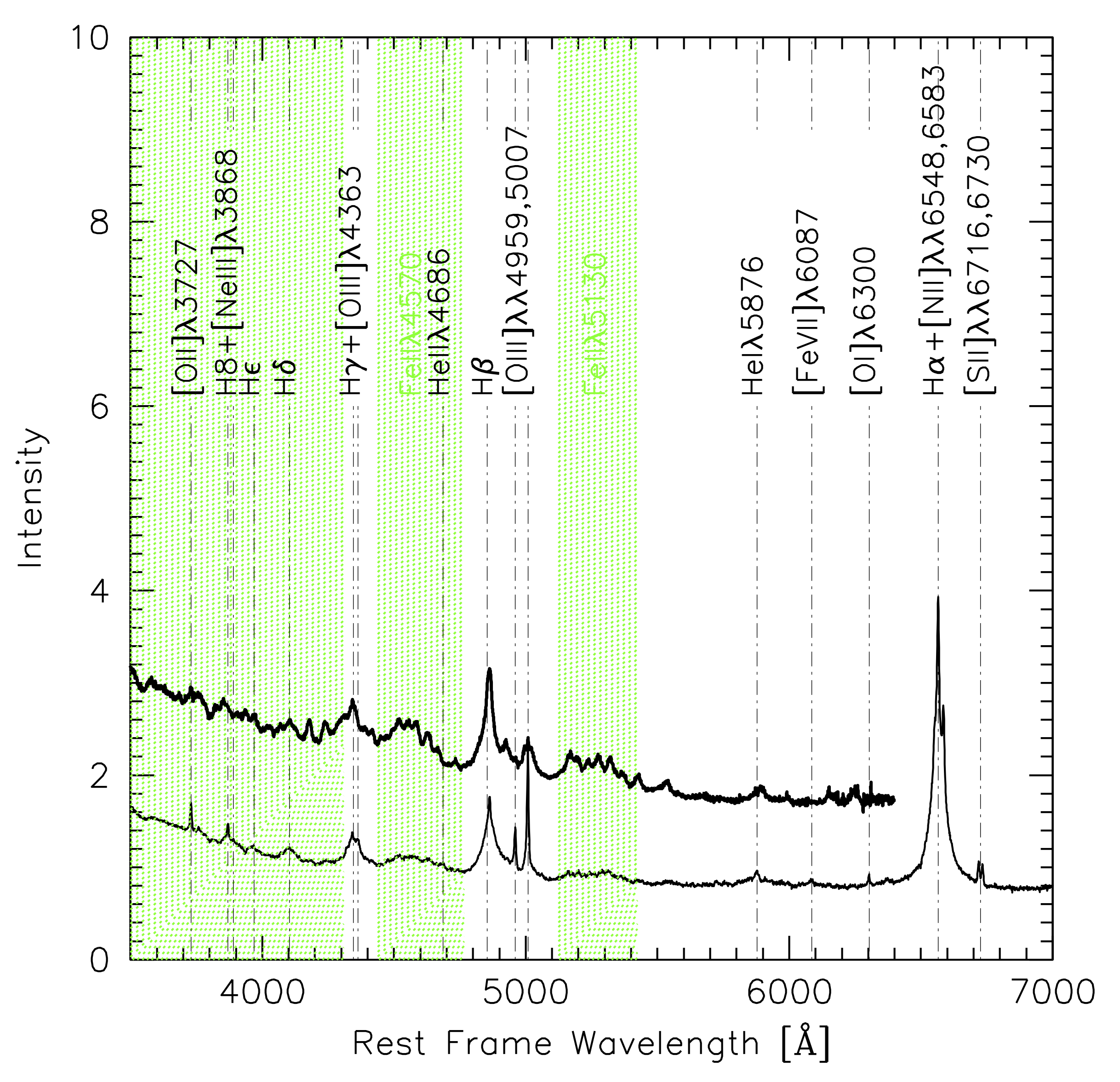}\hspace{-0.2cm}
\includegraphics[width=4.75 cm]{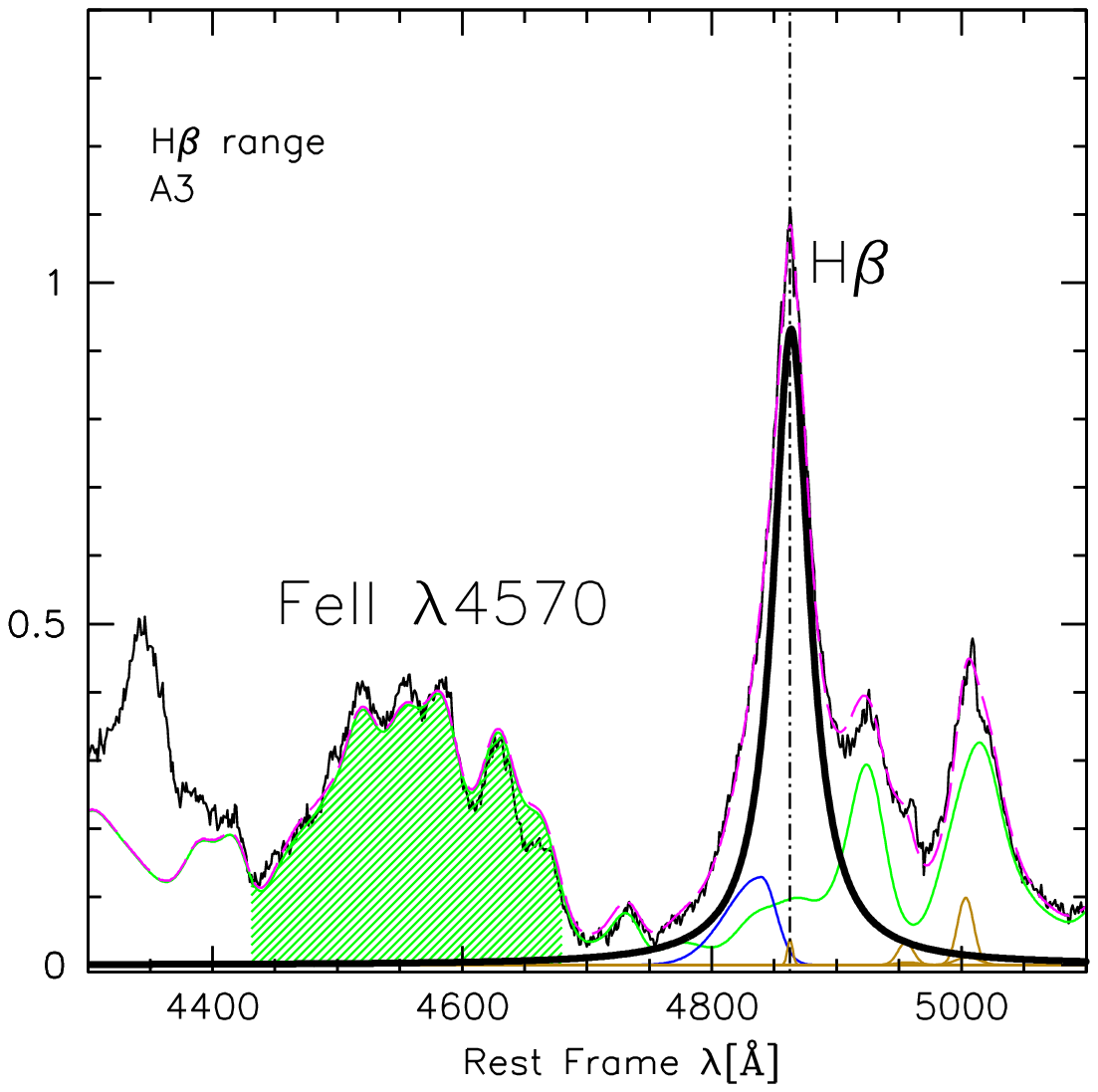}

\caption{\textbf{(Left and middle pane)l} 
 : xA quasar composite ultra-violet (UV) and optical spectrum (thick lines) compared to the median composite spectrum of \citet{vandenberketal01} for the Sloan Digital Sky Survey (SDSS). Spectral regions with prominent \feii\ are shaded green. \textbf{(Right)} continuum subtracted composite spectrum for spectral type A3 belonging to the xA sub-population~\cite{sulenticetal02}, with~emphasis on the \feii$\lambda$4570 blend measured for the derivation of \rfe, again shaded green. A model involving a Lorentzian, unshifted \hb\ component and a fainter component skewed and shifted to blue reproduces the \hb\ profile. \label{fig:spectra}}
\end{figure}

The main sequence (MS) is a classification scheme that organizes quasars according to their spectral properties, much like the main sequence of stars~\cite{sulenticetal00a,marzianietal01,shenho14,marzianietal18}. The MS is most commonly represented in the optical plane (Figure~\ref{fig:mslabeled}, see~\cite{wildyetal19} for an alternative), defined by two parameters:

\begin{itemize}
\item { R\(_\mathrm{FeII}\)}: the relative strength of \feii\ to \hb\ emission, measured as the ratio of the \feii\ complex (around 4570 \AA) to the broad \hb\ line:
 \(
 R_\mathrm{FeII} = {\text{\feii\ (4570 \AA)}}/{\text{H}\beta} 
 \)
(Figure~\ref{fig:spectra}). The \feii\ complex, integrated over the range 4434–4684 \AA\ (hereafter \feq\ for brevity) includes many multiplets, but~the strongest contributions around 4570 \AA\ come from multiplets 37 and 38~\cite{moore45}, which consist of lines from transition between terms $^{4}D $ and $^{4}P$ to $^{4}F$ (the F group of \citet{kovacevicetal10}). 
 Low \( R_\mathrm{FeII} \) values indicate weak global \feii\ emission, as the type-1 AGN \feii\ emission is changing intensity with respect to \hb\ (Figure~\ref{fig:spectra}), but~is otherwise self-similar over a broad range of \rfe\ and AGN luminosity (\cite{marzianietal18}, and references therein). High \( R_\mathrm{FeII} \) values are associated with strong \feii\ emitters and are linked to high Eddington ratio~\cite{borosongreen92,marzianietal01,boroson02,marzianietal03b,shenho14,sunshen15,pandaetal19}.
 
\item {FWHM of \hb}: the FWHM (in \kms) of the \hb\ line is a measure of the velocity of the broad-line region (BLR) gas~\cite{petersonwandel99,petersonetal04}, and is in the range of 1000–20,000 \kms\ \cite{sulenticetal00a}. 
 The FWHM is one of the most widely used ``virial broadening estimators'' used to infer black hole masses (e.g., ~\cite{vestergaardpeterson06,dallabontaetal20}). 
\end{itemize}

\vspace{-6pt}

Quasars can be divided into two broad populations based on their position in the optical plane, with~a sub-population of special~interest.

\begin{itemize}

\item {Population A:}
 narrower \hb\ lines (FWHM $<$ 4000 \kms), a~range of \feii\ emission (from weak to high \( R_\mathrm{FeII} \)), and often lower \civ\ and \oiii\ equivalent width~\cite{borosongreen92,sulenticetal02,dultzin-hacyanetal97}. Population A quasars are typically associated with high dimensionless accretion rates and higher Eddington ratios $L/L_{\text{Edd}} \approx 0.2$--$1.0$, compared to the lower values of $L/L_{\text{Edd}} \approx 0.01$--$0.2$ in Population B~\cite{marzianietal01,marzianietal03b,pandaetal19}. They are overwhelmingly radio-quiet. Radio-loud Narrow Line Seyfert 1~\cite{komossaetal06,abdoetal09,foschinietal15} are a tiny minority of optically selected Population A samples~\cite{zamfiretal08,gancietal19}. Meanwhile, { broad-absorption line} (BAL) QSOs belonging to Population A show more extreme high-velocity BAL troughs and a large BALnicity index~\cite{sulenticetal06a}. 

\item {Extreme Population A (\rfe~$\ge 1$):}
quasars located towards the high \(R_\mathrm{FeII}\) end of the MS (extreme Population A, or~xA for brevity) have been found to be characterized by extreme observational parameters among Pop. A sources and highest $\dot{m}$ above the Eddington limit (but still with bolometric luminosity \(L \sim L_\mathrm{\text{Edd}}\))~\cite{marzianisulentic14,duetal16a,negreteetal18}.

\item {{Population B:}} these quasars have broad \hb\ lines (FWHM $>$ 4000 \kms), weak \feii\ emission (low \( R_\mathrm{FeII} \)), and stronger \oiii\ emission~\cite{borosongreen92,sulenticetal02,marzianietal03a}. Population B quasars are located toward the low \(R_\mathrm{FeII}\) and broader \hb\ end, and associated with lower accretion rates and higher black hole masses, typically ranging from $10^{8.5}$ to $10^{9.5} M_{\odot}$~\cite{marzianietal03b,marziani23,terefemengistueetal23}. A defining feature is a strong redward asymmetry in the { low-ionization line} (LIL) profiles that is attributable to a black hole mass effect~\cite{marzianietal03b,marzianietal09,bonetal15,marziani23}. Population B quasars are more likely to be {radio-loud}, with~a probability of about 25\%, compared to the 3–4\% seen in Population A \citep{zamfiretal08}. Population B is also associated with less extreme BALs { and} lower density ($\log n_{\mathrm{H}} \approx 9.5$--$10$ [cm$^{-3}$]) in the BLR~\cite{marzianietal10}. 

\end{itemize}

The MS primarily highlights correlations between emission line properties and various physical properties of quasars, such as Eddington ratio, black hole mass, and \mbox{orientation~\cite{borosongreen92,marzianietal01,boroson02,marzianietal03b,shenho14,sunshen15,pandaetal19}}. Being the projection of a multidimensional space, the optical plane of the quasar MS is affected by several degeneracies. The limit $\text{FWHM(H}\beta) \approx 4000$~km/s, set at low $z$ and in moderate luminosity samples ($L \lesssim 10^{46}$ erg s$^{-1}$), is by itself luminosity and black hole mass dependent, if~it is interpreted as a well-defined limit in \lledd\ and if LILs are predominantly broadened by virial motions~\cite{marzianietal18a}. The distribution of data point of Figure~\ref{fig:mslabeled} applies to low-$z$ ($\lesssim$1) samples. If~a broad range of masses and luminosities is considered, the sequence is smeared into a wedge~\cite{shenho14}, although~the main trend associated with \rfe\ are~preserved. 

 The orientation of the quasar disk axis with respect to the observer’s line of sight can also affect the observed width of the emission lines~\cite{marzianietal01,shenho14}. Inclined quasars \mbox{($\theta \sim $ 45--60)} might exhibit broader lines due to the rotational motion of the BLR gas confined in a flattened geometry~\cite{wuhan01,bianzhao02,decarlietal11,marzianietal18}. Orientation and black hole mass affect the line width in the same way: an increase of mass and an increase of the angle between the line of sight and the accretion disk axis both lead to an increase in \hb\ FWHM~\cite{marzianietal18,marzianietal22a}. 
 The ratio \rfe\ ultimately depends on the \lledd, and possibly on orientation~\cite{marzianietal01,pandaetal19}, although~the dependence between \lledd\ and \rfe\, is indirect, with~\lledd\ affecting ionization parameter, metallicity, density, and column density of the line emitting gas in a way that is not fully clear as yet (\cite{pandaetal19,pandaetal20,pandaetal21,florisetal24a}, and references therein).

 The spectral appearance of quasars along the MS is retained~\cite{chenetal24} up to the Cosmic Noon, { a period around redshift $ z \sim 1$--$3$ (roughly from 5 to 2 billion years after the Big Bang), when star formation and black hole accretion reached their peak activity~\cite{madaudickinson14}. } A major survey carried out with the ISAAC instrument at the Very Large Telescope (VLT), covering 53 Hamburg-ESO (HE) quasars (hereafter HEMS), revealed very strong \feii\ emitters at $1 \lesssim z \lesssim 2$~\cite{sulenticetal04,sulenticetal06,marzianietal09}. Recent observations using ground-based infrared (IR) spectrometers and James Webb Space Telescope (JWST) have reinforced the stability and distinctiveness of the spectral features defining the MS~\cite{murayamaetal98,dodoricoetal23,yiangetal23,deconto-machadoetal24}: they confirm that the MS-defining properties are consistent across various luminosities and cosmic \mbox{epochs \mbox{\cite{marzianietal09,banadosetal16,banadosetal23,deconto-machadoetal23,onoratoetal24}}}. A higher prevalence of Population A sources at intermediate redshift reflects an evolution in accretion rate of quasars coming down from the Cosmic Noon to the present-day Universe \citep{cavalierevittorini00,hopkinsetal06,kellyetal10,hirschmannetal14}. 
 The main difference apparently resides in the broader emission lines of luminous quasars: broader because of the FWHM systematic increase in black hole mass \citep{marzianietal18a}, ascribed to a Malmquist-type bias in flux-limited samples~\cite{sulenticetal14}. 
 { Highly accreting quasars like I Zw 1 and H0557-385 may become undetectable at high redshift if their flux falls below a survey detection threshold. However, their high-luminosity counterparts—which exhibit similar \rfe values—would still be detectable, albeit with broader emission lines due to their larger \mbh.}

 \vspace{-3pt}

 \subsection{Identification from the Optical Rest~Frame}
 
\begin{figure}[H]
\includegraphics[width=13.5 cm]{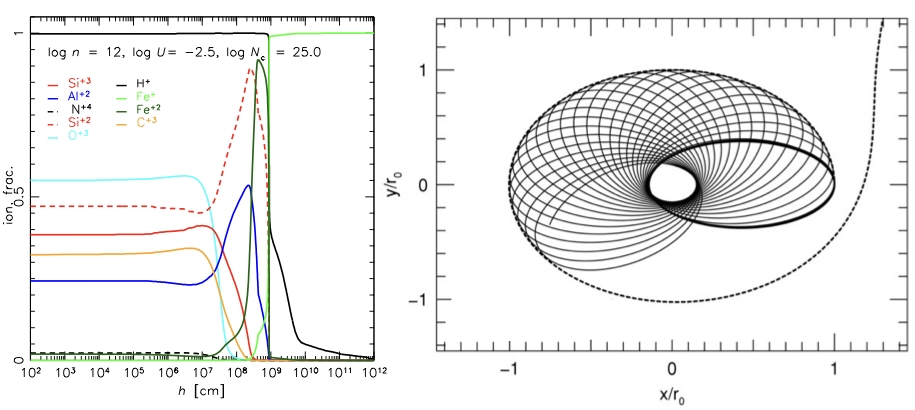}
\vspace{0cm}
\caption{(\textbf{{Left}
}) ionization structure as a function of the geometrical depth $h$ in a slab of emitting gas. The continuum is illuminating the slab from the left side ($h=0$) { giving rise to a fully-ionized zone, and a PIZ where hydrogen is partially ionized (black line) and almost all iron is singly-ionized (pale green)}. (\textbf{Right}) Different orbits for clouds of different column density, with the~same \mbh\ and \lledd. The highest column density clouds remain quasi-virialized with a precessional orbit due to the deviation from the inverse square law for their distance from the central continuum source. The lowest density clouds become unbound (dashed line). Adapted from Refs.~\cite{negreteetal12,netzermarziani10}. \label{fig:ionorbit}}
\end{figure}

 The optical singly-ionized iron emission is remarkably self-similar among type-1 quasars~\cite{marzianietal03a}, since significant \feii\ emission requires very similar conditions: in the photoionization context, a~partially ionized zone (PIZ) within the emitting gas with electron temperature $T_\mathrm{e} \sim 7000$K~\cite{kwankrolik81,reesetal89,netzer90,marzianietal10}, where Fe$^{+}$\ is the dominant ionization stage of iron (Figure~\ref{fig:ionorbit}). The relative intensity with respect to \hb\ is instead varying by a large factor along the MS, ranging from being undetectable to becoming so prominent that it overwhelms the emission lines within the H$\beta$ spectral range~\cite{phillips78a,borosongreen92,marzianietal03a}. 
 The selection criterion~\cite{marzianisulentic14,negreteetal18,duetal16a}
\begin{equation} 
 R_\mathrm{FeII} = {\text{\feii\ (4570 \AA)}}/{\text{H}\beta} \ge 1
 \end{equation}
isolates quasars with strongest \feii\ emission and other extreme properties at the tip of the quasar MS. xA sources exhibit distinct, and easily-recognizable spectra compared to typical quasars (Figure~\ref{fig:spectra}), due to the dominance of singly ionized iron emission across a wide range from the FUV to the NIR, and specifically around H$\beta$\ \citep{marzianietal03b,duetal16a}. 
 
The parameter \rfe\ might appear as obscure and irrelevant, but~strong \feii\ emission is actually the main cooling agent in the BLR~\cite{marinelloetal16}. \feii\ emission of xA sources can overwhelm other emission lines and create a pseudo-continuum, exceeding the emission of Ly$\alpha$\ and the strongest metal lines. Especially in the UV around the Mg{\sc ii} doublet, the contamination of the underlying accretion disk continuum creates difficulties even in assessing their relative contribution (e.g.,~\cite{kovacevicdojcinovicpopopvic15,sextonetal21,florisetal24a}). The optical \feii\ emission is not only a key SE indicator, but also remains remarkably stable over time in most cases (and at least within the temporal extension of monitoring campaigns \citep{duetal16,duetal18}), which is uncommon in other quasar spectra, especially in Pop. B sources where at low accretion rate the changing look phenomenon occurs frequently~\cite{komossaetal24}. 

\begin{figure}[H]
\includegraphics[width=1\linewidth]{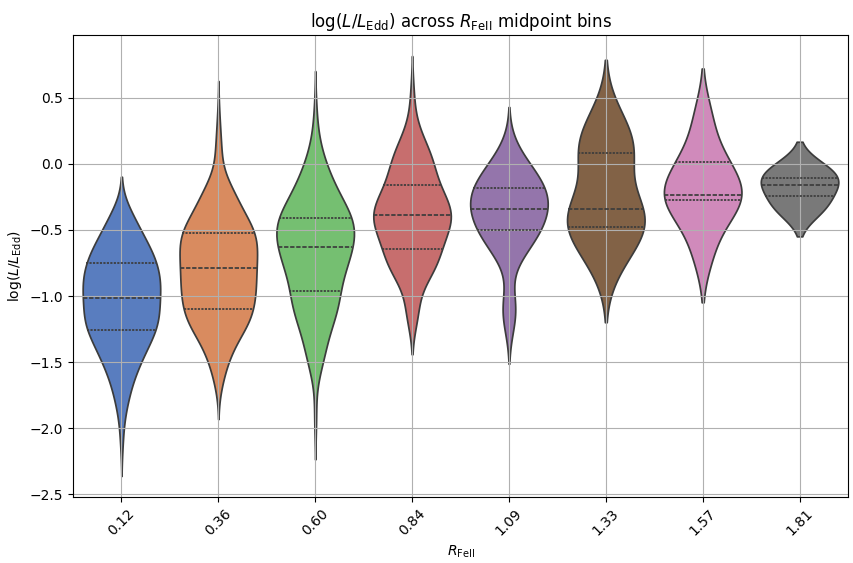}
\caption{Violin plot of the Eddington ratio \lledd\ as a function of the \feii\ prominence parameter \rfe. Horizontal lines within each range identify median and first and third quartile of the distribution. \label{fig:violin}}
\end{figure}

The criterion \rfe\ $\gtrsim 1$\ corresponds to the highest values of the Eddington ratio, close to 1 or slightly above 1 depending on different bolometric corrections and black hole mass estimates~\cite{marzianietal01,marzianisulentic14,sunshen15,duetal16a}. Figure~\ref{fig:violin} shows the distributions of the \lledd\ as a function of \rfe, in bins of $\delta$\rfe = 0.24, for~a large sample of almost 700 low-$z$ sources \citep{terefemengistueetal23}. The violin plots are well-suited to show the overall trend as well as the ample dispersion of the data in each bin. There is a steady increase in the median \lledd\ up to \rfe\ $\approx 0.84$; afterwards, the increase is shallower, confirming a threshold around \rfe\ $\approx 1$, beyond which, even if \rfe\ becomes significantly larger than 1, the \lledd\ is only weakly increasing. This behavior reminds the one of \lledd\ as a function of $\dot{m}$. The dispersion is in large part due to uncertainties in the \lledd\ estimates, so that the prominence of \feii\ could potentially serve as a more reliable proxy for the AGN accretion state. 

Other low ionization species are strongest in the spectra of xA quasars, in addition to singly-ionized iron: Ca{\sc ii}\ \cite{dultzinhacyanetal99,martinez-aldamaetal15,pandaetal20,martinez-aldamaetal21,pandaetal21}, as well as Si{\sc ii}$\lambda$1263 and $\lambda1816$\ \cite{lahaetal16}. In addition, the \civ\ equivalent width is minimized \citep{kinneyetal90,shemmerlieber15,plotkinetal15,marzianietal16a}, with~$W \lesssim 10$\ \AA\ qualifying most sources as weak lined quasars~\cite{diamond-stanicetal09,martinez-aldamaetal18,leungetal24}. Why does high Eddington ratio come with an extremely low ionization spectrum? This question was raised almost a quarter century ago. An answer given at that time was that emission line diagnostics suggest very high density and low ionization for the BLR \citep{negreteetal12}. High column density is also needed for strong \feii\ emission~\cite{collinjoly00,jolyetal08}, to~maintain the PIZ, where Fe$^{+}$\ is the dominant ionization stage of iron (Figure~\ref{fig:ionorbit}). It is as if the BLR of xA sources is somehow reduced to a dense, compact remnant, not unlike the Weigelt blobs of $\eta$ Carin\ae\ \citep{johanssonletokhov04,johanssonetal06}. Parenthetically, we note that large column density stabilizes the virial motion of the emitting gas against radiation pressure because the radiative acceleration is inversely proportional to the gas column density~\cite{marconietal09,netzermarziani10}. The right panel of Figure~\ref{fig:ionorbit} shows three conditions for an orbiting cloud originally in elliptical orbit, for~the same \lledd: at high $N_{\mathrm c}$, there is no noticeable effect (thick trace); for low $N_{\mathrm c}$\ the cloud becomes unbound; and at intermediate $N_{\mathrm c}$\ \mbox{($\sim$$10^{23}$ cm$^{-2}$)}, the cloud remains bound but the orbit undergoes a precessional motion. Large $N_{\mathrm c}$\ gas is expected { to be present at} the accretion disk~surface. 

\unskip

\unskip 

\subsection{Identification from UV Rest~Frame}

\begin{figure}[H]
\includegraphics[width=8 cm]{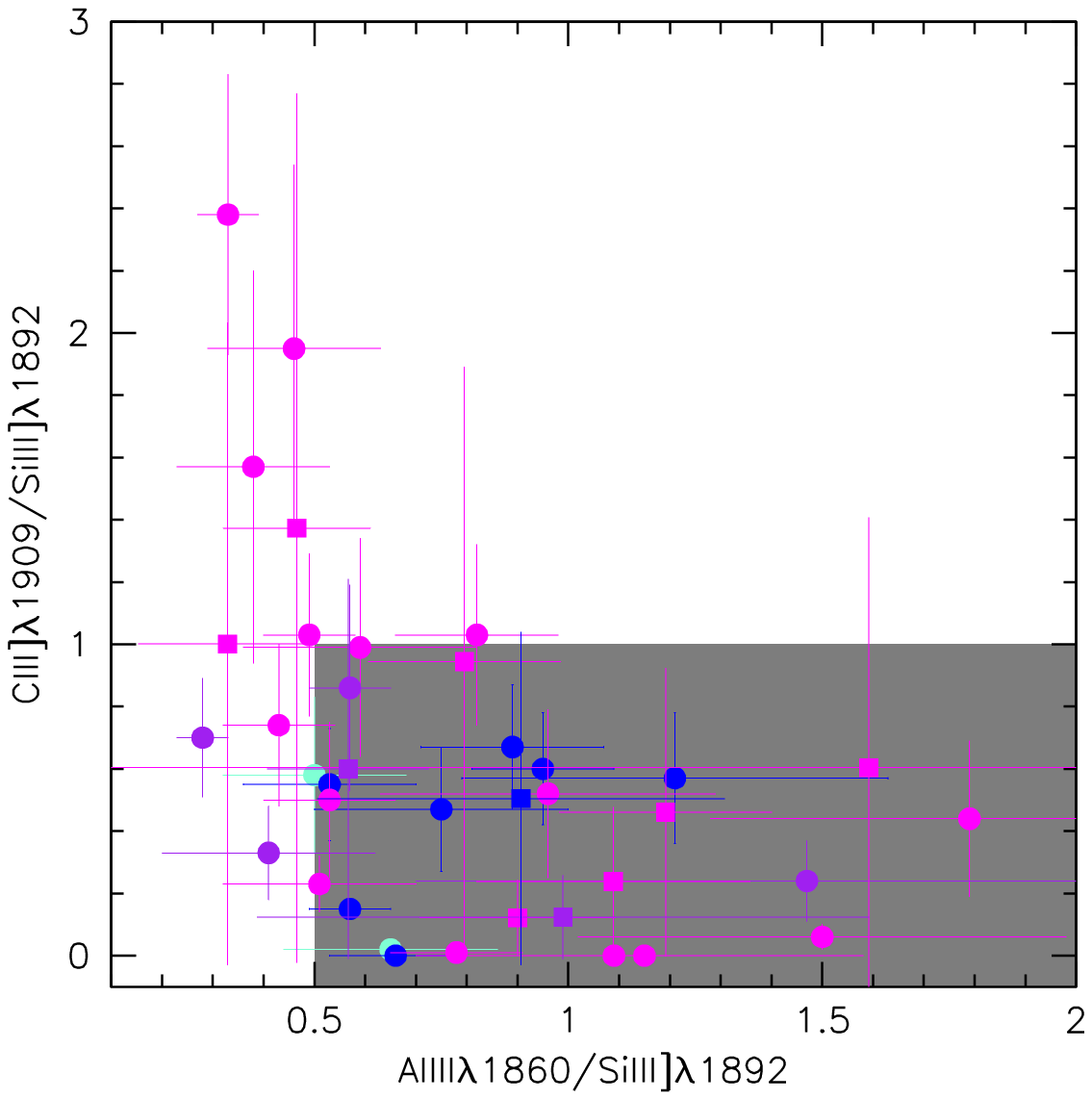}
\caption{Plane of intensity ratio \ciii/\siiii~vs. \aliii/\siiii. The data are coded according to their \rfe\ values. Cyan: A1 {(\rfe $<$ 0.5)}, Blue: A2 {($0.5\le$ \rfe\ $<$ 0.5)}; Magenta: A3 \mbox{($1\le$ rfe $<$ 1.5)}; Purple: A4 {($1.5\le$ \rfe\ $\lesssim$ 2)};. The shaded area { shows} the range of ratios used to identify the SE candidates. Circles represent the sources of \citet{marzianietal22}; squares are data from \citet{florisetal24} and Buendia Rios et~al., submitted. \label{fig:uvratios}}
\end{figure}

At redshift above $\approx 1$, observations of \hb\ become cumbersome because the line is red-shifted into the near IR. Dedicated instrumentation is needed, from~ground or space. Ground-based observations are affected by telluric absorption and sky emission. Thus, it would be expedient to have an UV selection criterion at hand. The condition \rfe\ $\gtrsim 1$\ is associated with strong Al{\sc iii} emission and weak C{\sc iii}]~\cite{hartigbaldwin86,bachevetal04,richardsetal11}. The weakness of C{\sc iii}] was one of the major aspects suggesting high density \citep{aokiyoshida99,marzianietal10}. In terms of the {Si III]/C III] ratio}, extreme Population A quasars show high values ($\gtrsim$1), while this ratio is more modest in the rest of Population A and is never as large in Population B~\cite{sulenticetal02,buendia-riosetal23}. The \mbox{following conditions}

\begin{equation} \mathrm{Si{\textsc{iii}}]}\lambda 1892 / \mathrm{Al{\textsc{ iii}}}\lambda 1860 > 0.5 \, \mathrm{and}\ \mathrm{Si{\textsc{iii}}]}\lambda 1892 / \mathrm{C{\textsc{iii}}]}\lambda 1909 < 1 \end{equation} 
define a criterion that is meant to be equivalent to \rfe\ $>$ 1. The correspondence between \rfe\ $>$ 1 and the UV conditions hold in $\approx$80\% of cases where both spectral ranges are available (Figure~\ref{fig:uvratios}; \citep{marzianietal14,marzianietal22,buendia-riosetal23}). Out of a sample of $\approx$30 xA optically-selected sources, two are located outside of the grey rectangle defining the area in agreement with the UV selection criteria; an additional four have \ciii\ too strong but \aliii/\siiii\ slightly lower than the threshold value 0.5 but consistent within the uncertainties. Data points in discordant location involve mainly borderline discrepancies, especially sources meeting the UV conditions and having \rfe\ close to 1 but somewhat less than 1 or viceversa, \rfe\ $>$ 1, and $\textrm{C}{\textsc{iii}}] \lambda 1909$ stronger than expected, possibly due to an undersubtraction of the $\textrm{Fe}{\textsc{iii}} \lambda 1914$, enhanced via Ly$\alpha$\ fluorescence~\cite{marzianietal10}. 

The optical and UV criteria can be supplemented by other indicators. xAs exhibit the highest {X-ray variability} along the sequence (e.g.,~\cite{ranietal17,rakshitstalin17,bolleretal21,parkeretal21,lietal19}), while {optical variability} is more frequent and occurs with greater amplitude in Population B quasars. Other UV intensity ratios can be used to reinforce the classification, typically $\mathrm{Si{\textsc{iv}}}+ \mathrm{O{\textsc{iv}}]}\lambda 1400$/$ \mathrm{C{\textsc{iv}}}\lambda1549$. In the most extreme quasars, this ratio tends toward unity. 

There are several caveats, mainly because both the UV and optical selection criteria are dependent on relative elemental abundances, and such dependence is not established. \rfe\ is highest in { conditions} of high density and column density, with~low or moderate ionization. Higher iron content makes it easier to achieve \rfe\ above 1, between $\approx$$1.5$--$2$, or even higher \citep{florisetal24}. The \rfe\ ratio is obviously dependent on metallicity, although~it is not yet universally accepted that highly super-solar abundances are needed to achieve values \mbox{\rfe\ $\gtrsim 1$.} Recent works suggest that \rfe\ $\gtrsim 1$ can be reached with solar \mbox{metallicity~\cite{templeetal21,zhangetal24},} albeit stretching other physical parameters to their limits. In addition, UV intermediate ionization line ratios are strongly dependent on density, as they are the ratios of transitions with different critical density~\cite{marzianietal20}, right in the domain of the densities expected for the BLR ($n_{\mathrm{H}} \sim 10^{9}$--$10^{11}$ cm$^{-3}$). Last, silicon is an $\alpha$\ element but aluminum is not, and there is a considerable uncertainty about its nucleosynthesis~\cite{pignatarietal10,thielemann19}.

In conclusion, the optical and UV criteria appear to be fairly effective in identifying SE candidates. A significant source of uncertainty in the interpretation of the selection criteria lies in the determination of \lledd, primarily due to uncertainties in mass estimates (which are influenced by orientation effects) and, even more critically, the bolometric correction. 
Therefore, we prefer to rely on empirical criteria based on the MS, acknowledging that they likely select only a limited sub-sample of SE candidates—excluding type-1 AGN that may be affected by reddening, embedded in dusty starbursts~\cite{farrahetal22}, or~ type-2. Obscured sources might well be highly accreting~\cite{miniuttietal13}. For~instance, the prototype type-2 AGN NGC 1068 shows strong \feii\ emission in polarized light \citep{antonuccimiller85,antonucci93}, and might have been classified as xA if the \feii\ had been observed in natural light.

A related issue is whether the selection identifies truly SE quasars. The optical and UV criteria focus on extreme properties observed of xA quasars: enhanced \feii\ emission, highest velocity outflows, highest chemical abundances, etc. The data do not yet rule out the possibility that an extreme sub-Eddington balance could be at play. Apart from the relation between $\dot{m}$, \lledd, and \rfe\ \cite{marzianisulentic14,duetal16a}, the evidence for SE accretion in quasars remains largely circumstantial and based on the large black hole masses at very high redshifts. Therefore, xA quasars, which represent a well-defined observational type, might be more appropriately considered SE candidates, even if, for~brevity, we refer to them simply as SE. 

\section{Properties of Super-Eddington Candidates in the Main Sequence~Context}
\label{ms}
\unskip

\subsection{Multi-Frequency~Correlates}
\label{multi}


 \citet{sulenticetal11} and \citet{fraix-burnetetal17} list several multifrequency parameters that are systematically different between Population A and B. xA sources can be thought as simply reaching the most extreme values in most observational parameters of Population A. One of the main distinctions is the {\civ\ blueshift} measured at half peak intensity ($\Delta v_{1/2}$ \civ) with respect to the quasar rest frame, which is typically largest in xA (in absolute value, $\gtrsim$1000 \kms) compared to the smaller and null values in the rest of Population A and Population B ($-250$ to $+70$ \kms~\cite{marzianietal96,corbinboroson96,sulenticetal07,richardsetal11}). xA quasars often show extremely weak \civ\ and \oiii\ emissions compared to Population B, where these lines are more prominent and with a narrower core close to rest frame~\cite{sulenticmarziani99,sulenticetal07}. The \oiii\ lines are found blueshifted by several hundred \kms\ among xA (the so-called ``blue outliers'' \cite{zamanovetal02,marzianietal03b,craccoetal16}), with~shifts that are apparently correlated with { those} of \civ\ \cite{zamanovetal02,vietrietal18,vietrietal20,deconto-machadoetal24}. Additionally, extreme Population A quasars tend to have {softer X-ray spectra} with large R\"ontgensatellit (ROSAT) photon indices ($\Gamma_\mathrm{soft} \gtrsim 2$; \(f(\epsilon) \propto \epsilon^{-\Gamma} \)) \citep{bolleretal16}, while Population B shows somewhat rarer instances of enhanced soft X-ray emission. A  $\Gamma_\mathrm{soft} \lesssim 2$\ indicates a flattish SED as observed in RL and Population B sources; a steeper slope $\Gamma_\mathrm{soft} \gtrsim 2$ is common among Population A sources, with~the most extreme values, $\Gamma_\mathrm{soft} \approx 3-4$,\ among xA AGN. Such large $\Gamma_\mathrm{soft} $ values diagnose a soft X-ray excess~\cite{bolleretal96,sulenticetal00a,grupe04,bolleretal16} due to the presence of a component superimposed on an ideal power law connecting the near UV emission and the X ray domain, the $\alpha_\mathrm{ox}$ parameter~\cite{tananbaumetal79}. The most widely accepted explanation is emission from a Compton-thick corona connected with the innermost accretion disk (\cite{walterfink93,wangnetzer03,petruccietal20}, and references therein)] and possibly even with the bare accretion disk (i.e., without~a Comptonizing corona \cite{doneetal12}).
 
 Together with \rfe\ and FWHM \hb, \civ\ shift and $\Gamma_\mathrm{soft} $ define the so-called 4DE1~\cite{sulenticetal00b,sulenticetal07,benschetal15}, one of the landmark achievements of the past 30 years of research on AGN~\cite{sulenticetal00b,sulenticmarziani15}. The 4DE1 scheme captured the main aspects of the type-1 AGN multifrequency phenomenology. In addition to FWHM \hb\ and  \rfe\ increasing with the Eddington ratio~\cite{marzianietal01,marzianietal03b,sunshen15,pandaetal19}, providing the source location in the MS optical plane, the 4DE1 includes two other parameters that are observationally ``orthogonal'', involving different frequency domains, and that are related to the main { parameters} of the accretion process in quasars (mass, Eddington ratio, viewing angle, and accretion mode). 

\subsection{Super-Eddington SED}

\begin{figure}[H]
\includegraphics[width=11 cm]{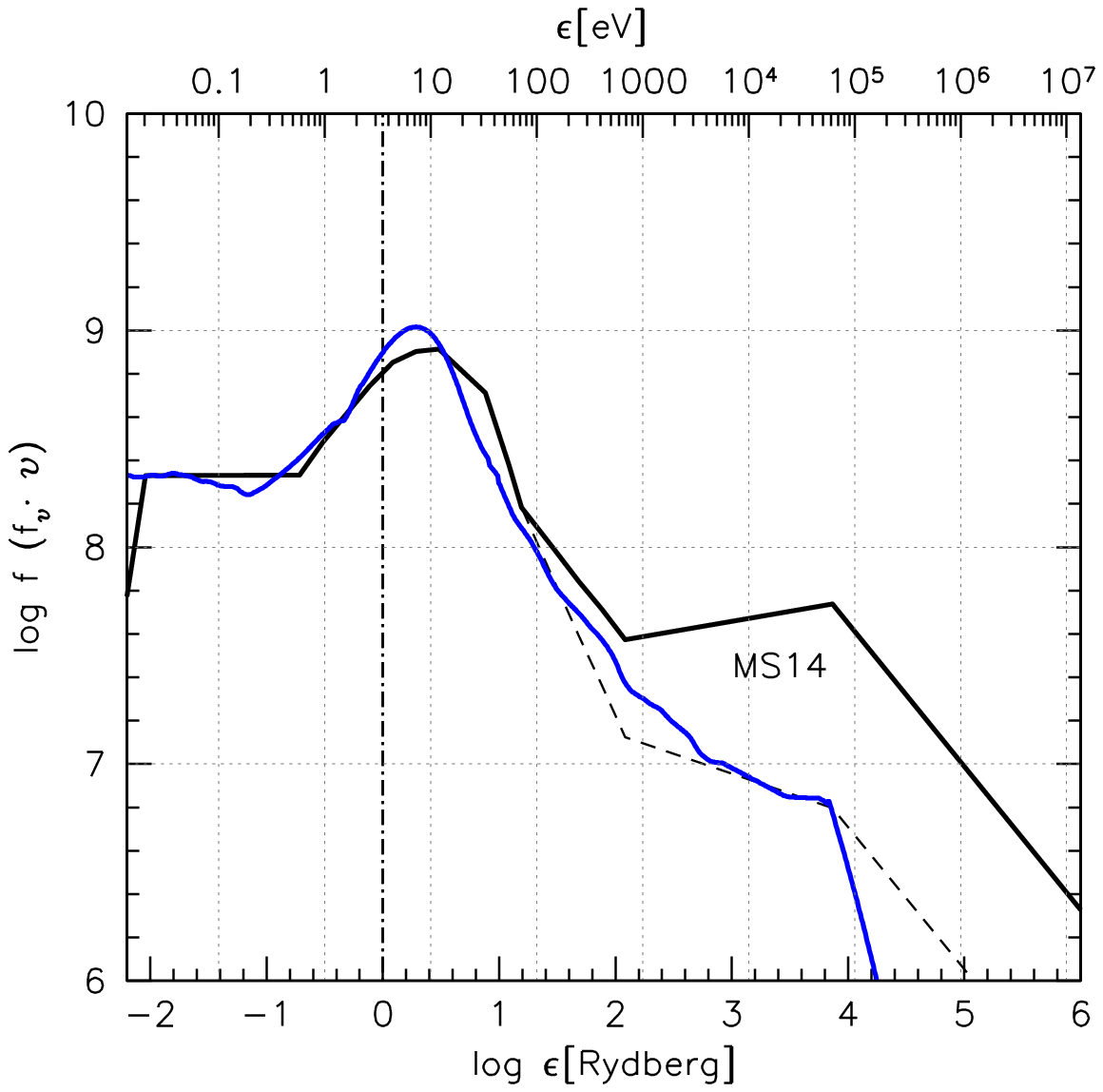}
\caption{Spectral energy distributions (SEDs) computed by Marziani and Sulentic (\cite{marzianisulentic14}, MS14) and Garnica~et~al. (submitted) are presented. In all cases, the sources used to construct the composite SEDs were selected based on the criterion \rfe\ $\ge 1$. The filled black line represents a composite SED with typical values across various spectral ranges, while the dashed line illustrates the extreme values~\cite{marzianisulentic14}. The new SED (blue), which is a median of more than 150 sources, shows good agreement with the extreme~composite. 
\label{fig:sedrfe1}}
\end{figure}

Figure~\ref{fig:sedrfe1} shows three SEDs obtained for sources satisyfing the criterion \rfe\ $\ge 1$ (see also \citet{pandamarziani23} for a contextualization of the SED along the main sequence).
A prominent big blue bump and a soft X-ray excess extending up to few keV has been a distinguishing feature of the SED of sources radiating at high Eddington ratio \mbox{\citep{shields78,czernyelvis87,walterfink93,laoretal97,grupeetal10,doneetal12,pandamarziani23}}. 
Slim disks manage to maintain surface luminosity at the local Eddington limit through radial advection, which alters the expected luminosity distribution. Under~conditions of high accretion rates, the emissivity decreases more slowly $(\epsilon(r) \propto r^{-2})$ than as seen in standard sub-Eddington disks ($ \propto r^{-3}$). In practice, the predicted SED should peak in the FUV right beyond the Lyman limit, with~a flat shape in correspondence of the maximum emission (\cite{kubotadone19}, Figure~\ref{fig:sedrfe1}). An observational test is difficult~\cite{capellupoetal15}, as the FUV is unobservable because of Galactic absorption.

On the high energy side, the behavior of the xA SED is even less clear. Up~to \mbox{$\approx 20$ keV,} the slope of the X-ray continuum was found to be steep, with~photon indexes larger than 2 and reaching the most extreme values for the highest \lledd\ \mbox{sources~\cite{wangetal13,marzianisulentic14,ferlandetal20,laurentietal21}}. Intriguingly, the photon index appears to correlated with the outflow dynamical \mbox{properties \citep{grunwaldetal23,boller23}}. 
On the one hand, a~sizable population of sources are weaker than expected from the slope $\alpha_\mathrm{ox}$\ trend with UV luminosity~\cite{laurentietal21}. On the other hand, beyond 20 keV, several xA sources show a flattening in their nuSTAR spectra. The origin of this flattening is not obvious. A cold Compton-reflection component is observed in Population B sources, and is consistent with scattering toward a geometrically thin, ``cold’’ \mbox{disk~\cite{haardtmaraschi91,haardtmaraschi93,nandrapounds94}}, where the Fe K$\alpha$\ line is observed at the rest-frame energies expected for low ionization { degrees} of iron. The accretion geometry is expected to be basically different for super- and sub-Eddington AGN \citep{karaetal16a,zhangetal24}; in the SE case, the Fe K$\alpha$ resonance line is shifted toward higher energies, since the X-ray reflection can only be due to a Compton-thick outflow encircling the funnel of an optically thick, geometrically thick accretion disk, and scattering toward the outer disk hampered by the inner disk geometry. The presence (or absence) of a reflection component might be at the origin of the difference between the SEDs of Figure~\ref{fig:sedrfe1}. 

On the low energy side, xA sources show a significant far infrared (FIR) excess with respect to the canonical \citet{mathewsferland87} SED, especially prominent for the highest \rfe\ values. Their host galaxies at low-$z$ are known to undergo vigorous star formation~\cite{xieetal21}, often qualifying them as luminous or ultra-luminous infrared galaxies. They are cold quasars in the interpretation of Kirkpatrick et al. (\cite{kirkpatricketal19}, \S\ \ref{evol}). 


\subsection{Chemical Properties and Broad Line~Region}
\label{chemicals}


Quasar metallicities are consistently reported to be high, typically around \linebreak \mbox{$5$--$10 Z_\odot$ \citep{hamannferland93,hamannferland99,xuetal18,shinetal13,sulenticetal14}}. Several diagnostic ratios which are easily measurable from the broad line spectrum of quasars show a fairly monotonic dependence on metallicity and other parameters over a wide range of physical conditions: the intensity ratio \civ/\heiiuv\ and \civ/\hb\ are sensitive to metallicity as well as to ionization parameter~\cite{sniegowskaetal21,florisetal24}; \civ/\ciii\ is in principle sensitive to the ionization parameter, albeit its use is not appropriate for xA sources because of the \ciii\ intrinsic weakness~\cite{marzianisulentic14}. The ratio \aliii/\heiiuv\ is monotonically dependent on metallicity; the ratios \siiv + \oiv/\civ\ and \siiv + \oiv/\heiiuv\ are known metallicity indicators~\cite{shields76,nagaoetal06,shinetal13,sulenticetal14,huangetal23}. A value of the \siiv+\oiv/\civ\ ratio close to $\approx 1$ as found for xA quasars suggests extreme metal abundances~\cite{garnicaetal22,huangetal23}. Ratios involving the semi-forbidden transitions are sensitive to density \citep{hamannferland93,nagaoetal06,shinetal13,huangetal23}. It is a safe approach to rely on UV resonance lines to estimate metallicity and physical properties such as ionization parameters and density (as done in several recent works~\cite{negreteetal12,sniegowskaetal21,garnicaetal22,florisetal24}), for~which atomic data are known with high precision, and radiative transfer may be better addressed by the escape probability formalism~\cite{hubenyetal01,dumontetal03}. In the PIZ {(Figure~\ref{fig:ionorbit})}, several non-local processes play crucial roles~\cite{kwankrolik81,netzer90,krolik99,osterbrockferland06,pradhannahar15}.

\begin{figure}[H]
\includegraphics[width=6.75cm]{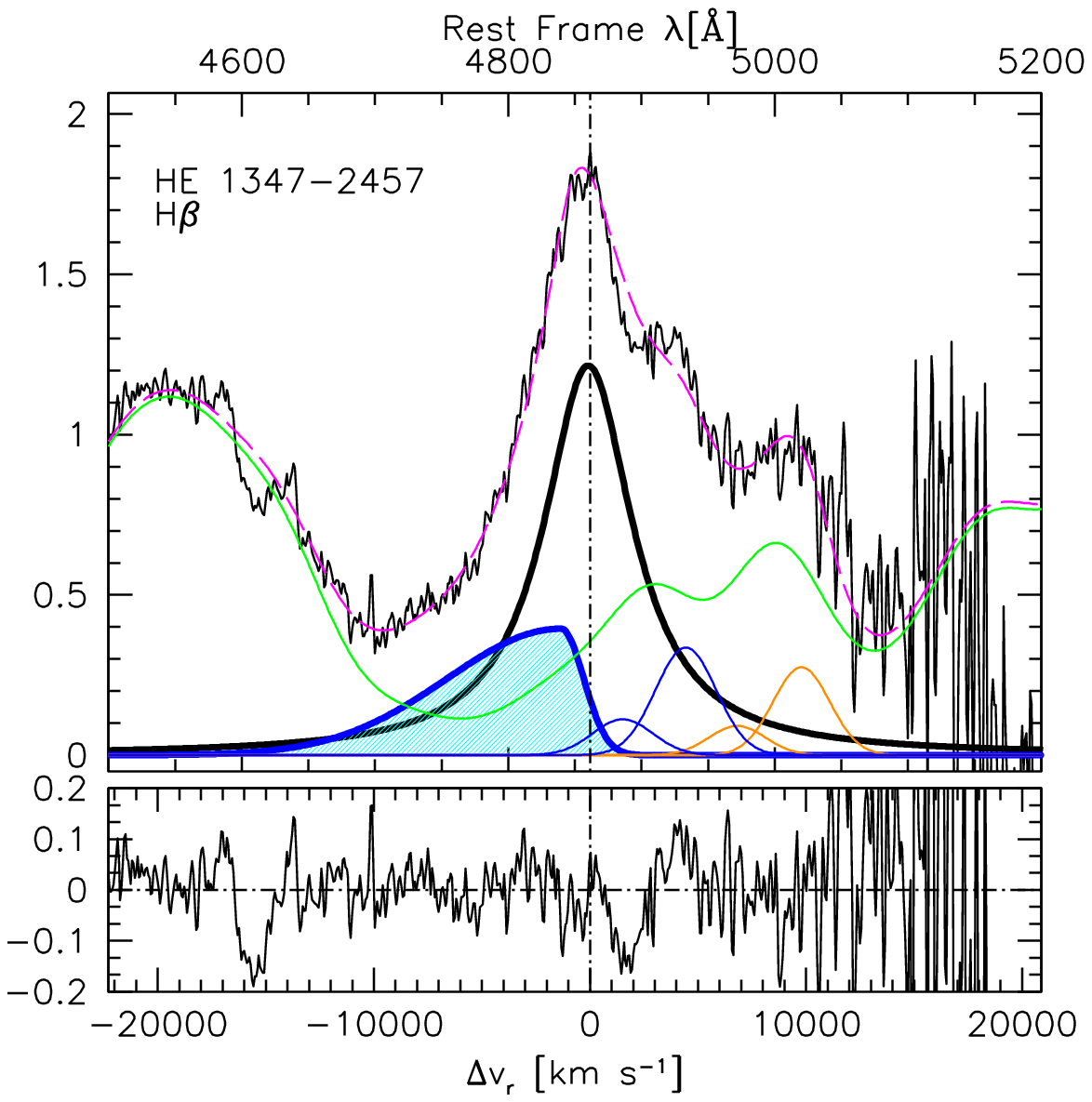}
\includegraphics[width=6.75cm]{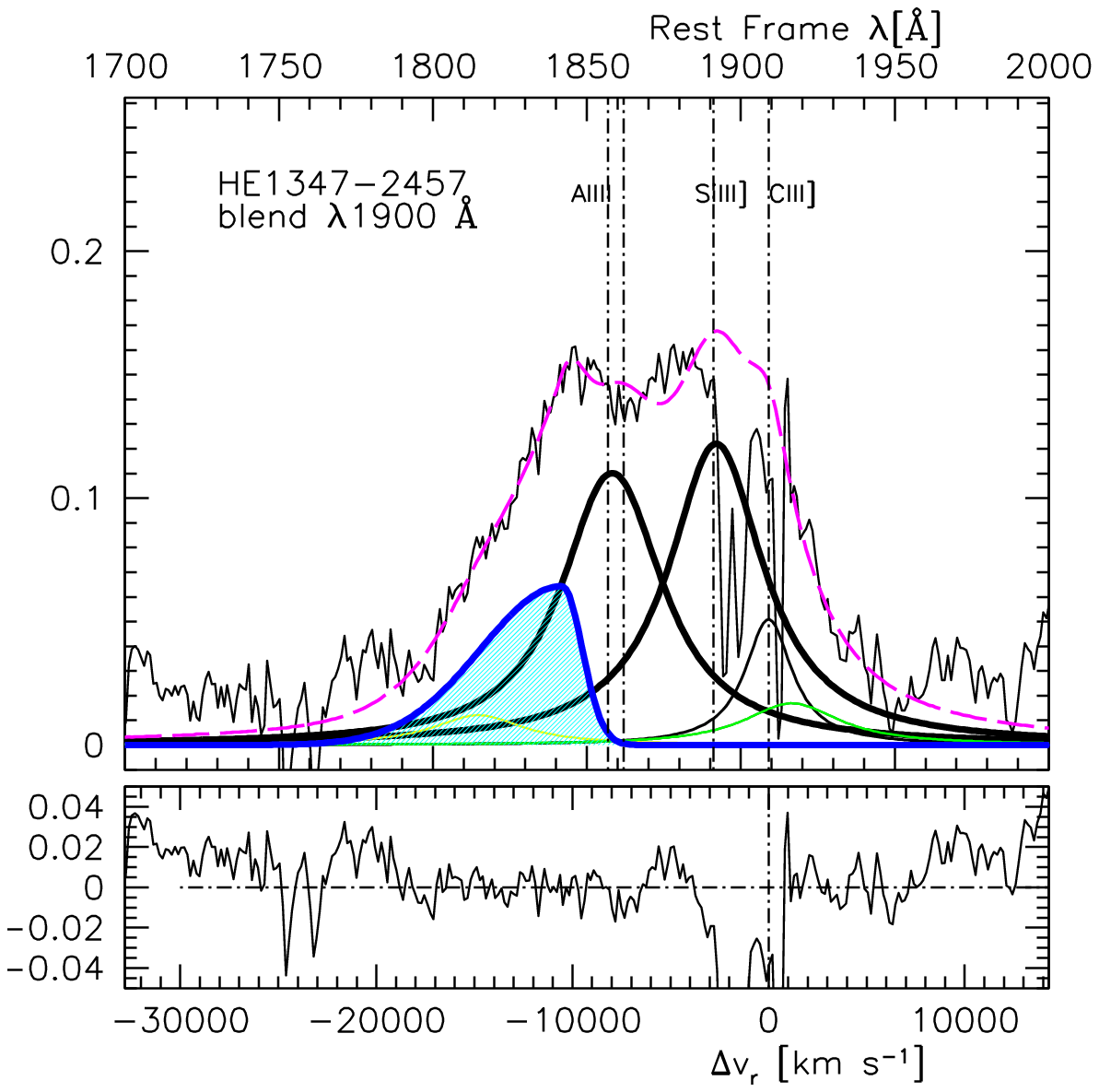}\\
\includegraphics[width=6.75cm]{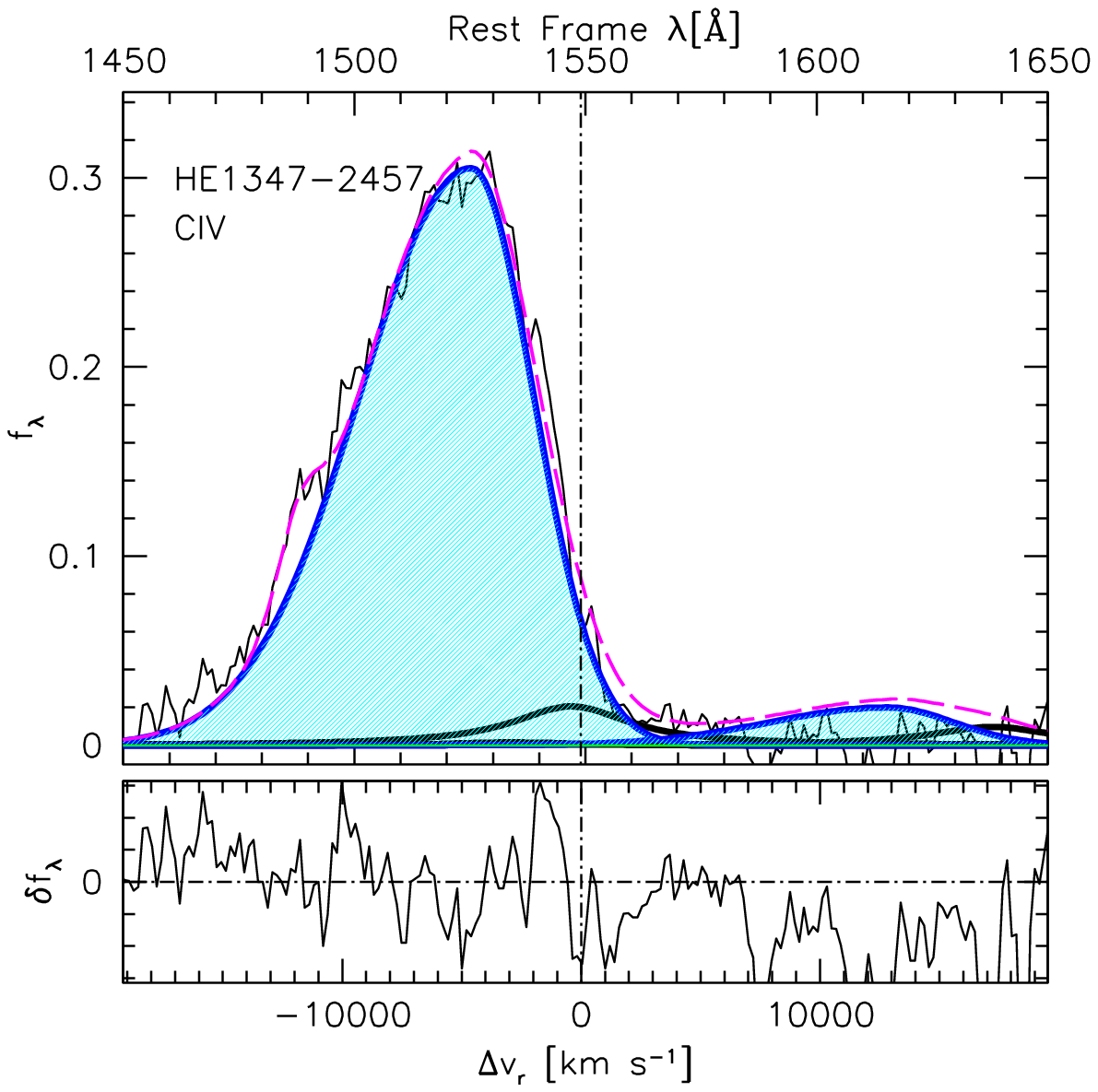}
\includegraphics[width=6.75cm]{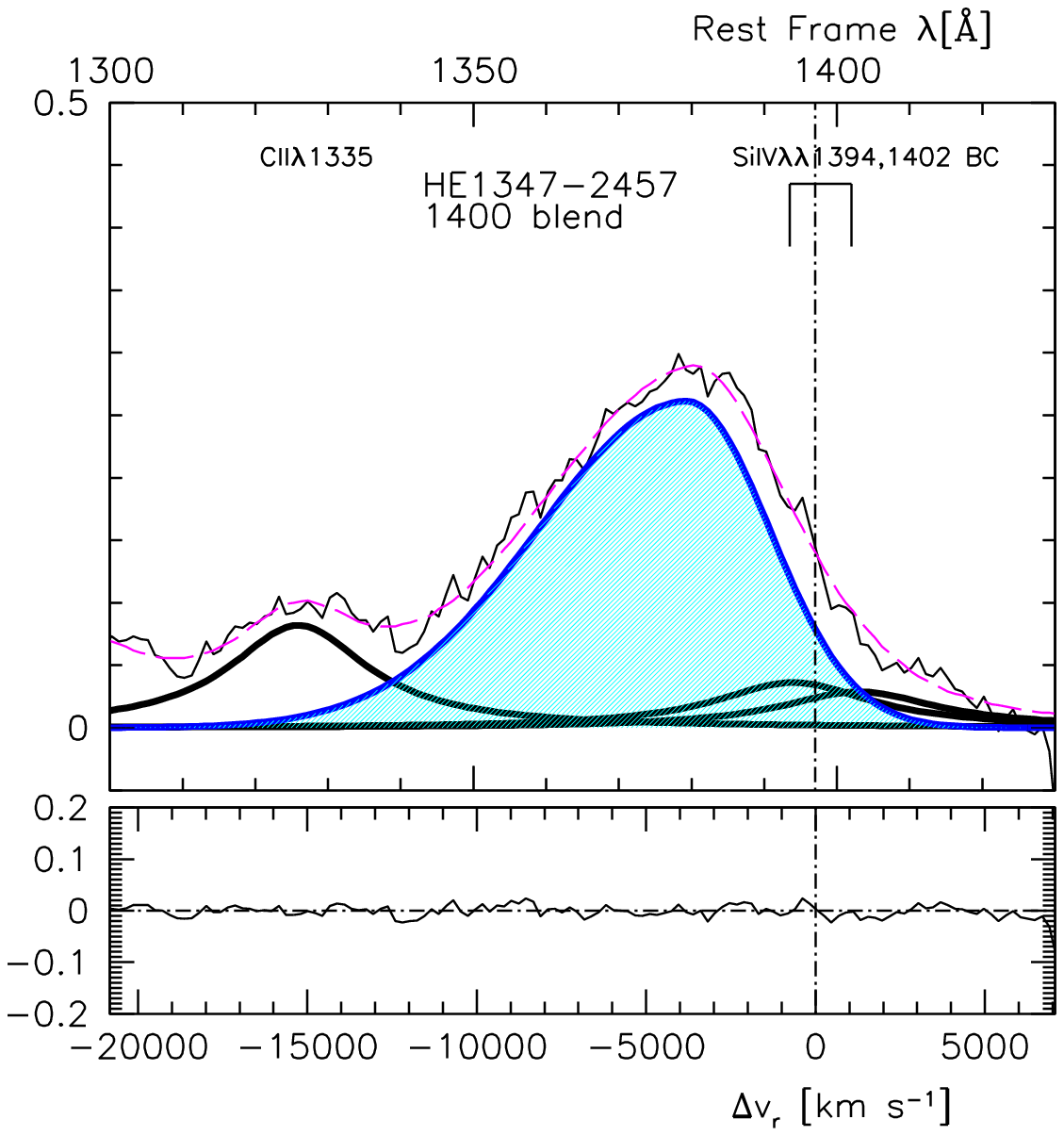}
\caption{{Rest-frame} 
 emission line profiles of \hb, 1,900\AA\ blend, \civ, and \siiv+\oiv, for~the HEMS quasar HE1347-2457, after~continuum subtraction. The dominance of the outflow emission is emphasized by the cyan shade of the blueshifted components, weakest in the \hb\ profile, but overwhelming in the high ionization lines. The quasar rest frame is identified by the dot-dashed vertical line. The virialized components are traced by the thick black lines, at~systemic redshift, being extremely weak in the case of the high-ionization \civ, \siiv, and \heiiuv\ lines. The \heiiuv\ line has been fit self consistently with \civ. The \hb\ line is strongly affected by \feii\ emission (green line), with~\rfe\ $\gtrsim 1$. \label{fig:outflows}}
\end{figure}

\begin{figure}[H]
\includegraphics[width=7.5cm]{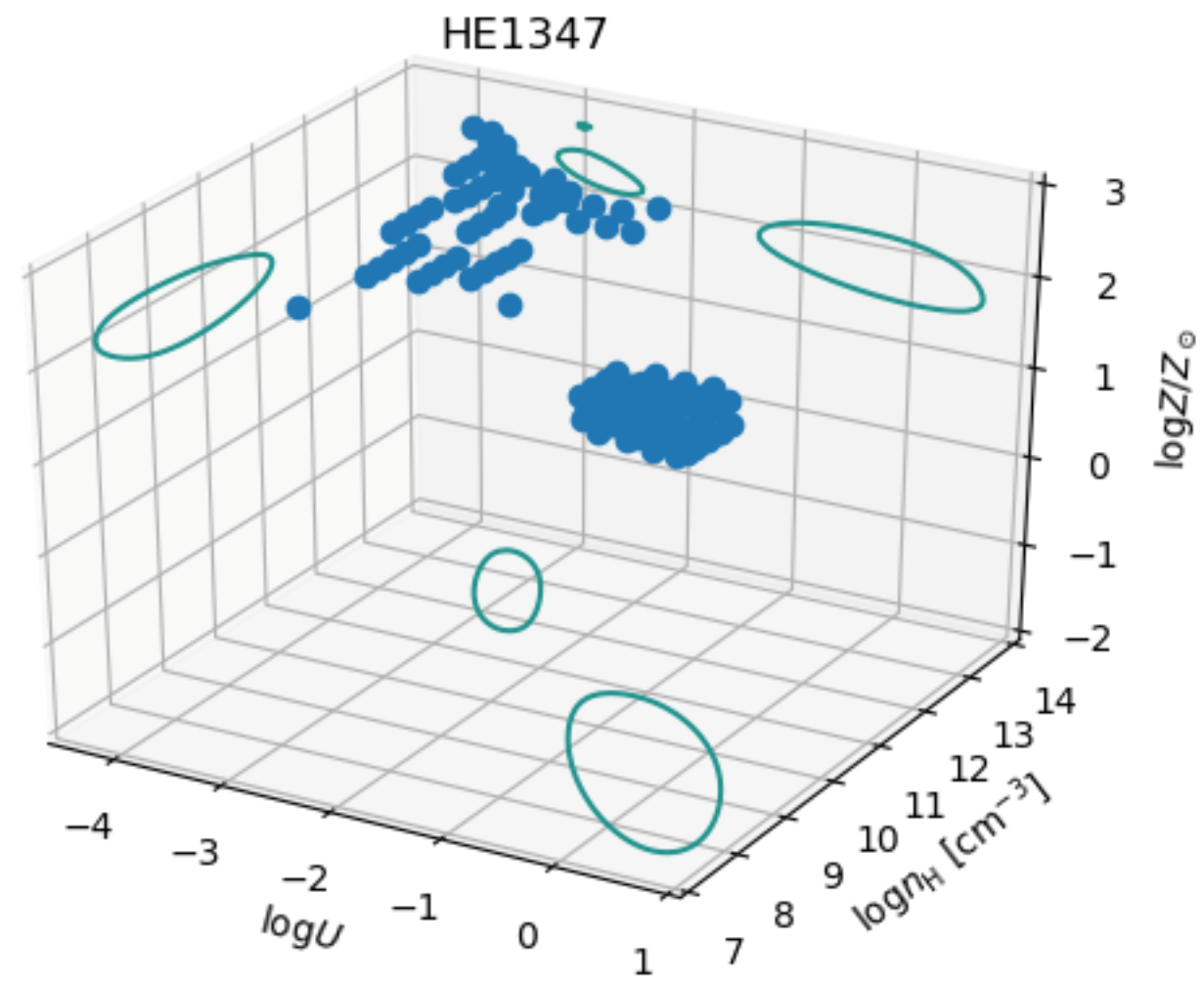}
 \caption{\textls[15]{Estimate of physical parameters for the HEMS quasar HE 1347-2457: metallicity $Z$, density, and ionization parameter. Dots represent photoionization simulations in the parameter space that are in agreement with the minimum $\chi^{2}$ within $1\sigma$ \ confidence limit. The seagreen lines trace the projection of the $1\sigma$ confidence limits in the three parameters after smoothing with a Gaussian kernel.} 
 \label{fig:3d}}
\end{figure}

The blends relevant for metallicity estimates in the optical and UV spectrum of the xA quasar HE 1347-2457 are shown in Figure~\ref{fig:outflows}. Figure~\ref{fig:3d} shows the results in the 3D parameter space ionization parameter $U$, Hydrogen density $n_\mathrm{H}$, and metallicity $Z$\ for HE 1347-2457. Measured broad line intensity ratios were compared with the prediction of arrays of {\tt CLOUDY} simulations to cover the entire parameter volume between $7 \le \log n_{\mathrm{H}} \le 14$\ [cm$^{-3}$. $-4.5 \le \log U \le 1$, $-2 \le \log Z \le 3$, with~a step of 0.25 or 0.3 dex, in search for the best model according to a minimum $\chi^{2}$ criterion. The derived metallicity value is very high, $Z \sim 100 Z_{\odot}$. The narrow projection on the planes $Z$-$U$ and $Z$-density suggest that the $Z$ \ uncertainty is small, and that the metallicity is well constrained, unlike the other physical parameters, which, however, have a minimum in correspondence of the expected low ionization solution, $\log U \sim -3$, $\log n_{\mathrm{H}} \sim 13$--$14$\ [cm$^{-3}$]. The high value of metallicity is supported by the high \siiv+\oiv/\civ\ ratio, as well as by the intense \aliii, and very prominent \feii\ (Figure~\ref{fig:outflows} \cite{sniegowskaetal21, garnicaetal22}). 
The high metallicity and large outflow velocity that can be inferred from \mbox{Figures~\ref{fig:outflows} and~\ref{fig:3d}} are consistent with a scenario in which the wind acceleration is driven by the scattering of AGN continuum radiation, through absorption and isotropic re-emission in resonant transitions of the most abundant metals~\cite{castoretal75,murrayetal95}. 

The spectra of xA quasars reveal that the gas in their BLR is consistently enriched in metals at low redshift as well. Metallicity levels reach up to 50 times the solar value in the most extreme xA PHL 1092, and are always at least 10 times solar~\cite{florisetal24}. Again, quasars such as I Zw 1 and PHL 1092 exhibit significant blueshifts in their { high-ionization lines} (HILs), particularly evident and easy to measure in the \civ\ line. The diagnostic ratios used to derive metallicities and the UV broad line weakness~\cite{marzianietal16a,martinez-aldamaetal18} provide a consistent picture of the physical conditions within the BLR, indicating a uniform structure among these SE candidates over several orders of magnitude in luminosity~\cite{sulenticetal00a,negreteetal12,shemmerlieber15,marzianietal16,chenetal24}. This uniformity suggests that the processes driving the enrichment associated with SE accretion are self-similar and produce consistent BLR characteristics across different xA quasars, with~low ionization and high density. 

The observed lack of redshift evolution in $Z$ (\cite{juarezetal09,yangetal21}, and references therein), the high $Z \gtrsim10$ inferred for the xA BLR, combined with the steep metallicity gradient from the BLR to the narrow-line region (NLR) and the stellar component \citep{xuetal18} suggest that metallicity enrichment occurs in situ \citep{collinzahn99,wangetal23,huangetal23,fanwu23}. This inference is further supported by the strong similarities between low-redshift xA quasars \citep{florisetal24} and those at the highest redshifts \citep{banadosetal16, dodoricoetal23,loiaconoetal24}. Given the relatively short ages at high redshift, there might not be sufficient time for efficient accretion of enriched material produced elsewhere across the host galaxies. Unless BLR $Z$\ estimates are wrong by two orders of magnitude, the  in situ origin of the enrichment is also consistent with the lower metallicity estimates for the NLR, which reach $Z \sim Z_{\odot}$ in the most extreme cases, and are usually between sub-solar and slightly super-solar~\cite{storchi-bergmannetal98,nagaoetal06b,xuetal18,armahetal23}. A direct comparison with the metallicity in the BLR of xA sources is, however, not appropriate: the NLR estimates refer to different AGN, some of which belong to Population B. Quantifying metallicity in the xA NLR is challenging, as the O$^{2+}$ lines are usually weak and blueshifted, and other lines are even weaker and difficult to measure. A characterization of the xA NLR metallicity is not yet available. 

\unskip

\subsection{Super-Eddington Accretion and Nuclear Star~Formation}

A starburst--AGN connection has been discussed from the 1980s through the 2000s, whereas several Seyfert galaxies were found in interacting systems with concomitant starbursts, many of them classified as luminous or ultra-luminous IRAS galaxies \mbox{(e.g.,~\cite{sandersetal88,rafanellimarziani92,marzianietal93,heckmanetal97,dultzin-hacyanetal03,king05,schweitzeretal06,wildetal10,rafanellietal11}).} The class of cold quasars (blue quasars detected in ultra-luminous IR galaxies because of their emission of cold dust at 250 $\upmu$m~\cite{kirkpatricketal19}) indicates that there can be a significant range of cosmic epochs where both star formation and AGN activity coexist. In addition, quasars that are hosted in luminous IRAS galaxies are often extreme \feii\ \mbox{emitters \citep{liparietal93,lipari94,haasetal03}}. This result led to propose an evolutionary unification scenario involving supermassive black holes, and starbursts with outflows to explain key observational properties of what we now call SE candidates~\cite{sandersetal88,dultzin-hacyanetal03a,liparietal06}. 

A more recent approach is still focused on explaining several aspects of nuclear activity as mainly associated with massive star formation, and  to evaluate the conditions in the inter-stellar medium (ISM) surrounding the black hole and their implications for the accretion process. Nuclear star formation (i.e., {\em {within or around the molecular torus} 
})~\cite{wangetal09sf,wangetal10sf,wangetal11sf,wangetal12sf} is expected to be present in most AGN with moderate or high accretion rates (a torus might be absent in low-luminosity AGN \cite{elitzurshlosman06,kangetal24}). The outer self-gravitating part of the disk and the torus provide a reservoir of atomic and molecular gas for the active nucleus fuelling, and star formation in the self-gravitating part of the accretion disk appears a condition naturally associated with high or SE accretion rate \citep{artymowiczetal93,lin97,collinzahn99,wangetal23,dittmanncantiello24,fabjetal24,liuetal24}. 

Stars formed within, or~embedded by the accretion disk are expected to follow evolutionary tracks different from the ones of stars evolving in a low-density environment. Accretion modified stars, embedded within the dense environment of the AGN accretion disk, can undergo substantial mass gain or experience powerful tidal forces, reshaping their structures, compositions, and lifespans~\cite{artymowiczetal93,nayakshinetal07}. Stars in close orbits around supermassive black holes may accrete disk material, leading to surface enrichment in heavy elements and altered thermal properties~\cite{alexanderlivio01a,alexanderlivio01b}. Such accretion-modified stars could become more massive and luminous, and may even exhibit enhanced rotation due to the angular momentum transfer from the disk~\cite{davieslin20}. Stars might even become ``immortal'', meaning they accrete gas at a rate that balances their mass loss through stellar winds~\cite{jermynetal22,dittmannetal23,nasimetal23}. The disk/star interactions not only influence the stars themselves but may also contribute to AGN variability and feedback processes, as accretion-modified stars can fuel further activity within the nucleus through stellar winds or episodic mass loss back into the accretion flow. Over~time, the stars in AGN disks can inject elements like nitrogen (the main product of the CNO cycle \cite{clayton68}) into the surrounding medium, and  yield peculiar enrichment patterns \citep{napolitanoetal24}. These stars can also contribute to phenomena like supernov\ae\ (SN\ae) and gamma-ray bursts (GRBs) within the AGN disk. 

The connection between nuclear star formation and AGN properties is still sketchy, in part because data cannot resolve nuclear star formation, but~also because a full stellar evolution theory for accretion-modified stars in a dense environment is still to be developed~\cite{dittmanncantiello24}. 
 At present, the evidence of star formation as a source of infrared and radio emission in xA sources is associated with the host galaxies: several xA sources tend to follow the FIR---radio correlation of star forming galaxies~\cite{zhouetal02,sanietal10,caccianigaetal15,gancietal19}. The modest radio-to-optical ratio places xA sources in the radio-intermediate domain~\cite{zamfiretal08}. Given the resolution limits, especially severe in the FIR, it has been impossible to probe the extent of nuclear star formation. The origin of the relatively weak radio emission of radio-quiet and radio-intermediate quasars remains unclear~\cite{laorbehar08,laoretal19}. The inner X-ray corona and the wind could also produce significant nuclear radio emission~\cite{panessaetal19,yangetal20,chenetal23}. 
 
It is the heavy enrichment of the BLR emitting gas that makes it necessary to postulate a close association between SE activity and nuclear star formation. Gas enrichment suggested by significant \feii\ emission in high (and even the highest) redshift \mbox{quasars~\cite{elstonetal94,marzianietal09,dodoricoetal23,banadosetal16,yiangetal23}} implies that the enrichment occurs very proximate to or within the accretion disk~\cite{fanwu23}. 
 Core-collapse supernov\ae\ from massive progenitors have a significant yield in iron~\cite{chieffilimongi13}, along with Al and Si. This makes it possible for highly accreting quasars in Population A not only to exhibit powerful metal-rich winds~\cite{wangetal09a,wangetal10sf,wangetal11sf,wangetal12sf}, but~also to enhance the low- and intermediate-ionization lines that are especially prominent in xA sources.



\subsection{Evolution Along the Quasar Main Sequence}
\label{evol}

\begin{figure}[H]
\includegraphics[width=1\linewidth]{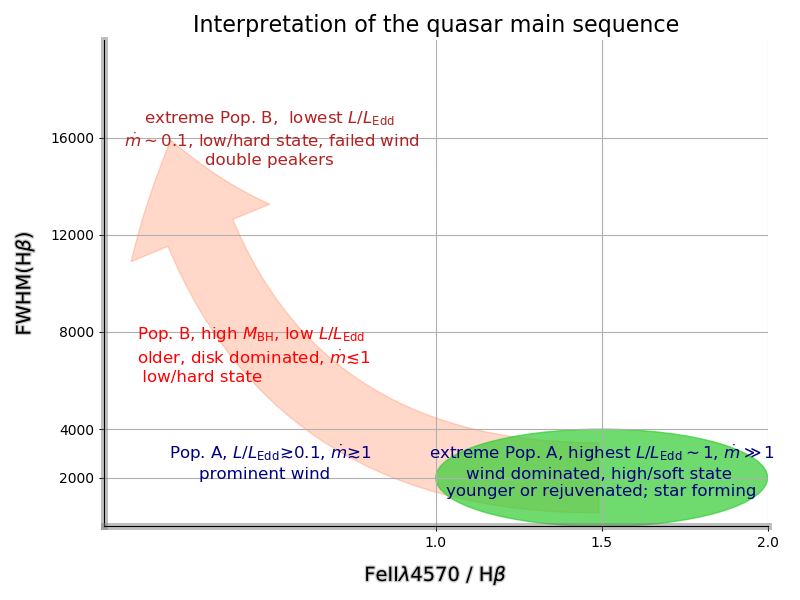}
\caption{{Evolutionary} 
 interpretation of the optical plane FWHM \hb\ vs. \rfe\ of the quasar main sequence. \label{fig:evol}}
\end{figure}

The scheme of Figure~\ref{fig:evol} (see also Figure~6 of \citet{marzianietal24}) illustrates a possible evolutionary path, emphasizing the transition from highly accreting, young quasars to massive, low Eddington ratio systems. Population A quasars are characterized by lower black hole (BH) masses ($M_\mathrm{BH}$), higher \lledd, strong outflows, and low equivalent width \oiii\ and \civ\ HILs. These are typically young or rejuvenated, highly accreting systems, some of which host powerful winds. At~the lower-right corner of the sequence, the extreme Population A (\rfe\ strong NLSy1s at low luminosity) includes younger, highly accreting systems with strong winds, low equivalent width emission lines, and significant blueshifts in both \civ\ and \oiii\ lines. {  VLTI/MATISSE and GRAVITY observations of H0557-385 and I Zw 1 revealed that the intense radiation field might by at the origin of a flattening in the molecular torus \citep{drewesetal25}. The K, L, and M band results indicate that the dust structures are compact, and exhibit no significant polar elongation. Notably, in H0557-385, the dust sizes between 3.5 and 8 { $\upmu$m} remain constant at approximately 3 to 10 times the sublimation radii, suggesting a direct view of the wind launching region. The absence of polar elongation implies that any winds are likely launched in an equatorial direction or are influenced by the AGN continuum radiation pressure. If~equatorial outflows dominate, this could reduce the amount of gas available for accretion, potentially lowering the accretion rate over time, or~leading to a stationary configuration contributing to the low ionization part of the BLR~\cite{czernyhryniewicz11}. } 

On the other hand, Population B quasars have higher \mbh, lower \lledd, weaker outflows, and high W \oiii\ lines, representing more evolved, older systems with less active accretion~\cite{fraix-burnetetal17}. Powerful radio-loud (jetted \cite{bertonetal15,padovani17}) sources are located in the upper-left quadrant, associated with older, less active quasars with weaker winds and higher \oiii\ emission. At~the upper left end of the MS, we encounter sources with $\dot{m}$\ so low that they are at the verge of the inefficient radiative domain. In several cases, \hb\ emission is showing a relativistic disk profile, meaning that the emission is associated with a bare, small-size (perhaps truncated) accretion disk~\cite{stratevaetal03,terefemengistueetal24}. These systems are starving black holes~\cite{marzianietal22a}. 
 
 As the quasar evolves and feedback processes (such as winds and other outflows) deplete the surrounding material, the quasar transitions toward lower \lledd\ ratios, with~a gradual shift into Population B as the black hole accretion rate decreases and the quasar begins to receive less accretion material. In this phase, outflows weaken, and the quasar \oiii\ emission becomes more prominent~\cite{marzianietal16}. Overall, the sequence encapsulates the idea that quasars evolve from highly accreting (possibly even at SE rate), young, wind-dominated systems in Population A to older, massive, low-Eddington ratio systems in Population B, with~ feedback and accretion rate playing crucial roles in this transition. In other words, the MS shape of Figure~\ref{fig:mslabeled} can be well understood as a temporal sequence with black hole mass as a rough clock of accretion history~\cite{fraix-burnetetal17}. 
 
 Massive stars embedded in the accretion disk may be a scenario possible only for young or rejuvenated sources~\cite{wu09a}. More evolved systems might involve embedded black holes with extreme mass ratio or neutron stars, yielding gravitational wave chirps that could be detected with LISA up to high redshift~\cite{donofriomarziani18}. 

 \subsection{Accretion Modes}

Quasars can be classified into distinct accretion modes based on their position in the mass-luminosity diagram, in addition to their placement along the MS~\cite{marzianietal03b,shen13,donofriomarziani21,wushen22}. In the plane luminosity vs. black hole mass, type 1 AGN occupy a relatively narrow strip constrained between \lledd\ $\approx 0.01$ and $\approx 1.$ Inefficient radiators are found at or below the lower end of the Eddington ratio range, characterized by low accretion rates and jet-like outflows \citep{reesetal82,narayanyi94,narayanyi95}. They are rare in optically selected samples of type 1 AGN, since a cold disk~\cite{shakurasunyaev73} might evaporate into an X-ray hot plasma if \lledd\ $\ll 0.01$ \citep{kangetal24}. Population B sources occupy the middle Eddington ratio range, $0.01 \lesssim $\lledd$\lesssim 0.1$--$0.2$, where the accretion disk can be arguably modeled using the classical Shakura--Sunyaev framework~\cite{shakurasunyaev73}. These quasars exhibit moderate accretion rates and relatively stable disks, with~limited wind activity (also due to an over-ionization of the inner disk region exposed to the AGN continuum~\cite{murrayetal95,progaetal00,progakallman04}). The change in structure with increasing dimensionless accretion rate is well-summarized in Figure~1 of \citet{giustiniproga19}, where all the main accretion modes are sketched. Table~\ref{tab:modes} provides a slightly modified scheme, in which the onset of an optically thick, geometrically thick region occurs at \lledd\ $\sim 0.1$--$0.2$, where we see a transition between Pop. A and B properties~\cite{marzianietal03b,blandfordetal19}. Above~this Eddington ratio, outflows leading to a change in emission line parameters of high ionization lines become frequent, and \feii\ emission becomes more prominent. It is tempting to think that this critical Eddington ratio is accompanied by a structure change in the inner accretion disk, and a shielded region may develop \citep{wangetal14a,marzianietal18}, yielding the ideal conditions for the production of the low ionization spectrum emitted in Pop. A and extreme Pop. A. At~the same time, radiative forces are higher, and the wind acceleration might become more efficient: low-column density gas is easily ablated away \citep{netzermarziani10}. Toward the high end of the Eddington ratio range, we enter in the domain of xA sources and we find sources which are even more wind-dominated \citep{sulenticetal07,richardsetal11,coatmanetal16,sulenticetal17,vietrietal18}. These quasars exhibit significant outflows, as evidenced by the pronounced blue shifts in high ionization lines like \civ, with~radial velocities of the outflow that can reach $\gtrsim$1000 \kms, up~to 5000 \kms\ at the highest luminosity (Section~\ref{winds}; Figure~\ref{fig:outflows}). These are among the outflows that may play a role in regulating the growth of the black hole and the evolution of galaxies~\cite{dimatteoetal05,hopkinselvis10,fabian12}. 

\begin{table}[H] 
\caption{Populations in the luminosity--black hole mass diagrams of type-1~AGN.
\label{tab:modes}}
\newcolumntype{C}{>{\centering\arraybackslash}X}
\begin{tabularx}{\textwidth}{Cm{5cm}<{\centering}C}
\toprule
\textbf{Population}	& \textbf{Eddington Ratio Range}	& \textbf{Prevalence}\\
\midrule
Inefficient radiators		& \lledd\ $ \lesssim\ 10^{-3}$--$10^{-2}$	& not included in Figure~\ref{fig:mslabeled} \\
Population B		 & $10^{-3}$--$10^{-2} \lesssim$ \lledd\ $ \lesssim 0.1$--$0.3 $			& $\approx 0.46$\\
Population A		 & $0.1$--$0.3 \lesssim$ \lledd\ $\lesssim $ 1			& $\approx$ 0.43\\
extreme Pop. A & $1 \lesssim$\lledd\ $\lesssim $ a few &$\approx$ 0.10 \\
highly super-Eddington & \lledd\ $\gg$ a few & not existent? \\
\bottomrule
\end{tabularx}
\end{table}


\section{Coexistence of a Virialized System and Powerful~Outflows}
\unskip

\subsection{A Virialized Low-Ionization System} 
\label{virial} 

The HIL and LIL profiles in Population A and xA reflects a dichotomy in the physical and dynamical conditions of the BLR. 
An impressive, extreme example is shown in Figure~\ref{fig:outflows}, which emphasizes the profile differences between \civ\ and \hb: while \hb\ retains a mostly symmetric profile, roughly consistent with the rest frame, most of the \civ\ flux is due to a blueshifted excess. The line component which is shaded blue is ascribed to an outflow, most likely a disk wind driven by radiation pressure~\cite{murrayetal95,murraychiang97,progaetal00,progakallman04}. Outflows in AGN are very pervasive~\cite{marzianietal22a} and there will be always some low column density gas that will be expelled from the gravitational sphere of influence of the central black hole~\cite{ferlandetal09,marconietal09,risalitielvis10,netzermarziani10}, no matter how low \lledd\ might be. However, it is only above a threshold \lledd\ $\approx 0.1$--$0.2$ \ we start seeing prominent outflows in \civ\ and other high ionization lines like \siiv\ and \heiiuv\ \cite{sulenticetal07,richardsetal11} such as the one shown in Figure~\ref{fig:outflows}. 
The figure also shows that a symmetric, unshifted component is dominating the emission of \hb\ and is still prominent in the intermediate ionization line \aliii, although~it is overwhelmed in \civ. The line profiles of \hb, \mgii, and \aliii\ are well modeled by a Lorentzian function, and are consistent with an extended disk profile even if broadened by turbulent motions of the emitting gas~\cite{sulenticetal02,goadetal12,marzianietal13,popovicetal19}, with~negligible relativistic effects~\cite{chenhalpern89}. 
 
This distinction between HILs and LILs is especially pronounced for xA quasars, where the high accretion rates lead to stronger winds for the HILs, while the LILs remain tied to the more virialized region near the disk~\cite{marzianietal14,dultzinetal20}: the \civ\ and \hb\ lines appear to behave independently of one another~\cite{capellupoetal15,mejia-restrepoetal16}. 

\begin{figure}[H]
\includegraphics[width=1\linewidth]{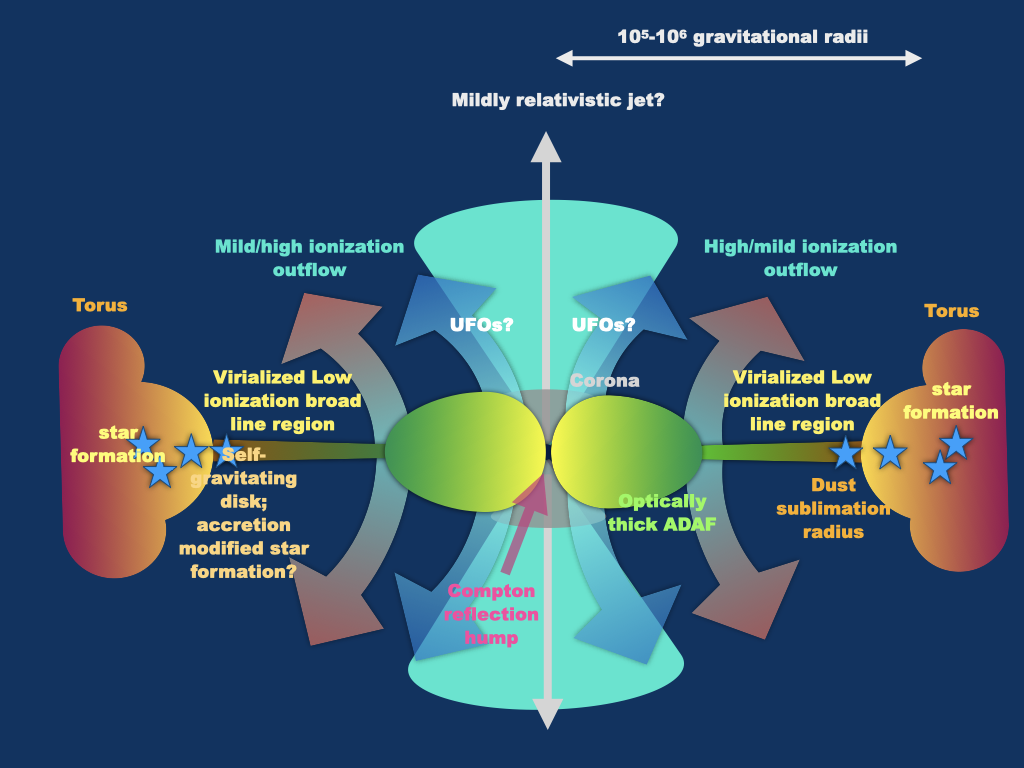}
\caption{{A tentative} 
 sketch of the inner working of a super-Eddington quasar. The placement and the configuration of several elements is not drawn to scale, and is largely hypothetical. { The scheme is meant to emphasize the existence of a shielded low-ionization region, along with the possibility of concomitant star formation in the outer regions of the accretion disk and in the torus.}
\label{fig:agn}}
\end{figure}

The sketch of Figure~\ref{fig:agn} illustrates the structure and components of SE AGN focusing on the regions surrounding the supermassive black hole (from 1 to $10^5$--$10^6$) gravitational radii. At~the center, the SMBH is enveloped by an ADAF. A mildly relativistic jet may emerge from the black hole along the spin axis, indicating the outflow of material along the polar axis~\cite{collinsouffrinetal88,marzianietal96,elvis00}. The sketch also highlights the presence of a corona above and below the accretion flow, which is responsible for high-energy X-ray emission. The geometry of the corona is actually highly uncertain, with~a lamppost geometry usually preferred over a more symmetric configuration~\cite{cackettetal07,uttleyetal14}. A ``special'' Compton reflection hump representing the interaction of high-energy photons with Compton-thick outflowing material from the { very} edge of { a} geometrically-thick accretion disk { might be also present}: in this case, the emission of Fe K$\alpha$\ is seen displaced by (0.1--0.2) $c$~\cite{karaetal16,zhangetal24}. { Ultra-fast outflows} (UFOs \cite{tombesietal10}) are indicated as the inner arrows extending outward from the central engine. The location and the geometry is not fully clear as yet~\cite{tombesietal12,tombesietal13,lahaetal21}, but~they are believed to be on a scale of {$10^2$--$10^4$} gravitational radii, perhaps closer to the SMBH axis than the mildly ionized outflows, which might be associated with the outer ADAF but also with the part of the disk not dominated by radiation pressure (shown as a flatter section in between the ADAF and the torus), shielded (at least in part) by the coronal X-ray radiation. The flattening shown in Figure~\ref{fig:agn} may not actually occur; however, the thick disk geometry allows for a shielded region (region II of \citet{wangetal14a}). Among~the mildly ionized outflow, there is an interesting relation between the amplitude of the line blueshift and the ionization potential of the parent ionic species: highest blueshifts in \civ, $\gtrsim$1000 \kms\ correspond to moderate blueshifts in \aliii\ $\approx 0.2$\ the ones of \civ\ \cite{marzianietal22}, and of a few hundred \kms\ in \mgii\ \cite{marzianietal13}. The decreasing velocity could be a consequence of an increase in the launching radius of a radially-stratified wind. Extrapolating the trend to higher energy, one should ascribe the X-ray high speed wind to the innermost region exposed to funnel radiation. 

 The sketch in Figure~\ref{fig:agn} highlights an additional layer of complexity in the outer regions of the accretion disk: the potential for star formation within the self-gravitating portion of the disk~\cite{linpringle87,artymowiczetal93,collinzahn99}. Stars that form within the disk, or~are captured from elsewhere, follow evolutionary paths that are influenced by the accretion of disk material. Farther out, a~torus of dust and gas surrounds the AGN, with~the dust sublimation radius marking the point beyond which dust can survive without being sublimated by the AGN intense radiation. Dust-driven outflows might originate around this transition radius~\cite{czernyetal17,naddafetal21,naddafetal21a,choietal22}. Star-forming regions are indicated where the disk may become self-gravitating. The sketch is meant to suggest that these star-forming regions could affect the accretion process, potentially influencing the overall evolution of the system, and even the development of two main structural components of the active nucleus, the torus and the accretion disk. 
  
 \subsection{Powerful Outflows: A Major Factor in Galactic Evolution?}
\label{galev}

\label{winds}

Recent surveys—such as the Wide-field Infrared Survey Explorer (WISE)/SDSS \mbox{selected} hyper-luminous (WISSH)~\cite{bischettietal17} and the Hamburg-ESO main sequence \linebreak (HEMS)~\cite{marzianietal09}—have considered the most luminous quasars at the Cosmic Noon. 
The large blueshifts observed in the \civ\ line are indicative of strong winds generated during their high-accretion phases, producing outflows capable of escaping the BLR and enriching the surrounding regions \citep{vietrietal18, vietrietal20,deconto-machadoetal24}. Albeit rare, these sources are of extremely high luminosity, \mbox{$L \gtrsim 10^{47}$ erg s$^{-1}$}. The estimate of the mechanical parameters of the mildly ionized gas outflows traced by \civ\ and \oiii\ emission permits { testing} whether AGN may cause a significant feedback effect at least when their outflows are most powerful. The mass outflow rate {($\dot{M}$)} and the kinetic power {($\dot{E}_{\mathrm{kin}}$)} have been estimated adopting a simple biconical geometry where the emitting gas is in radial outflow~\cite{canodiazetal12,marzianietal16,marzianietal17,vietrietal18}. {In this framework, the relations are formally identical for \oiiionly\ and \civ, and namely:}
 $ M_{\textrm{ion}} \propto L^\mathrm{out} Z^{-1} n_\mathrm{H}^{-1}$, where $L$ is the outflow-emitted line luminosity, $n_\mathrm{H}$\ the electron density and $Z$\ the metallicity. The mass outflow rate ($\dot{M}^{\textrm{out}}_\mathrm{ion}$) at a radius $r$ and with an outflow velocity $v_{\mathrm{out}}$, might be written as:
$
\dot{M}_\textrm{ion}^\textrm{out} \propto L^\textrm{out}\ v_{\mathrm{out}}\ r^{-1} Z^{-1} n_\textrm{H}^{-1}
$.
The kinetic power, ${\dot{E}_{\mathrm{kin}}}$, is then given by $\dot{E}_{\mathrm{out}}\sim \frac{1}{2}\dot{M}_{\mathrm{out}}v_{\mathrm{out}}^2$, which leads to
$ \dot{E}_{\mathrm{kin}} \propto L^{\mathrm{out}} v_{\mathrm{out}}^3\ r^{-1} Z^{-1}n_\mathrm{H}^{-1}$. 
The values obtained scale with line luminosity and can therefore increase by a factor $\sim$$10^{3}$\ going from moderately to very luminous quasars. The increase in radial velocity is more modest, a~factor $\sim$$5$, due to the weak luminosity dependence of the terminal velocity for radiation-driven winds \citep{laorbrandt02,sulenticetal17}. 

\begin{figure}[H]
\includegraphics[width=1\linewidth]{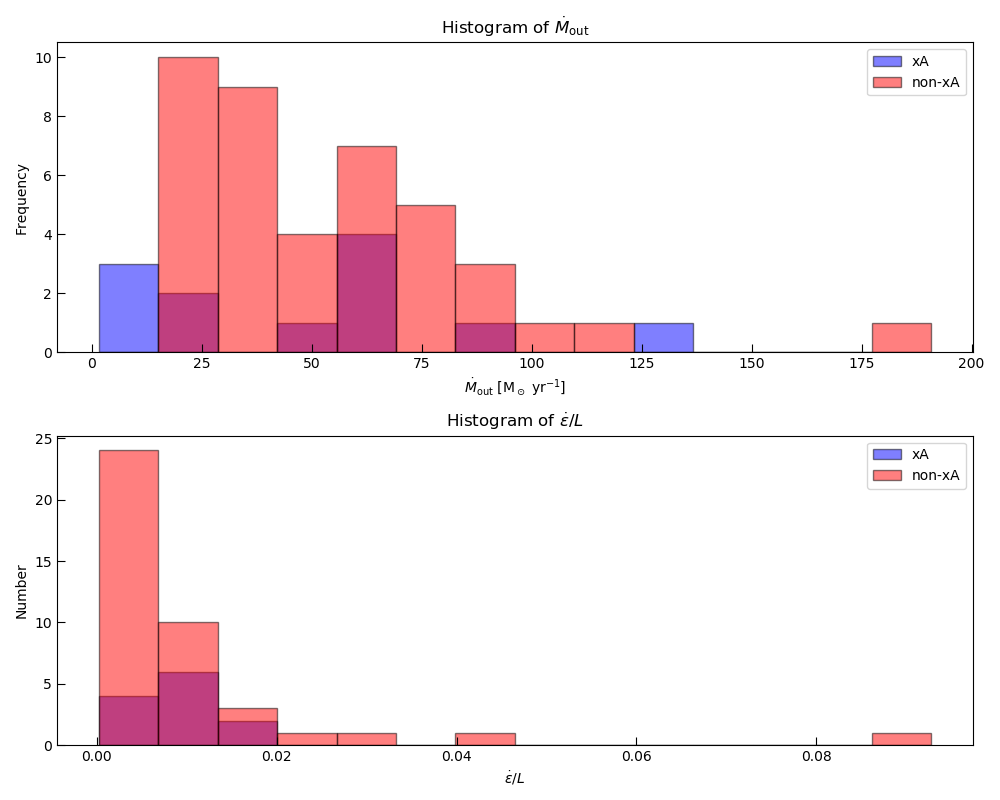}
\caption{Distribution
 of mass outflow rate (top) and ratio between the kinetic power $\dot{\epsilon}$\ and the bolometric luminosity (bottom) for sources with $L \gtrsim 10^{47}$ erg s$^{-1}$. \label{fig:xAcomp}}
\end{figure}

Figure~\ref{fig:outflows} shows an example of xA quasar where the outflow component is clearly dominating the emission in the high ionization lines \civ, \heiiuv, \siiv, and \oiv. Figure~\ref{fig:xAcomp} shows the distribution of mass outflow rate in M$_{\odot}$ yr$^{-1}$ and of the ratio between kinetic power and bolometric luminosity derived from the \civ\ lines applying the equations written above, for~a sample of high-luminosity quasars \mbox{$L \gtrsim 10^{47}$ erg s$^{-1}$}, mainly from the HEMS survey~\cite{sulenticetal04,sulenticetal06,marzianietal09} and from the recent works \mbox{by \citet{deconto-machadoetal23,deconto-machadoetal24}.} 
There is no obvious systematic difference between the xA sources and the rest of the sample, as confirmed by Kolmogorov--Smirnov and $U$-Mann--Whitney tests. Significant outflows are detected also in Population B, even if the most powerful are restricted to Population A~\cite{deconto-machadoetal23,deconto-machadoetal24}. The \civ\ outflows are related to the larger scale outflows traced by \oiii\ \cite{zamanovetal02,vietrietal18,deconto-machadoetal23,deconto-machadoetal24}. It is tempting to think that the emitting gas might be the one expelled from the active nucleus~\cite{villar-martinetal24}. At~a low redshift, the low \oiii\ equivalent width and its large FWHM are consistent with a compact NLR and hence with a more straightforward physical connection between the \civ\ and the \oiii\ emitting gas: in principle, the \oiii\ could be emitted by gas further downstream in the nuclear outflow~\cite{zamanovetal02}. For~quasars at high luminosity, the \oiii\ emission is spread over kpc-sized \mbox{distances \citep{harrison2014,harrisonetal14}}. The extensive size of the \oiii\ emitting regions in high-luminosity \mbox{quasars~\cite{kakkadetal20,singhaetal22}} may suggest that the emission arises predominantly or even solely from radiative feedback from the AGN on the host galaxy gas.

 Even if the $Z$\ is measured from diagnostic ratios, the uncertainties are at least \mbox{$\sim$$0.3$ dex}. The geometry of the outflow introduces a factor of 3, based on the comparison between spherical symmetry and a flat layer. The 1$\sigma$ uncertainty in the emitting radius is approximately $\pm 30$\%, considering only the uncertainty of the scaling law parameter, while the uncertainty in the outflow velocity is typically around 30\%. Quadratically propagating these uncertainties results in a total factor of $\approx 5$ at the 1$\sigma$ confidence level in the kinetic power~\cite{deconto-machadoetal24}. The energetics remains extremely high even if conservative lower limits are applied, with~$\dot{M}_{\mathrm out} \sim 100 $M$_{\odot}$ { yr$^{{-1}}$}, with~typical kinetics powers around $\sim$$1$\%\ the bolometric luminosity, due to the extreme line luminosities reaching $\sim$$10^{45}$ erg s$^{-1}$\ for \civ. Figure~\ref{fig:xAcomp} shows that the xA quasars have mass outflow rates and kinetic--power ratios to luminosity comparable to non-xA quasars, but~that xAs are not {\em} extreme (the equivalent width of \civ\ is among the lowest~\cite{plotkinetal15,martinez-aldamaetal18,chenetal24}), even if they are expected to have extreme radiative output per unit black hole mass. The implication is that there is likely a broader population of quasars beyond the SE candidates—essentially encompassing all of Population A—that is capable of exerting significant feedback effects on their host galaxies.
 
The \civ\ resonance line is representative of the mildly ionized outflows originating from the accretion disk \citep{murrayetal95,murraychiang97,risalitielvis10}. This is expected to involve a large fraction of the mass load of the AGN wind, but~is not sampling gas in higher ionization stages, specifically the one associated with the UFOs). UFOs are present in jetted AGNs~\cite{tombesietal14}, and detected in about half { of} nearby AGNs, although there are hints { of} a larger prevalence in highly accreting quasars~\cite{goffordetal15,mesticietal24}. An UV line-driven disk wind could be at the origin of UFOs in AGN~\cite{mizumotoetal21}: this appears to be at least the case of the high-Eddington ratio quasar PDS 456 \citep{matzeuetal17}. Acceleration of UFOs likely needs an additional mechanism to yield velocities $\sim$0.1--0.3 $c$, such as intense magnetic fields~\cite{fukumuraetal14,fukumuraetal18} or thermal pressure gradient~\cite{begelmanetal83}. The mass load and the mass outflow rates of the UFO are comparable or superior to the one of the outflows traced by \civ\ \cite{tombesietal12,gianollietal24}. UFOs and mildly ionized winds are likely to be associated with coexisting phenomena, i.e.,~UFOs might not be entirely due to gas downstream or upstream of the BLR outflowing gas. Due to the multiphase nature of quasar outflows, the mass outflow rate and dynamical parameters derived from the \civ\ line may represent only a fraction of the total outflow. If~so, it is likely that the combined effect of UFOs and mildly ionized winds could result in a more significant and widespread impact~\cite{tombesietal10,mesticietal24}.

\unskip 

\section{Super-Eddington Quasars for~Cosmology} 
\label{cosmo}
\unskip


\subsection{Distance~Indicators}
\label{mu}
 
 The contrast between the behavior of HILs and LILs in Population A and extreme Population A quasars highlights the multiphase nature of quasar outflows and the complex, dichotomic structure of the BLR. The potential to use quasars as reliable cosmic distance indicators is grounded in the well-established understanding that their emitting regions form a virialized low-ionization system~\cite{petersonwandel99}, governed by the gravitational force of the central supermassive black hole, even in the case of the highest radiative output~\cite{negreteetal12}. Capitalizing on this result, we can consider three basic elements to derive a cosmological luminosity estimators: (1) SE candidates tend to converge toward a limiting value of \lledd; (2) their optical and UV spectrum is self similar, and easy to recognize, which in turns implies that the radius of the emitting regions should scale $\propto L^\frac{1}{2}$, lest significant differences are introduced in the ionization of the line emitting gas; and (3) there is a measure that accurately reflects the virial motion around the central black hole. If~we consider the three factors together, it is possible to rewrite $M_\mathrm{\mathrm{BH}}$\ in terms of luminosity and virial broadening  {\citep{marzianisulentic14}:}
 
 \begin{equation}
 L \propto \mathrm{FWHM}^{4} 
 \end{equation} 

 The derivation of the Faber-Jackson~\cite{tullyfisher77} law followed exactly the same line of reasoning. The caveat is that \lledd\ needs to be constant (in observational terms, have a small scatter around a well-defined value). Paradoxically, this assumption seems to be more easy to achieve for super-Eddington accretors than for galaxies, which are much larger and more complex bodies, each with its own evolutionary path~\cite{donofrioetal16a}. 

Virial broadening estimators have been sought for decades \citep{vestergaardpeterson06,marzianietal06,marzianisulentic12,shen13,runnoeetal13,brothertonetal15,savicetal18,afanasievetal19,savicetal20,dallabontaetal20,marzianietal22}. They allow for the least-inaccurate measurement of the central black hole mass, following the deceptively simple virial mass definition
\begin{equation} M_\mathrm{\mathrm{BH}} = \mathcal{F}\frac{ r_\mathrm{BLR}\cdot \mathrm{FWHM}}{G}, 
 \end{equation}
 where the factor $\mathcal{F}$ incorporates all dependences on geometry, dynamics, and accretion mode. The HI Balmer line \hb\ seems to be a good choice (Section~\ref{virial};~\cite{vestergaardpeterson06,marzianietal19}) to construct the Hubble diagram~\cite{marzianisulentic14,dultzinetal20,marzianietal21}. At~a second stance, the UV \aliii\ doublet~\cite{marzianietal22} could be a good tool to extend the Hubble diagram to redshifts beyond $\approx 1.4$, where \hb\ can be observed only with IR spectrometers~\cite{czernyetal21}. Preliminary applications have been carried out employing several samples where the virial broadening estimator has been \hb\ FWHM and $\sigma$\ only or \hb\ and \aliii\ jointly~\cite{marzianisulentic14,dultzinetal20,czernyetal21,marzianietal21,donofrioetal24}. Despite high dispersion (presently around 0.3 dex in the luminosity estimates), this method provides a promising approach for using quasars as cosmological standard candles, especially for measuring the matter density parameter ($\Omega_\mathrm{\mathrm{M}}$)~\cite{dultzinetal20,czernyetal21}. The main reason resides in the ability of quasars to sample a redshift range around $z \approx 2$: at the corresponding cosmic epochs, the repulsive effect of  dark energy was too low, and the main factor in the Universe dynamics was the attractive gravitational force (see Figure~6 of \citet{marzianisulentic14}, which shows how differences in cosmological models become quite large, $\approx 0.2 $ dex in luminosity, around $z \approx 2$). 



\subsection{Grown with the Wind, or~Saving $\Lambda$-Cold Dark~Matter}
\label{growth}

Some high-$z$\ quasars with very massive black hole ($\sim$$2 \cdot 10^{9} M_{\odot}$) would have been classified as an unremarkable Pop. B quasar if observed at $z \lesssim 1$. Two examples---{ along with others at $z \approx 5$ \citep{wolfetal24}}---are PJ308-21 at $z \approx 6.234$~\cite{loiaconoetal24} and { ULAS} J1120+0641, whose black hole mass is (1.35 $\pm 0.04) \times 10^{9}$ M$_\mathrm{\odot}$\ based on the Mg{\sc ii}$\lambda$2800 line, Eddington \mbox{ratio $\approx$ 0.40,} and X-ray properties indistinguishable from other bright quasars at $1 < z < 5 $ with similar Eddington rates~\cite{bosmanetal24}. JWST discovered a population of fairly massive black holes even at the highest $z$, for~example UHZ1 with a mass of $\sim$$10^{7.5}$ M$_\odot$\ at redshift $z \approx 10.3$~\cite{bogdanetal24}. These newly discovered, extremely massive black holes present a challenge to current growth models, with~potential implications for cosmology~\cite{melia24,melia24a}, unless a SE accretion phase is considered (e.g.,~\cite{huskoetal24}, and references therein). 

{ On the one hand,} there are no convincing cases with \lledd\ $\gg 1$, in agreement with the solutions computed for SE ADAFs. Even if the balance between infall, radiation trapping, the feedback exerted by radiation forces, magnetic field, and mass loss due to winds~\cite{ohsugaetal05,ogawaetal17} is not fully understood nor modeled in general relativistic radiation magnetohydrodynamics simulations \citep{inayoshietal20}, models do not exclude a SE regime with \mbox{$\dot{m}\gtrsim 10^{3}$} and radiative efficiency as low as $\eta \sim 0.01$~\cite{jiangetal14,sadowskietal15,jiangetal19} that would make possible a rapid black hole growth. 
 xA quasars with \rfe\ $\gg1$ \ like PG0043+052 and Mark 231 have been found up to $z\approx 6.5$~\cite{elstonetal94,murayamaetal98,chenetal24}. Convincing cases with best virial mass estimates range from $10^{6}$ to $10^9$\ solar masses: from the case of I Zw 1, PHL 1092, PDS 456, Mark \mbox{231 \citep{huangetal19,marzianietal19,marinelloetal20b,lietal24}} to the most luminous quasars at the Cosmic Noon (e.g., HE 0359—3959 \cite{marzianietal09,deconto-machadoetal24,martinez-aldamaetal17}). 
 
{ On the other hand,} there are several basic unknowns. The first question is whether accretion is truly SE, or~ a form of sub-Eddington balance resulting in a radiative output around \lledd\ $\sim$ 1. A continuity argument suggests caution, as there is no clear discontinuity between Pop. A and extreme Pop. A, unlike the distinct transition observed between Pop. B and Pop. A. The extreme Pop. A in this respect might be just an extreme once the threshold in Eddington ratio at $\approx$$0.1$--$0.2$ has been passed. The mass outflow rates also raise pressing problems. If~the outflow rates estimated in Section~\ref{winds} apply to both sub- and super-Eddington regimes, they may become comparable to the mass accretion rate in the sub-Eddington case, slowing the growth of the black holes. In addition, there is an environmental caveat to consider. ADAF models deal mostly with the innermost few hundreds gravitational radii, and assume that an arbitrary large amount is reaching the disk. On larger scales, the interplay between radiation forces and gravity is not so obvious, especially in the regions beyond the Bondi radius (the radius at which the gravitational pull of the central mass dominates over the thermal pressure of the gas, allowing the gas to move inward and accrete; see, e.g.,~\cite{mushanoetal24}). It is not clear whether a large amount of matter in excess to hundreds of solar masses per year can be effectively funneled down to the Bondi radius of the black hole~\cite{milosavlijevicetal09,inayoshietal20}. 

We can make a few elementary considerations on the black hole growth with mass loss due to a wind. The rate of change of the black hole mass \(M(t)\) is the difference between the accretion rate and the mass loss rate due to winds: \( {dM}/{dt} = \dot{M}_\mathrm{\text{acc}} - \dot{M}_\mathrm{\text{wind}}, \) where \(\dot{M}_\mathrm{\text{acc}}\) is the accretion rate of the black hole and \(\dot{M}_\mathrm{\text{wind}}\) is mass loss rate due to winds. The accretion rate can be expressed in terms of the Eddington accretion rate \(\dot{M}_\mathrm{\text{Edd}}\) and the dimensionless accretion rate \(\dot{m}\): \( \dot{M}_\mathrm{\text{acc}} = (1-\eta) \dot{m} \cdot \dot{M}_\mathrm{\text{Edd}} \)
and the mass loss rate due to winds as a fraction \(f_\mathrm{\text{wind}}\) of the accretion rate:
\( \dot{M}_\mathrm{\text{wind}} = f_\mathrm{\text{wind}} \cdot \dot{M}_\mathrm{\text{acc}}, \) so that we can write:
\begin{equation} {dM}/{dt} = \dot{m} \cdot {4 \pi G m_\mathrm{p}}/{ \sigma_T c} \left(1 - \eta)(1 - f_\mathrm{\text{wind}}\right) M. \end{equation}

The solution of this differential equation describes the exponential growth of the black hole mass \(M(t)\) considering both accretion and mass loss due to winds is:
 \( M(t) = M_0 \exp\left({t}/{t_\mathrm{\textrm{growth}}} \right), \) where the characteristic growth timescale \(t_\mathrm{\text{growth}}\) is given by \( t_\mathrm{\text{growth}} = { \sigma_T c}/[{\dot{m} \cdot 4 \pi G m_\mathrm{p} \left(1 - \eta)(1 - f_\mathrm{\text{wind}}\right)}] = {t_\mathrm{\text{Edd}}}/{\dot{m} \left(1 - \eta)(1 - f_\mathrm{\text{wind}}\right)}. \)
 
\begin{figure}[H]
\includegraphics[width=1\linewidth]{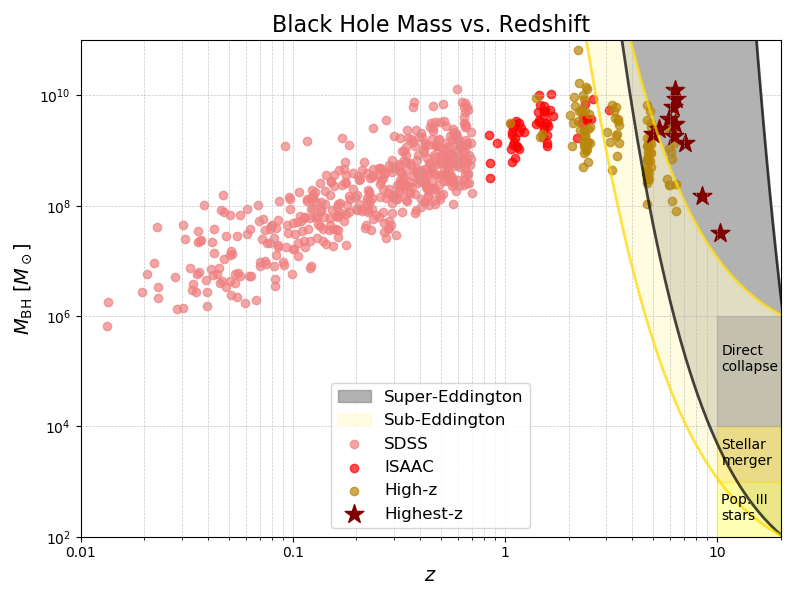}
\caption{Black hole mass estimates for several samples of quasars as a function of redshift, with~a focus on the highest masses at high redshift. The coral data points are from an \hb\ SDSS-sample~\cite{zamfiretal10}, the red ones are the data of the HEMS survey~\cite{marzianietal09}, and the dark golden are several high-redshift \mbh\ estimates presented in~\cite{marzianisulentic12}. The brick-brown stars mostly refer to the most recent observations made possible by JWST~\cite{yueetal24,bosmanetal24,stoneetal24,kokorevetal23,mazzucchellietal23,ubleretal23,bogdanetal24}. The shaded areas show the range permitted for black hole growth starting from Pop. III star progenitors (\mbh~$\sim 10^{2}$M$_{\odot}$) and direct collapse (\mbh~$\sim 10^{5}$M$_{\odot}$), for~the sub-Eddington (yellow) and super-Eddington regime. See text for~details. \label{fig:mz}}
\end{figure}

Figure~\ref{fig:mz} shows black hole mass estimates as a function of $z$, assuming standard $\Lambda$---cold dark matter cosmology ($H_{0} =70$ \kms\ Mpc$^{{-1}}$, $\Omega_{\mathrm{M}} = 0.3$,\ $\Omega_{\Lambda} = 0.7$) (\cite{donofrioburigana09}, and references therein). The samples include \hb\ based measurements for low-$z$ quasars from the SDSS~\cite{zamfiretal10}, the HEMS survey~\cite{marzianietal09}, and several samples of high-$z$\ quasars shown \mbox{by (\cite{marzianisulentic12}, c.f.~\cite{trakhtenbrotnetzer12}).} With~respect to \citet{marzianisulentic12}, the figure adds some recent \mbh\ estimates (brown stars) for the highest redshift quasars, in the range $6 \lesssim z \lesssim 10$. The yellow shaded area represents Eddingon limited growth from seeds between $10^{2}$ M$_{\odot}$ (Pop. III stars) and $10^{6}$ M$_{\odot}$ (intermediate mass black holes (IMBHs) produced by direct collapse) formed at $z \approx 20$, assuming $f_\mathrm{\text{wind}} = 0.5$. The new observations fall outside the permitted area; the Eddington limited growth is too slow to account for the newest observations, but~was still able to explain the highest \mbh\ up to $z \approx 5$. A modest super-Eddington regime ($\dot{m} \sim 10^{2}$) is able to explain the highest masses at the highest redshift, since the observed mass outflow rate becomes negligible compared to the accretion rate. This toy scheme is not considering feedback effects which may slow the growth, and an SE phase of $\sim$$10^{7}$ yr is still necessary to explain a \mbh\ $\sim 5 \cdot 10^{7}$M$_{\odot}$ at $z \approx 10$, when the Universe was just half a billion years old, assuming direct collapse with $M_{0} \sim 10^{5} - 10^{6}$M$_{\odot}$. 

{Systematic searches for IMBHs in the local Universe remain elusive. ~\cite{mapelli20,greeneetal20,davisetal24}. These objects, with~masses ranging from approximately 100 to 100,000 M$\odot$, are still mostly candidates as definitive dynamical confirmation remains challenging. IMBHs are thought to form through processes such as the merger of smaller black holes or the direct collapse of massive stars in low-metallicity environments~\cite{ebisuzakietal01,mapelli16}. Another possible pathway, particularly relevant at high redshifts, involves runaway collisions of massive stars in dense stellar clusters, which could rapidly produce an IMBH of $10^4 - 10^{5}$ M$_{\odot}$\ in just a few million years~\cite{portegiesetal02,portegiesetal04}. At~low redshift, a~handful of super-Eddington IMBH candidates have been identified, with~some exhibiting X-ray brightness or luminosities approaching the Eddington limit~\cite{reinesetal13,chilingarianetal18}. However, they appear to be a tiny minority within the \citet{chilingarianetal18} sample, and their true nature remains uncertain. It is still unclear whether these low-$z$ candidates serve as valid analogs of the IMBHs expected at the highest redshifts, where a much higher fraction of rapidly accreting black holes is anticipated. } 

\unskip 

\section{Analogies Between Stellar Mass and Supermassive Black~Holes}
\label{stars}
\unskip

 \subsection{Super-Eddington AGN and Relativistic~Jets}

The possibility of developing a relativistic jet { probably} rests on the magnetic field strength: strong magnetic fields lead to a magnetically arrested disk (MAD). The defining feature of a MAD is the presence of a large-scale, ordered magnetic field that is sufficiently strong to exert significant pressure on the inflowing gas. As the gas moves inward toward the black hole, magnetic flux is advected inward and accumulates near the event horizon. When the magnetic pressure near the black hole becomes comparable to or exceeds the ram pressure of the inflowing gas, the accretion flow is arrested~\cite{narayanetal03}. The strong magnetic field near the black hole can efficiently extract rotational energy from the black hole via the Blandford--Znajek process~\cite{blandfordznajek77}. However, it is not obvious if the thick geometry is preserved~\cite{mckinneyetal15}, nor if the SE regime and a MAD may coexist, or~lead to a rapid spin-down of even a maximally rotating supermassive black hole \citep{ricarteetal23,lowelletal24}. Hydrodynamic simulations of SE accretion disk produce only mildly relativistic jets, $v \approx 0.5c$, close to the disk axis~\cite{sadowskinarayan15,yangetal23,czernyetal23}. 

\textls[-10]{From an observational point of view, do we have evidence of sources that are unambiguously jetted, meeting a strict selection criterion for being radio loud? The criterion requiring a radio-to-optical specific flux ratio in excess of \mbox{$\log R_\mathrm{K} = f_\nu(6 \mathrm{cm})/f_\nu(g) \approx 1.8$\ \cite{zamfiretal08,gancietal19}}} is a restrictive albeit sufficient condition for the identification of true radio-loud AGN, i.e.,~with radio emission due to a powerful relativistic jet~\cite{bertonetal15,padovani17}. There is no doubt that several type 1 RL AGN show significant \feii\ emission. If~we apply the radio-to-optical specific flux ratio $\log f_{\nu,\mathrm{1.4GHz}} / f_{\nu,\mathrm{5100 }\AA} \gtrsim 1.8$~\cite{zamfiretal08,zamfiretal10}, along with the SE criterion \rfe\ $\gtrsim 1$ accepted at face value, SE radio loud NLSy1s~\cite{komossaetal06,foschinietal15} may be a very rare phenomenon, if~present at all: among the sources listed in the newly published catalog \mbox{by \citet{paliyaetal24},} they are just $\approx$0.06\%. At~redshift $z \gtrsim$ 2, a~recent comparative analysis between radio-quiet and jetted sources did not reveal any jetted quasar with \mbox{\rfe\ $\gtrsim 1$ \citep{deconto-machadoetal23,deconto-machadoetal24}}. Nor were they detected in the HEMS survey (\cite{marzianietal09}, and references therein). This might not be surprising as a large part of the accretion power is funnelled into the mechanical power of the jet, which usually exceeds its radiative power by a large factor ($\lesssim 10$ \cite{foschinietal24}). 

\subsection{Soft and Hard States of Stellar Mass Black~Holes}

Galactic black hole candidates frequently experience X-ray nova outbursts, where they transition from a quiescent state to a highly active state ({e.g.,~\cite{lewinetal97,remillard2006}}). During~these outbursts, a~given source can undergo dramatic changes in its accretion rate over the course of a few months. This evolution typically involves the source shifting from low to high accretion rates and then returning to a lower rate as the outburst subsides. The changes in the accretion flow structure and associated emission are well captured in a hardness--intensity diagram, which plots the source X-ray luminosity against its hardness ratio—a measure of the proportion of high-energy (hard) versus low-energy (soft) X-rays \mbox{\cite{fenderetal04,homanbelloni05,belloni10}}. In this diagram, high hardness ratios indicate a state where the system is dominated by a hot, extended inner flow or corona and is typically associated with the presence of strong, steady jets {\cite{koljonenetal18}.} This configuration corresponds to the low/hard state of the black hole binary. Conversely, low hardness ratios signify that the cold, geometrically thin accretion disk extends closer to the black hole’s innermost stable circular orbit, leading to a dominance of soft X-ray emission and a reduction or complete suppression of the jet, known as the high/soft state {\cite{doneetal07}}. Ultra-luminous X-ray sources (ULXs) are best cantidates for SE accretors, as  they emit X-rays at luminosities exceeding \(10^{39}\) erg/s, significantly higher than the Eddington limit for stellar-mass black holes (e.g.,~\cite{gladstoneetal09,middletonetal13}). Following the discovery of pulsating ULXs that must contain neutron stars~\cite{bachettietal14}, it is now thought that accreting binaries exceeding super-Eddington constitute the majority of the ULX population~\cite{kaaretetal17}. A fraction of the ULX (the most luminous ones) might be still more easily explained by intermediate-mass black holes~\cite{mckenzieetal23}, especially if there is a physical limit to the accretion luminosity per unit black hole mass. 

It is instructive to focus the attention on a prototypical case, GX 339-4~\cite{makishimaetal86}. GX 339-4 is a well-known black hole X-ray binary that exhibits a wide range of variability across different wavelengths, most notably in the X-ray. GX 339-4 is known for its recurrent outbursts, thought to be caused by increases in the mass accretion rate, possibly due to changes in the accretion disk structure~\cite{dunnetal10}. GX 339-4 undergoes transitions between several distinct spectral states, primarily defined in X-rays: in the low-hard state, GX 339-4 is characterized by a hard X-ray spectrum, dominated by a power-law component with a high-energy cutoff. The emission is likely produced by a hot, optically thin corona. In this state, the system shows strong radio emission from a compact jet~\cite{corbeletal00}. The source transitions to a soft state when the accretion rate increases. It shows a dominant thermal component from the accretion disk, with~the X-ray spectrum softening. The radio jet is usually quenched or much weaker in this state~\cite{homanetal05,fenderetal09}. During~transitions between hard and soft state, GX 339-4 often exhibits quasi-periodic oscillations and rapid variability, suggesting changes in the inner disk configuration and the corona~\cite{bellonietal05,mottaetal11}, in general reproducible by general-relativistic magneto-hydrodynamic simulations~\cite{dihingiaetal23}. The source often follows a $q$-shaped hysteresis track in the hardness-intensity diagram, transitioning from the low to high state and back~\cite{fenderetal04,homanbelloni05}. The decline phase does not retrace the same path, as required for a true hysteresis cycle. In the low/hard state, optical and IR emissions are often dominated by synchrotron radiation from the compact jet~\cite{corbelfender02}. GX 339–4 produces steady, compact radio jets, showing a correlation between the radio and X-ray flux, which follow a power-law relation~\cite{corbeletal03,galloetal03}. During~the transition to the high/soft state, the radio jet is often suppressed, and the radio flux drops significantly.


\subsection{Hysteresis Cycles in Binary Stars and AGN}

In the context of binary stars, hysteresis cycles occur when the path followed by the system during an increase in accretion rate differs from the path followed during a decrease. This results in different spectral and luminosity states for the same accretion rate, depending on whether the rate is increasing or decreasing. In other words, the transition from one state to another (e.g., from~low/hard to high/soft) does not follow the same path as the reverse transition (e.g., from~high/soft back to low/hard). This means that the system exhibits a ``loop'' or ``cycle'' in the diagram plotting one state variable against another (such as luminosity vs. spectral hardness), rather than a single 1D path \citep{esin1997,maccarone2003,remillard2006,done2007}.

This cyclical evolution of accretion states and jet formation is characteristic of galactic black hole binaries. In contrast, similar evolutionary patterns in AGN would unfold over much longer timescales (in the order of thousands to millions of years) if the instability associated with a change in viscosity parameter has to propagate over the full disk~\cite{siemiginowskaetal96}. Nevertheless, there are several intriguing analogies between SE accretion in stellar-mass black holes and supermassive black holes: (a) a soft state at high accretion rate contrasts with a hard state at low accretion rate; (b) excess radiation pressure in super-Eddington regimes can drive powerful outflows or winds, expelling material from the accretion disk---massive outflows and winds from the disks of SMBH can be observed as blueshifted emission lines, broad absorption lines, or UFOs; (c) the accretion disk around stellar-mass black holes can become geometrically thick and advective, trapping radiation and allowing continued accretion despite high radiation pressures;
(d) high variability in X-ray and optical emissions can indicate SE accretion rates---quasars exhibit high variability in their X-ray emission and super soft spectrum, albeit the optical variability of xA sources is apparently of low amplitude~\cite{duetal16}; (e) SE super-massive black holes that are relativistically jetted are very rare, if~they exist at all. 

\begin{figure}[H]
\includegraphics[width=10 cm]{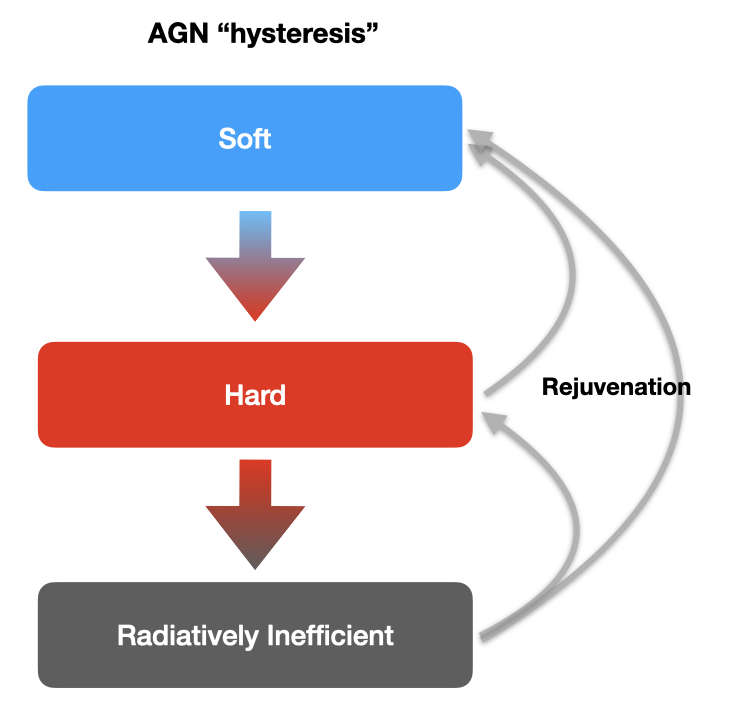}
\caption{{A straight} 
 evolutionary path from an initial soft state of high \lledd and low \mbh may be followed by a ``hard'' (disk dominated) state with higher \mbh and lower \lledd, down to inefficient accretion if $\dot{m} \ll\ $1. The conditions might be reversed in a process that is conceptually similar to the hysteresis cycle in binary stars if the nuclear activity is restarted or enhanced following an increase of accretion rate. The most extreme forms of ``rejuvenation'' may give rise to a changing look~phenomenology. 
\label{fig:hyster}}
\end{figure}

The analogies between super-Eddington accretion in stellar-mass and supermassive black holes illustrate similar underlying physical processes and phenomena despite the vast differences in scale. The 4DE1 scheme captured the main changes between a soft and hard state in type 1 AGN, as well as the following: 
(a) \rfe\ is a tracer of softness, and reaches the highest values in correspondence of SE candidates~\cite{marzianisulentic14,duetal16a}; (b) the soft X-ray photon index $\Gamma_\mathrm{soft}$ plays a role equivalent to the hardness ratio in X-ray binaries: a flat $\Gamma_\mathrm{soft} \lesssim 2$\ indicated a flattish SED as observed in RL and Population B sources, while a steeper slope $\Gamma_\mathrm{soft} \gtrsim 2$ is common among Population A sources, with~the most extreme values $\Gamma_\mathrm{soft} \approx 3-4$\ in the cases of highest \lledd. The \civ\ blueshifts represents the prominence of the accretion disk outflow, which is stronger in the high/soft state~\cite{sulenticetal07}. The 4DE1 optical relations have been successfully scaled down to stellar-mass sources, predicting the location of two accreting white { dwarfs} in the \rfe\---FWHM \hb\ plane~\cite{zamanovmarziani02}. This said, the details of the transition between a soft and hard state are as yet very poorly known. It would imply the transition from a Population B sources to a Population A or xA, if~we consider SE candidates, giving rise to some changing look phenomenology (Figure~\ref{fig:hyster} and the next section). Apparently, the most extensively monitored sources, such as NGC 5548, remain confined into the B spectral types~\cite{bonetal18}. In the survey of changing look candidates along the quasar MS carried out by \citet{pandasniegowska24}, Population B sources stay Population B (apart from a few, unconvincing cases). Some transient events, however, most likely associated with TDEs~\cite{komossagreiner99}, have yielded a surge in luminosity leading to SE properties~\cite{karaetal16,zhangetal24}. 

\subsection{Transient Super-Eddington AGN}

Changing-look AGN are a class of active galactic nuclei (AGN) that exhibit dramatic, rapid changes in their spectral features, often switching between states of high and low activity, and vice versa~\cite{komossaetal24,pandasniegowska24}. These transitions can occur over timescales as short as a few weeks to a few months, which are much shorter than the typical dynamical timescales expected for AGN~\cite{sniegowskaetal20}. Transient phenomena such as tidal disruption events (TDE) might induce a SE state \citep{karaetal16a,daietal18,petrushevskaetal23,zhangetal24}. 
 In passing, we note that turn-on transients offer a straightforward test of the main parameter associated with the E1 MS: a transition from high \feii\ to lower \feii\ with increasing luminosity might not be possible, if~\feii\ is governed by \lledd\ \citep{petrushevskaetal23}. 

{ A most remarkable case of a changing look AGN is provided by \mbox{NGC 1566~\cite{ilicetal20,oknyanskyetal20,oknyanskyetal21,oknyansky22}:} with increasing luminosity, \feii\ emission, which was undetectable in the lower luminosity state, appears in the spectrum as if out of nowhere. However, the NGC 1566 high state \rfe\ is well below 1. The source may qualify as a repeated transition between type-2, type-1 Pop. B, and Pop. A~\cite{xuetal24}, but~super-Eddington luminosities are not involved: in the scheme of Table~\ref{tab:modes}, the change of NGC 1566 would correspond to a change between the radiatively inefficient and the Pop. A state. Over~time, cases of transition to a stable SE state might be detected.} The available data are still insufficient to characterize any hysteresis cycle in AGN definitively. However, AGN are observed to vary more frequently and with larger amplitudes than previously thought~\cite{ulrichetal97}, and suitable data to detect a transition to a super-Eddington regime (or its turn-off) may become available in due time via the extensive photometric monitoring surveys that have been carried out (from Catalina~\cite{drakeetal09} to the ZTF~\cite{grahametal19,mascietal19,bellmetal19}, to~the PanSTARRs~\cite{chambersetal16}, and All Sky Automated Survey for SuperNovae, ASAS-SN \cite{kochaneketal17}) and, from~2025, surveys that will be conducted by~the Vera Rubin telescope~\cite{izevicetal19}. 




\section{Conclusions}
\label{conclusion}


 The main sequence correlations have allowed for the identification of super-Eddington candidates through well-defined optical and UV criteria. These sources, powering quasars and less luminous AGN, are observed over a broad range of black hole masses from the present-day Universe to the earliest cosmic epochs. The SE regime appears to be closely associated with the evolution of a nuclear star forming system, intimately connected to the development of structures in the AGN, in particular the obscuring torus, and the self-gravitation regions of the accretion disk where in situ star formation is possible. This scenario has two main supporting elements: (1) the explanation of the enrichment of the broad line emitting gas over short timescales and (2) the increase of the black hole mass as fast as needed to preserve the current $\Lambda$-cold dark matter (CDM) model.
 



Several open questions remain regarding super-Eddington accretion, concerning the ultimate limits of mass accretion rates, and essentially addressing the following question: how fast can black holes really grow? Does a highly SE regime really exist, and { which form may it take~\cite{pognanetal20}}? We believe the observational evidence supports a positive answer to the second question, even if it remains largely circumstantial. The powerful winds generated in these systems can significantly influence not only the growth of the black hole but also the evolution of the surrounding galaxy. 
Accurate estimates of wind parameters and feedback effects over the full range of AGN black hole masses are essential for providing a \mbox{conclusive answer.} 

The observed high metallicities suggest efficient metal enrichment processes that need to be understood in the context of nuclear or circum-nuclear star formation. A huge effort in theoretical modeling of accretion-modified stars, nuclear stellar systems, and advection-dominated accretion flows, as well as in the collection of dedicated observational data, are required to elucidate these processes. The connection between low ionization spectra and the super-Eddington phenomenology is another area that requires further exploration, as the virialized low emitting regions might be providing a viable avenue for the use of super-Eddington quasars as cosmological standard candles. 
\label{open}




\vspace{6pt}

\authorcontributions{Writing---original draft preparation, P.M.; writing---review and editing, P.M., A.F., and K.G.L. 
	 A.d.O., A.D.-M., C.A.N., T.B.-R., and D.D. contributed to the several papers on which this review is based in part. A.d.O., A.d.M., and T.B.-R. contributed with a critical reading and comments. C.A.N. also contributed by signaling helpful references. All authors have read and agreed to the published version of the manuscript.}

\funding{A.F. 
 was funded by the European Union ERC-2022-STG - BOOTES -101076343. The views and opinions expressed are, however, those of the author(s) only and do not necessarily reflect those of the European Union or the European Research Council Executive Agency. Neither the European Union nor the granting authority can be held responsible for them. A.d.O. and A.D.M. acknowledge financial support from the Spanish MCIU through project PID2022-140871NB-C21 by “ERDF A way of mak-
ing Europe”, and the Severo Ochoa grant CEX2021- 515001131-S funded by MCIN/AEI/10.13039/501100011033.
}



\dataavailability{This review is based entirely on published data or on data that are expected to be published soon. } 


\conflictsofinterest{{} 
} 



\abbreviations{Abbreviations}{
The following abbreviations are used in this manuscript:\\

\noindent 
\begin{tabular}{@{}ll}
4DE1 & 4D Eigenvector 1 \\
ADAF & Advection Dominated Accretion Flow \\
AGN & Active Galactic Nucleus/i \\
BH & Black Hole \\
BAL & Broad Absorption Line\\ 
BLR & Broad Line Region \\
E1 & Eigenvector 1\\
FIR & Far Infrared \\
FWHM & Full Width Half Maximum \\
HE & Hamburg-ESO\\
HEMS & Hamburg-ESO Main Sequence \\
HIL & High Ionization Line\\
ISM & Inter-Stellar Medium \\
JWST & James Webb Space Telescope \\
LIL & Low Ionization Line\\
LISA & Laser Interferometer Space Antenna \\
MDPI & Multidisciplinary Digital Publishing Institute\\
MS & Main Sequence\\
NLR & Narrow Line Region \\
NLSy1 & Narrow Line Seyfert 1 \\
PIZ & Partially Ionized Zone \\
ROSAT & R\"ontgensatellit \\
SDSS & Sloan Digital Sky Survey\\
SE & Super-Eddington \\
SMBH & Super-Massive Black Hole \\
SNR & SuperNova Remnant\\
UFO & Ultra Fast Outflow\\
UV & Ultra-Violet \\
ULX& Ultra-Luminous X-ray\\
VLT & Very Large Telescope\\
WISE & Wide-field Infrared Survey Explorer \\
\end{tabular}
}



\begin{adjustwidth}{-\extralength}{0cm}
\printendnotes[custom] 

\reftitle{References}

\PublishersNote{}
\end{adjustwidth}

\begin{thebibliography}{999}

\bibitem[{Peterson}(1997)]{peterson97}
{Peterson}, B.M.
\newblock {\em {An Introduction to Active Galactic Nuclei}}; Cambridge University Press: {Cambridge, UK}, 1997.

\bibitem[{Elvis}(2000)]{elvis00}
{Elvis}, M.
\newblock {A Structure for Quasars}.
\newblock {\em \apj} {\bf 2000}, {\em 545},~63--76.
\newblock {\url{https://doi.org/10.1086/317778}}.

\bibitem[{Frank} {et~al.}(2002){Frank}, {King}, and {Raine}]{franketal02}
{Frank}, J.; {King}, A.; {Raine}, D.J.
\newblock {\em Accretion Power in Astrophysics, {Third} 
 Edition}; Cambridge University Press: Cambridge, UK, 2002.

\bibitem[{D'Onofrio} {et~al.}(2012){D'Onofrio}, {Marziani}, and {
 Sulentic}]{donofrioetal12}
{D'Onofrio}, M.; {Marziani}, P.; { Sulentic}, J.W. (Eds.)
\newblock {\em Fifty Years of Quasars From Early Observations and Ideas to
 Future Research}; Astrophysics and Space Science Library; Springer: Berlin/Heidelberg, Germany, 2012; Volume 386.

\bibitem[{Netzer}(2013)]{netzer13}
{Netzer}, H.
\newblock {\emph{The Physics and Evolution of Active Galactic Nuclei}};
 Cambridge University Press: Cambridge, UK, 2013.

\bibitem[{Event Horizon Telescope Collaboration} {et~al.}(2019){Event
 Horizon Telescope Collaboration}, {Akiyama}, {Alberdi}, {Alef}, {Asada},
 {Azulay}, {Baczko}, {Ball}, {Balokovi{\'c}}, {Barrett}, {Bintley},
 {Blackburn}, {Boland}, {Bouman}, {Bower}, {Bremer}, {Brinkerink},
 {Brissenden}, {Britzen}, {Broderick}, {Broguiere}, {Bronzwaer}, {Byun},
 {Carlstrom}, {Chael}, {Chan}, {Chatterjee}, {Chatterjee}, {Chen}, {Chen},
 {Cho}, {Christian}, {Conway}, {Cordes}, {Crew}, {Cui}, {Davelaar}, {De
 Laurentis}, {Deane}, {Dempsey}, {Desvignes}, {Dexter}, {Doeleman}, {Eatough},
 {Falcke}, {Fish}, {Fomalont}, {Fraga-Encinas}, {Freeman}, {Friberg}, {Fromm},
 {G{\'o}mez}, {Galison}, {Gammie}, {Garc{\'\i}a}, {Gentaz}, {Georgiev},
 {Goddi}, {Gold}, {Gu}, {Gurwell}, {Hada}, {Hecht}, {Hesper}, {Ho}, {Ho},
 {Honma}, {Huang}, {Huang}, {Hughes}, {Ikeda}, {Inoue}, {Issaoun}, {James},
 {Jannuzi}, {Janssen}, {Jeter}, {Jiang}, {Johnson}, {Jorstad}, {Jung},
 {Karami}, {Karuppusamy}, {Kawashima}, {Keating}, {Kettenis}, {Kim}, {Kim},
 {Kim}, {Kino}, {Koay}, {Koch}, {Koyama}, {Kramer}, {Kramer}, {Krichbaum},
 {Kuo}, {Lauer}, {Lee}, {Li}, {Li}, {Lindqvist}, {Liu}, {Liuzzo}, {Lo},
 {Lobanov}, {Loinard}, {Lonsdale}, {Lu}, {MacDonald}, {Mao}, {Markoff},
 {Marrone}, {Marscher}, {Mart{\'\i}-Vidal}, {Matsushita}, {Matthews},
 {Medeiros}, {Menten}, {Mizuno}, {Mizuno}, {Moran}, {Moriyama},
 {Moscibrodzka}, {M{\"u}ller}, {Nagai}, {Nagar}, {Nakamura}, {Narayan},
 {Narayanan}, {Natarajan}, {Neri}, {Ni}, {Noutsos}, {Okino}, {Olivares},
 {Oyama}, {{\"O}zel}, {Palumbo}, {Patel}, {Pen}, {Pesce}, {Pi{\'e}tu},
 {Plambeck}, {PopStefanija}, {Porth}, {Prather}, {Preciado-L{\'o}pez},
 {Psaltis}, {Pu}, {Ramakrishnan}, {Rao}, {Rawlings}, {Raymond}, {Rezzolla},
 {Ripperda}, {Roelofs}, {Rogers}, {Ros}, {Rose}, {Roshanineshat}, {Rottmann},
 {Roy}, {Ruszczyk}, {Ryan}, {Rygl}, {S{\'a}nchez}, {S{\'a}nchez-Arguelles},
 {Sasada}, {Savolainen}, {Schloerb}, {Schuster}, {Shao}, {Shen}, {Small},
 {Sohn}, {SooHoo}, {Tazaki}, {Tiede}, {Tilanus}, {Titus}, {Toma}, {Torne},
 {Trent}, {Trippe}, {Tsuda}, {van Bemmel}, {van Langevelde}, {van Rossum},
 {Wagner}, {Wardle}, {Weintroub}, {Wex}, {Wharton}, {Wielgus}, {Wong}, {Wu},
 {Young}, {Young}, {Younsi}, {Yuan}, {Yuan}, {Zensus}, {Zhao}, {Zhao}, {Zhu},
 {Farah}, {Meyer-Zhao}, {Michalik}, {Nadolski}, {Nishioka}, {Pradel},
 {Primiani}, {Souccar}, {Vertatschitsch}, and {Yamaguchi}]{eht19}
{Event Horizon Telescope Collaboration}; {Akiyama}, K.; {Alberdi}, A.; {Alef},
 W.; {Asada}, K.; {Azulay}, R.; {Baczko}, A.K.; {Ball}, D.; {Balokovi{\'c}},
 M.; {Barrett}, J.; et~al.
\newblock {First M87 Event Horizon Telescope Results. IV. Imaging the Central
 Supermassive Black Hole}.
\newblock {\em \apjl} {\bf 2019}, {\em 875},~L4.
\newblock {\url{https://doi.org/10.3847/2041-8213/ab0e85}}.

\bibitem[{Urry} and {Padovani}(1995)]{urrypadovani95}
{Urry}, C.M.; {Padovani}, P.
\newblock {Unified Schemes for Radio-Loud Active Galactic Nuclei}.
\newblock {\em PASP} {\bf 1995}, {\em 107},~803.
\newblock {\url{https://doi.org/10.1086/133630}}.

\bibitem[{Dultzin-Hacyan} {et~al.}(1999){Dultzin-Hacyan}, {Krongold},
 {Fuentes-Guridi}, and {Marziani}]{dultzin-hacyanetal99}
{Dultzin-Hacyan}, D.; {Krongold}, Y.; {Fuentes-Guridi}, I.; {Marziani}, P.
\newblock {The Close Environment of Seyfert Galaxies and Its Implication for
 Unification Models}.
\newblock {\em ApJ} {\bf 1999}, {\em 513},~L111--L114.
\newblock {\url{https://doi.org/10.1086/311925}}.

\bibitem[{Bianchi} {et~al.}(2008){Bianchi}, {Corral}, {Panessa}, {Barcons},
 {Matt}, {Bassani}, {Carrera}, and {Jim{\'e}nez-Bail{\'o}n}]{bianchietal08}
{Bianchi}, S.; {Corral}, A.; {Panessa}, F.; {Barcons}, X.; {Matt}, G.;
 {Bassani}, L.; {Carrera}, F.J.; {Jim{\'e}nez-Bail{\'o}n}, E.
\newblock {NGC 3147: A `true' type 2 Seyfert galaxy without the broad-line
 region}.
\newblock {\em \mnras} {\bf 2008}, {\em 385},~195--199.
\newblock {\url{https://doi.org/10.1111/j.1365-2966.2007.12625.x}}.

\bibitem[{Villarroel} and {Korn}(2014)]{villarroelkorn14}
{Villarroel}, B.; {Korn}, A.J.
\newblock {The different neighbours around Type-1 and Type-2 active galactic
 nuclei}.
\newblock {\em Nat. Phys.} {\bf 2014}, {\em 10},~417--420.
\newblock {\url{https://doi.org/10.1038/nphys2951}}.

\bibitem[{Shangguan} and {Ho}(2019)]{shangguanho19}
{Shangguan}, J.; {Ho}, L.C.
\newblock {Testing the Evolutionary Link between Type 1 and Type 2 Quasars with
 Measurements of the Interstellar Medium}.
\newblock {\em \apj} {\bf 2019}, {\em 873},~90.
\newblock {\url{https://doi.org/10.3847/1538-4357/ab0555}}.

\bibitem[{Zou} {et~al.}(2019){Zou}, {Yang}, {Brandt}, and {Xue}]{zouetal19}
{Zou}, F.; {Yang}, G.; {Brandt}, W.N.; {Xue}, Y.
\newblock {The Host-galaxy Properties of Type 1 versus Type 2 Active Galactic
 Nuclei}.
\newblock {\em \apj} {\bf 2019}, {\em 878},~11.
\newblock {\url{https://doi.org/10.3847/1538-4357/ab1eb1}}.

\bibitem[{L{\'o}pez-Navas} {et~al.}(2023){L{\'o}pez-Navas}, {Ar{\'e}valo},
 {Bernal}, {Graham}, {Hern{\'a}ndez-Garc{\'\i}a}, {Lira}, and
 {S{\'a}nchez-S{\'a}ez}]{lopez-navasetal23}
{L{\'o}pez-Navas}, E.; {Ar{\'e}valo}, P.; {Bernal}, S.; {Graham}, M.J.;
 {Hern{\'a}ndez-Garc{\'\i}a}, L.; {Lira}, P.; {S{\'a}nchez-S{\'a}ez}, P.
\newblock {The Type 1 and Type 2 AGN dichotomy according to their ZTF optical
 variability}.
\newblock {\em \mnras} {\bf 2023}, {\em 518},~1531--1542.
\newblock {\url{https://doi.org/10.1093/mnras/stac3174}}.

\bibitem[Storey-Fisher {et~al.}(2024)Storey-Fisher, Hogg, Rix, Eilers,
 Fabbian, Blanton, and Alonso]{storey-fisheretal24}
Storey-Fisher, K.; Hogg, D.W.; Rix, H.W.; Eilers, A.C.; Fabbian, G.; Blanton,
 M.R.; Alonso, D.
\newblock Quaia, the Gaia-unWISE Quasar Catalog: An All-sky Spectroscopic
 Quasar Sample.
\newblock {\em Astrophys. J.} {\bf 2024}, {\em 964},~69.
\newblock {\url{https://doi.org/10.3847/1538-4357/ad1328}}.

\bibitem[Margon(1984)]{margon1984}
Margon, B.
\newblock The properties of SS 433.
\newblock {\em Annu. Rev. Astron. Astrophys.} {\bf 1984}, {\em
 22},~507--536.
\newblock {\url{https://doi.org/10.1146/annurev.aa.22.090184.002451}}.

\bibitem[Greiner {et~al.}(2001)Greiner, Cuby, and McCaughrean]{greiner2001}
Greiner, J.; Cuby, J.G.; McCaughrean, M.J.
\newblock The microquasar GRS 1915+105: Multi-wavelength observations and
 modeling.
\newblock {\em Nature} {\bf 2001}, {\em 414},~522--525.
\newblock {\url{https://doi.org/10.1038/35107019}}.

\bibitem[{Teerikorpi}(2005)]{teerikorpi05}
{Teerikorpi}, P.
\newblock {The distance scale and Eddington efficiency of luminous quasars}.
\newblock {\em arXiv} {\bf 2005}, arXiv:astro-ph/0510382.

\bibitem[{Paczy{\'n}sky} and {Wiita}(1980)]{paczynskywiita80}
{Paczy{\'n}sky}, B.; {Wiita}, P.J.
\newblock {Thick Accretion Disks and Supercritical Luminosities}.
\newblock {\em \aap} {\bf 1980}, {\em 88},~23.

\bibitem[{Paczynski} and {Abramowicz}(1982)]{paczynskiabramowicz82}
{Paczynski}, B.; {Abramowicz}, M.A.
\newblock {A model of a thick disk with equatorial accretion}.
\newblock {\em \apj} {\bf 1982}, {\em 253},~897--907.
\newblock {\url{https://doi.org/10.1086/159689}}.

\bibitem[{Balbus} and {Hawley}(1991)]{bh91}
{Balbus}, S.A.; {Hawley}, J.F.
\newblock {A powerful local shear instability in weakly magnetized disks. I -
 Linear analysis. II---Nonlinear evolution}.
\newblock {\em \apj} {\bf 1991}, {\em 376},~214--233.
\newblock {\url{https://doi.org/10.1086/170270}}.

\bibitem[{Balbus} and {Hawley}(1998)]{bh98}
{Balbus}, S.A.; {Hawley}, J.F.
\newblock {Instability, turbulence, and enhanced transport in accretion disks}.
\newblock {\em Rev. Mod. Phys.} {\bf 1998}, {\em 70},~1--53.
\newblock {\url{https://doi.org/10.1103/RevModPhys.70.1}}.

\bibitem[{Shakura} and {Sunyaev}(1973)]{shakurasunyaev73}
{Shakura}, N.I.; {Sunyaev}, R.A.
\newblock {Black holes in binary systems. Observational appearance.}
\newblock {\em \aap} {\bf 1973}, {\em 24},~337--355.

\bibitem[{Di Matteo} {et~al.}(2005){Di Matteo}, {Springel}, and
 {Hernquist}]{dimatteoetal05}
{Di Matteo}, T.; {Springel}, V.; {Hernquist}, L.
\newblock {Energy input from quasars regulates the growth and activity of black
 holes and their host galaxies}.
\newblock {\em \nat} {\bf 2005}, {\em 433},~604--607.
\newblock {\url{https://doi.org/10.1038/nature03335}}.

\bibitem[{King}(2005)]{king05}
{King}, A.
\newblock {The AGN-Starburst Connection, Galactic Superwinds, and
 M$_{BH}$-{\ensuremath{\sigma}}}.
\newblock {\em \apjl} {\bf 2005}, {\em 635},~L121--L123.
\newblock {\url{https://doi.org/10.1086/499430}}.

\bibitem[{King} and {Muldrew}(2016)]{kingmuldrew16}
{King}, A.; {Muldrew}, S.I.
\newblock {Black hole winds II: Hyper-Eddington winds and feedback}.
\newblock {\em \mnras} {\bf 2016}, {\em 455},~1211--1217.
\newblock {\url{https://doi.org/10.1093/mnras/stv2347}}.

\bibitem[{Marziani} and {Sulentic}(2014)]{marzianisulentic14}
{Marziani}, P.; {Sulentic}, J.W.
\newblock {Highly accreting quasars: Sample definition and possible
 cosmological implications}.
\newblock {\em \mnras} {\bf 2014}, {\em 442},~1211--1229.
\newblock {\url{https://doi.org/10.1093/mnras/stu951}}.

\bibitem[{Dultzin} {et~al.}(2020){Dultzin}, {Marziani}, {de Diego},
 {Negrete}, {Del Olmo}, {Mart{\'\i}nez-Aldama}, {D'Onofrio}, {Bon}, {Bon}, and
 {Stirpe}]{dultzinetal20}
{Dultzin}, D.; {Marziani}, P.; {de Diego}, J.A.; {Negrete}, C.A.; {Del Olmo},
 A.; {Mart{\'\i}nez-Aldama}, M.L.; {D'Onofrio}, M.; {Bon}, E.; {Bon}, N.;
 {Stirpe}, G.M.
\newblock {Extreme quasars as distance indicators in cosmology}.
\newblock {\em Front. Astron. Space Sci.} {\bf 2020}, {\em
 6},~80. 
\newblock {\url{https://doi.org/10.3389/fspas.2019.00080}}.

\bibitem[{Czerny} {et~al.}(2021){Czerny}, {Mart{\'\i}nez-Aldama},
 {Wojtkowska}, {Zaja{\v{c}}ek}, {Marziani}, {Dultzin}, {Naddaf}, {Panda},
 {Prince}, {Przyluski}, {Ralowski}, and {{\'S}niegowska}]{czernyetal21}
{Czerny}, B.; {Mart{\'\i}nez-Aldama}, M.L.; {Wojtkowska}, G.; {Zaja{\v{c}}ek},
 M.; {Marziani}, P.; {Dultzin}, D.; {Naddaf}, M.H.; {Panda}, S.; {Prince}, R.;
 {Przyluski}, R.; et~al.
\newblock {Dark Energy Constraintsfrom Quasar Observations}.
\newblock {\em Acta Phys. Pol. A} {\bf 2021}, {\em 139},~389--393.
\newblock {\url{https://doi.org/10.12693/APhysPolA.139.389}}.

\bibitem[{Marziani} {et~al.}(2021){Marziani}, {Dultzin}, {del Olmo},
 {D'Onofrio}, {de Diego}, {Stirpe}, {Bon}, {Bon}, {Czerny}, {Perea}, {Panda},
 {Loli Martinez-Aldama}, and {Negrete}]{marzianietal21}
{Marziani}, P.; {Dultzin}, D.; {del Olmo}, A.; {D'Onofrio}, M.; {de Diego},
 J.A.; {Stirpe}, G.M.; {Bon}, E.; {Bon}, N.; {Czerny}, B.; {Perea}, J.;
 et~al.
\newblock {The quasar main sequence and its potential for cosmology}.
\newblock In \emph{{IAU Symposium S356: Nuclear Activity in Galaxies Across Cosmic Time, 
Proceedings of the International Astronomical Union, Ababa, Ethiopia, 7--11 October} 2019}; 
 {Povi{\'c}}, M., {Marziani}, P., {Masegosa}, J., {Netzer}, H., {Negu}, S.H., {Tessema}, S.B., Eds.; 
 {Cambridge University Press: Cambridge, UK}, 2021; pp. 66--71.
\newblock {\url{https://doi.org/10.1017/S1743921320002598}}.

\bibitem[{Sulentic} {et~al.}(2000){Sulentic}, {Marziani}, and
 {Dultzin-Hacyan}]{sulenticetal00a}
{Sulentic}, J.W.; {Marziani}, P.; {Dultzin-Hacyan}, D.
\newblock {Phenomenology of Broad Emission Lines in Active Galactic Nuclei}.
\newblock {\em ARA\&A} {\bf 2000}, {\em 38},~521--571.
\newblock {\url{https://doi.org/10.1146/annurev.astro.38.1.521}}.

\bibitem[{Du} {et~al.}(2016){Du}, {Wang}, {Hu}, {Ho}, {Li}, and
 {Bai}]{duetal16a}
{Du}, P.; {Wang}, J.M.; {Hu}, C.; {Ho}, L.C.; {Li}, Y.R.; {Bai}, J.M.
\newblock {The Fundamental Plane of the Broad-line Region in Active Galactic
 Nuclei}.
\newblock {\em \apjl} {\bf 2016}, {\em 818},~L14.
\newblock {\url{https://doi.org/10.3847/2041-8205/818/1/L14}}.

\bibitem[{Paczynski}(1990)]{paczynski90}
{Paczynski}, B.
\newblock {Super-Eddington Winds from Neutron Stars}.
\newblock {\em \apj} {\bf 1990}, {\em 363},~218.
\newblock {\url{https://doi.org/10.1086/169332}}.

\bibitem[{Lupi} {et~al.}(2024){Lupi},{Quadri},{Volonteri},{Colpi},{Regan}]{lupietal24}  Lupi, A; Quadri, G.;  {Volonteri}, M.;  Colpi, M.; {Regan}, J. A.
\newblock {Sustained super-Eddington accretion in high-redshift quasars}.
\newblock {\em \aap} {\bf 2024}, {\em 686},~A256.
\newblock {\url{	https://doi.org/10.1051/0004-6361/202348788}}.




\bibitem[{Yang} and {Yuan}(2024)]{yangyuan24}
{Yang}, H.; {Yuan}, F.
\newblock {Wind from the Hot Accretion Flow and Super-Eddington Accretion
 Flow}.
\newblock {\em arXiv} {\bf 2024}, arXiv:2408.16595.
\newblock {\url{https://doi.org/10.48550/arXiv.2408.16595}}.

\bibitem[{Wang} {et~al.}(2013){Wang}, {Du}, {Valls-Gabaud}, {Hu}, and
 {Netzer}]{wangetal13}
{Wang}, J.M.; {Du}, P.; {Valls-Gabaud}, D.; {Hu}, C.; {Netzer}, H.
\newblock {Super-Eddington Accreting Massive Black Holes as Long-Lived
 Cosmological Standards}.
\newblock {\em Phys. Rev. Lett.} {\bf 2013}, {\em 110},~081301.
\newblock {\url{https://doi.org/10.1103/PhysRevLett.110.081301}}.

\bibitem[{D'Onofrio} {et~al.}(2016){D'Onofrio}, {Rampazzo}, and
 {Zaggia}]{donofrioetal16a}
{D'Onofrio}, M.; {Rampazzo}, R.; {Zaggia}, S. (Eds.)
\newblock {\em {From the Realm of the Nebulae to Populations of Galaxies}}; 
Astrophysics and Space Science Library; {Springer: Cham, Switzerland}, 2016; Volume 435.
\newblock {\url{https://doi.org/10.1007/978-3-319-31006-0}}.

\bibitem[{D'Onofrio} {et~al.}(2017){D'Onofrio}, {Cariddi}, {Chiosi},
 {Chiosi}, and {Marziani}]{donofrioetal17}
{D'Onofrio}, M.; {Cariddi}, S.; {Chiosi}, C.; {Chiosi}, E.; {Marziani}, P.
\newblock {On the Origin of the Fundamental Plane and Faber-Jackson Relations:
 Implications for the Star Formation Problem}.
\newblock {\em \apj} {\bf 2017}, {\em 838},~163.
\newblock {\url{https://doi.org/10.3847/1538-4357/aa6540}}.

\bibitem[{D'Onofrio} and {Marziani}(2018)]{donofriomarziani18}
{D'Onofrio}, M.; {Marziani}, P.
\newblock {A multimessenger view of galaxies and quasars from now to
 mid-century}.
\newblock {\em Front. Astron. Space Sci.} {\bf 2018}, {\em
 5},~31.
\newblock {\url{https://doi.org/10.3389/fspas.2018.00031}}.

\bibitem[{Davis} and {Laor}(2011)]{davislaor11}
{Davis}, S.W.; {Laor}, A.
\newblock {The Radiative Efficiency of Accretion Flows in Individual Active
 Galactic Nuclei}.
\newblock {\em \apj} {\bf 2011}, {\em 728},~98.
\newblock {\url{https://doi.org/10.1088/0004-637X/728/2/98}}.

\bibitem[Panda {et~al.}(2018)Panda, Czerny, Adhikari, Hryniewicz, Wildy,
 Kuraszkiewicz, and {\'{S}}niegowska]{pandaetal18}
Panda, S.; Czerny, B.; Adhikari, T.P.; Hryniewicz, K.; Wildy, C.;
 Kuraszkiewicz, J.; {\'{S}}niegowska, M.
\newblock Modeling of the Quasar Main Sequence in the Optical Plane.
\newblock {\em Astrophys. J.} {\bf 2018}, {\em 866},~115.
\newblock {\url{https://doi.org/10.3847/1538-4357/aae209}}.

\bibitem[{Kubota} and {Done}(2019)]{kubotadone19}
{Kubota}, A.; {Done}, C.
\newblock {Modelling the spectral energy distribution of super-Eddington
 quasars}.
\newblock {\em \mnras} {\bf 2019}, {\em 489},~524--533.
\newblock {\url{https://doi.org/10.1093/mnras/stz2140}}.

\bibitem[{Jiang} {et~al.}(2019){Jiang}, {Stone}, and {Davis}]{jiangetal19}
{Jiang}, Y.F.; {Stone}, J.M.; {Davis}, S.W.
\newblock {Super-Eddington Accretion Disks around Supermassive Black Holes}.
\newblock {\em \apj} {\bf 2019}, {\em 880},~67.
\newblock {\url{https://doi.org/10.3847/1538-4357/ab29ff}}.

\bibitem[{Inayoshi} {et~al.}(2016){Inayoshi}, {Haiman}, and
 {Ostriker}]{inayoshietal16}
{Inayoshi}, K.; {Haiman}, Z.; {Ostriker}, J.P.
\newblock {Hyper-Eddington accretion flows on to massive black holes}.
\newblock {\em \mnras} {\bf 2016}, {\em 459},~3738--3755.
\newblock {\url{https://doi.org/10.1093/mnras/stw836}}.

\bibitem[{Toyouchi} {et~al.}(2021){Toyouchi}, {Inayoshi}, {Hosokawa}, and
 {Kuiper}]{toyouchietal21}
{Toyouchi}, D.; {Inayoshi}, K.; {Hosokawa}, T.; {Kuiper}, R.
\newblock {Super-Eddington Mass Growth of Intermediate-Mass Black Holes
 Embedded in Dusty Circumnuclear Disks}.
\newblock {\em \apj} {\bf 2021}, {\em 907},~74.
\newblock {\url{https://doi.org/10.3847/1538-4357/abcfc2}}.

\bibitem[{Abramowicz} {et~al.}(1988){Abramowicz}, {Czerny}, {Lasota}, and
 {Szuszkiewicz}]{abramowiczetal88}
{Abramowicz}, M.A.; {Czerny}, B.; {Lasota}, J.P.; {Szuszkiewicz}, E.
\newblock {Slim accretion disks}.
\newblock {\em \apj} {\bf 1988}, {\em 332},~646--658.
\newblock {\url{https://doi.org/10.1086/166683}}.

\bibitem[{Sirko} and {Goodman}(2003)]{sirkogoodman03}
{Sirko}, E.; {Goodman}, J.
\newblock {Spectral energy distributions of marginally self-gravitating
 quasi-stellar object discs}.
\newblock {\em \mnras} {\bf 2003}, {\em 341},~501--508.
\newblock {\url{https://doi.org/10.1046/j.1365-8711.2003.06431.x}}.

\bibitem[{Thompson} {et~al.}(2005){Thompson}, {Quataert}, and
 {Murray}]{thompson05}
{Thompson}, T.A.; {Quataert}, E.; {Murray}, N.
\newblock {Radiation Pressure-supported Starburst Disks and Active Galactic
 Nucleus Fueling}.
\newblock {\em \apj} {\bf 2005}, {\em 630},~167--185.
\newblock {\url{https://doi.org/10.1086/431923}}.

\bibitem[{S{\k{a}}dowski}(2009)]{sadowski09}
{S{\k{a}}dowski}, A.
\newblock {Slim Disks Around Kerr Black Holes Revisited}.
\newblock {\em \apjs} {\bf 2009}, {\em 183},~171--178.
\newblock {\url{https://doi.org/10.1088/0067-0049/183/2/171}}.

\bibitem[{S{\k{a}}dowski} {et~al.}(2014){S{\k{a}}dowski}, {Narayan},
 {McKinney}, and {Tchekhovskoy}]{sadowskietal14}
{S{\k{a}}dowski}, A.; {Narayan}, R.; {McKinney}, J.C.; {Tchekhovskoy}, A.
\newblock {Numerical simulations of super-critical black hole accretion flows
 in general relativity}.
\newblock {\em \mnras} {\bf 2014}, {\em 439},~503--520.
\newblock {\url{https://doi.org/10.1093/mnras/stt2479}}.

\bibitem[{Abramowicz}(2005)]{abramowicz05}
{Abramowicz}, M.A.
\newblock {Super-Eddington black hole accretion: Polish doughnuts and slim disks}.
\newblock In \emph{Growing Black Holes: Accretion in a
 Cosmological Context}; {Merloni}, A., {Nayakshin}, S., {Sunyaev}, R.A., Eds.;
 {Springer: Berlin/Heidelberg, Germany}, 2005; pp. 257--273.
\newblock {\url{https://doi.org/10.1007/11403913_49}}.

\bibitem[{Abramowicz} and {Straub}(2014)]{abramowiczstaub14}
{Abramowicz}, M.A.; {Straub}, O.
\newblock {Accretion discs}.
\newblock {\em Scholarpedia} {\bf 2014}, {\em 9},~2408.
\newblock {\url{https://doi.org/10.4249/scholarpedia.2408}}.

\bibitem[{Madau}(1988)]{madau88}
{Madau}, P.
\newblock {Thick Accretion Disks around Black Holes and the UV/Soft X-Ray
 Excess in Quasars}.
\newblock {\em \apj} {\bf 1988}, {\em 327},~116.
\newblock {\url{https://doi.org/10.1086/166175}}.

\bibitem[{Calvani} {et~al.}(1989){Calvani}, {Marziani}, and
 {Padovani}]{calvanietal89}
{Calvani}, M.; {Marziani}, P.; {Padovani}, P.
\newblock {(Thick) accretion discs---Where are they?}
\newblock In Proceedings of the General Relativity and Gravitational Physics,
 1989; pp. 102--123.

\bibitem[{Wang} and {Zhou}(1999)]{wangzhou99}
{Wang}, J.M.; {Zhou}, Y.Y.
\newblock {Self-Similar Solution of Optically Thick Advection-dominated Flows}.
\newblock {\em \apj} {\bf 1999}, {\em 516},~420--424.
\newblock {\url{https://doi.org/10.1086/307080}}.

\bibitem[{Jin} {et~al.}(2017){Jin}, {Done}, and {Ward}]{jinetal17}
{Jin}, C.; {Done}, C.; {Ward}, M.
\newblock {Super-Eddington QSO RX J0439.6-5311 - I. Origin of the soft X-ray
 excess and structure of the inner accretion flow}.
\newblock {\em \mnras} {\bf 2017}, {\em 468},~3663--3681.
\newblock {\url{https://doi.org/10.1093/mnras/stx718}}.

\bibitem[{Wang} {et~al.}(2014){Wang}, {Qiu}, {Du}, and {Ho}]{wangetal14a}
{Wang}, J.M.; {Qiu}, J.; {Du}, P.; {Ho}, L.C.
\newblock {Self-shadowing Effects of Slim Accretion Disks in Active Galactic
 Nuclei: The Diverse Appearance of the Broad-line Region}.
\newblock {\em \apj} {\bf 2014}, {\em 797},~65.
\newblock {\url{https://doi.org/10.1088/0004-637X/797/1/65}}.

\bibitem[{Mineshige} {et~al.}(2000){Mineshige}, {Kawaguchi}, {Takeuchi}, and
 {Hayashida}]{mineshigeetal00}
{Mineshige}, S.; {Kawaguchi}, T.; {Takeuchi}, M.; {Hayashida}, K.
\newblock {Slim-Disk Model for Soft X-Ray Excess and Variability of Narrow-Line
 Seyfert 1 Galaxies}.
\newblock {\em \pasj} {\bf 2000}, {\em 52},~499--508.

\bibitem[{Madau} {et~al.}(2014){Madau}, {Haardt}, and {Dotti}]{madauetal14}
{Madau}, P.; {Haardt}, F.; {Dotti}, M.
\newblock {Super-critical Growth of Massive Black Holes from Stellar-Mass
 Seeds}.
\newblock {\em \apjl} {\bf 2014}, {\em 784},~L38.
\newblock {\url{https://doi.org/10.1088/2041-8205/784/2/L38}}.

\bibitem[{Inayoshi} {et~al.}(2020){Inayoshi}, {Visbal}, and
 {Haiman}]{inayoshietal20}
{Inayoshi}, K.; {Visbal}, E.; {Haiman}, Z.
\newblock {The Assembly of the First Massive Black Holes}.
\newblock {\em \araa} {\bf 2020}, {\em 58},~27--97.
\newblock {\url{https://doi.org/10.1146/annurev-astro-120419-014455}}.

\bibitem[{Ohsuga} {et~al.}(2005){Ohsuga}, {Mori}, {Nakamoto}, and
 {Mineshige}]{ohsugaetal05}
{Ohsuga}, K.; {Mori}, M.; {Nakamoto}, T.; {Mineshige}, S.
\newblock {Supercritical Accretion Flows around Black Holes: Two-dimensional,
 Radiation Pressure-dominated Disks with Photon Trapping}.
\newblock {\em \apj} {\bf 2005}, {\em 628},~368--381.
\newblock {\url{https://doi.org/10.1086/430728}}.

\bibitem[{Ogawa} {et~al.}(2017){Ogawa}, {Mineshige}, {Kawashima}, {Ohsuga},
 and {Hashizume}]{ogawaetal17}
{Ogawa}, T.; {Mineshige}, S.; {Kawashima}, T.; {Ohsuga}, K.; {Hashizume}, K.
\newblock {Radiation hydrodynamic simulations of a super-Eddington accretor as
 a model for ultra-luminous sources}.
\newblock {\em \pasj} {\bf 2017}, {\em 69},~33.
\newblock {\url{https://doi.org/10.1093/pasj/psx006}}.

\bibitem[{Pognan} {et~al.}(2020){Pognan}, {Trakhtenbrot}, {Sbarrato},
 {Schawinski}, and {Bertemes}]{pognanetal20}
{Pognan}, Q.; {Trakhtenbrot}, B.; {Sbarrato}, T.; {Schawinski}, K.; {Bertemes},
 C.
\newblock {Searching for super-Eddington quasars using a photon trapping
 accretion disc model}.
\newblock {\em \mnras} {\bf 2020}, {\em 492},~4058--4079.
\newblock {\url{https://doi.org/10.1093/mnras/staa078}}.

\bibitem[{Marziani} {et~al.}(2001){Marziani}, {Sulentic}, {Zwitter},
 {Dultzin-Hacyan}, and {Calvani}]{marzianietal01}
{Marziani}, P.; {Sulentic}, J.W.; {Zwitter}, T.; {Dultzin-Hacyan}, D.;
 {Calvani}, M.
\newblock {Searching for the Physical Drivers of the Eigenvector 1 Correlation
 Space}.
\newblock {\em ApJ} {\bf 2001}, {\em 558},~553--560.
\newblock {\url{https://doi.org/10.1086/322286}}.

\bibitem[{Shen} and {Ho}(2014)]{shenho14}
{Shen}, Y.; {Ho}, L.C.
\newblock {The diversity of quasars unified by accretion and orientation}.
\newblock {\em \nat} {\bf 2014}, {\em 513},~210--213.
\newblock {\url{https://doi.org/10.1038/nature13712}}.

\bibitem[{Marziani} {et~al.}(2018){Marziani}, {Dultzin}, {Sulentic}, {Del
 Olmo}, {Negrete}, {Mart{\'\i}nez-Aldama}, {D'Onofrio}, {Bon}, {Bon}, and
 {Stirpe}]{marzianietal18}
{Marziani}, P.; {Dultzin}, D.; {Sulentic}, J.W.; {Del Olmo}, A.; {Negrete},
 C.A.; {Mart{\'\i}nez-Aldama}, M.L.; {D'Onofrio}, M.; {Bon}, E.; {Bon}, N.;
 {Stirpe}, G.M.
\newblock {A main sequence for quasars}.
\newblock {\em Front. Astron. Space Sci.} {\bf 2018}, {\em
 5},~6.
\newblock {\url{https://doi.org/10.3389/fspas.2018.00006}}.

\bibitem[{Wildy} {et~al.}(2019){Wildy}, {Czerny}, and {Panda}]{wildyetal19}
{Wildy}, C.; {Czerny}, B.; {Panda}, S.
\newblock {Quasar main sequence: A line or a plane}.
\newblock {\em \aap} {\bf 2019}, {\em 632},~A41.
\newblock {\url{https://doi.org/10.1051/0004-6361/201935620}}.

\bibitem[{Moore}(1945)]{moore45}
{Moore}, C.E.
\newblock {A Multiplet Table of Astrophysical Interest. Revised Edition. Part I---Table of Multiplets}.
\newblock {\em Contrib. Princet. Univ. Obs.} {\bf
 1945}, {\em 20},~1--110.

\bibitem[{Kova{\v c}evi{\'c}} {et~al.}(2010){Kova{\v c}evi{\'c}},
 {Popovi{\'c}}, and {Dimitrijevi{\'c}}]{kovacevicetal10}
{Kova{\v c}evi{\'c}}, J.; {Popovi{\'c}}, L.{\v C}.; {Dimitrijevi{\'c}}, M.S.
\newblock {Analysis of Optical Fe II Emission in a Sample of Active Galactic
 Nucleus Spectra}.
\newblock {\em \apjs} {\bf 2010}, {\em 189},~15--36.
\newblock {\url{https://doi.org/10.1088/0067-0049/189/1/15}}.

\bibitem[{Boroson} and {Green}(1992)]{borosongreen92}
{Boroson}, T.A.; {Green}, R.F.
\newblock {The Emission-Line Properties of Low-Redshift Quasi-stellar Objects}.
\newblock {\em \apjs} {\bf 1992}, {\em 80},~109.
\newblock {\url{https://doi.org/10.1086/191661}}.

\bibitem[{Boroson}(2002)]{boroson02}
{Boroson}, T.A.
\newblock {Black Hole Mass and Eddington Ratio as Drivers for the Observable
 Properties of Radio-loud and Radio-quiet QSOs}.
\newblock {\em \apj} {\bf 2002}, {\em 565},~78--85.
\newblock {\url{https://doi.org/10.1086/324486}}.

\bibitem[{Marziani} {et~al.}(2003){Marziani}, {Zamanov}, {Sulentic}, and
 {Calvani}]{marzianietal03b}
{Marziani}, P.; {Zamanov}, R.K.; {Sulentic}, J.W.; {Calvani}, M.
\newblock {Searching for the physical drivers of eigenvector 1: Influence of
 black hole mass and Eddington ratio}.
\newblock {\em MNRAS} {\bf 2003}, {\em 345},~1133--1144.
\newblock {\url{https://doi.org/10.1046/j.1365-2966.2003.07033.x}}.

\bibitem[{Sun} and {Shen}(2015)]{sunshen15}
{Sun}, J.; {Shen}, Y.
\newblock {Dissecting the Quasar Main Sequence: Insight from Host Galaxy
 Properties}.
\newblock {\em \apjl} {\bf 2015}, {\em 804},~L15.
\newblock {\url{https://doi.org/10.1088/2041-8205/804/1/L15}}.

\bibitem[{Panda} {et~al.}(2019){Panda}, {Marziani}, and
 {Czerny}]{pandaetal19}
{Panda}, S.; {Marziani}, P.; {Czerny}, B.
\newblock {The Quasar Main Sequence Explained by the Combination of Eddington
 Ratio, Metallicity, and Orientation}.
\newblock {\em \apj} {\bf 2019}, {\em 882},~79.
\newblock {\url{https://doi.org/10.3847/1538-4357/ab3292}}.

\bibitem[{Peterson} and {Wandel}(1999)]{petersonwandel99}
{Peterson}, B.M.; {Wandel}, A.
\newblock {Keplerian Motion of Broad-Line Region Gas as Evidence for
 Supermassive Black Holes in Active Galactic Nuclei}.
\newblock {\em \apjl} {\bf 1999}, {\em 521},~L95--L98.
\newblock {\url{https://doi.org/10.1086/312190}}.

\bibitem[{Peterson} {et~al.}(2004){Peterson}, {Ferrarese}, {Gilbert},
 {Kaspi}, {Malkan}, {Maoz}, {Merritt}, {Netzer}, {Onken}, {Pogge},
 {Vestergaard}, and {Wandel}]{petersonetal04}
{Peterson}, B.M.; {Ferrarese}, L.; {Gilbert}, K.M.; {Kaspi}, S.; {Malkan},
 M.A.; {Maoz}, D.; {Merritt}, D.; {Netzer}, H.; {Onken}, C.A.; {Pogge}, R.W.;
 et~al.
\newblock {Central Masses and Broad-Line Region Sizes of Active Galactic
 Nuclei. II. A Homogeneous Analysis of a Large Reverberation-Mapping
 Database}.
\newblock {\em ApJ} {\bf 2004}, {\em 613},~682--699.
\newblock {\url{https://doi.org/10.1086/423269}}.

\bibitem[{Vestergaard} and {Peterson}(2006)]{vestergaardpeterson06}
{Vestergaard}, M.; {Peterson}, B.M.
\newblock {Determining Central Black Hole Masses in Distant Active Galaxies and
 Quasars. II. Improved Optical and UV Scaling Relationships}.
\newblock {\em ApJ} {\bf 2006}, {\em 641},~689--709.
\newblock {\url{https://doi.org/10.1086/500572}}.

\bibitem[{Dalla Bont{\`a}} {et~al.}(2020){Dalla Bont{\`a}}, {Peterson},
 {Bentz}, {Brandt}, {Ciroi}, {De Rosa}, {Fonseca Alvarez}, {Grier}, {Hall},
 {Hern{\'a}ndez Santisteban}, {Ho}, {Homayouni}, {Horne}, {Kochanek}, {Li},
 {Morelli}, {Pizzella}, {Pogge}, {Schneider}, {Shen}, {Trump}, and
 {Vestergaard}]{dallabontaetal20}
{Dalla Bont{\`a}}, E.; {Peterson}, B.M.; {Bentz}, M.C.; {Brandt}, W.N.;
 {Ciroi}, S.; {De Rosa}, G.; {Fonseca Alvarez}, G.; {Grier}, C.J.; {Hall},
 P.B.; {Hern{\'a}ndez Santisteban}, J.V.; et~al.
\newblock {The Sloan Digital Sky Survey Reverberation Mapping Project:
 Estimating Masses of Black Holes in Quasars with Single-epoch Spectroscopy}.
\newblock {\em \apj} {\bf 2020}, {\em 903},~112.
\newblock {\url{https://doi.org/10.3847/1538-4357/abbc1c}}.

\bibitem[{Sulentic} {et~al.}(2002){Sulentic}, {Marziani}, {Zamanov},
 {Bachev}, {Calvani}, and {Dultzin-Hacyan}]{sulenticetal02}
{Sulentic}, J.W.; {Marziani}, P.; {Zamanov}, R.; {Bachev}, R.; {Calvani}, M.;
 {Dultzin-Hacyan}, D.
\newblock {Average Quasar Spectra in the Context of Eigenvector 1}.
\newblock {\em ApJL} {\bf 2002}, {\em 566},~L71--L75.
\newblock {\url{https://doi.org/10.1086/339594}}.

\bibitem[{Dultzin-Hacyan} {et~al.}(1997){Dultzin-Hacyan}, {Sulentic},
 {Marziani}, {Calvani}, and {Moles}]{dultzin-hacyanetal97}
{Dultzin-Hacyan}, D.; {Sulentic}, J.; {Marziani}, P.; {Calvani}, M.; {Moles},
 M.
\newblock {A Correlation Analysis for Emission Lines in 52 AGN}.
\newblock In \emph{Proceedings of the IAU Colloq. 159: Emission Lines in Active Galaxies: New Methods and Technique}; 
{Peterson}, B.M., {Cheng}, F.Z., {Wilson}, A.S., Eds.; 
Astronomical Society of the Pacific Conference Series; 
{Astronomical Society of the Pacific: San Francisco, CA, USA}, 1997; Volume 113, p. 262.

\bibitem[{Komossa} {et~al.}(2006){Komossa}, {Voges}, {Xu}, {Mathur},
 {Adorf}, {Lemson}, {Duschl}, and {Grupe}]{komossaetal06}
{Komossa}, S.; {Voges}, W.; {Xu}, D.; {Mathur}, S.; {Adorf}, H.M.; {Lemson},
 G.; {Duschl}, W.J.; {Grupe}, D.
\newblock {Radio-loud Narrow-Line Type 1 Quasars}.
\newblock {\em \aj} {\bf 2006}, {\em 132},~531--545.
\newblock {\url{https://doi.org/10.1086/505043}}.

\bibitem[{Abdo} {et~al.}(2009){Abdo}, {Ackermann}, {Ajello}, {Baldini},
 {Ballet}, {Barbiellini}, {Bastieri}, {Bechtol}, {Bellazzini}, {Berenji},
 {Bloom}, {Bonamente}, {Borgland}, {Bregeon}, {Brez}, {Brigida}, {Bruel},
 {Burnett}, {Caliandro}, {Cameron}, {Caraveo}, {Casandjian}, {Cecchi},
 {{\v{C}}elik}, {Chekhtman}, {Cheung}, {Chiang}, {Ciprini}, {Claus},
 {Cohen-Tanugi}, {Conrad}, {Cutini}, {Dermer}, {de Palma}, {Silva}, {Drell},
 {Dubois}, {Dumora}, {Farnier}, {Favuzzi}, {Fegan}, {Focke}, {Foschini},
 {Frailis}, {Fukazawa}, {Fusco}, {Gargano}, {Gehrels}, {Germani}, {Giebels},
 {Giglietto}, {Giordano}, {Giroletti}, {Glanzman}, {Godfrey}, {Grenier},
 {Grove}, {Guillemot}, {Guiriec}, {Hayashida}, {Hays}, {Horan}, {Hughes},
 {J{\'o}hannesson}, {Johnson}, {Johnson}, {Kadler}, {Kamae}, {Katagiri},
 {Kataoka}, {Kerr}, {Kn{\"o}dlseder}, {Kuss}, {Lande}, {Latronico}, {Longo},
 {Loparco}, {Lott}, {Lovellette}, {Lubrano}, {Makeev}, {Mazziotta},
 {McConville}, {McEnery}, {Meurer}, {Michelson}, {Mitthumsiri}, {Mizuno},
 {Monte}, {Monzani}, {Morselli}, {Moskalenko}, {Murgia}, {Nolan}, {Norris},
 {Nuss}, {Ohsugi}, {Omodei}, {Orlando}, {Ormes}, {Pelassa}, {Pepe}, {Persic},
 {Pesce-Rollins}, {Piron}, {Porter}, {Rain{\`o}}, {Rando}, {Razzano},
 {Rochester}, {Rodriguez}, {Ryde}, {Sadrozinski}, {Sambruna}, {Sander}, {Saz
 Parkinson}, {Scargle}, {Sgr{\`o}}, {Smith}, {Spandre}, {Spinelli},
 {Strickman}, {Suson}, {Tagliaferri}, {Takahashi}, {Takahashi}, {Tanaka},
 {Thayer}, {Thayer}, {Thompson}, {Tibaldo}, {Tibolla}, {Torres}, {Tosti},
 {Tramacere}, {Uchiyama}, {Usher}, {Vasileiou}, {Vilchez}, {Vitale}, {Waite},
 {Wang}, {Winer}, {Wood}, {Ylinen}, {Ziegler}, {Fermi/LAT Collaboration},
 {Ghisellini}, {Maraschi}, and {Tavecchio}]{abdoetal09}
{Abdo}, A.A.; {Ackermann}, M.; {Ajello}, M.; {Baldini}, L.; {Ballet}, J.;
 {Barbiellini}, G.; {Bastieri}, D.; {Bechtol}, K.; {Bellazzini}, R.;
 {Berenji}, B.; et~al.
\newblock {Radio-Loud Narrow-Line Seyfert 1 as a New Class of Gamma-Ray Active
 Galactic Nuclei}.
\newblock {\em \apjl} {\bf 2009}, {\em 707},~L142--L147.
\newblock {\url{https://doi.org/10.1088/0004-637X/707/2/L142}}.

\bibitem[{Foschini} {et~al.}(2015){Foschini}, {Berton}, {Caccianiga},
 {Ciroi}, {Cracco}, {Peterson}, {Angelakis}, {Braito}, {Fuhrmann}, {Gallo},
 {Grupe}, {J{\"a}rvel{\"a}}, {Kaufmann}, {Komossa}, {Kovalev},
 {L{\"a}hteenm{\"a}ki}, {Lisakov}, {Lister}, {Mathur}, {Richards}, {Romano},
 {Sievers}, {Tagliaferri}, {Tammi}, {Tibolla}, {Tornikoski}, {Vercellone}, {La
 Mura}, {Maraschi}, and {Rafanelli}]{foschinietal15}
{Foschini}, L.; {Berton}, M.; {Caccianiga}, A.; {Ciroi}, S.; {Cracco}, V.;
 {Peterson}, B.M.; {Angelakis}, E.; {Braito}, V.; {Fuhrmann}, L.; {Gallo}, L.;
 et~al.
\newblock {Properties of flat-spectrum radio-loud narrow-line Seyfert 1
 galaxies}.
\newblock {\em \aap} {\bf 2015}, {\em 575},~A13.
\newblock {\url{https://doi.org/10.1051/0004-6361/201424972}}.

\bibitem[{Zamfir} {et~al.}(2008){Zamfir}, {Sulentic}, and
 {Marziani}]{zamfiretal08}
{Zamfir}, S.; {Sulentic}, J.W.; {Marziani}, P.
\newblock {New insights on the QSO radio-loud/radio-quiet dichotomy: SDSS
 spectra in the context of the 4D eigenvector1 parameter space}.
\newblock {\em MNRAS} {\bf 2008}, {\em 387},~856--870.
\newblock {\url{https://doi.org/10.1111/j.1365-2966.2008.13290.x}}.

\bibitem[{Ganci} {et~al.}(2019){Ganci}, {Marziani}, {D'Onofrio}, {del Olmo},
 {Bon}, {Bon}, and {Negrete}]{gancietal19}
{Ganci}, V.; {Marziani}, P.; {D'Onofrio}, M.; {del Olmo}, A.; {Bon}, E.; {Bon},
 N.; {Negrete}, C.A.
\newblock {Radio loudness along the quasar main sequence}.
\newblock {\em \aap} {\bf 2019}, {\em 630},~A110.
\newblock {\url{https://doi.org/10.1051/0004-6361/201936270}}.

\bibitem[{Sulentic} {et~al.}(2006){Sulentic}, {Dultzin-Hacyan}, {Marziani},
 {Bongardo}, {Braito}, {Calvani}, and {Zamanov}]{sulenticetal06a}
{Sulentic}, J.W.; {Dultzin-Hacyan}, D.; {Marziani}, P.; {Bongardo}, C.;
 {Braito}, V.; {Calvani}, M.; {Zamanov}, R.
\newblock {Low Redshift BAL QSOs in the Eigenvector 1 Context}.
\newblock {\em Rev. Mex. Astron. Y Astrofis.} {\bf 2006}, {\em
 42},~23--39.

\bibitem[{Negrete} {et~al.}(2018){Negrete}, {Dultzin}, {Marziani},
 {Esparza}, {Sulentic}, {del Olmo}, {Mart{\'{\i}}nez-Aldama}, {Garc{\'{\i}}a
 L{\'o}pez}, {D'Onofrio}, {Bon}, and {Bon}]{negreteetal18}
{Negrete}, C.A.; {Dultzin}, D.; {Marziani}, P.; {Esparza}, D.; {Sulentic},
 J.W.; {del Olmo}, A.; {Mart{\'{\i}}nez-Aldama}, M.L.; {Garc{\'{\i}}a
 L{\'o}pez}, A.; {D'Onofrio}, M.; {Bon}, N.; et~al.
\newblock {Highly accreting quasars: The SDSS low-redshift catalog}.
\newblock {\em \aap} {\bf 2018}, {\em 620},~A118.
\newblock {\url{https://doi.org/10.1051/0004-6361/201833285}}.

\bibitem[{Marziani} {et~al.}(2003){Marziani}, {Sulentic}, {Zamanov},
 {Calvani}, {Dultzin-Hacyan}, {Bachev}, and {Zwitter}]{marzianietal03a}
{Marziani}, P.; {Sulentic}, J.W.; {Zamanov}, R.; {Calvani}, M.;
 {Dultzin-Hacyan}, D.; {Bachev}, R.; {Zwitter}, T.
\newblock {An Optical Spectroscopic Atlas of Low-Redshift Active Galactic
 Nuclei}.
\newblock {\em ApJS} {\bf 2003}, {\em 145},~199--211.
\newblock {\url{https://doi.org/10.1086/346025}}.

\bibitem[{Marziani}(2023)]{marziani23}
{Marziani}, P.
\newblock {Accretion/Ejection Phenomena and Emission-Line Profile (A)symmetries
 in Type-1 Active Galactic Nuclei}.
\newblock {\em Symmetry} {\bf 2023}, {\em 15},~1859.
\newblock {\url{https://doi.org/10.3390/sym15101859}}.

\bibitem[{Mengistue} {et~al.}(2023){Mengistue}, {Del Olmo}, {Marziani},
 {Povi{\'c}}, {Mart{\'\i}nez-Carballo}, {Perea}, and
 {M{\'a}rquez}]{terefemengistueetal23}
{Mengistue}, S.T.; {Del Olmo}, A.; {Marziani}, P.; {Povi{\'c}}, M.;
 {Mart{\'\i}nez-Carballo}, M.A.; {Perea}, J.; {M{\'a}rquez}, I.
\newblock {Optical and near-UV spectroscopic properties of low-redshift jetted
 quasars in the main sequence context}.
\newblock {\em \mnras} {\bf 2023}, {\em 525},~4474--4496.
\newblock {\url{https://doi.org/10.1093/mnras/stad2467}}.

\bibitem[{Marziani} {et~al.}(2009){Marziani}, {Sulentic}, {Stirpe},
 {Zamfir}, and {Calvani}]{marzianietal09}
{Marziani}, P.; {Sulentic}, J.W.; {Stirpe}, G.M.; {Zamfir}, S.; {Calvani}, M.
\newblock {VLT/ISAAC spectra of the H{$\beta$} region in intermediate-redshift
 quasars. III. H{$\beta$} broad-line profile analysis and inferences about BLR
 structure}.
\newblock {\em A\&Ap} {\bf 2009}, {\em 495},~83--112.
\newblock {\url{https://doi.org/10.1051/0004-6361:200810764}}.

\bibitem[{Bon} {et~al.}(2015){Bon}, {Bon}, {Marziani}, and
 {Jovanovi{\'c}}]{bonetal15}
{Bon}, N.; {Bon}, E.; {Marziani}, P.; {Jovanovi{\'c}}, P.
\newblock {Gravitational redshift of emission lines in the AGN spectra}.
\newblock {\em \apss} {\bf 2015}, {\em 360},~7.
\newblock {\url{https://doi.org/10.1007/s10509-015-2555-5}}.

\bibitem[{Marziani} {et~al.}(2010){Marziani}, {Sulentic}, {Negrete},
 {Dultzin}, {Zamfir}, and {Bachev}]{marzianietal10}
{Marziani}, P.; {Sulentic}, J.W.; {Negrete}, C.A.; {Dultzin}, D.; {Zamfir}, S.;
 {Bachev}, R.
\newblock {Broad-line region physical conditions along the quasar eigenvector 1
 sequence}.
\newblock {\em \mnras} {\bf 2010}, {\em 409},~1033--1048.
\newblock {\url{https://doi.org/10.1111/j.1365-2966.2010.17357.x}}.

\bibitem[{Marziani} {et~al.}(2018){Marziani}, {del Olmo}, {D'Onofrio},
 {Dultzin}, {Negrete}, {Mart{\'{\i}}nez-Aldama}, {Bon}, {Bon}, and
 {Stirpe}]{marzianietal18a}
{Marziani}, P.; {del Olmo}, A.; {D'Onofrio}, M.; {Dultzin}, D.; {Negrete},
 C.A.; {Mart{\'{\i}}nez-Aldama}, M.L.; {Bon}, E.; {Bon}, N.; {Stirpe}, G.M.
\newblock {Narrow-line Seyfert 1s: What is wrong in a name?}
\newblock In Proceedings of the Revisiting Narrow-Line Seyfert 1 Galaxies and
 Their Place in the Universe, Padova Botanical Garden, {Padua}, Italy, 9--13 April 2018; id.2. SISSA/ISAS, Volume {PoS(NLS1-2018)}
 . Available online: \url{https://pos.sissa.it/cgi-bin/reader/conf.cgi?confid=328} {(accessed on 1 December 2020)}
.

\bibitem[{Wu} and {Han}(2001)]{wuhan01}
{Wu}, X.B.; {Han}, J.L.
\newblock {Inclinations and Black Hole Masses of Seyfert 1 Galaxies}.
\newblock {\em \apjl} {\bf 2001}, {\em 561},~L59--L62.
\newblock {\url{https://doi.org/10.1086/324408}}.

\bibitem[{Bian} and {Zhao}(2002)]{bianzhao02}
{Bian}, W.; {Zhao}, Y.
\newblock {Masses, accretion rates and inclinations of AGNs}.
\newblock {\em \aap} {\bf 2002}, {\em 395},~465--473.
\newblock {\url{https://doi.org/10.1051/0004-6361:20021319}}.

\bibitem[{Decarli} {et~al.}(2011){Decarli}, {Dotti}, and
 {Treves}]{decarlietal11}
{Decarli}, R.; {Dotti}, M.; {Treves}, A.
\newblock {Geometry and inclination of the broad-line region in blazars}.
\newblock {\em \mnras} {\bf 2011}, {\em 413},~39--46.
\newblock {\url{https://doi.org/10.1111/j.1365-2966.2010.18102.x}}.

\bibitem[{Marziani} {et~al.}(2022){Marziani}, {Bon}, {Bon}, {D'Onofrio},
 {Punsly}, {{\'S}niegowska}, {Czerny}, {Panda}, {Martnez Aldama}, {del Olmo},
 {Deconto--Machado}, {Negrete}, {Dultzin}, {Buendia}, and
 {Garnica}]{marzianietal22a}
{Marziani}, P.; {Bon}, E.; {Bon}, N.; {D'Onofrio}, M.; {Punsly}, B.;
 {{\'S}niegowska}, M.; {Czerny}, B.; {Panda}, S.; {Martnez Aldama}, M.L.; {del
 Olmo}, A.; et~al.
\newblock {The main sequence of quasars: The taming of the extremes}.
\newblock {\em Astron. Nachrichten} {\bf 2022}, {\em 343},~e210082.
\newblock {\url{https://doi.org/10.1002/asna.20210082}}.

\bibitem[{Panda} {et~al.}(2020){Panda}, {Ma{\l}ek}, {{\'S}niegowska}, and
 {Czerny}]{pandaetal20}
{Panda}, S.; {Ma{\l}ek}, K.; {{\'S}niegowska}, M.; {Czerny}, B.
\newblock {Strong FeII emission in NLS1s: An unsolved mystery}.
\newblock In \emph{Challenges in Panchromatic Modelling with Next Generation Facilities}; 
{Boquien}, M., {Lusso}, E., {Gruppioni}, C., {Tissera}, P., Eds.; 
IAU Symposium; 
{Cambridge University Press: Cambridge, UK}, 2020; Volume 341, pp. 297--298.
\newblock {\url{https://doi.org/10.1017/S174392131900187X}}.

\bibitem[{Panda}(2021)]{pandaetal21}
{{Panda}, S.} 
\newblock {The CaFe project: Optical Fe II and near-infrared Ca II triplet
 emission in active galaxies: Simulated EWs and the co-dependence of cloud
 size and metal content}.
\newblock {\em \aap} {\bf 2021}, {\em 650},~A154.
\newblock {\url{https://doi.org/10.1051/0004-6361/202140393}}.

\bibitem[{Floris} {et~al.}(2024){Floris}, {Pandey}, {Czerny}, {Martinez
 Aldama}, {Panda}, {Marziani}, and {Prince}]{florisetal24a}
{Floris}, A.; {Pandey}, A.; {Czerny}, B.; {Martinez Aldama}, M.L.; {Panda}, S.;
 {Marziani}, P.; {Prince}, R.
\newblock {Dark and bright sides of the Broad Line Region clouds as seen in the
 FeII emission of SDSS RM 102}.
\newblock {\em arXiv} {\bf 2024}, arXiv:2408.17323.
\newblock {\url{https://doi.org/10.48550/arXiv.2408.17323}}.

\bibitem[{Chen} {et~al.}(2024){Chen}, {Luo}, {Brandt}, {Zuo}, {Dix}, {Ha},
 {Matthews}, {Paul}, {Plotkin}, and {Shemmer}]{chenetal24}
{Chen}, Y.; {Luo}, B.; {Brandt}, W.N.; {Zuo}, W.; {Dix}, C.; {Ha}, T.;
 {Matthews}, B.; {Paul}, J.D.; {Plotkin}, R.M.; {Shemmer}, O.
\newblock {Rest-frame Optical Spectroscopy of Ten z {\ensuremath{\sim}} 2 Weak
 Emission-line Quasars}.
\newblock {\em \apj} {\bf 2024}, {\em 972},~191.
\newblock {\url{https://doi.org/10.3847/1538-4357/ad5f89}}.

\bibitem[{Madau} and {Dickinson}(2014)]{madaudickinson14}
{Madau}, P.; {Dickinson}, M.
\newblock {Cosmic Star-Formation History}.
\newblock {\em \araa} {\bf 2014}, {\em 52},~415--486.
\newblock {\url{https://doi.org/10.1146/annurev-astro-081811-125615}}.

\bibitem[{Sulentic} {et~al.}(2004){Sulentic}, {Stirpe}, {Marziani},
 {Zamanov}, {Calvani}, and {Braito}]{sulenticetal04}
{Sulentic}, J.W.; {Stirpe}, G.M.; {Marziani}, P.; {Zamanov}, R.; {Calvani}, M.;
 {Braito}, V.
\newblock {VLT/ISAAC spectra of the H{$\beta$} region in intermediate redshift
 quasars}.
\newblock {\em A\&Ap} {\bf 2004}, {\em 423},~121--132.
\newblock {\url{https://doi.org/10.1051/0004-6361:20035912}}.

\bibitem[{Sulentic} {et~al.}(2006){Sulentic}, {Repetto}, {Stirpe},
 {Marziani}, {Dultzin-Hacyan}, and {Calvani}]{sulenticetal06}
{Sulentic}, J.W.; {Repetto}, P.; {Stirpe}, G.M.; {Marziani}, P.;
 {Dultzin-Hacyan}, D.; {Calvani}, M.
\newblock {VLT/ISAAC spectra of the H{$\beta$} region in intermediate-redshift
 quasars. II. Black hole mass and Eddington ratio}.
\newblock {\em A\&Ap} {\bf 2006}, {\em 456},~929--939.
\newblock {\url{https://doi.org/10.1051/0004-6361:20054153}}.

\bibitem[{Murayama} {et~al.}(1998){Murayama}, {Taniguchi}, {Evans},
 {Sanders}, {Ohyama}, {Kawara}, and {Arimoto}]{murayamaetal98}
{Murayama}, T.; {Taniguchi}, Y.; {Evans}, A.S.; {Sanders}, D.B.; {Ohyama}, Y.;
 {Kawara}, K.; {Arimoto}, N.
\newblock {Near-Infrared Spectroscopy of the High-Redshift Quasar S4 0636+68 at
 Z = 3.2}.
\newblock {\em \aj} {\bf 1998}, {\em 115},~2237--2243.
\newblock {\url{https://doi.org/10.1086/300352}}.

\bibitem[{D'Odorico} {et~al.}(2023){D'Odorico}, {Ba{\~n}ados}, {Becker},
 {Bischetti}, {Bosman}, {Cupani}, {Davies}, {Farina}, {Ferrara}, {Feruglio},
 {Mazzucchelli}, {Ryan-Weber}, {Schindler}, {Sodini}, {Venemans}, {Walter},
 {Chen}, {Lai}, {Zhu}, {Bian}, {Campo}, {Carniani}, {Cristiani}, {Davies},
 {Decarli}, {Drake}, {Eilers}, {Fan}, {Gaikwad}, {Gallerani}, {Greig},
 {Haehnelt}, {Hennawi}, {Keating}, {Kulkarni}, {Mesinger}, {Meyer},
 {Neeleman}, {Onoue}, {Pallottini}, {Qin}, {Rojas-Ruiz}, {Satyavolu},
 {Sebastian}, {Tripodi}, {Wang}, {Wolfson}, {Yang}, and
 {Zanchettin}]{dodoricoetal23}
{D'Odorico}, V.; {Ba{\~n}ados}, E.; {Becker}, G.D.; {Bischetti}, M.; {Bosman},
 S.E.I.; {Cupani}, G.; {Davies}, R.; {Farina}, E.P.; {Ferrara}, A.;
 {Feruglio}, C.; et~al.
\newblock {XQR-30: The ultimate XSHOOTER quasar sample at the reionization
 epoch}.
\newblock {\em \mnras} {\bf 2023}, {\em 523},~1399--1420.
\newblock {\url{https://doi.org/10.1093/mnras/stad1468}}.

\bibitem[{Yang} {et~al.}(2023){Yang}, {Wang}, {Fan}, {Hennawi}, {Barth},
 {Ba{\~n}ados}, {Sun}, {Liu}, {Cai}, {Jiang}, {Li}, {Onoue}, {Schindler},
 {Shen}, {Wu}, {Bhowmick}, {Bieri}, {Blecha}, {Bosman}, {Champagne}, {Colina},
 {Connor}, {Costa}, {Davies}, {Decarli}, {De Rosa}, {Drake}, {Egami},
 {Eilers}, {Evans}, {Farina}, {Habouzit}, {Haiman}, {Jin}, {Jun}, {Kakiichi},
 {Khusanova}, {Kulkarni}, {Loiacono}, {Lupi}, {Mazzucchelli}, {Pan},
 {Rojas-Ruiz}, {Strauss}, {Tee}, {Trakhtenbrot}, {Trebitsch}, {Venemans},
 {Vestergaard}, {Volonteri}, {Walter}, {Xie}, {Yue}, {Zhang}, {Zhang}, and
 {Zou}]{yiangetal23}
{Yang}, J.; {Wang}, F.; {Fan}, X.; {Hennawi}, J.F.; {Barth}, A.J.;
 {Ba{\~n}ados}, E.; {Sun}, F.; {Liu}, W.; {Cai}, Z.; {Jiang}, L.; et~al.
\newblock {A SPectroscopic Survey of Biased Halos in the Reionization Era
 (ASPIRE): A First Look at the Rest-frame Optical Spectra of z > 6.5 Quasars
 Using JWST}.
\newblock {\em \apjl} {\bf 2023}, {\em 951},~L5.
\newblock {\url{https://doi.org/10.3847/2041-8213/acc9c8}}.

\bibitem[{Deconto--Machado} {et~al.}(2024){Deconto--Machado}, {del Olmo}, and
 {Marziani}]{deconto-machadoetal24}
{Deconto--Machado}, A.; {del Olmo}, A.; {Marziani}, P.
\newblock {Exploring the links between quasar winds and radio emission along
 the main sequence at high redshift}.
\newblock {\em \aap} {\bf 2024}, {\em 691},~A15.
\newblock {\url{https://doi.org/10.1051/0004-6361/202449976}}.

\bibitem[{Ba{\~n}ados} {et~al.}(2016){Ba{\~n}ados}, {Venemans}, {Decarli},
 {Farina}, {Mazzucchelli}, {Walter}, {Fan}, {Stern}, {Schlafly}, {Chambers},
 {Rix}, {Jiang}, {McGreer}, {Simcoe}, {Wang}, {Yang}, {Morganson}, {De Rosa},
 {Greiner}, {Balokovi{\'c}}, {Burgett}, {Cooper}, {Draper}, {Flewelling},
 {Hodapp}, {Jun}, {Kaiser}, {Kudritzki}, {Magnier}, {Metcalfe}, {Miller},
 {Schindler}, {Tonry}, {Wainscoat}, {Waters}, and {Yang}]{banadosetal16}
{Ba{\~n}ados}, E.; {Venemans}, B.P.; {Decarli}, R.; {Farina}, E.P.;
 {Mazzucchelli}, C.; {Walter}, F.; {Fan}, X.; {Stern}, D.; {Schlafly}, E.;
 {Chambers}, K.C.; et~al.
\newblock {The Pan-STARRS1 Distant z{\,}$>${\,}5.6 Quasar Survey: More than 100
 Quasars within the First Gyr of the Universe}.
\newblock {\em \apjs} {\bf 2016}, {\em 227},~11.
\newblock {\url{https://doi.org/10.3847/0067-0049/227/1/11}}.

\bibitem[{Ba{\~n}ados} {et~al.}(2023){Ba{\~n}ados}, {Schindler}, {Venemans},
 {Connor}, {Decarli}, {Farina}, {Mazzucchelli}, {Meyer}, {Stern}, {Walter},
 {Fan}, {Hennawi}, {Khusanova}, {Morrell}, {Nanni}, {Noirot}, {Pensabene},
 {Rix}, {Simon}, {Verdoes Kleijn}, {Xie}, {Yang}, and {Connor}]{banadosetal23}
{Ba{\~n}ados}, E.; {Schindler}, J.T.; {Venemans}, B.P.; {Connor}, T.;
 {Decarli}, R.; {Farina}, E.P.; {Mazzucchelli}, C.; {Meyer}, R.A.; {Stern},
 D.; {Walter}, F.; et~al.
\newblock {The Pan-STARRS1 z > 5.6 Quasar Survey. II. Discovery of 55 Quasars
 at 5.6 < z < 6.5}.
\newblock {\em \apjs} {\bf 2023}, {\em 265},~29.
\newblock {\url{https://doi.org/10.3847/1538-4365/acb3c7}}.

\bibitem[{Deconto--Machado} {et~al.}(2023){Deconto--Machado}, {del Olmo
 Orozco}, {Marziani}, {Perea}, and {Stirpe}]{deconto-machadoetal23}
{Deconto--Machado}, A.; {del Olmo Orozco}, A.; {Marziani}, P.; {Perea}, J.;
 {Stirpe}, G.M.
\newblock {High-redshift quasars along the main sequence}.
\newblock {\em \aap} {\bf 2023}, {\em 669},~A83.
\newblock {\url{https://doi.org/10.1051/0004-6361/202243801}}.

\bibitem[{Onorato} {et~al.}(2024){Onorato}, {Hennawi}, {Schindler}, {Yang},
 {Wang}, {Barth}, {Ba{\~n}ados}, {Eilers}, {Bosman}, {Davies}, {Venemans},
 {Mazzucchelli}, {Belladitta}, {Vito}, {Farina}, {Andika}, {Fan}, {Walter},
 {Decarli}, {Onoue}, and {Nanni}]{onoratoetal24}
{Onorato}, S.; {Hennawi}, J.F.; {Schindler}, J.T.; {Yang}, J.; {Wang}, F.;
 {Barth}, A.J.; {Ba{\~n}ados}, E.; {Eilers}, A.C.; {Bosman}, S.E.I.; {Davies},
 F.B.; et~al.
\newblock {Optical and near-infrared spectroscopy of quasars at $z>6.5$: Public
 data release and composite spectrum}.
\newblock {\em arXiv} {\bf 2024}, arXiv:2406.07612.
\newblock {\url{https://doi.org/10.48550/arXiv.2406.07612}}.

\bibitem[{Cavaliere} and {Vittorini}(2000)]{cavalierevittorini00}
{Cavaliere}, A.; {Vittorini}, V.
\newblock {The Fall of the Quasar Population}.
\newblock {\em \apj} {\bf 2000}, {\em 543},~599--610.
\newblock {\url{https://doi.org/10.1086/317155}}.

\bibitem[{Hopkins} {et~al.}(2006){Hopkins}, {Hernquist}, {Cox}, {Di Matteo},
 {Robertson}, and {Springel}]{hopkinsetal06}
{Hopkins}, P.F.; {Hernquist}, L.; {Cox}, T.J.; {Di Matteo}, T.; {Robertson},
 B.; {Springel}, V.
\newblock {A Unified, Merger-driven Model of the Origin of Starbursts, Quasars,
 the Cosmic X-Ray Background, Supermassive Black Holes, and Galaxy Spheroids}.
\newblock {\em \apjs} {\bf 2006}, {\em 163},~1--49.
\newblock {\url{https://doi.org/10.1086/499298}}.

\bibitem[{Kelly} {et~al.}(2010){Kelly}, {Vestergaard}, {Fan}, {Hopkins},
 {Hernquist}, and {Siemiginowska}]{kellyetal10}
{Kelly}, B.C.; {Vestergaard}, M.; {Fan}, X.; {Hopkins}, P.; {Hernquist}, L.;
 {Siemiginowska}, A.
\newblock {Constraints on Black Hole Growth, Quasar Lifetimes, and Eddington
 Ratio Distributions from the SDSS Broad-line Quasar Black Hole Mass
 Function}.
\newblock {\em \apj} {\bf 2010}, {\em 719},~1315--1334.
\newblock {\url{https://doi.org/10.1088/0004-637X/719/2/1315}}.

\bibitem[{Hirschmann} {et~al.}(2014){Hirschmann}, {Dolag}, {Saro},
 {Bachmann}, {Borgani}, and {Burkert}]{hirschmannetal14}
{Hirschmann}, M.; {Dolag}, K.; {Saro}, A.; {Bachmann}, L.; {Borgani}, S.;
 {Burkert}, A.
\newblock {Cosmological simulations of black hole growth: AGN luminosities and
 downsizing}.
\newblock {\em \mnras} {\bf 2014}, {\em 442},~2304--2324.
\newblock {\url{https://doi.org/10.1093/mnras/stu1023}}.

\bibitem[{Sulentic} {et~al.}(2014){Sulentic}, {Marziani}, {del Olmo},
 {Dultzin}, {Perea}, and {Alenka Negrete}]{sulenticetal14}
{Sulentic}, J.W.; {Marziani}, P.; {del Olmo}, A.; {Dultzin}, D.; {Perea}, J.;
 {Alenka Negrete}, C.
\newblock {GTC spectra of z {$\approx$} 2.3 quasars: Comparison with local
 luminosity analogs}.
\newblock {\em \aap} {\bf 2014}, {\em 570},~A96.
\newblock {\url{https://doi.org/10.1051/0004-6361/201423975}}.

\bibitem[{Vanden Berk} {et~al.}(2001){Vanden Berk}, {Richards}, {Bauer},
 {Strauss}, {Schneider}, {Heckman}, {York}, {Hall}, and
 {Fan}]{vandenberketal01}
{Vanden Berk}, D.E.; {Richards}, G.T.; {Bauer}, A.; {Strauss}, M.A.;
 {Schneider}, D.P.; {Heckman}, T.M.; {York}, D.G.; {Hall}, P.B.; {Fan}, X.e.a.
\newblock {Composite Quasar Spectra from the Sloan Digital Sky Survey}.
\newblock {\em AJ} {\bf 2001}, {\em 122},~549--564.
\newblock {\url{https://doi.org/10.1086/321167}}.

\bibitem[{Kwan} and {Krolik}(1981)]{kwankrolik81}
{Kwan}, J.; {Krolik}, J.H.
\newblock {The formation of emission lines in quasars and Seyfert nuclei}.
\newblock {\em \apj} {\bf 1981}, {\em 250},~478--507.
\newblock {\url{https://doi.org/10.1086/159395}}.

\bibitem[{Rees} {et~al.}(1989){Rees}, {Netzer}, and {Ferland}]{reesetal89}
{Rees}, M.J.; {Netzer}, H.; {Ferland}, G.J.
\newblock {Small dense broad-line regions in active nuclei}.
\newblock {\em \apj} {\bf 1989}, {\em 347},~640--655.
\newblock {\url{https://doi.org/10.1086/168155}}.

\bibitem[{Netzer}(1990)]{netzer90}
{Netzer}, H.
\newblock {AGN emission lines.}
\newblock In \emph{Active Galactic Nuclei}; {Blandford, R.D., Netzer, H., Woltjer, L., Courvoisier, T.J.-L., Mayor, M.}, Eds.; {Springer: Berlin/Heidelberg, Germany}, 1990; pp. 57--160.

\bibitem[{Phillips}(1978)]{phillips78a}
{Phillips}, M.M.
\newblock {Permitted fe II Emission in Seyfert 1 Galaxies and QSOs I.
 Observations}.
\newblock {\em ApJS} {\bf 1978}, {\em 38},~187.
\newblock {\url{https://doi.org/10.1086/190553}}.

\bibitem[{Marinello} {et~al.}(2016){Marinello}, {Rodriguez-Ardila},
 {Garcia-Rissmann}, {Sigut}, and {Pradhan}]{marinelloetal16}
{Marinello}, A.O.M.; {Rodriguez-Ardila}, A.; {Garcia-Rissmann}, A.; {Sigut},
 T.A.A.; {Pradhan}, A.K.
\newblock {The FeII emission in active galactic nuclei: Excitation mechanisms
 and location of the emitting region}.
\newblock {\em ApJ} {\bf 2016}, {\em 820},~116.

\bibitem[{Kova{\v c}evi{\'c}-Doj{\v c}inovi{\'c}} and
 {Popovi{\'c}}(2015)]{kovacevicdojcinovicpopopvic15}
{Kova{\v c}evi{\'c}-Doj{\v c}inovi{\'c}}, J.; {Popovi{\'c}}, L.{\v C}.
\newblock {The Connections Between the UV and Optical Fe ii Emission Lines in
 Type 1 AGNs}.
\newblock {\em \apjs} {\bf 2015}, {\em 221},~35.
\newblock {\url{https://doi.org/10.1088/0067-0049/221/2/35}}.

\bibitem[{Sexton} {et~al.}(2021){Sexton}, {Matzko}, {Darden}, {Canalizo},
 and {Gorjian}]{sextonetal21}
{Sexton}, R.O.; {Matzko}, W.; {Darden}, N.; {Canalizo}, G.; {Gorjian}, V.
\newblock {Bayesian AGN Decomposition Analysis for SDSS spectra: A correlation
 analysis of [O III] {\ensuremath{\lambda}}5007 outflow kinematics with AGN
 and host galaxy properties}.
\newblock {\em \mnras} {\bf 2021}, {\em 500},~2871--2895.
\newblock {\url{https://doi.org/10.1093/mnras/staa3278}}.

\bibitem[{Du} {et~al.}(2016){Du}, {Lu}, {Hu}, {Qiu}, {Li}, {Huang}, {Wang},
 {Bai}, {Bian}, {Yuan}, {Ho}, {Wang}, and {SEAMBH Collaboration}]{duetal16}
{Du}, P.; {Lu}, K.X.; {Hu}, C.; {Qiu}, J.; {Li}, Y.R.; {Huang}, Y.K.; {Wang},
 F.; {Bai}, J.M.; {Bian}, W.H.; {Yuan}, Y.F.; et~al.
\newblock {Supermassive Black Holes with High Accretion Rates in Active
 Galactic Nuclei. VI. Velocity-resolved Reverberation Mapping of the
 H{$\beta$} Line}.
\newblock {\em \apj} {\bf 2016}, {\em 820},~27.
\newblock {\url{https://doi.org/10.3847/0004-637X/820/1/27}}.

\bibitem[{Du} {et~al.}(2018){Du}, {Zhang}, {Wang}, {Huang}, {Zhang}, {Lu},
 {Hu}, {Li}, {Bai}, {Bian}, {Yuan}, {Ho}, {Wang}, and {SEAMBH
 Collaboration}]{duetal18}
{Du}, P.; {Zhang}, Z.X.; {Wang}, K.; {Huang}, Y.K.; {Zhang}, Y.; {Lu}, K.X.;
 {Hu}, C.; {Li}, Y.R.; {Bai}, J.M.; {Bian}, W.H.; et~al.
\newblock {Supermassive Black Holes with High Accretion Rates in Active
 Galactic Nuclei. IX. 10 New Observations of Reverberation Mapping and
 Shortened H{$\beta$} Lags}.
\newblock {\em \apj} {\bf 2018}, {\em 856},~6.
\newblock {\url{https://doi.org/10.3847/1538-4357/aaae6b}}.

\bibitem[{Komossa} {et~al.}(2024){Komossa}, {Grupe}, {Marziani}, {Popovic},
 {Marceta-Mandic}, {Bon}, {Ilic}, {Kovacevic}, {Kraus}, {Haiman}, {Petrecca},
 {De Cicco}, {Dimitrijevic}, {Sreckovic}, {Kovacevic Dojcinovic},
 {Pannikkote}, {Bon}, {Gupta}, and {Iacob}]{komossaetal24}
{Komossa}, S.; {Grupe}, D.; {Marziani}, P.; {Popovic}, L.C.; {Marceta-Mandic},
 S.; {Bon}, E.; {Ilic}, D.; {Kovacevic}, A.B.; {Kraus}, A.; {Haiman}, Z.;
 et~al.
\newblock {The extremes of AGN variability: Outbursts, deep fades, changing
 looks, exceptional spectral states, and semi-periodicities}.
\newblock {\em arXiv} {\bf 2024}, arXiv:2408.00089.
\newblock {\url{https://doi.org/10.48550/arXiv.2408.00089}}.

\bibitem[{Dultzin-Hacyan} {et~al.}(1999){Dultzin-Hacyan}, {Taniguchi}, and
 {Uranga}]{dultzinhacyanetal99}
{Dultzin-Hacyan}, D.; {Taniguchi}, Y.; {Uranga}, L.
\newblock {Where is the Ca II Triplet Emitting Region in AGN?}
\newblock In \emph{Structure and Kinematics of Quasar Broad Line
 Regions}; {Gaskell}, C.M., {Brandt}, W.N., {Dietrich}, M., {Dultzin-Hacyan},
 D., {Eracleous}, M., Eds.; {Astronomical Society of the Pacific Conference Series}; Astronomical Society of the Pacific (ASP): {San Francisco, CA, USA}, 1999; Volume 175, p. 303.

\bibitem[{Mart{\'{\i}}nez-Aldama} {et~al.}(2015){Mart{\'{\i}}nez-Aldama},
 {Dultzin}, {Marziani}, {Sulentic}, {Bressan}, {Chen}, and
 {Stirpe}]{martinez-aldamaetal15}
{Mart{\'{\i}}nez-Aldama}, M.L.; {Dultzin}, D.; {Marziani}, P.; {Sulentic},
 J.W.; {Bressan}, A.; {Chen}, Y.; {Stirpe}, G.M.
\newblock {O I and Ca II Observations in Intermediate Redshift Quasars}.
\newblock {\em ApJS} {\bf 2015}, {\em 217},~3.
\newblock {\url{https://doi.org/10.1088/0067-0049/217/1/3}}.

\bibitem[{Mart{\'\i}nez-Aldama} {et~al.}(2021){Mart{\'\i}nez-Aldama},
 {Panda}, {Czerny}, {Marinello}, {Marziani}, and
 {Dultzin}]{martinez-aldamaetal21}
{Mart{\'\i}nez-Aldama}, M.L.; {Panda}, S.; {Czerny}, B.; {Marinello}, M.;
 {Marziani}, P.; {Dultzin}, D.
\newblock {The CaFe Project: Optical FeII and Near-Infrared Ca II triplet
 emission in active galaxies. II. The driver(s) of the Ca II and Fe II and its
 potential use as a chemical clock}.
\newblock {\em arXiv} {\bf 2021}, arXiv:2101.06999.


\bibitem[{Laha} {et~al.}(2016){Laha}, {Keenan}, {Ferland}, {Ramsbottom}, and
 {Aggarwal}]{lahaetal16}
{Laha}, S.; {Keenan}, F.P.; {Ferland}, G.J.; {Ramsbottom}, C.A.; {Aggarwal},
 K.M.
\newblock {Ultraviolet Emission Lines of Si II in Quasars: Investigating the
 ``Si II Disaster''}.
\newblock {\em \apj} {\bf 2016}, {\em 825},~28.
\newblock {\url{https://doi.org/10.3847/0004-637X/825/1/28}}.

\bibitem[{Kinney} {et~al.}(1990){Kinney}, {Rivolo}, and
 {Koratkar}]{kinneyetal90}
{Kinney}, A.L.; {Rivolo}, A.R.; {Koratkar}, A.P.
\newblock {A study of the Baldwin effect in the IUE data set}.
\newblock {\em \apj} {\bf 1990}, {\em 357},~338--345.
\newblock {\url{https://doi.org/10.1086/168924}}.

\bibitem[{Shemmer} and {Lieber}(2015)]{shemmerlieber15}
{Shemmer}, O.; {Lieber}, S.
\newblock {Weak Emission-line Quasars in the Context of a Modified Baldwin
 Effect}.
\newblock {\em \apj} {\bf 2015}, {\em 805},~124.
\newblock {\url{https://doi.org/10.1088/0004-637X/805/2/124}}.

\bibitem[{Plotkin} {et~al.}(2015){Plotkin}, {Shemmer}, {Trakhtenbrot},
 {Anderson}, {Brandt}, {Fan}, {Gallo}, {Lira}, {Luo}, {Richards}, {Schneider},
 {Strauss}, and {Wu}]{plotkinetal15}
{Plotkin}, R.M.; {Shemmer}, O.; {Trakhtenbrot}, B.; {Anderson}, S.F.; {Brandt},
 W.N.; {Fan}, X.; {Gallo}, E.; {Lira}, P.; {Luo}, B.; {Richards}, G.T.;
 et~al.
\newblock {Detection of Rest-frame Optical Lines from X-shooter Spectroscopy of
 Weak Emission Line Quasars}.
\newblock {\em \apj} {\bf 2015}, {\em 805},~123.
\newblock {\url{https://doi.org/10.1088/0004-637X/805/2/123}}.

\bibitem[{Marziani} {et~al.}(2016){Marziani}, {Mart{\'{\i}}nez Carballo},
 {Sulentic}, {Del Olmo}, {Stirpe}, and {Dultzin}]{marzianietal16a}
{Marziani}, P.; {Mart{\'{\i}}nez Carballo}, M.A.; {Sulentic}, J.W.; {Del Olmo},
 A.; {Stirpe}, G.M.; {Dultzin}, D.
\newblock {The most powerful quasar outflows as revealed by the Civ
 {$\lambda$}1549 resonance line}.
\newblock {\em \apss} {\bf 2016}, {\em 361},~29.
\newblock {\url{https://doi.org/10.1007/s10509-015-2611-1}}.

\bibitem[{Diamond-Stanic} {et~al.}(2009){Diamond-Stanic}, {Fan}, {Brandt},
 {Shemmer}, {Strauss}, {Anderson}, {Carilli}, {Gibson}, {Jiang}, {Kim},
 {Richards}, {Schmidt}, {Schneider}, {Shen}, {Smith}, {Vestergaard}, and
 {Young}]{diamond-stanicetal09}
{Diamond-Stanic}, A.M.; {Fan}, X.; {Brandt}, W.N.; {Shemmer}, O.; {Strauss},
 M.A.; {Anderson}, S.F.; {Carilli}, C.L.; {Gibson}, R.R.; {Jiang}, L.; {Kim},
 J.S.; et~al.
\newblock {High-redshift SDSS Quasars with Weak Emission Lines}.
\newblock {\em \apj} {\bf 2009}, {\em 699},~782--799.
\newblock {\url{https://doi.org/10.1088/0004-637X/699/1/782}}.

\bibitem[{Mart{\'\i}nez-Aldama} {et~al.}(2018){Mart{\'\i}nez-Aldama}, {del
 Olmo}, {Marziani}, {Sulentic}, {Negrete}, {Dultzin}, {D'Onofrio}, and
 {Perea}]{martinez-aldamaetal18}
{Mart{\'\i}nez-Aldama}, M.L.; {del Olmo}, A.; {Marziani}, P.; {Sulentic}, J.W.;
 {Negrete}, C.A.; {Dultzin}, D.; {D'Onofrio}, M.; {Perea}, J.
\newblock {Extreme quasars at high redshift}.
\newblock {\em \aap} {\bf 2018}, {\em 618},~A179.
\newblock {\url{https://doi.org/10.1051/0004-6361/201833541}}.

\bibitem[{Leung} {et~al.}(2024){Leung}, {Finkelstein},
 {P{\'e}rez-Gonz{\'a}lez}, {Morales}, {Taylor}, {Barro}, {Kocevski}, {Akins},
 {Carnall}, {Ch{\'a}vez Ortiz}, {Cleri}, {Cullen}, {Donnan}, {Dunlop},
 {Ellis}, {Grogin}, {Hirschmann}, {Koekemoer}, {Kokorev}, {Lucas}, {McLeod},
 {Papovich}, and {Yung}]{leungetal24}
 {Leung, G.C.K.; Finkelstein, S.L.; Pérez-González, P.G.; Morales, A.M.; Taylor, A.J.; Barro, G.; Kocevski, D.D.; Akins, H.B.; Carnall, A.C.; Chávez Ortiz, Ó.A.; et al.}
\newblock {{Exploring the Nature of Little Red Dots: Constraints on AGN and
 Stellar Contributions from PRIMER MIRI Imaging}}.
\newblock {\em arXiv} {\bf 2024}, arXiv:2411.12005.
\newblock {\url{https://doi.org/10.48550/arXiv.2411.12005}}.

\bibitem[{Negrete} {et~al.}(2012){Negrete}, {Dultzin}, {Marziani}, and
 {Sulentic}]{negreteetal12}
{Negrete}, A.; {Dultzin}, D.; {Marziani}, P.; {Sulentic}, J.
\newblock {BLR Physical Conditions in Extreme Population A Quasars: A Method to
 Estimate Central Black Hole Mass at High Redshift}.
\newblock {\em ApJ} {\bf 2012}, {\em 757},~62.

\bibitem[{Collin} and {Joly}(2000)]{collinjoly00}
{Collin}, S.; {Joly}, M.
\newblock {The Fe II problem in NLS1s}.
\newblock {\em NAR} {\bf 2000}, {\em 44},~531--537.
\newblock {\url{https://doi.org/10.1016/S1387-6473(00)00093-2}}.

\bibitem[{Joly} {et~al.}(2008){Joly}, {V{\'e}ron-Cetty}, and
 {V{\'e}ron}]{jolyetal08}
{Joly}, M.; {V{\'e}ron-Cetty}, M.; {V{\'e}ron}, P.
\newblock {Fe II emission in AGN}.
\newblock In \emph{Revista Mexicana de Astronomia y Astrofisica}; Conference Series; 
{Instituto de Astronomía: Distrito Federal, México}, 2008; Volume~32, pp. 59--61.

\bibitem[{Johansson} and {Letokhov}(2004)]{johanssonletokhov04}
{Johansson}, S.; {Letokhov}, V.S.
\newblock {Anomalous Fe II Spectral Effects and High H I Ly{$\alpha$}
 Temperature in Gas Blobs Near {$\eta$} Carinae}.
\newblock {\em Astron. Lett.} {\bf 2004}, {\em 30},~58--63.
\newblock {\url{https://doi.org/10.1134/1.1647477}}.

\bibitem[{Johansson} {et~al.}(2006){Johansson}, {Hartman}, and
 {Letokhov}]{johanssonetal06}
{Johansson}, S.; {Hartman}, H.; {Letokhov}, V.S.
\newblock {Resonance-enhanced two-photon ionization (RETPI) of Si II and an
 anomalous, variable intensity of the {\ensuremath{\lambda}}1892 Si III] line
 in the Weigelt blobs of {\ensuremath{\eta}} Carinae}.
\newblock {\em \aap} {\bf 2006}, {\em 452},~253--256.
\newblock {\url{https://doi.org/10.1051/0004-6361:20054375}}.

\bibitem[{Marconi} {et~al.}(2009){Marconi}, {Axon}, {Maiolino}, {Nagao},
 {Pietrini}, {Risaliti}, {Robinson}, and {Torricelli}]{marconietal09}
{Marconi}, A.; {Axon}, D.J.; {Maiolino}, R.; {Nagao}, T.; {Pietrini}, P.;
 {Risaliti}, G.; {Robinson}, A.; {Torricelli}, G.
\newblock {On the Observed Distributions of Black Hole Masses and Eddington
 Ratios from Radiation Pressure Corrected Virial Indicators}.
\newblock {\em \apjl} {\bf 2009}, {\em 698},~L103--L107.
\newblock {\url{https://doi.org/10.1088/0004-637X/698/2/L103}}.

\bibitem[{Netzer} and {Marziani}(2010)]{netzermarziani10}
{Netzer}, H.; {Marziani}, P.
\newblock {The Effect of Radiation Pressure on Emission-line Profiles and Black
 Hole Mass Determination in Active Galactic Nuclei}.
\newblock {\em \apj} {\bf 2010}, {\em 724},~318--328.
\newblock {\url{https://doi.org/10.1088/0004-637X/724/1/318}}.

\bibitem[{Hartig} and {Baldwin}(1986)]{hartigbaldwin86}
{Hartig}, G.F.; {Baldwin}, J.A.
\newblock {The emission-line regions in broad absorption line quasars}.
\newblock {\em \apj} {\bf 1986}, {\em 302},~64--80.
\newblock {\url{https://doi.org/10.1086/163974}}.

\bibitem[{Bachev} {et~al.}(2004){Bachev}, {Marziani}, {Sulentic}, {Zamanov},
 {Calvani}, and {Dultzin-Hacyan}]{bachevetal04}
{Bachev}, R.; {Marziani}, P.; {Sulentic}, J.W.; {Zamanov}, R.; {Calvani}, M.;
 {Dultzin-Hacyan}, D.
\newblock {Average Ultraviolet Quasar Spectra in the Context of Eigenvector 1:
 A Baldwin Effect Governed by the Eddington Ratio?}
\newblock {\em ApJ} {\bf 2004}, {\em 617},~171--183.
\newblock {\url{https://doi.org/10.1086/425210}}.

\bibitem[{Richards} {et~al.}(2011){Richards}, {Kruczek}, {Gallagher},
 {Hall}, {Hewett}, {Leighly}, {Deo}, {Kratzer}, and {Shen}]{richardsetal11}
{Richards}, G.T.; {Kruczek}, N.E.; {Gallagher}, S.C.; {Hall}, P.B.; {Hewett},
 P.C.; {Leighly}, K.M.; {Deo}, R.P.; {Kratzer}, R.M.; {Shen}, Y.
\newblock {Unification of Luminous Type 1 Quasars through C IV Emission}.
\newblock {\em \aj} {\bf 2011}, {\em 141},~167.
\newblock {\url{https://doi.org/10.1088/0004-6256/141/5/167}}.

\bibitem[{Aoki} and {Yoshida}(1999)]{aokiyoshida99}
{Aoki}, K.; {Yoshida}, M.
\newblock {The Correlation between SI III[ {$\lambda$}1892/CIII]
 {$\lambda$}1909 and FeII {$\lambda$} 4500/H{$\beta$} in Low Redshift QSOs}.
\newblock In \emph{Proceedings of the Quasars and Cosmology}; 
{Ferland}, G., {Baldwin}, J., Eds.; 
 {Astronomical Society of the Pacific Conference Series}; 
 {Astronomical Society of the Pacific: San Francisco, CA, USA}, 1999; Volume 162, p. 385.

\bibitem[{Buendia-Rios} {et~al.}(2023){Buendia-Rios}, {Negrete}, {Marziani},
 and {Dultzin}]{buendia-riosetal23}
{Buendia-Rios}, T.M.; {Negrete}, C.A.; {Marziani}, P.; {Dultzin}, D.
\newblock {Statistical analysis of Al III and C III] emission lines as virial
 black hole mass estimators in quasars}.
\newblock {\em \aap} {\bf 2023}, {\em 669},~A135.
\newblock {\url{https://doi.org/10.1051/0004-6361/202244177}}.

\bibitem[{Marziani} {et~al.}(2014){Marziani}, {Sulentic}, {Negrete},
 {Dultzin}, {D'Onofrio}, {Del Olmo}, and {Martinez-Aldama}]{marzianietal14}
{Marziani}, P.; {Sulentic}, J.W.; {Negrete}, C.A.; {Dultzin}, D.; {D'Onofrio},
 M.; {Del Olmo}, A.; {Martinez-Aldama}, M.L.
\newblock {Low- and high-z highly accreting quasars in the 4D Eigenvector 1
 context}.
\newblock {\em Astron. Rev.} {\bf 2014}, {\em 9},~6--25.

\bibitem[{Marziani} {et~al.}(2022){Marziani}, {Olmo}, {Negrete}, {Dultzin},
 {Piconcelli}, {Vietri}, {Mart{\'\i}nez-Aldama}, {D'Onofrio}, {Bon}, {Bon},
 {Deconto Machado}, {Stirpe}, and {Buendia Rios}]{marzianietal22}
{Marziani}, P.; {Olmo}, A.d.; {Negrete}, C.A.; {Dultzin}, D.; {Piconcelli}, E.;
 {Vietri}, G.; {Mart{\'\i}nez-Aldama}, M.L.; {D'Onofrio}, M.; {Bon}, E.;
 {Bon}, N.; et~al.
\newblock {The Intermediate-ionization Lines as Virial Broadening Estimators
 for Population A Quasars}.
\newblock {\em \apjs} {\bf 2022}, {\em 261},~30.
\newblock {\url{https://doi.org/10.3847/1538-4365/ac6fd6}}.

\bibitem[{Rani} {et~al.}(2017){Rani}, {Stalin}, and {Rakshit}]{ranietal17}
{Rani}, P.; {Stalin}, C.S.; {Rakshit}, S.
\newblock {X-ray flux variability of active galactic nuclei observed using
 NuSTAR}.
\newblock {\em \mnras} {\bf 2017}, {\em 466}, \mbox{3309--3322}.
\newblock {\url{https://doi.org/10.1093/mnras/stw3228}}.

\bibitem[{Rakshit} and {Stalin}(2017)]{rakshitstalin17}
{Rakshit}, S.; {Stalin}, C.S.
\newblock {Optical Variability of Narrow-line and Broad-line Seyfert 1
 Galaxies}.
\newblock {\em \apj} {\bf 2017}, {\em 842},~96.
\newblock {\url{https://doi.org/10.3847/1538-4357/aa72f4}}.

\bibitem[{Boller} {et~al.}(2021){Boller}, {Liu}, {Weber}, {Arcodia},
 {Dauser}, {Wilms}, {Nandra}, {Buchner}, {Merloni}, {Freyberg}, {Krumpe}, and
 {Waddell}]{bolleretal21}
{Boller}, T.; {Liu}, T.; {Weber}, P.; {Arcodia}, R.; {Dauser}, T.; {Wilms}, J.;
 {Nandra}, K.; {Buchner}, J.; {Merloni}, A.; {Freyberg}, M.J.; et~al.
\newblock {Extreme ultra-soft X-ray variability in an eROSITA observation of
 the narrow-line Seyfert 1 galaxy 1H 0707-495}.
\newblock {\em \aap} {\bf 2021}, {\em 647},~A6.
\newblock {\url{https://doi.org/10.1051/0004-6361/202039316}}.

\bibitem[{Parker} {et~al.}(2021){Parker}, {Alston}, {H{\"a}rer}, {Igo},
 {Joyce}, {Buisson}, {Chainakun}, {Fabian}, {Jiang}, {Kosec}, {Matzeu},
 {Pinto}, {Xu}, and {Zaidouni}]{parkeretal21}
{Parker}, M.L.; {Alston}, W.N.; {H{\"a}rer}, L.; {Igo}, Z.; {Joyce}, A.;
 {Buisson}, D.J.K.; {Chainakun}, P.; {Fabian}, A.C.; {Jiang}, J.; {Kosec}, P.;
 et~al.
\newblock {The nature of the extreme X-ray variability in the NLS1 1H
 0707-495}.
\newblock {\em \mnras} {\bf 2021}, {\em 508},~1798--1816.
\newblock {\url{https://doi.org/10.1093/mnras/stab2434}}.

\bibitem[{Li} {et~al.}(2019){Li}, {Sun}, {Wang}, {He}, and {Xue}]{lietal19}
{Li}, J.; {Sun}, M.; {Wang}, T.; {He}, Z.; {Xue}, Y.
\newblock {On the origin of the dramatic spectral variability of WPVS 007}.
\newblock {\em \mnras} {\bf 2019}, {\em 487},~4592--4602.
\newblock {\url{https://doi.org/10.1093/mnras/stz1393}}.

\bibitem[{Floris} {et~al.}(2024){Floris}, {Marziani}, {Panda}, {Sniegowska},
 {D'Onofrio}, {Deconto--Machado}, {Del Olmo}, and {Czerny}]{florisetal24}
{Floris}, A.; {Marziani}, P.; {Panda}, S.; {Sniegowska}, M.; {D'Onofrio}, M.;
 {Deconto--Machado}, A.; {Del Olmo}, A.; {Czerny}, B.
\newblock {Chemical abundances along the quasar main sequence}.
\newblock {\em arXiv} {\bf 2024}, arXiv:2405.04456.
\newblock {\url{https://doi.org/10.48550/arXiv.2405.04456}}.

\bibitem[{Temple} {et~al.}(2021){Temple}, {Ferland}, {Rankine}, {Chatzikos},
 and {Hewett}]{templeetal21}
{Temple}, M.J.; {Ferland}, G.J.; {Rankine}, A.L.; {Chatzikos}, M.; {Hewett},
 P.C.
\newblock {High-ionization emission-line ratios from quasar broad-line regions:
 metallicity or density?}
\newblock {\em \mnras} {\bf 2021}, {\em 505},~3247--3259.
\newblock {\url{https://doi.org/10.1093/mnras/stab1610}}.

\bibitem[{Zhang} {et~al.}(2024){Zhang}, {Thomsen}, {Dai}, {Reynolds},
 {Garc{\'\i}a}, {Kara}, {Connors}, {Masterson}, {Yao}, and
 {Dauser}]{zhangetal24}
{Zhang}, Z.; {Thomsen}, L.L.; {Dai}, L.; {Reynolds}, C.S.; {Garc{\'\i}a}, J.A.;
 {Kara}, E.; {Connors}, R.; {Masterson}, M.; {Yao}, Y.; {Dauser}, T.
\newblock {Modeling X-Ray Multi-Reflection in Super-Eddington Winds}.
\newblock {\em arXiv} {\bf 2024}, arXiv:2407.08596.
\newblock {\url{https://doi.org/10.48550/arXiv.2407.08596}}.

\bibitem[{Marziani} {et~al.}(2020){Marziani}, {del Olmo}, {Perea},
 {D'Onofrio}, and {Panda}]{marzianietal20}
{Marziani}, P.; {del Olmo}, A.; {Perea}, J.; {D'Onofrio}, M.; {Panda}, S.
\newblock {Broad UV Emission Lines in Type-1 Active Galactic Nuclei: A Note on
 Spectral Diagnostics and the Excitation Mechanism}.
\newblock {\em Atoms} {\bf 2020}, {\em 8},~94.
\newblock {\url{https://doi.org/10.3390/atoms8040094}}.

\bibitem[{Pignatari} {et~al.}(2010){Pignatari}, {Gallino}, {Heil},
 {Wiescher}, {K{\"a}ppeler}, {Herwig}, and {Bisterzo}]{pignatarietal10}
{Pignatari}, M.; {Gallino}, R.; {Heil}, M.; {Wiescher}, M.; {K{\"a}ppeler}, F.;
 {Herwig}, F.; {Bisterzo}, S.
\newblock {The Weak s-Process in Massive Stars and its Dependence on the
 Neutron Capture Cross Sections}.
\newblock {\em \apj} {\bf 2010}, {\em 710},~1557--1577.
\newblock {\url{https://doi.org/10.1088/0004-637X/710/2/1557}}.

\bibitem[{Thielemann}(2019)]{thielemann19}
{Thielemann}, F.K.
\newblock {Explosive Nucleosynthesis: What We Learned and What We Still Do Not Understand}.
\newblock In Proceedings of the Nuclei in the Cosmos XV, {Assergi, Italy, 24--29 June} 2019; 
Volume 219, pp. 125--134.
\newblock {\url{https://doi.org/10.1007/978-3-030-13876-9_21}}.

\bibitem[{Farrah} {et~al.}(2022){Farrah}, {Efstathiou}, {Afonso},
 {Bernard-Salas}, {Cairns}, {Clements}, {Croker}, {Hatziminaoglou}, {Joyce},
 {Lacy}, {Lebouteiller}, {Lieblich}, {Lonsdale}, {Oliver}, {Pearson}, {Petty},
 {Pitchford}, {Rigopoulou}, {Rowan-Robinson}, {Runburg}, {Spoon}, {Verma}, and
 {Wang}]{farrahetal22}
{Farrah}, D.; {Efstathiou}, A.; {Afonso}, J.; {Bernard-Salas}, J.; {Cairns},
 J.; {Clements}, D.L.; {Croker}, K.; {Hatziminaoglou}, E.; {Joyce}, M.;
 {Lacy}, M.; et~al.
\newblock {Stellar and black hole assembly in z < 0.3 infrared-luminous
 mergers: Intermittent starbursts versus super-Eddington accretion}.
\newblock {\em \mnras} {\bf 2022}, {\em 513},~4770--4786.
\newblock {\url{https://doi.org/10.1093/mnras/stac980}}.

\bibitem[{Miniutti} {et~al.}(2013){Miniutti}, {Saxton},
 {Rodr{\'\i}guez-Pascual}, {Read}, {Esquej}, {Colless}, {Dobbie}, and
 {Spolaor}]{miniuttietal13}
{Miniutti}, G.; {Saxton}, R.D.; {Rodr{\'\i}guez-Pascual}, P.M.; {Read}, A.M.;
 {Esquej}, P.; {Colless}, M.; {Dobbie}, P.; {Spolaor}, M.
\newblock {A high Eddington-ratio, true Seyfert 2 galaxy candidate:
 implications for broad-line region models}.
\newblock {\em \mnras} {\bf 2013}, {\em 433},~1764--1777.
\newblock {\url{https://doi.org/10.1093/mnras/stt850}}.

\bibitem[{Antonucci} and {Miller}(1985)]{antonuccimiller85}
{Antonucci}, R.R.J.; {Miller}, J.S.
\newblock {Spectropolarimetry and the nature of NGC 1068}.
\newblock {\em \apj} {\bf 1985}, {\em 297},~621--632.
\newblock {\url{https://doi.org/10.1086/163559}}.

\bibitem[{Antonucci}(1993)]{antonucci93}
{Antonucci}, R.
\newblock {Unified models for active galactic nuclei and quasars}.
\newblock {\em \araa} {\bf 1993}, {\em 31},~473--521.
\newblock {\url{https://doi.org/10.1146/annurev.aa.31.090193.002353}}.

\bibitem[{Sulentic} {et~al.}(2011){Sulentic}, {Marziani}, and
 {Zamfir}]{sulenticetal11}
{Sulentic}, J.; {Marziani}, P.; {Zamfir}, S.
\newblock {The Case for Two Quasar Populations}.
\newblock {\em Balt. Astron.} {\bf 2011}, {\em 20},~427--434.

\bibitem[Fraix-Burnet {et~al.}(2017)Fraix-Burnet, Marziani, D'Onofrio, and
 Dultzin]{fraix-burnetetal17}
Fraix-Burnet, D.; Marziani, P.; D'Onofrio, M.; Dultzin, D.
\newblock The Phylogeny of Quasars and the Ontogeny of Their Central Black
 Holes.
\newblock {\em Front. Astron. Space Sci.} {\bf 2017}, {\em
 4},~1.
\newblock {\url{https://doi.org/10.3389/fspas.2017.00001}}.

\bibitem[{Marziani} {et~al.}(1996){Marziani}, {Sulentic}, {Dultzin-Hacyan},
 {Calvani}, and {Moles}]{marzianietal96}
{Marziani}, P.; {Sulentic}, J.W.; {Dultzin-Hacyan}, D.; {Calvani}, M.; {Moles},
 M.
\newblock {Comparative Analysis of the High- and Low-Ionization Lines in the
 Broad-Line Region of Active Galactic Nuclei}.
\newblock {\em ApJS} {\bf 1996}, {\em 104},~37.
\newblock {\url{https://doi.org/10.1086/192291}}.

\bibitem[{Corbin} and {Boroson}(1996)]{corbinboroson96}
{Corbin}, M.R.; {Boroson}, T.A.
\newblock {Combined Ultraviolet and Optical Spectra of 48 Low-Redshift QSOs and
 the Relation of the Continuum and Emission-Line Properties}.
\newblock {\em \apjs} {\bf 1996}, {\em 107},~69.
\newblock {\url{https://doi.org/10.1086/192355}}.

\bibitem[{Sulentic} {et~al.}(2007){Sulentic}, {Bachev}, {Marziani},
 {Negrete}, and {Dultzin}]{sulenticetal07}
{Sulentic}, J.W.; {Bachev}, R.; {Marziani}, P.; {Negrete}, C.A.; {Dultzin}, D.
\newblock {C IV {\ensuremath{\lambda}}1549 as an Eigenvector 1 Parameter for
 Active Galactic Nuclei}.
\newblock {\em \apj} {\bf 2007}, {\em 666},~757--777.
\newblock {\url{https://doi.org/10.1086/519916}}.

\bibitem[{Sulentic} and {Marziani}(1999)]{sulenticmarziani99}
{Sulentic}, J.W.; {Marziani}, P.
\newblock {The Intermediate-Line Region in Active Galactic Nuclei: A Region
 ``Pr{\ae}ter Necessitatem''?}
\newblock {\em ApJL} {\bf 1999}, {\em 518},~L9--L12.
\newblock {\url{https://doi.org/10.1086/312060}}.

\bibitem[{Zamanov} {et~al.}(2002){Zamanov}, {Marziani}, {Sulentic},
 {Calvani}, {Dultzin-Hacyan}, and {Bachev}]{zamanovetal02}
{Zamanov}, R.; {Marziani}, P.; {Sulentic}, J.W.; {Calvani}, M.;
 {Dultzin-Hacyan}, D.; {Bachev}, R.
\newblock {Kinematic Linkage between the Broad- and Narrow-Line-emitting Gas in
 Active Galactic Nuclei}.
\newblock {\em ApJL} {\bf 2002}, {\em 576},~L9--L13.
\newblock {\url{https://doi.org/10.1086/342783}}.

\bibitem[{Cracco} {et~al.}(2016){Cracco}, {Ciroi}, {Berton}, {Di Mille},
 {Foschini}, {La Mura}, and {Rafanelli}]{craccoetal16}
{Cracco}, V.; {Ciroi}, S.; {Berton}, M.; {Di Mille}, F.; {Foschini}, L.; {La
 Mura}, G.; {Rafanelli}, P.
\newblock {A spectroscopic analysis of a sample of narrow-line Seyfert 1
 galaxies selected from the Sloan Digital Sky Survey}.
\newblock {\em \mnras} {\bf 2016}, {\em 462},~1256--1280.
\newblock {\url{https://doi.org/10.1093/mnras/stw1689}}.

\bibitem[{Vietri} {et~al.}(2018){Vietri}, {Piconcelli}, {Bischetti},
 {Duras}, {Martocchia}, {Bongiorno}, {Marconi}, {Zappacosta}, {Bisogni},
 {Bruni}, {Brusa}, {Comastri}, {Cresci}, {Feruglio}, {Giallongo}, {La Franca},
 {Mainieri}, {Mannucci}, {Ricci}, {Sani}, {Testa}, {Tombesi}, {Vignali}, and
 {Fiore}]{vietrietal18}
{Vietri}, G.; {Piconcelli}, E.; {Bischetti}, M.; {Duras}, F.; {Martocchia}, S.;
 {Bongiorno}, A.; {Marconi}, A.; {Zappacosta}, L.; {Bisogni}, S.; {Bruni}, G.;
 et~al.
\newblock {The WISSH quasars project. IV. Broad line region versus
 kiloparsec-scale winds}.
\newblock {\em A\&A} {\bf 2018}, {\em 617},~A81.
\newblock {\url{https://doi.org/10.1051/0004-6361/201732335}}.

\bibitem[{Vietri} {et~al.}(2020){Vietri}, {Mainieri}, {Kakkad}, {Netzer},
 {Perna}, {Circosta}, {Harrison}, {Zappacosta}, {Husemann}, {Padovani},
 {Bischetti}, {Bongiorno}, {Brusa}, {Carniani}, {Cicone}, {Comastri},
 {Cresci}, {Feruglio}, {Fiore}, {Lanzuisi}, {Mannucci}, {Marconi},
 {Piconcelli}, {Puglisi}, {Salvato}, {Schramm}, {Schulze}, {Scholtz},
 {Vignali}, and {Zamorani}]{vietrietal20}
{Vietri}, G.; {Mainieri}, V.; {Kakkad}, D.; {Netzer}, H.; {Perna}, M.;
 {Circosta}, C.; {Harrison}, C.M.; {Zappacosta}, L.; {Husemann}, B.;
 {Padovani}, P.; et~al.
\newblock {SUPER. III. Broad line region properties of AGNs at z
 {\ensuremath{\sim}} 2}.
\newblock {\em \aap} {\bf 2020}, {\em 644},~A175.
\newblock {\url{https://doi.org/10.1051/0004-6361/202039136}}.

\bibitem[{Boller} {et~al.}(2016){Boller}, {Freyberg}, {Tr{\"u}mper},
 {Haberl}, {Voges}, and {Nandra}]{bolleretal16}
{Boller}, T.; {Freyberg}, M.J.; {Tr{\"u}mper}, J.; {Haberl}, F.; {Voges}, W.;
 {Nandra}, K.
\newblock {Second ROSAT all-sky survey (2RXS) source catalogue}.
\newblock {\em \aap} {\bf 2016}, {\em 588},~A103.
\newblock {\url{https://doi.org/10.1051/0004-6361/201525648}}.

\bibitem[{Boller} {et~al.}(1996){Boller}, {Brandt}, and
 {Fink}]{bolleretal96}
{Boller}, T.; {Brandt}, W.N.; {Fink}, H.
\newblock {Soft X-ray properties of narrow-line Seyfert 1 galaxies.}
\newblock {\em \aap} {\bf 1996}, {\em 305},~53.

\bibitem[{Grupe}(2004)]{grupe04}
{Grupe}, D. A Complete Sample of Soft X-Ray-selected AGNs. II. Statistical Analysis.
\newblock {\em \aj} {\bf 2004}, {\em 127},~1799--1810.
\newblock {\url{https://doi.org/10.1086/382516}}.

\bibitem[{Tananbaum} {et~al.}(1979){Tananbaum}, {Avni}, {Branduardi},
 {Elvis}, {Fabbiano}, {Feigelson}, {Giacconi}, {Henry}, {Pye}, {Soltan}, and
 {Zamorani}]{tananbaumetal79}
{Tananbaum}, H.; {Avni}, Y.; {Branduardi}, G.; {Elvis}, M.; {Fabbiano}, G.;
 {Feigelson}, E.; {Giacconi}, R.; {Henry}, J.P.; {Pye}, J.P.; {Soltan}, A.;
 et~al.
\newblock {X-ray studies of quasars with the Einstein Observatory.}
\newblock {\em \apjl} {\bf 1979}, {\em 234},~L9--L13.
\newblock {\url{https://doi.org/10.1086/183100}}.

\bibitem[{Walter} and {Fink}(1993)]{walterfink93}
{Walter}, R.; {Fink}, H.H.
\newblock {The ultraviolet to soft X-ray bump of Seyfert 1 type active galactic
 nuclei.}
\newblock {\em \aap} {\bf 1993}, {\em 274},~105.

\bibitem[{Wang} and {Netzer}(2003)]{wangnetzer03}
{Wang}, J.M.; {Netzer}, H.
\newblock {Extreme slim accretion disks and narrow line Seyfert 1 galaxies: The
 nature of the soft X-ray hump}.
\newblock {\em \aap} {\bf 2003}, {\em 398},~927--936.
\newblock {\url{https://doi.org/10.1051/0004-6361:20021511}}.

\bibitem[{Petrucci} {et~al.}(2020){Petrucci}, {Gronkiewicz}, {Rozanska},
 {Belmont}, {Bianchi}, {Czerny}, {Matt}, {Malzac}, {Middei}, {De Rosa},
 {Ursini}, and {Cappi}]{petruccietal20}
{Petrucci}, P.O.; {Gronkiewicz}, D.; {Rozanska}, A.; {Belmont}, R.; {Bianchi},
 S.; {Czerny}, B.; {Matt}, G.; {Malzac}, J.; {Middei}, R.; {De Rosa}, A.;
 et~al.
\newblock {Radiation spectra of warm and optically thick coronae in AGNs}.
\newblock {\em \aap} {\bf 2020}, {\em 634},~A85.
\newblock {\url{https://doi.org/10.1051/0004-6361/201937011}}.

\bibitem[{Done} {et~al.}(2012){Done}, {Davis}, {Jin}, {Blaes}, and
 {Ward}]{doneetal12}
{Done}, C.; {Davis}, S.W.; {Jin}, C.; {Blaes}, O.; {Ward}, M.
\newblock {Intrinsic disc emission and the soft X-ray excess in active galactic
 nuclei}.
\newblock {\em \mnras} {\bf 2012}, {\em 420},~1848--1860.
\newblock {\url{https://doi.org/10.1111/j.1365-2966.2011.19779.x}}.

\bibitem[{Sulentic} {et~al.}(2000){Sulentic}, {Marziani}, {Zwitter},
 {Dultzin-Hacyan}, and {Calvani}]{sulenticetal00b}
{Sulentic}, J.W.; {Marziani}, P.; {Zwitter}, T.; {Dultzin-Hacyan}, D.;
 {Calvani}, M.
\newblock {The Demise of the Classical Broad-Line Region in the Luminous Quasar
 PG 1416-129}.
\newblock {\em ApJL} {\bf 2000}, {\em 545},~L15--L18.
\newblock {\url{https://doi.org/10.1086/317330}}.

\bibitem[{Bensch} {et~al.}(2015){Bensch}, {del Olmo}, {Sulentic}, {Perea},
 and {Marziani}]{benschetal15}
{Bensch}, K.; {del Olmo}, A.; {Sulentic}, J.; {Perea}, J.; {Marziani}, P.
\newblock {Measures of the Soft X-ray Excess as an Eigenvector 1 Parameter for
 Active Galactic Nuclei}.
\newblock {\em J. Astrophys. Astron.} {\bf 2015}, {\em
 36},~467--474.
\newblock {\url{https://doi.org/10.1007/s12036-015-9355-8}}.

\bibitem[{Sulentic} and {Marziani}(2015)]{sulenticmarziani15}
{Sulentic}, J.; {Marziani}, P.
\newblock {Quasars in the 4D Eigenvector 1 Context: A stroll down memory lane}.
\newblock {\em Front. Astron. Space Sci.} {\bf 2015}, {\em
 2},~6.
\newblock {\url{https://doi.org/10.3389/fspas.2015.00006}}.

\bibitem[{Panda} and {Marziani}(2023)]{pandamarziani23}
{Panda}, S.; {Marziani}, P.
\newblock {High Eddington quasars as discovery tools: Current state and
 challenges}.
\newblock {\em Front. Astron. Space Sci.} {\bf 2023}, {\em
 10},~1130103.
\newblock {\url{https://doi.org/10.3389/fspas.2023.1130103}}.

\bibitem[{Shields}(1978)]{shields78}
{Shields}, G.A.
\newblock {Thermal continuum from acretion disks in quasars}.
\newblock {\em \nat} {\bf 1978}, {\em 272},~706--708.
\newblock {\url{https://doi.org/10.1038/272706a0}}.

\bibitem[{Czerny} and {Elvis}(1987)]{czernyelvis87}
{Czerny}, B.; {Elvis}, M.
\newblock {Constraints on Quasar Accretion Disks from the
 Optical/Ultraviolet/Soft X-Ray Big Bump}.
\newblock {\em \apj} {\bf 1987}, {\em 321},~305.
\newblock {\url{https://doi.org/10.1086/165630}}.

\bibitem[{Laor} {et~al.}(1997){Laor}, {Fiore}, {Elvis}, {Wilkes}, and
 {McDowell}]{laoretal97}
{Laor}, A.; {Fiore}, F.; {Elvis}, M.; {Wilkes}, B.J.; {McDowell}, J.C.
\newblock {The Soft X-Ray Properties of a Complete Sample of Optically Selected
 Quasars. II. Final Results}.
\newblock {\em \apj} {\bf 1997}, {\em 477},~93--113.
\newblock {\url{https://doi.org/10.1086/303696}}.

\bibitem[{Grupe} {et~al.}(2010){Grupe}, {Komossa}, {Leighly}, and
 {Page}]{grupeetal10}
{Grupe}, D.; {Komossa}, S.; {Leighly}, K.M.; {Page}, K.L.
\newblock {The Simultaneous Optical-to-X-Ray Spectral Energy Distribution of
 Soft X-Ray Selected Active Galactic Nuclei Observed by Swift}.
\newblock {\em \apjs} {\bf 2010}, {\em 187},~64--106.
\newblock {\url{https://doi.org/10.1088/0067-0049/187/1/64}}.

\bibitem[{Capellupo} {et~al.}(2015){Capellupo}, {Netzer}, {Lira},
 {Trakhtenbrot}, and {Mej{\'\i}a-Restrepo}]{capellupoetal15}
{Capellupo}, D.M.; {Netzer}, H.; {Lira}, P.; {Trakhtenbrot}, B.;
 {Mej{\'\i}a-Restrepo}, J.
\newblock {Active galactic nuclei at z {\ensuremath{\sim}} 1.5 - I. Spectral
 energy distribution and accretion discs}.
\newblock {\em \mnras} {\bf 2015}, {\em 446},~3427--3446.
\newblock {\url{https://doi.org/10.1093/mnras/stu2266}}.

\bibitem[{Ferland} {et~al.}(2020){Ferland}, {Done}, {Jin}, {Landt}, and
 {Ward}]{ferlandetal20}
{Ferland}, G.J.; {Done}, C.; {Jin}, C.; {Landt}, H.; {Ward}, M.J.
\newblock {State-of-the-art AGN SEDs for photoionization models: BLR
 predictions confront the observations}.
\newblock {\em \mnras} {\bf 2020}, {\em 494},~5917--5922.
\newblock {\url{https://doi.org/10.1093/mnras/staa1207}}.

\bibitem[{Laurenti} {et~al.}(2021){Laurenti}, {Luminari}, {Tombesi},
 {Vagnetti}, {Middei}, and {Piconcelli}]{laurentietal21}
{Laurenti}, M.; {Luminari}, A.; {Tombesi}, F.; {Vagnetti}, F.; {Middei}, R.;
 {Piconcelli}, E.
\newblock {Location and energetics of the ultra-fast outflow in PG 1448+273}.
\newblock {\em \aap} {\bf 2021}, {\em 645},~A118.
\newblock {\url{https://doi.org/10.1051/0004-6361/202039409}}.

\bibitem[{Gr{\"u}nwald} {et~al.}(2023){Gr{\"u}nwald}, {Boller}, {Rakshit},
 {Buchner}, {Dauser}, {Freyberg}, {Liu}, {Salvato}, and
 {Schichtel}]{grunwaldetal23}
{Gr{\"u}nwald}, G.; {Boller}, T.; {Rakshit}, S.; {Buchner}, J.; {Dauser}, T.;
 {Freyberg}, M.; {Liu}, T.; {Salvato}, M.; {Schichtel}, A.
\newblock {The first look at narrow-line Seyfert 1 galaxies with eROSITA}.
\newblock {\em \aap} {\bf 2023}, {\em 669},~A37.
\newblock {\url{https://doi.org/10.1051/0004-6361/202244620}}.

\bibitem[{Boller}(2023)]{boller23}
{Boller}, T.
\newblock {Unraveling the enigmatic soft x-ray excess: Current understanding
 and future perspectives}.
\newblock {\em Astron. Nachrichten} {\bf 2023}, {\em 344},~e20230105.
\newblock {\url{https://doi.org/10.1002/asna.20230105}}.

\bibitem[{Haardt} and {Maraschi}(1991)]{haardtmaraschi91}
{Haardt}, F.; {Maraschi}, L.
\newblock {A two-phase model for the X-ray emission from Seyfert galaxies}.
\newblock {\em \apjl} {\bf 1991}, {\em 380},~L51--L54.
\newblock {\url{https://doi.org/10.1086/186171}}.

\bibitem[{Haardt} and {Maraschi}(1993)]{haardtmaraschi93}
{Haardt}, F.; {Maraschi}, L.
\newblock {X-Ray Spectra from Two-Phase Accretion Disks}.
\newblock {\em \apj} {\bf 1993}, {\em 413},~507.
\newblock {\url{https://doi.org/10.1086/173020}}.

\bibitem[{Nandra} and {Pounds}(1994)]{nandrapounds94}
{Nandra}, K.; {Pounds}, K.A.
\newblock {GINGA observations of the X-ray spectra of Seyfert galaxies.}
\newblock {\em \mnras} {\bf 1994}, {\em 268},~405--429.
\newblock {\url{https://doi.org/10.1093/mnras/268.2.405}}.

\bibitem[{Kara} {et~al.}(2016){Kara}, {Miller}, {Reynolds}, and
 {Dai}]{karaetal16a}
{Kara}, E.; {Miller}, J.M.; {Reynolds}, C.; {Dai}, L.
\newblock {Relativistic reverberation in the accretion flow of a tidal
 disruption event}.
\newblock {\em \nat} {\bf 2016}, {\em 535},~388--390.
\newblock {\url{https://doi.org/10.1038/nature18007}}.

\bibitem[{Mathews} and {Ferland}(1987)]{mathewsferland87}
{Mathews}, W.G.; {Ferland}, G.J.
\newblock {What heats the hot phase in active nuclei?}
\newblock {\em ApJ} {\bf 1987}, {\em 323},~456--467.
\newblock {\url{https://doi.org/10.1086/165843}}.

\bibitem[{Xie} {et~al.}(2021){Xie}, {Ho}, {Zhuang}, and
 {Shangguan}]{xieetal21}
{Xie}, Y.; {Ho}, L.C.; {Zhuang}, M.Y.; {Shangguan}, J.
\newblock {The Infrared Emission and Vigorous Star Formation of Low-redshift
 Quasars}.
\newblock {\em \apj} {\bf 2021}, {\em 910},~124.
\newblock {\url{https://doi.org/10.3847/1538-4357/abe404}}.

\bibitem[Kirkpatrick {et~al.}(2019)Kirkpatrick, Pope, Roebuck, Yan, Alberts,
 Herter, and Armus]{kirkpatricketal19}
Kirkpatrick, A.; Pope, A.; Roebuck, E.; Yan, L.; Alberts, S.; Herter, T.;
 Armus, L.
\newblock Cold Quasars: A New Phase in Quasar Evolution.
\newblock {\em Astrophys. J.} {\bf 2019}, {\em 879},~41.
\newblock {\url{https://doi.org/10.3847/1538-4357/ab20cc}}.

\bibitem[{Hamann} and {Ferland}(1993)]{hamannferland93}
{Hamann}, F.; {Ferland}, G.
\newblock {The Chemical Evolution of QSOs and the Implications for Cosmology
 and Galaxy Formation}.
\newblock {\em \apj} {\bf 1993}, {\em 418},~11.
\newblock {\url{https://doi.org/10.1086/173366}}.

\bibitem[{Hamann} and {Ferland}(1999)]{hamannferland99}
{Hamann}, F.; {Ferland}, G.
\newblock {Elemental Abundances in Quasistellar Objects: Star Formation and
 Galactic Nuclear Evolution at High Redshifts}.
\newblock {\em ARA\&A} {\bf 1999}, {\em 37},~487--531.
\newblock {\url{https://doi.org/10.1146/annurev.astro.37.1.487}}.

\bibitem[{Xu} {et~al.}(2018){Xu}, {Bian}, {Shen}, {Zuo}, {Fan}, and
 {Zhu}]{xuetal18}
{Xu}, F.; {Bian}, F.; {Shen}, Y.; {Zuo}, W.; {Fan}, X.; {Zhu}, Z.
\newblock {The evolution of chemical abundance in quasar broad line region}.
\newblock {\em \mnras} {\bf 2018}, {\em 480},~345--357.
\newblock {\url{https://doi.org/10.1093/mnras/sty1763}}.

\bibitem[{Shin} {et~al.}(2013){Shin}, {Woo}, {Nagao}, and {Kim}]{shinetal13}
{Shin}, J.; {Woo}, J.H.; {Nagao}, T.; {Kim}, S.C.
\newblock {The Chemical Properties of Low-redshift QSOs}.
\newblock {\em \apj} {\bf 2013}, {\em 763},~58.
\newblock {\url{https://doi.org/10.1088/0004-637X/763/1/58}}.

\bibitem[{{\'S}niegowska} {et~al.}(2021){{\'S}niegowska}, {Marziani},
 {Czerny}, {Panda}, {Mart{\'\i}nez-Aldama}, {del Olmo}, and
 {D'Onofrio}]{sniegowskaetal21}
{{\'S}niegowska}, M.; {Marziani}, P.; {Czerny}, B.; {Panda}, S.;
 {Mart{\'\i}nez-Aldama}, M.L.; {del Olmo}, A.; {D'Onofrio}, M.
\newblock {High Metal Content of Highly Accreting Quasars}.
\newblock {\em \apj} {\bf 2021}, {\em 910},~115.
\newblock {\url{https://doi.org/10.3847/1538-4357/abe1c8}}.

\bibitem[{Shields}(1976)]{shields76}
{Shields}, G.A.
\newblock {The abundance of nitrogen in QSOs.}
\newblock {\em \apj} {\bf 1976}, {\em 204},~330--336.
\newblock {\url{https://doi.org/10.1086/154176}}.

\bibitem[{Nagao} {et~al.}(2006){Nagao}, {Marconi}, and
 {Maiolino}]{nagaoetal06}
{Nagao}, T.; {Marconi}, A.; {Maiolino}, R.
\newblock {The evolution of the broad-line region among SDSS quasars}.
\newblock {\em A\&Ap} {\bf 2006}, {\em 447},~157--172.
\newblock {\url{https://doi.org/10.1051/0004-6361:20054024}}.

\bibitem[{Huang} {et~al.}(2023){Huang}, {Lin}, and {Shields}]{huangetal23}
{Huang}, J.; {Lin}, D.N.C.; {Shields}, G.
\newblock {Metal enrichment due to embedded stars in AGN discs}.
\newblock {\em \mnras} {\bf 2023}, {\em 525},~5702--5718.
\newblock {\url{https://doi.org/10.1093/mnras/stad2642}}.

\bibitem[{Garnica} {et~al.}(2022){Garnica}, {Negrete}, {Marziani},
 {Dultzin}, {{\'S}niegowska}, and {Panda}]{garnicaetal22}
{Garnica}, K.; {Negrete}, C.A.; {Marziani}, P.; {Dultzin}, D.;
 {{\'S}niegowska}, M.; {Panda}, S.
\newblock {High metal content of highly accreting quasars: Analysis of an
 extended sample}.
\newblock {\em \aap} {\bf 2022}, {\em 667},~A105.
\newblock {\url{https://doi.org/10.1051/0004-6361/202142837}}.

\bibitem[{Hubeny} {et~al.}(2001){Hubeny}, {Blaes}, {Krolik}, and
 {Agol}]{hubenyetal01}
{Hubeny}, I.; {Blaes}, O.; {Krolik}, J.H.; {Agol}, E.
\newblock {Non-LTE Models and Theoretical Spectra of Accretion Disks in Active
 Galactic Nuclei. IV. Effects of Compton Scattering and Metal Opacities}.
\newblock {\em \apj} {\bf 2001}, {\em 559},~680--702.
\newblock {\url{https://doi.org/10.1086/322344}}.

\bibitem[{Dumont} {et~al.}(2003){Dumont}, {Collin}, {Paletou}, {Coup{\'e}},
 {Godet}, and {Pelat}]{dumontetal03}
{Dumont}, A.M.; {Collin}, S.; {Paletou}, F.; {Coup{\'e}}, S.; {Godet}, O.;
 {Pelat}, D.
\newblock {Escape probability methods versus ``exact'' transfer for modelling
 the X-ray spectrum of Active Galactic Nuclei and X-ray binaries}.
\newblock {\em \aap} {\bf 2003}, {\em 407},~13--30.
\newblock {\url{https://doi.org/10.1051/0004-6361:20030890}}.

\bibitem[{Krolik}(1999)]{krolik99}
{Krolik}, J.H.
\newblock {\em {Active Galactic Nuclei: From the Central Black Hole to the Galactic Environment}}; 
{Princeton University Press: Princeton, NJ, USA}, 1999.

\bibitem[{Osterbrock} and {Ferland}(2006)]{osterbrockferland06}
{Osterbrock}, D.E.; {Ferland}, G.J.
\newblock {\em Astrophysics of Gaseous Nebulae and Active Galactic Nuclei};
 University Science Books: Mill Valley, CA, 2006.

\bibitem[{Pradhan} and {Nahar}(2015)]{pradhannahar15}
{Pradhan}, A.K.; {Nahar}, S.N.
\newblock {\em {Atomic Astrophysics and Spectroscopy}}; Cambridge University
 Press: Cambridge, UK, 2015.

\bibitem[{Castor} {et~al.}(1975){Castor}, {Abbott}, and
 {Klein}]{castoretal75}
{Castor}, J.I.; {Abbott}, D.C.; {Klein}, R.I.
\newblock {Radiation-driven winds in Of stars}.
\newblock {\em \apj} {\bf 1975}, {\em 195},~157--174.
\newblock {\url{https://doi.org/10.1086/153315}}.

\bibitem[{Murray} {et~al.}(1995){Murray}, {Chiang}, {Grossman}, and
 {Voit}]{murrayetal95}
{Murray}, N.; {Chiang}, J.; {Grossman}, S.A.; {Voit}, G.M.
\newblock {Accretion Disk Winds from Active Galactic Nuclei}.
\newblock {\em \apj} {\bf 1995}, {\em 451},~498.
\newblock {\url{https://doi.org/10.1086/176238}}.

\bibitem[{Marziani} {et~al.}(2016){Marziani}, {Sulentic}, {Stirpe},
 {Dultzin}, {Del Olmo}, and {Mart{\'{\i}}nez-Carballo}]{marzianietal16}
{Marziani}, P.; {Sulentic}, J.W.; {Stirpe}, G.M.; {Dultzin}, D.; {Del Olmo},
 A.; {Mart{\'{\i}}nez-Carballo}, M.A.
\newblock {Blue outliers among intermediate redshift quasars}.
\newblock {\em \apss} {\bf 2016}, {\em 361},~3.
\newblock {\url{https://doi.org/10.1007/s10509-015-2590-2}}.

\bibitem[{Juarez} {et~al.}(2009){Juarez}, {Maiolino}, {Mujica}, {Pedani},
 {Marinoni}, {Nagao}, {Marconi}, and {Oliva}]{juarezetal09}
{Juarez}, Y.; {Maiolino}, R.; {Mujica}, R.; {Pedani}, M.; {Marinoni}, S.;
 {Nagao}, T.; {Marconi}, A.; {Oliva}, E.
\newblock {The metallicity of the most distant quasars}.
\newblock {\em A\&Ap} {\bf 2009}, {\em 494},~L25--L28.
\newblock {\url{https://doi.org/10.1051/0004-6361:200811415}}.

\bibitem[Yang {et~al.}(2021)Yang, Wang, Fan, Barth, Hennawi, Nanni, Bian,
 Davies, Farina, Schindler, Ba{\~n}ados, Decarli, Eilers, Green, Guo, Jiang,
 Li, Venemans, Walter, Wu, and Yue]{yangetal21}
Yang, J.; Wang, F.; Fan, X.; Barth, A.J.; Hennawi, J.F.; Nanni, R.; Bian, F.;
 Davies, F.B.; Farina, E.P.; Schindler, J.T.; et~al.
\newblock Probing Early Supermassive Black Hole Growth and Quasar Evolution
 with Near-infrared Spectroscopy of 37 Reionization-era Quasars at $6.3 < z <
 7.64$.
\newblock {\em Astrophys. J.} {\bf 2021}, {\em 923},~262.
\newblock {\url{https://doi.org/10.3847/1538-4357/ac2b32}}.

\bibitem[{Collin} and {Zahn}(1999)]{collinzahn99}
{Collin}, S.; {Zahn}, J.P.
\newblock {Star formation and evolution in accretion disks around massive black
 holes.}
\newblock {\em A\&Ap} {\bf 1999}, {\em 344},~433--449.

\bibitem[{Wang} {et~al.}(2023){Wang}, {Zhai}, {Li}, {Songsheng}, {Ho},
 {Chen}, {Liu}, {Du}, and {Yuan}]{wangetal23}
{Wang}, J.M.; {Zhai}, S.; {Li}, Y.R.; {Songsheng}, Y.Y.; {Ho}, L.C.; {Chen},
 Y.J.; {Liu}, J.R.; {Du}, P.; {Yuan}, Y.F.
\newblock {Star Formation in Self-gravitating Disks in Active Galactic Nuclei.
 III. Efficient Production of Iron and Infrared Spectral Energy
 Distributions}.
\newblock {\em \apj} {\bf 2023}, {\em 954},~84.
\newblock {\url{https://doi.org/10.3847/1538-4357/acdf48}}.

\bibitem[{Fan} and {Wu}(2023)]{fanwu23}
{Fan}, X.; {Wu}, Q.
\newblock {In Situ Star Formation in Accretion Disks and Explanation of
 Correlation between the Black Hole Mass and Metallicity in Active Galactic
 Nuclei}.
\newblock {\em \apj} {\bf 2023}, {\em 944},~159.
\newblock {\url{https://doi.org/10.3847/1538-4357/acb532}}.

\bibitem[{Loiacono} {et~al.}(2024){Loiacono}, {Decarli}, {Mignoli},
 {Farina}, {Ba{\~n}ados}, {Bosman}, {Eilers}, {Schindler}, {Strauss},
 {Vestergaard}, {Wang}, {Blecha}, {Carilli}, {Comastri}, {Connor}, {Costa},
 {Dotti}, {Fan}, {Gilli}, {Jun}, {Liu}, {Lupi}, {Marshall}, {Mazzucchelli},
 {Meyer}, {Neeleman}, {Overzier}, {Pensabene}, {Riechers}, {Trakhtenbrot},
 {Trebitsch}, {Venemans}, {Walter}, and {Yang}]{loiaconoetal24}
{Loiacono}, F.; {Decarli}, R.; {Mignoli}, M.; {Farina}, E.P.; {Ba{\~n}ados},
 E.; {Bosman}, S.; {Eilers}, A.C.; {Schindler}, J.T.; {Strauss}, M.A.;
 {Vestergaard}, M.; et~al.
\newblock {A quasar-galaxy merger at z {\ensuremath{\sim}} 6.2: Black hole mass
 and quasar properties from the NIRSpec spectrum}.
\newblock {\em \aap} {\bf 2024}, {\em 685},~A121.
\newblock {\url{https://doi.org/10.1051/0004-6361/202348535}}.

\bibitem[{Storchi-Bergmann} {et~al.}(1998){Storchi-Bergmann}, {Schmitt},
 {Calzetti}, and {Kinney}]{storchi-bergmannetal98}
{Storchi-Bergmann}, T.; {Schmitt}, H.R.; {Calzetti}, D.; {Kinney}, A.L.
\newblock {Chemical Abundance Calibrations for the Narrow-Line Region of Active
 Galaxies}.
\newblock {\em \aj} {\bf 1998}, {\em 115},~909--914.
\newblock {\url{https://doi.org/10.1086/300242}}.

\bibitem[{Nagao} {et~al.}(2006){Nagao}, {Maiolino}, and
 {Marconi}]{nagaoetal06b}
{Nagao}, T.; {Maiolino}, R.; {Marconi}, A.
\newblock {Gas metallicity in the narrow-line regions of high-redshift active galactic nuclei}.
\newblock {\em \aap} {\bf 2006}, {\em 447},~863--876.
\newblock {\url{https://doi.org/10.1051/0004-6361:20054127}}.

\bibitem[{Armah} {et~al.}(2023){Armah}, {Riffel}, {Dors}, {Oh}, {Koss},
 {Ricci}, {Trakhtenbrot}, {Valerdi}, {Riffel}, and {Krabbe}]{armahetal23}
{Armah}, M.; {Riffel}, R.; {Dors}, O.L.; {Oh}, K.; {Koss}, M.J.; {Ricci}, C.;
 {Trakhtenbrot}, B.; {Valerdi}, M.; {Riffel}, R.A.; {Krabbe}, A.C.
\newblock {Oxygen abundances in the narrow line regions of Seyfert galaxies and
 the metallicity-luminosity relation}.
\newblock {\em \mnras} {\bf 2023}, {\em 520},~1687--1703.
\newblock {\url{https://doi.org/10.1093/mnras/stad217}}.

\bibitem[{Sanders} {et~al.}(1988){Sanders}, {Soifer}, {Elias}, {Madore},
 {Matthews}, {Neugebauer}, and {Scoville}]{sandersetal88}
{Sanders}, D.B.; {Soifer}, B.T.; {Elias}, J.H.; {Madore}, B.F.; {Matthews}, K.;
 {Neugebauer}, G.; {Scoville}, N.Z.
\newblock {Ultraluminous infrared galaxies and the origin of quasars}.
\newblock {\em \apj} {\bf 1988}, {\em 325},~74--91.
\newblock {\url{https://doi.org/10.1086/165983}}.

\bibitem[{Rafanelli} and {Marziani}(1992)]{rafanellimarziani92}
{Rafanelli}, P.; {Marziani}, P.
\newblock {The complex nature of the interacting system NGC 7592}.
\newblock {\em AJ} {\bf 1992}, {\em 103},~743--756.
\newblock {\url{https://doi.org/10.1086/116098}}.

\bibitem[{Marziani} {et~al.}(1993){Marziani}, {Sulentic}, {Calvani},
 {Perez}, {Moles}, and {Penston}]{marzianietal93}
{Marziani}, P.; {Sulentic}, J.W.; {Calvani}, M.; {Perez}, E.; {Moles}, M.;
 {Penston}, M.V.
\newblock {The peculiar Balmer line profiles of OQ 208}.
\newblock {\em ApJ} {\bf 1993}, {\em 410},~56--67.
\newblock {\url{https://doi.org/10.1086/172724}}.

\bibitem[{Heckman} {et~al.}(1997){Heckman}, {Gonz{\'a}lez-Delgado},
 {Leitherer}, {Meurer}, {Krolik}, {Wilson}, {Koratkar}, and
 {Kinney}]{heckmanetal97}
{Heckman}, T.M.; {Gonz{\'a}lez-Delgado}, R.; {Leitherer}, C.; {Meurer}, G.R.;
 {Krolik}, J.; {Wilson}, A.S.; {Koratkar}, A.; {Kinney}, A.
\newblock {A Powerful Nuclear Starburst in the Seyfert Galaxy Markarian 477:
 Implications for the Starburst-Active Galactic Nucleus Connection}.
\newblock {\em \apj} {\bf 1997}, {\em 482},~114--132.
\newblock {\url{https://doi.org/10.1086/304139}}.

\bibitem[{Dultzin-Hacyan} {et~al.}(2003){Dultzin-Hacyan}, {Krongold}, and
 {Marziani}]{dultzin-hacyanetal03}
{Dultzin-Hacyan}, D.; {Krongold}, Y.; {Marziani}, P.
\newblock {The Environment of AGN and AN Evolutionary Scheme}.
\newblock In \emph{Revista Mexicana de Astronomia y Astrofisica}; 
 Conference Series; 
{Reyes-Ruiz}, M., {V{\'a}zquez-Semadeni}, E., Eds.; 
 {Instituto de Astronomía: Distrito Federal, México}, 2003; 
Volume~18, p. 147.

\bibitem[{Schweitzer} {et~al.}(2006){Schweitzer}, {Lutz}, {Sturm},
 {Contursi}, {Tacconi}, {Lehnert}, {Dasyra}, {Genzel}, {Veilleux}, {Rupke},
 {Kim}, {Baker}, {Netzer}, {Sternberg}, {Mazzarella}, and
 {Lord}]{schweitzeretal06}
{Schweitzer}, M.; {Lutz}, D.; {Sturm}, E.; {Contursi}, A.; {Tacconi}, L.J.;
 {Lehnert}, M.D.; {Dasyra}, K.M.; {Genzel}, R.; {Veilleux}, S.; {Rupke}, D.;
 et~al.
\newblock {Spitzer Quasar and ULIRG Evolution Study (QUEST). I. The Origin of
 the Far-Infrared Continuum of QSOs}.
\newblock {\em \apj} {\bf 2006}, {\em 649},~79--90.
\newblock {\url{https://doi.org/10.1086/506510}}.

\bibitem[{Wild} {et~al.}(2010){Wild}, {Heckman}, and {Charlot}]{wildetal10}
{Wild}, V.; {Heckman}, T.; {Charlot}, S.
\newblock {Timing the starburst-AGN connection}.
\newblock {\em \mnras} {\bf 2010}, {\em 405},~933--947.
\newblock {\url{https://doi.org/10.1111/j.1365-2966.2010.16536.x}}.

\bibitem[{Rafanelli} {et~al.}(2011){Rafanelli}, {La Mura}, {Bindoni},
 {Ciroi}, {Cracco}, {Di Mille}, and {Vaona}]{rafanellietal11}
{Rafanelli}, P.; {La Mura}, G.; {Bindoni}, D.; {Ciroi}, S.; {Cracco}, V.; {Di
 Mille}, F.; {Vaona}, L.
\newblock {The Starburst---AGN Connection: A Critical Review}.
\newblock {\em Balt. Astron.} {\bf 2011}, {\em 20},~419--426.
\newblock {\url{https://doi.org/10.1515/astro-2017-0313}}.

\bibitem[{Lipari} {et~al.}(1993){Lipari}, {Terlevich}, and
 {Macchetto}]{liparietal93}
{Lipari}, S.; {Terlevich}, R.; {Macchetto}, F.
\newblock {Extreme optical Fe II emission in luminous IRAS active galactic
 nuclei}.
\newblock {\em \apj} {\bf 1993}, {\em 406},~451--456.
\newblock {\url{https://doi.org/10.1086/172456}}.

\bibitem[{Lipari}(1994)]{lipari94}
{Lipari}, S.
\newblock {Galaxies with Extreme Infrared and Fe II Emission. II. IRAS
 07598+6508: A Starburst/Young Broad Absorption Line QSO}.
\newblock {\em \apj} {\bf 1994}, {\em 436},~102.
\newblock {\url{https://doi.org/10.1086/174884}}.

\bibitem[{Haas} {et~al.}(2003){Haas}, {Klaas}, {M{\"u}ller}, {Bertoldi},
 {Camenzind}, {Chini}, {Krause}, {Lemke}, {Meisenheimer}, {Richards}, and
 {Wilkes}]{haasetal03}
{Haas}, M.; {Klaas}, U.; {M{\"u}ller}, S.A.H.; {Bertoldi}, F.; {Camenzind}, M.;
 {Chini}, R.; {Krause}, O.; {Lemke}, D.; {Meisenheimer}, K.; {Richards}, P.J.;
 et~al.
\newblock {The ISO view of Palomar-Green quasars}.
\newblock {\em \aap} {\bf 2003}, {\em 402},~87--111.
\newblock {\url{https://doi.org/10.1051/0004-6361:20030110}}.

\bibitem[{Krongold} {et~al.}(2003){Krongold}, {Dultzin-Hacyan}, and
 {Marziani}]{dultzin-hacyanetal03a}
{Krongold}, Y.; {Dultzin-Hacyan}, D.; {Marziani}, P.
\newblock {An Evolutionary Sequence for AGN}.
\newblock In Proceedings of the Active Galactic Nuclei: From Central Engine to Host Galaxy; 
 {Collin}, S., {Combes}, F., {Shlosman}, I., Eds.; 
 {Astronomical Society of the Pacific Conference Series}; 
{Astronomical Society of the Pacific: San Francisco, CA, USA}, 2003; 
Volume 290, p. 523.

\bibitem[{L{\'\i}pari} and {Terlevich}(2006)]{liparietal06}
{L{\'\i}pari}, S.L.; {Terlevich}, R.J.
\newblock {Evolutionary unification in composite active galactic nuclei}.
\newblock {\em \mnras} {\bf 2006}, {\em 368},~1001--1015.
\newblock {\url{https://doi.org/10.1111/j.1365-2966.2006.10215.x}}.

\bibitem[{Wang} {et~al.}(2009){Wang}, {Yan}, {Li}, {Chen}, {Xiang}, {Hu},
 {Ge}, and {Zhang}]{wangetal09sf}
{Wang}, J.M.; {Yan}, C.S.; {Li}, Y.R.; {Chen}, Y.M.; {Xiang}, F.; {Hu}, C.;
 {Ge}, J.Q.; {Zhang}, S.
\newblock {Evolution of Gaseous Disk Viscosity Driven by Supernova Explosions
 in Star-Forming Galaxies at High Redshift}.
\newblock {\em \apjl} {\bf 2009}, {\em 701},~L7--L11.
\newblock {\url{https://doi.org/10.1088/0004-637X/701/1/L7}}.

\bibitem[{Wang} {et~al.}(2010){Wang}, {Deng}, and {Wei}]{wangetal10sf}
{Wang}, J.; {Deng}, J.S.; {Wei}, J.Y.
\newblock {Ongoing star formation in AGN host galaxy discs: A view from
 core-collapse supernovae}.
\newblock {\em \mnras} {\bf 2010}, {\em 405},~2529--2533.
\newblock {\url{https://doi.org/10.1111/j.1365-2966.2010.16629.x}}.

\bibitem[{Wang} {et~al.}(2011){Wang}, {Ge}, {Hu}, {Baldwin}, {Li},
 {Ferland}, {Xiang}, {Yan}, and {Zhang}]{wangetal11sf}
{Wang}, J.M.; {Ge}, J.Q.; {Hu}, C.; {Baldwin}, J.A.; {Li}, Y.R.; {Ferland},
 G.J.; {Xiang}, F.; {Yan}, C.S.; {Zhang}, S.
\newblock {Star Formation in Self-gravitating Disks in Active Galactic Nuclei.
 I. Metallicity Gradients in Broad-line Regions}.
\newblock {\em \apj} {\bf 2011}, {\em 739},~3.
\newblock {\url{https://doi.org/10.1088/0004-637X/739/1/3}}.

\bibitem[{Wang} {et~al.}(2012){Wang}, {Du}, {Baldwin}, {Ge}, {Hu}, and
 {Ferland}]{wangetal12sf}
{Wang}, J.M.; {Du}, P.; {Baldwin}, J.A.; {Ge}, J.Q.; {Hu}, C.; {Ferland}, G.J.
\newblock {Star Formation in Self-gravitating Disks in Active Galactic Nuclei.
 II. Episodic Formation of Broad-line Regions}.
\newblock {\em \apj} {\bf 2012}, {\em 746},~137.
\newblock {\url{https://doi.org/10.1088/0004-637X/746/2/137}}.

\bibitem[{Elitzur} and {Shlosman}(2006)]{elitzurshlosman06}
{Elitzur}, M.; {Shlosman}, I.
\newblock {The AGN-obscuring Torus: The End of the ``Doughnut'' Paradigm?}
\newblock {\em \apjl} {\bf 2006}, {\em 648},~L101--L104.
\newblock {\url{https://doi.org/10.1086/508158}}.

\bibitem[{Kang} {et~al.}(2024){Kang}, {Done}, {Hagen}, {Temple},
 {Silverman}, {Li}, and {Liu}]{kangetal24}
{Kang}, J.L.; {Done}, C.; {Hagen}, S.; {Temple}, M.J.; {Silverman}, J.D.; {Li},
 J.; {Liu}, T.
\newblock {Systematic collapse of the accretion disc in AGN confirmed by UV
 photometry and broad line spectra}.
\newblock {\em arXiv} {\bf 2024}, arXiv:2410.06730.
\newblock {\url{https://doi.org/10.48550/arXiv.2410.06730}}.

\bibitem[{Artymowicz} {et~al.}(1993){Artymowicz}, {Lin}, and
 {Wampler}]{artymowiczetal93}
{Artymowicz}, P.; {Lin}, D.N.C.; {Wampler}, E.J.
\newblock {Star Trapping and Metallicity Enrichment in Quasars and Active
 Galactic Nuclei}.
\newblock {\em \apj} {\bf 1993}, {\em 409},~592.
\newblock {\url{https://doi.org/10.1086/172690}}.

\bibitem[{Lin}(1997)]{lin97}
{Lin}, D.N.C.
\newblock {Star/Disk Interaction in the Nuclei of Active Galaxies}.
\newblock In \emph{Proceedings of the IAU Colloq. 159: Emission Lines in Active Galaxies: New Methods and Techniques}; 
 {Peterson}, B.M., {Cheng}, F.Z., {Wilson}, A.S., Eds.; 
 {Astronomical Society of the Pacific Conference Series};
{Astronomical Society of the Pacific: San Francisco, CA, USA} 1997; 
 Volume 113, p.~64.

\bibitem[{Dittmann} and {Cantiello}(2024)]{dittmanncantiello24}
{Dittmann}, A.J.; {Cantiello}, M.
\newblock {A Semi-Analytical Model for Stellar Evolution in AGN Disks}.
\newblock {\em arXiv} {\bf 2024}, arXiv:2409.02981.
\newblock {\url{https://doi.org/10.48550/arXiv.2409.02981}}.

\bibitem[{Fabj} {et~al.}(2024){Fabj}, {Dittmann}, {Cantiello}, {Perna}, and
 {Samsing}]{fabjetal24}
{Fabj}, G.; {Dittmann}, A.J.; {Cantiello}, M.; {Perna}, R.; {Samsing}, J.
\newblock {Mapping the Outcomes of Stellar Evolution in the Disks of Active
 Galactic Nuclei}.
\newblock {\em arXiv} {\bf 2024}, arXiv:2408.16050.
\newblock {\url{https://doi.org/10.48550/arXiv.2408.16050}}.

\bibitem[{Liu} {et~al.}(2024){Liu}, {Wang}, and {Wang}]{liuetal24}
{Liu}, J.R.; {Wang}, Y.L.; {Wang}, J.M.
\newblock {Accretion-modified Stars in Accretion Disks of Active Galactic
 Nuclei: Observational Characteristics in Different Regions of the Disks}.
\newblock {\em \apj} {\bf 2024}, {\em 969},~37.
\newblock {\url{https://doi.org/10.3847/1538-4357/ad463a}}.

\bibitem[{Nayakshin} {et~al.}(2007){Nayakshin}, {Cuadra}, and
 {Springel}]{nayakshinetal07}
{Nayakshin}, S.; {Cuadra}, J.; {Springel}, V.
\newblock {Simulations of star formation in a gaseous disc around Sgr A*---A
 failed active galactic nucleus}.
\newblock {\em \mnras} {\bf 2007}, {\em 379},~21--33.
\newblock {\url{https://doi.org/10.1111/j.1365-2966.2007.11938.x}}.

\bibitem[{Alexander} and {Livio}(2001)]{alexanderlivio01a}
{Alexander}, T.; {Livio}, M.
\newblock {Tidal Scattering of Stars on Supermassive Black Holes in Galactic
 Centers}.
\newblock {\em \apjl} {\bf 2001}, {\em 560},~L143--L146.
\newblock {\url{https://doi.org/10.1086/324324}}.

\bibitem[{Alexander} and {Livio}(2004)]{alexanderlivio01b}
{Alexander}, T.; {Livio}, M.
\newblock {Orbital Capture of Stars by a Massive Black Hole via Exchanges with
 Compact Remnants}.
\newblock {\em \apjl} {\bf 2004}, {\em 606},~L21--L24.
\newblock {\url{https://doi.org/10.1086/421112}}.

\bibitem[{Davies} and {Lin}(2020)]{davieslin20}
{Davies}, M.B.; {Lin}, D.N.C.
\newblock {Making massive stars in the Galactic Centre via accretion on to
 low-mass stars within an accretion disc}.
\newblock {\em \mnras} {\bf 2020}, {\em 498},~3452--3456.
\newblock {\url{https://doi.org/10.1093/mnras/staa2590}}.

\bibitem[{Jermyn} {et~al.}(2022){Jermyn}, {Dittmann}, {McKernan}, {Ford},
 and {Cantiello}]{jermynetal22}
{Jermyn}, A.S.; {Dittmann}, A.J.; {McKernan}, B.; {Ford}, K.E.S.; {Cantiello},
 M.
\newblock {Effects of an Immortal Stellar Population in AGN Disks}.
\newblock {\em \apj} {\bf 2022}, {\em 929},~133.
\newblock {\url{https://doi.org/10.3847/1538-4357/ac5d40}}.

\bibitem[{Dittmann} {et~al.}(2023){Dittmann}, {Jermyn}, and
 {Cantiello}]{dittmannetal23}
{Dittmann}, A.J.; {Jermyn}, A.S.; {Cantiello}, M.
\newblock {The Influence of Disk Composition on the Evolution of Stars in the
 Disks of Active Galactic Nuclei}.
\newblock {\em \apj} {\bf 2023}, {\em 946},~56.
\newblock {\url{https://doi.org/10.3847/1538-4357/acacf2}}.

\bibitem[{Nasim} {et~al.}(2023){Nasim}, {Fabj}, {Caban}, {Secunda}, {Ford},
 {McKernan}, {Bellovary}, {Leigh}, and {Lyra}]{nasimetal23}
{Nasim}, S.S.; {Fabj}, G.; {Caban}, F.; {Secunda}, A.; {Ford}, K.E.S.;
 {McKernan}, B.; {Bellovary}, J.M.; {Leigh}, N.W.C.; {Lyra}, W.
\newblock {Aligning Retrograde Nuclear Cluster Orbits with an Active Galactic
 Nucleus Accretion Disc}.
\newblock {\em \mnras} {\bf 2023}, {\em 522},~5393--5401.
\newblock {\url{https://doi.org/10.1093/mnras/stad1295}}.

\bibitem[{Clayton}(1968)]{clayton68}
{Clayton}, D.D.
\newblock {\em {Principles of Stellar Evolution and Nucleosynthesis}}; 
{The University of Chicago Press: Chicago, IL, USA}, 1968.

\bibitem[{Napolitano} {et~al.}(2024){Napolitano}, {Castellano},
 {Pentericci}, {Vignali}, {Gilli}, {Fontana}, {Santini}, {Treu},
 {Calabr{\`o}}, {Llerena}, {Piconcelli}, {Zappacosta}, {Mascia}, {Bergamini},
 {Bakx}, {Dickinson}, {Glazebrook}, {Henry}, {Leethochawalit}, {Mazzolari},
 {Merlin}, {Morishita}, {Nanayakkara}, {Paris}, {Puccetti}, {Roberts-Borsani},
 {Rojas Ruiz}, {Vanzella}, {Vito}, {Vulcani}, {Wang}, {Yoon}, and
 {Zavala}]{napolitanoetal24}
{Napolitano}, L.; {Castellano}, M.; {Pentericci}, L.; {Vignali}, C.; {Gilli},
 R.; {Fontana}, A.; {Santini}, P.; {Treu}, T.; {Calabr{\`o}}, A.; {Llerena},
 M.; et~al.
\newblock {The dual nature of GHZ9: Coexisting AGN and star formation activity
 in a remote X-ray source at z = 10.145}.
\newblock {\em arXiv} {\bf 2024}, arXiv:2410.18763.
\newblock {\url{https://doi.org/10.48550/arXiv.2410.18763}}.

\bibitem[{Zhou} {et~al.}(2002){Zhou}, {Wang}, {Zhou}, {Cheng-Li}, and
 {Dong}]{zhouetal02}
{Zhou}, H.Y.; {Wang}, T.G.; {Zhou}, Y.Y.; {Li}, C.; {Dong}, X.B.
\newblock {Discovery of a Radio-loud Superstrong Fe II-emitting Quasar}.
\newblock {\em \apj} {\bf 2002}, {\em 581},~96--102.
\newblock {\url{https://doi.org/10.1086/344201}}.

\bibitem[{Sani} {et~al.}(2010){Sani}, {Lutz}, {Risaliti}, {Netzer}, {Gallo},
 {Trakhtenbrot}, {Sturm}, and {Boller}]{sanietal10}
{Sani}, E.; {Lutz}, D.; {Risaliti}, G.; {Netzer}, H.; {Gallo}, L.C.;
 {Trakhtenbrot}, B.; {Sturm}, E.; {Boller}, T.
\newblock {Enhanced star formation in narrow-line Seyfert 1 active galactic
 nuclei revealed by Spitzer}.
\newblock {\em MNRAS} {\bf 2010}, {\em 403},~1246--1260.
\newblock {\url{https://doi.org/10.1111/j.1365-2966.2009.16217.x}}.

\bibitem[{Caccianiga} {et~al.}(2015){Caccianiga}, {Ant{\'o}n}, {Ballo},
 {Foschini}, {Maccacaro}, {Della Ceca}, {Severgnini}, {March{\~a}}, {Mateos},
 and {Sani}]{caccianigaetal15}
{Caccianiga}, A.; {Ant{\'o}n}, S.; {Ballo}, L.; {Foschini}, L.; {Maccacaro},
 T.; {Della Ceca}, R.; {Severgnini}, P.; {March{\~a}}, M.J.; {Mateos}, S.;
 {Sani}, E.
\newblock {WISE colours and star formation in the host galaxies of radio-loud
 narrow-line Seyfert 1}.
\newblock {\em \mnras} {\bf 2015}, {\em 451},~1795--1805.
\newblock {\url{https://doi.org/10.1093/mnras/stv939}}.

\bibitem[{Laor} and {Behar}(2008)]{laorbehar08}
{Laor}, A.; {Behar}, E.
\newblock {On the origin of radio emission in radio-quiet quasars}.
\newblock {\em \mnras} {\bf 2008}, {\em 390},~847--862.
\newblock {\url{https://doi.org/10.1111/j.1365-2966.2008.13806.x}}.

\bibitem[{Laor} {et~al.}(2019){Laor}, {Baldi}, and {Behar}]{laoretal19}
{Laor}, A.; {Baldi}, R.D.; {Behar}, E.
\newblock {What drives the radio slopes in radio-quiet quasars?}
\newblock {\em \mnras} {\bf 2019}, {\em 482},~5513--5523.
\newblock {\url{https://doi.org/10.1093/mnras/sty3098}}.

\bibitem[{Panessa} {et~al.}(2019){Panessa}, {Baldi}, {Laor}, {Padovani},
 {Behar}, and {McHardy}]{panessaetal19}
{Panessa}, F.; {Baldi}, R.D.; {Laor}, A.; {Padovani}, P.; {Behar}, E.;
 {McHardy}, I.
\newblock {The origin of radio emission from radio-quiet active galactic
 nuclei}.
\newblock {\em Nat. Astron.} {\bf 2019}, {\em 3},~387--396.
\newblock {\url{https://doi.org/10.1038/s41550-019-0765-4}}.

\bibitem[{Yang} {et~al.}(2020){Yang}, {Yao}, {Yang}, {Ho}, {An}, {Wang},
 {Baan}, {Gu}, {Liu}, {Yang}, and {Joshi}]{yangetal20}
{Yang}, X.; {Yao}, S.; {Yang}, J.; {Ho}, L.C.; {An}, T.; {Wang}, R.; {Baan},
 W.A.; {Gu}, M.; {Liu}, X.; {Yang}, X.; et~al.
\newblock {Radio Activity of Supermassive Black Holes with Extremely High
 Accretion Rates}.
\newblock {\em \apj} {\bf 2020}, {\em 904},~200.
\newblock {\url{https://doi.org/10.3847/1538-4357/abb775}}.

\bibitem[{Chen} {et~al.}(2023){Chen}, {Laor}, {Behar}, {Baldi}, and
 {Gelfand}]{chenetal23}
{Chen}, S.; {Laor}, A.; {Behar}, E.; {Baldi}, R.D.; {Gelfand}, J.D.
\newblock {The radio emission in radio-quiet quasars: The VLBA perspective}.
\newblock {\em \mnras} {\bf 2023}, {\em 525},~164--182.
\newblock {\url{https://doi.org/10.1093/mnras/stad2289}}.

\bibitem[{Elston} {et~al.}(1994){Elston}, {Thompson}, and
 {Hill}]{elstonetal94}
{Elston}, R.; {Thompson}, K.L.; {Hill}, G.J.
\newblock {Detection of strong iron emission from quasars at redshift z > 3}.
\newblock {\em \nat} {\bf 1994}, {\em 367},~250--251.
\newblock {\url{https://doi.org/10.1038/367250a0}}.

\bibitem[{Chieffi} and {Limongi}(2013)]{chieffilimongi13}
{Chieffi}, A.; {Limongi}, M.
\newblock {Pre-supernova Evolution of Rotating Solar Metallicity Stars in the
 Mass Range 13--120 M $_{{\ensuremath{\odot}}}$ and their Explosive Yields}.
\newblock {\em \apj} {\bf 2013}, {\em 764},~21.
\newblock {\url{https://doi.org/10.1088/0004-637X/764/1/21}}.

\bibitem[{Wang} {et~al.}(2009){Wang}, {Zhou}, {Yuan}, {Lu}, {Dong}, and
 {Shan}]{wangetal09a}
{Wang}, T.; {Zhou}, H.; {Yuan}, W.; {Lu}, H.L.; {Dong}, X.; {Shan}, H.
\newblock {Metal-Enriched Outflows in the Ultraluminous Infrared Quasar
 Q1321+058}.
\newblock {\em \apj} {\bf 2009}, {\em 702},~851--861.
\newblock {\url{https://doi.org/10.1088/0004-637X/702/2/851}}.

\bibitem[{Marziani} {et~al.}(2024){Marziani}, {Floris}, {Deconto--Machado},
 {Panda}, {Sniegowska}, {Garnica}, {Dultzin}, {D'Onofrio}, {Del Olmo}, {Bon},
 and {Bon}]{marzianietal24}
{Marziani}, P.; {Floris}, A.; {Deconto--Machado}, A.; {Panda}, S.; {Sniegowska},
 M.; {Garnica}, K.; {Dultzin}, D.; {D'Onofrio}, M.; {Del Olmo}, A.; {Bon}, E.;
 et~al.
\newblock {From Sub-Solar to Super-Solar Chemical Abundances along the Quasar
 Main Sequence}.
\newblock {\em Physics} {\bf 2024}, {\em 6},~216--236.
\newblock {\url{https://doi.org/10.3390/physics6010016}}.

\bibitem[{Drewes} {et~al.}(2025){Drewes}, {Leftley}, {H{\"o}nig},
 {Tristram}, and {Kishimoto}]{drewesetal25}
{Drewes}, F.; {Leftley}, J.H.; {H{\"o}nig}, S.F.; {Tristram}, K.R.W.;
 {Kishimoto}, M.
\newblock {I Zw 1 and H0557-385: The dusty tori of two high Eddington AGNs
 observed in the MATISSE LM-Bands}.
\newblock {\em \mnras} {\bf 2025}, \emph{537}, 1369--1384.
\newblock {\url{https://doi.org/10.1093/mnras/staf110}}.

\bibitem[{Czerny} and {Hryniewicz}(2011)]{czernyhryniewicz11}
{Czerny}, B.; {Hryniewicz}, K.
\newblock {The origin of the broad line region in active galactic nuclei}.
\newblock {\em \aap} {\bf 2011}, {\em 525},~L8.
\newblock {\url{https://doi.org/10.1051/0004-6361/201016025}}.

\bibitem[{Berton} {et~al.}(2015){Berton}, {Foschini}, {Ciroi}, {Cracco}, {La
 Mura}, {Lister}, {Mathur}, {Peterson}, {Richards}, and
 {Rafanelli}]{bertonetal15}
{Berton}, M.; {Foschini}, L.; {Ciroi}, S.; {Cracco}, V.; {La Mura}, G.;
 {Lister}, M.L.; {Mathur}, S.; {Peterson}, B.M.; {Richards}, J.L.;
 {Rafanelli}, P.
\newblock {Parent population of flat-spectrum radio-loud narrow-line Seyfert 1
 galaxies}.
\newblock {\em \aap} {\bf 2015}, {\em 578},~A28.
\newblock {\url{https://doi.org/10.1051/0004-6361/201525691}}.

\bibitem[{Padovani}(2017)]{padovani17}
{Padovani}, P.
\newblock {Active Galactic Nuclei at All Wavelengths and from All Angles}.
\newblock {\em Front. Astron. Space Sci.} {\bf 2017}, {\em
 4},~35.
\newblock {\url{https://doi.org/10.3389/fspas.2017.00035}}.

\bibitem[{Strateva} {et~al.}(2003){Strateva}, {Strauss}, {Hao}, {Schlegel},
 {Hall}, {Gunn}, {Li}, {Ivezi{\'c}}, {Richards}, {Zakamska}, {Voges},
 {Anderson}, {Lupton}, {Schneider}, {Brinkmann}, and {Nichol}]{stratevaetal03}
{Strateva}, I.V.; {Strauss}, M.A.; {Hao}, L.; {Schlegel}, D.J.; {Hall}, P.B.;
 {Gunn}, J.E.; {Li}, L.; {Ivezi{\'c}}, {\v Z}.; {Richards}, G.T.; {Zakamska},
 N.L.; et~al.
\newblock {Double-peaked Low-Ionization Emission Lines in Active Galactic
 Nuclei}.
\newblock {\em AJ} {\bf 2003}, {\em 126},~1720--1749.
\newblock {\url{https://doi.org/10.1086/378367}}.

\bibitem[{Mengistue} {et~al.}(2024){Mengistue}, {Marziani}, {del Olmo},
 {Povi{\'c}}, {Perea}, and {Deconto Machado}]{terefemengistueetal24}
{Mengistue}, S.T.; {Marziani}, P.; {del Olmo}, A.; {Povi{\'c}}, M.; {Perea},
 J.; {Deconto Machado}, A.
\newblock {Quasar 3C 47: Extreme Population B jetted source with double-peaked
 profiles}.
\newblock {\em \aap} {\bf 2024}, {\em 685},~A116.
\newblock {\url{https://doi.org/10.1051/0004-6361/202348800}}.

\bibitem[{Wu}(2009)]{wu09a}
{Wu}, Q.
\newblock {Observational Evidence for Young Radio Galaxies Is Triggered by
 Accretion Disk Instability}.
\newblock {\em \apjl} {\bf 2009}, {\em 701},~L95--L99.
\newblock {\url{https://doi.org/10.1088/0004-637X/701/2/L95}}.

\bibitem[{Shen}(2013)]{shen13}
{Shen}, Y.
\newblock {The mass of quasars}.
\newblock {\em Bull. Astron. Soc. India} {\bf 2013}, {\em
 41},~61--115.

\bibitem[{D'Onofrio} {et~al.}(2021){D'Onofrio}, {Marziani}, and
 {Chiosi}]{donofriomarziani21}
{D'Onofrio}, M.; {Marziani}, P.; {Chiosi}, C.
\newblock {Past, Present and Future of the Scaling Relations of Galaxies and
 Active Galactic Nuclei}.
\newblock {\em Front. Astron. Space Sci.} {\bf 2021}, {\em
 8},~157.
\newblock {\url{https://doi.org/10.3389/fspas.2021.694554}}.

\bibitem[Wu and Shen(2022)]{wushen22}
Wu, Q.; Shen, Y.
\newblock A Catalog of Quasar Properties from Sloan Digital Sky Survey Data
 Release 16.
\newblock {\em Astrophys. J. Suppl. Ser.} {\bf 2022}, {\em
 263},~42.
\newblock {\url{https://doi.org/10.3847/1538-4365/ac9ead}}.

\bibitem[{Rees} {et~al.}(1982){Rees}, {Begelman}, {Blandford}, and
 {Phinney}]{reesetal82}
{Rees}, M.J.; {Begelman}, M.C.; {Blandford}, R.D.; {Phinney}, E.S.
\newblock {Ion-supported tori and the origin of radio jets}.
\newblock {\em \nat} {\bf 1982}, {\em 295},~17--21.
\newblock {\url{https://doi.org/10.1038/295017a0}}.

\bibitem[{Narayan} and {Yi}(1994)]{narayanyi94}
{Narayan}, R.; {Yi}, I.
\newblock {Advection-dominated Accretion: A Self-similar Solution}.
\newblock {\em \apjl} {\bf 1994}, {\em 428},~L13.
\newblock {\url{https://doi.org/10.1086/187381}}.

\bibitem[{Narayan} and {Yi}(1995)]{narayanyi95}
{Narayan}, R.; {Yi}, I.
\newblock {Advection-dominated Accretion: Self-Similarity and Bipolar
 Outflows}.
\newblock {\em \apj} {\bf 1995}, {\em 444},~231.
\newblock {\url{https://doi.org/10.1086/175599}}.

\bibitem[{Proga} {et~al.}(2000){Proga}, {Stone}, and {Kallman}]{progaetal00}
{Proga}, D.; {Stone}, J.M.; {Kallman}, T.R.
\newblock {Dynamics of Line-Driven Disk Winds in Active Galactic Nuclei}.
\newblock {\em \apj} {\bf 2000}, {\em 543},~686--696.
\newblock {\url{https://doi.org/10.1086/317154}}.

\bibitem[{Proga} and {Kallman}(2004)]{progakallman04}
{Proga}, D.; {Kallman}, T.R.
\newblock {Dynamics of Line-Driven Disk Winds in Active Galactic Nuclei. II.
 Effects of Disk Radiation}.
\newblock {\em \apj} {\bf 2004}, {\em 616},~688--695.
\newblock {\url{https://doi.org/10.1086/425117}}.

\bibitem[{Giustini} and {Proga}(2019)]{giustiniproga19}
{Giustini}, M.; {Proga}, D.
\newblock {A global view of the inner accretion and ejection flow around super
 massive black holes. Radiation-driven accretion disk winds in a physical
 context}.
\newblock {\em \aap} {\bf 2019}, {\em 630},~A94.
\newblock {\url{https://doi.org/10.1051/0004-6361/201833810}}.

\bibitem[{Blandford} {et~al.}(2019){Blandford}, {Meier}, and
 {Readhead}]{blandfordetal19}
{Blandford}, R.; {Meier}, D.; {Readhead}, A.
\newblock {Relativistic Jets from Active Galactic Nuclei}.
\newblock {\em \araa} {\bf 2019}, {\em 57},~467--509.
\newblock {\url{https://doi.org/10.1146/annurev-astro-081817-051948}}.

\bibitem[{Coatman} {et~al.}(2016){Coatman}, {Hewett}, {Banerji}, and
 {Richards}]{coatmanetal16}
{Coatman}, L.; {Hewett}, P.C.; {Banerji}, M.; {Richards}, G.T.
\newblock {C iv emission-line properties and systematic trends in quasar black
 hole mass estimates}.
\newblock {\em \mnras} {\bf 2016}, {\em 461},~647--665.
\newblock {\url{https://doi.org/10.1093/mnras/stw1360}}.

\bibitem[{Sulentic} {et~al.}(2017){Sulentic}, {del Olmo}, {Marziani},
 {Mart{\'{\i}}nez-Carballo}, {D'Onofrio}, {Dultzin}, {Perea},
 {Mart{\'{\i}}nez-Aldama}, {Negrete}, {Stirpe}, and {Zamfir}]{sulenticetal17}
{Sulentic}, J.W.; {del Olmo}, A.; {Marziani}, P.; {Mart{\'{\i}}nez-Carballo},
 M.A.; {D'Onofrio}, M.; {Dultzin}, D.; {Perea}, J.; {Mart{\'{\i}}nez-Aldama},
 M.L.; {Negrete}, C.A.; {Stirpe}, G.M.; et~al.
\newblock {What does CIV{$\lambda$}1549 tell us about the physical driver of
 the Eigenvector quasar sequence?}
\newblock {\em \aap} {\bf 2017}, {\em 608},~A122.
\newblock {\url{https://doi.org/10.1051/0004-6361/201630309}}.

\bibitem[{Hopkins} and {Elvis}(2010)]{hopkinselvis10}
{Hopkins}, P.F.; {Elvis}, M.
\newblock {Quasar feedback: More bang for your buck}.
\newblock {\em \mnras} {\bf 2010}, {\em 401},~7--14.
\newblock {\url{https://doi.org/10.1111/j.1365-2966.2009.15643.x}}.

\bibitem[{Fabian}(2012)]{fabian12}
{Fabian}, A.C.
\newblock {Observational Evidence of Active Galactic Nuclei Feedback}.
\newblock {\em \araa} {\bf 2012}, {\em 50},~455--489.
\newblock {\url{https://doi.org/10.1146/annurev-astro-081811-125521}}.

\bibitem[{Murray} and {Chiang}(1997)]{murraychiang97}
{Murray}, N.; {Chiang}, J.
\newblock {Disk Winds and Disk Emission Lines}.
\newblock {\em \apj} {\bf 1997}, {\em 474},~91.
\newblock {\url{https://doi.org/10.1086/303443}}.

\bibitem[{Ferland} {et~al.}(2009){Ferland}, {Hu}, {Wang}, {Baldwin},
 {Porter}, {van Hoof}, and {Williams}]{ferlandetal09}
{Ferland}, G.J.; {Hu}, C.; {Wang}, J.; {Baldwin}, J.A.; {Porter}, R.L.; {van
 Hoof}, P.A.M.; {Williams}, R.J.R.
\newblock {Implications of Infalling Fe II-Emitting Clouds in Active Galactic
 Nuclei: Anisotropic Properties}.
\newblock {\em \apjl} {\bf 2009}, {\em 707},~L82--L86.
\newblock {\url{https://doi.org/10.1088/0004-637X/707/1/L82}}.

\bibitem[{Risaliti} and {Elvis}(2010)]{risalitielvis10}
{Risaliti}, G.; {Elvis}, M.
\newblock {A non-hydrodynamical model for acceleration of line-driven winds in
 active galactic nuclei}.
\newblock {\em \aap} {\bf 2010}, {\em 516},~A89.
\newblock {\url{https://doi.org/10.1051/0004-6361/200912579}}.

\bibitem[{Goad} {et~al.}(2012){Goad}, {Korista}, and {Ruff}]{goadetal12}
{Goad}, M.R.; {Korista}, K.T.; {Ruff}, A.J.
\newblock {The broad emission-line region: The confluence of the outer
 accretion disc with the inner edge of the dusty torus}.
\newblock {\em \mnras} {\bf 2012}, {\em 426},~3086--3111.
\newblock {\url{https://doi.org/10.1111/j.1365-2966.2012.21808.x}}.

\bibitem[{Marziani} {et~al.}(2013){Marziani}, {Sulentic}, {Plauchu-Frayn},
 and {del Olmo}]{marzianietal13}
{Marziani}, P.; {Sulentic}, J.W.; {Plauchu-Frayn}, I.; {del Olmo}, A.
\newblock {Low-Ionization Outflows in High Eddington Ratio Quasars}.
\newblock {\em ApJ} {\bf 2013}, {\em 764},

\bibitem[{Popovi{\'c}} {et~al.}(2019){Popovi{\'c}},
 {Kova{\v{c}}evi{\'c}-Doj{\v{c}}inovi{\'c}}, and {Mar{\v{c}}eta-Mand
 i{\'c}}]{popovicetal19}
{Popovi{\'c}}, L.{\v{C}}.; {Kova{\v{c}}evi{\'c}-Doj{\v{c}}inovi{\'c}}, J.;
 {Mar{\v{c}}eta-Mand i{\'c}}, S.
\newblock {The structure of the Mg II broad line emitting region in Type 1
 AGNs}.
\newblock {\em \mnras} {\bf 2019}, {\em 484},~3180--3197.
\newblock {\url{https://doi.org/10.1093/mnras/stz157}}.

\bibitem[{Chen} and {Halpern}(1989)]{chenhalpern89}
{Chen}, K.; {Halpern}, J.P.
\newblock {Structure of line-emitting accretion disks in active galactic nuclei---ARP 102B}.
\newblock {\em \apj} {\bf 1989}, {\em 344},~115--124.
\newblock {\url{https://doi.org/10.1086/167782}}.

\bibitem[{Mej{\'{\i}}a-Restrepo} {et~al.}(2016){Mej{\'{\i}}a-Restrepo},
 {Trakhtenbrot}, {Lira}, {Netzer}, and {Capellupo}]{mejia-restrepoetal16}
{Mej{\'{\i}}a-Restrepo}, J.E.; {Trakhtenbrot}, B.; {Lira}, P.; {Netzer}, H.;
 {Capellupo}, D.M.
\newblock {Active galactic nuclei at z\~{}1.5: II. Black Hole Mass estimation
 by means of broad emission lines}.
\newblock {\em \mnras} {\bf 2016}, {\em 460},

\bibitem[{Collin-Souffrin} {et~al.}(1988){Collin-Souffrin}, {Dyson},
 {McDowell}, and {Perry}]{collinsouffrinetal88}
{Collin-Souffrin}, S.; {Dyson}, J.E.; {McDowell}, J.C.; {Perry}, J.J.
\newblock {The environment of active galactic nuclei. I---A two-component broad
 emission line model}.
\newblock {\em MNRAS} {\bf 1988}, {\em 232},~539--550.

\bibitem[{Cackett} {et~al.}(2007){Cackett}, {Horne}, and
 {Winkler}]{cackettetal07}
{Cackett}, E.M.; {Horne}, K.; {Winkler}, H.
\newblock {Testing thermal reprocessing in active galactic nuclei accretion
 discs}.
\newblock {\em \mnras} {\bf 2007}, {\em 380},~669--682.
\newblock {\url{https://doi.org/10.1111/j.1365-2966.2007.12098.x}}.

\bibitem[{Uttley} {et~al.}(2014){Uttley}, {Cackett}, {Fabian}, {Kara}, and
 {Wilkins}]{uttleyetal14}
{Uttley}, P.; {Cackett}, E.M.; {Fabian}, A.C.; {Kara}, E.; {Wilkins}, D.R.
\newblock {X-ray reverberation around accreting black holes}.
\newblock {\em Astron. Astrophys. Rev.} {\bf 2014}, {\em 22},~72.
\newblock {\url{https://doi.org/10.1007/s00159-014-0072-0}}.

\bibitem[{Kara} {et~al.}(2016){Kara}, {Alston}, {Fabian}, {Cackett},
 {Uttley}, {Reynolds}, and {Zoghbi}]{karaetal16}
{Kara}, E.; {Alston}, W.N.; {Fabian}, A.C.; {Cackett}, E.M.; {Uttley}, P.;
 {Reynolds}, C.S.; {Zoghbi}, A.
\newblock {A global look at X-ray time lags in Seyfert galaxies}.
\newblock {\em \mnras} {\bf 2016}, {\em 462},~511--531.
\newblock {\url{https://doi.org/10.1093/mnras/stw1695}}.

\bibitem[{Tombesi} {et~al.}(2010){Tombesi}, {Sambruna}, {Reeves}, {Braito},
 {Ballo}, {Gofford}, {Cappi}, and {Mushotzky}]{tombesietal10}
{Tombesi}, F.; {Sambruna}, R.M.; {Reeves}, J.N.; {Braito}, V.; {Ballo}, L.;
 {Gofford}, J.; {Cappi}, M.; {Mushotzky}, R.F.
\newblock {Discovery of Ultra-fast Outflows in a Sample of Broad-line Radio
 Galaxies Observed with Suzaku}.
\newblock {\em \apj} {\bf 2010}, {\em 719},~700--715.
\newblock {\url{https://doi.org/10.1088/0004-637X/719/1/700}}.

\bibitem[{Tombesi} {et~al.}(2012){Tombesi}, {Cappi}, {Reeves}, and
 {Braito}]{tombesietal12}
{Tombesi}, F.; {Cappi}, M.; {Reeves}, J.N.; {Braito}, V.
\newblock {Evidence for ultrafast outflows in radio-quiet AGNs - III. Location
 and energetics}.
\newblock {\em \mnras} {\bf 2012}, {\em 422},~L1--L5.
\newblock {\url{https://doi.org/10.1111/j.1745-3933.2012.01221.x}}.

\bibitem[{Tombesi} {et~al.}(2013){Tombesi}, {Cappi}, {Reeves}, {Nemmen},
 {Braito}, {Gaspari}, and {Reynolds}]{tombesietal13}
{Tombesi}, F.; {Cappi}, M.; {Reeves}, J.N.; {Nemmen}, R.S.; {Braito}, V.;
 {Gaspari}, M.; {Reynolds}, C.S.
\newblock {Unification of X-ray winds in Seyfert galaxies: From ultra-fast
 outflows to warm absorbers}.
\newblock {\em \mnras} {\bf 2013}, {\em 430},~1102--1117.
\newblock {\url{https://doi.org/10.1093/mnras/sts692}}.

\bibitem[{Laha} {et~al.}(2021){Laha}, {Reynolds}, {Reeves}, {Kriss},
 {Guainazzi}, {Smith}, {Veilleux}, and {Proga}]{lahaetal21}
{Laha}, S.; {Reynolds}, C.S.; {Reeves}, J.; {Kriss}, G.; {Guainazzi}, M.;
 {Smith}, R.; {Veilleux}, S.; {Proga}, D.
\newblock {Ionized outflows from active galactic nuclei as the essential
 elements of feedback}.
\newblock {\em Nat. Astron.} {\bf 2021}, {\em 5},~13--24.
\newblock {\url{https://doi.org/10.1038/s41550-020-01255-2}}.

\bibitem[{Lin} and {Pringle}(1987)]{linpringle87}
{Lin}, D.N.C.; {Pringle}, J.E.
\newblock {A viscosity prescription for a self-gravitating accretion disc}.
\newblock {\em \mnras} {\bf 1987}, {\em 225},~607--613.
\newblock {\url{https://doi.org/10.1093/mnras/225.3.607}}.

\bibitem[{Czerny} {et~al.}(2017){Czerny}, {Li}, {Hryniewicz}, {Panda},
 {Wildy}, {Sniegowska}, {Wang}, {Sredzinska}, and {Karas}]{czernyetal17}
{Czerny}, B.; {Li}, Y.R.; {Hryniewicz}, K.; {Panda}, S.; {Wildy}, C.;
 {Sniegowska}, M.; {Wang}, J.M.; {Sredzinska}, J.; {Karas}, V.
\newblock {Failed Radiatively Accelerated Dusty Outflow Model of the Broad Line
 Region in Active Galactic Nuclei. I. Analytical Solution}.
\newblock {\em \apj} {\bf 2017}, {\em 846},~154.
\newblock {\url{https://doi.org/10.3847/1538-4357/aa8810}}.

\bibitem[{Naddaf} {et~al.}(2023){Naddaf}, {Martinez-Aldama}, {Marziani},
 {Panda}, {Sniegowska}, and {Czerny}]{naddafetal21}
{Naddaf}, M.H.; {Martinez-Aldama}, M.L.; {Marziani}, P.; {Panda}, S.;
 {Sniegowska}, M.; {Czerny}, B.
\newblock {Dust-driven wind as a model of broad absorption line quasars}.
\newblock {\em \aap} {\bf 2023}, {\em 675},~A43.
\newblock {\url{https://doi.org/10.1051/0004-6361/202245698}}.

\bibitem[{Naddaf} {et~al.}(2021){Naddaf}, {Czerny}, and
 {Szczerba}]{naddafetal21a}
{Naddaf}, M.H.; {Czerny}, B.; {Szczerba}, R.
\newblock {The Picture of BLR in 2.5D FRADO: Dynamics and Geometry}.
\newblock {\em \apj} {\bf 2021}, {\em 920},~30.
\newblock {\url{https://doi.org/10.3847/1538-4357/ac139d}}.

\bibitem[{Choi} {et~al.}(2022{\natexlab{a}}){Choi}, {Leighly}, {Terndrup},
 {Dabbieri}, {Gallagher}, and {Richards}]{choietal22}
{Choi, H.; Leighly, K.M.; Terndrup, D.M.; Dabbieri, C.; Gallagher, S.C.; Richards, G.T.} 
\newblock {The Physical Properties of Low-redshift FeLoBAL Quasars. I.
 Spectral-synthesis Analysis of the Broad Absorption-line (BAL) Outflows Using
 SimBAL}.
\newblock {\em \apj} {\bf 2022}, {\em 937},~74.
\newblock {\url{https://doi.org/10.3847/1538-4357/ac61d9}}.


\bibitem[{Bischetti} {et~al.}(2017){Bischetti}, {Piconcelli}, {Vietri},
 {Bongiorno}, {Fiore}, {Sani}, {Marconi}, {Duras}, {Zappacosta}, {Brusa},
 {Comastri}, {Cresci}, {Feruglio}, {Giallongo}, {La Franca}, {Mainieri},
 {Mannucci}, {Martocchia}, {Ricci}, {Schneider}, {Testa}, and
 {Vignali}]{bischettietal17}
{Bischetti}, M.; {Piconcelli}, E.; {Vietri}, G.; {Bongiorno}, A.; {Fiore}, F.;
 {Sani}, E.; {Marconi}, A.; {Duras}, F.; {Zappacosta}, L.; {Brusa}, M.;
 et~al.
\newblock {The WISSH quasars project. I. Powerful ionised outflows in
 hyper-luminous quasars}.
\newblock {\em \aap} {\bf 2017}, {\em 598},~A122.
\newblock {\url{https://doi.org/10.1051/0004-6361/201629301}}.

\bibitem[{Cano-D{\'{\i}}az} {et~al.}(2012){Cano-D{\'{\i}}az}, {Maiolino},
 {Marconi}, {Netzer}, {Shemmer}, and {Cresci}]{canodiazetal12}
{Cano-D{\'{\i}}az}, M.; {Maiolino}, R.; {Marconi}, A.; {Netzer}, H.; {Shemmer},
 O.; {Cresci}, G.
\newblock {Observational evidence of quasar feedback quenching star formation
 at high redshift}.
\newblock {\em \aap} {\bf 2012}, {\em 537},~L8.
\newblock {\url{https://doi.org/10.1051/0004-6361/201118358}}.

\bibitem[{Marziani} {et~al.}(2017){Marziani}, {Negrete}, {Dultzin},
 {Martinez-Aldama}, {Del Olmo}, {Esparza}, {Sulentic}, {D'Onofrio}, {Stirpe},
 {Bon}, and {Bon}]{marzianietal17}
{Marziani}, P.; {Negrete}, C.A.; {Dultzin}, D.; {Martinez-Aldama}, M.L.; {Del
 Olmo}, A.; {Esparza}, D.; {Sulentic}, J.W.; {D'Onofrio}, M.; {Stirpe}, G.M.;
 {Bon}, E.; et~al.
\newblock {Highly accreting quasars: A tool for cosmology?}
\newblock \emph{Proc. Int. Astron. Union} {\bf 2017}, \emph{324}, 245--246.
\newblock {\url{https://doi.org/10.1017/S1743921316012655}}.

\bibitem[{Laor} and {Brandt}(2002)]{laorbrandt02}
{Laor}, A.; {Brandt}, W.N.
\newblock {The Luminosity Dependence of Ultraviolet Absorption in Active
 Galactic Nuclei}.
\newblock {\em \apj} {\bf 2002}, {\em 569},~641--654.
\newblock {\url{https://doi.org/10.1086/339476}}.

\bibitem[{Villar Mart{\'\i}n} {et~al.}(2024){Villar Mart{\'\i}n}, {L{\'o}pez
 Cob{\'a}}, {Cazzoli}, {P{\'e}rez Montero}, and {Cabrera
 Lavers}]{villar-martinetal24}
{Villar Mart{\'\i}n}, M.; {L{\'o}pez Cob{\'a}}, C.; {Cazzoli}, S.; {P{\'e}rez
 Montero}, E.; {Cabrera Lavers}, A.
\newblock {AGN feedback can produce metal enrichment on galaxy scales}.
\newblock {\em \aap} {\bf 2024}, {\em 690},~A397.
\newblock {\url{https://doi.org/10.1051/0004-6361/202449621}}.

\bibitem[Harrison {et~al.}(2014)Harrison, Alexander, Mullaney, and
 Swinbank]{harrison2014}
Harrison, C.M.; Alexander, D.M.; Mullaney, J.R.; Swinbank, A.M.
\newblock Energetic galaxy-wide outflows in high-redshift quasars driven by AGN activity.
\newblock {\em Mon. Not. R. Astron. Soc.} {\bf 2014},
 {\em 426},~1073--1096.
\newblock {{\url{https://doi.org/10.1111/j.1365-2966.2012.21723.x}.} 
}

\bibitem[{Harrison} {et~al.}(2014){Harrison}, {Alexander}, {Mullaney}, and
 {Swinbank}]{harrisonetal14}
{Harrison}, C.M.; {Alexander}, D.M.; {Mullaney}, J.R.; {Swinbank}, A.M.
\newblock {Kiloparsec-scale outflows are prevalent among luminous AGN: Outflows
 and feedback in the context of the overall AGN population}.
\newblock {\em \mnras} {\bf 2014}, {\em 441},~3306--3347.
\newblock {\url{https://doi.org/10.1093/mnras/stu515}}.

\bibitem[{Kakkad} {et~al.}(2020){Kakkad}, {Mainieri}, {Vietri}, {Carniani},
 {Harrison}, {Perna}, {Scholtz}, {Circosta}, {Cresci}, {Husemann},
 {Bischetti}, {Feruglio}, {Fiore}, {Marconi}, {Padovani}, {Brusa}, {Cicone},
 {Comastri}, {Lanzuisi}, {Mannucci}, {Menci}, {Netzer}, {Piconcelli},
 {Puglisi}, {Salvato}, {Schramm}, {Silverman}, {Vignali}, {Zamorani}, and
 {Zappacosta}]{kakkadetal20}
{Kakkad}, D.; {Mainieri}, V.; {Vietri}, G.; {Carniani}, S.; {Harrison}, C.M.;
 {Perna}, M.; {Scholtz}, J.; {Circosta}, C.; {Cresci}, G.; {Husemann}, B.;
 et~al.
\newblock {SUPER. II. Spatially resolved ionised gas kinematics and scaling
 relations in z {\textasciitilde} 2 AGN host galaxies}.
\newblock {\em \aap} {\bf 2020}, {\em 642},~A147.
\newblock {\url{https://doi.org/10.1051/0004-6361/202038551}}.

\bibitem[{Singha} {et~al.}(2022){Singha}, {Husemann}, {Urrutia}, {O'Dea},
 {Scharw{\"a}chter}, {Gaspari}, {Combes}, {Nevin}, {Terrazas},
 {P{\'e}rez-Torres}, {Rose}, {Davis}, {Tremblay}, {Neumann},
 {Smirnova-Pinchukova}, and {Baum}]{singhaetal22}
{Singha}, M.; {Husemann}, B.; {Urrutia}, T.; {O'Dea}, C.P.; {Scharw{\"a}chter},
 J.; {Gaspari}, M.; {Combes}, F.; {Nevin}, R.; {Terrazas}, B.A.;
 {P{\'e}rez-Torres}, M.; et~al.
\newblock {The Close AGN Reference Survey (CARS). Locating the [O III] wing
 component in luminous local Type 1 AGN}.
\newblock {\em \aap} {\bf 2022}, {\em 659},~A123.
\newblock {\url{https://doi.org/10.1051/0004-6361/202040122}}.

\bibitem[{Tombesi} {et~al.}(2014){Tombesi}, {Tazaki}, {Mushotzky}, {Ueda},
 {Cappi}, {Gofford}, {Reeves}, and {Guainazzi}]{tombesietal14}
{Tombesi}, F.; {Tazaki}, F.; {Mushotzky}, R.F.; {Ueda}, Y.; {Cappi}, M.;
 {Gofford}, J.; {Reeves}, J.N.; {Guainazzi}, M.
\newblock {Ultrafast outflows in radio-loud active galactic nuclei}.
\newblock {\em \mnras} {\bf 2014}, {\em 443},~2154--2182.
\newblock {\url{https://doi.org/10.1093/mnras/stu1297}}.

\bibitem[{Gofford} {et~al.}(2015){Gofford}, {Reeves}, {McLaughlin},
 {Braito}, {Turner}, {Tombesi}, and {Cappi}]{goffordetal15}
{Gofford}, J.; {Reeves}, J.N.; {McLaughlin}, D.E.; {Braito}, V.; {Turner},
 T.J.; {Tombesi}, F.; {Cappi}, M.
\newblock {The Suzaku view of highly ionized outflows in AGN - II. Location,
 energetics and scalings with bolometric luminosity}.
\newblock {\em \mnras} {\bf 2015}, {\em 451},~4169--4182.
\newblock {\url{https://doi.org/10.1093/mnras/stv1207}}.

\bibitem[{Mestici} {et~al.}(2024){Mestici}, {Tombesi}, {Gaspari},
 {Piconcelli}, and {Panessa}]{mesticietal24}
{Mestici}, S.; {Tombesi}, F.; {Gaspari}, M.; {Piconcelli}, E.; {Panessa}, F.
\newblock {Unified properties of supermassive black hole winds in radio-quiet
 and radio-loud AGN}.
\newblock {\em \mnras} {\bf 2024}, {\em 532},~3036--3055.
\newblock {\url{https://doi.org/10.1093/mnras/stae1617}}.

\bibitem[{Mizumoto} {et~al.}(2021){Mizumoto}, {Nomura}, {Done}, {Ohsuga},
 and {Odaka}]{mizumotoetal21}
{Mizumoto}, M.; {Nomura}, M.; {Done}, C.; {Ohsuga}, K.; {Odaka}, H.
\newblock {UV line-driven disc wind as the origin of UltraFast Outflows in
 AGN}.
\newblock {\em \mnras} {\bf 2021}, {\em 503},~1442--1458.
\newblock {\url{https://doi.org/10.1093/mnras/staa3282}}.

\bibitem[{Matzeu} {et~al.}(2017){Matzeu}, {Reeves}, {Braito}, {Nardini},
 {McLaughlin}, {Lobban}, {Tombesi}, and {Costa}]{matzeuetal17}
{Matzeu}, G.A.; {Reeves}, J.N.; {Braito}, V.; {Nardini}, E.; {McLaughlin},
 D.E.; {Lobban}, A.P.; {Tombesi}, F.; {Costa}, M.T.
\newblock {Evidence for a radiatively driven disc-wind in PDS 456?}
\newblock {\em \mnras} {\bf 2017}, {\em 472},~L15--L19.
\newblock {\url{https://doi.org/10.1093/mnrasl/slx129}}.

\bibitem[{Fukumura} {et~al.}(2014){Fukumura}, {Tombesi}, {Kazanas},
 {Shrader}, {Behar}, and {Contopoulos}]{fukumuraetal14}
{Fukumura}, K.; {Tombesi}, F.; {Kazanas}, D.; {Shrader}, C.; {Behar}, E.;
 {Contopoulos}, I.
\newblock {Stratified Magnetically Driven Accretion-disk Winds and Their
 Relations to Jets}.
\newblock {\em \apj} {\bf 2014}, {\em 780},~120.
\newblock {\url{https://doi.org/10.1088/0004-637X/780/2/120}}.

\bibitem[{Fukumura} {et~al.}(2018){Fukumura}, {Kazanas}, {Shrader}, {Behar},
 {Tombesi}, and {Contopoulos}]{fukumuraetal18}
{Fukumura}, K.; {Kazanas}, D.; {Shrader}, C.; {Behar}, E.; {Tombesi}, F.;
 {Contopoulos}, I.
\newblock {Variable Nature of Magnetically Driven Ultra-fast Outflows}.
\newblock {\em \apjl} {\bf 2018}, {\em 864},~L27.
\newblock {\url{https://doi.org/10.3847/2041-8213/aadd10}}.

\bibitem[{Begelman} {et~al.}(1983){Begelman}, {McKee}, and
 {Shields}]{begelmanetal83}
{Begelman}, M.C.; {McKee}, C.F.; {Shields}, G.A.
\newblock {Compton heated winds and coronae above accretion disks. I.
 Dynamics.}
\newblock {\em \apj} {\bf 1983}, {\em 271},~70--88.
\newblock {\url{https://doi.org/10.1086/161178}}.

\bibitem[{Gianolli} {et~al.}(2024){Gianolli}, {Bianchi}, {Petrucci},
 {Brusa}, {Chartas}, {Lanzuisi}, {Matzeu}, {Parra}, {Ursini}, {Behar},
 {Bischetti}, {Comastri}, {Costantini}, {Cresci}, {Dadina}, {De Marco}, {De
 Rosa}, {Fiore}, {Gaspari}, {Gilli}, {Giustini}, {Guainazzi}, {King},
 {Kraemer}, {Kriss}, {Krongold}, {La Franca}, {Longinotti}, {Luminari},
 {Maiolino}, {Marconi}, {Mathur}, {Matt}, {Mehdipour}, {Merloni}, {Middei},
 {Miniutti}, {Nardini}, {Panessa}, {Perna}, {Piconcelli}, {Ponti}, {Ricci},
 {Serafinelli}, {Tombesi}, {Vignali}, and {Zappacosta}]{gianollietal24}
{Gianolli}, V.E.; {Bianchi}, S.; {Petrucci}, P.O.; {Brusa}, M.; {Chartas}, G.;
 {Lanzuisi}, G.; {Matzeu}, G.A.; {Parra}, M.; {Ursini}, F.; {Behar}, E.;
 et~al.
\newblock {Supermassive Black Hole Winds in X-rays: SUBWAYS. III. A population
 study on ultra-fast outflows}.
\newblock {\em \aap} {\bf 2024}, {\em 687},~A235.
\newblock {\url{https://doi.org/10.1051/0004-6361/202348908}}.

\bibitem[{Tully} and {Fisher}(1977)]{tullyfisher77}
{Tully}, R.B.; {Fisher}, J.R.
\newblock {A new method of determining distances to galaxies}.
\newblock {\em \aap} {\bf 1977}, {\em 54},~661--673.

\bibitem[{Marziani} {et~al.}(2006){Marziani}, {Dultzin-Hacyan}, and
 {Sulentic}]{marzianietal06}
{Marziani}, P.; {Dultzin-Hacyan}, D.; {Sulentic}, J.W.
\newblock {Accretion onto Supermassive Black Holes in Quasars: Learning from Optical/UV Observations}. 
In {\em New Developments in Black Hole Research};
 {Kreitler}, P.V., Ed.; Nova Press: New York, NY, USA, 2006; p. 123.

\bibitem[{Marziani} and {Sulentic}(2012)]{marzianisulentic12}
{Marziani}, P.; {Sulentic}, J.W.
\newblock {Estimating black hole masses in quasars using broad optical and UV
 emission lines}.
\newblock {\em NARev} {\bf 2012}, {\em 56},~49--63.
\newblock {\url{https://doi.org/10.1016/j.newar.2011.09.001}}.

\bibitem[{Runnoe} {et~al.}(2013){Runnoe}, {Shang}, and
 {Brotherton}]{runnoeetal13}
{Runnoe}, J.C.; {Shang}, Z.; {Brotherton}, M.S.
\newblock {The orientation dependence of quasar spectral energy distributions}.
\newblock {\em \mnras} {\bf 2013}, {\em 435},~3251--3261.
\newblock {\url{https://doi.org/10.1093/mnras/stt1528}}.

\bibitem[{Brotherton} {et~al.}(2015){Brotherton}, {Runnoe}, {Shang}, and
 {DiPompeo}]{brothertonetal15}
{Brotherton}, M.S.; {Runnoe}, J.C.; {Shang}, Z.; {DiPompeo}, M.A.
\newblock {Bias in C IV-based quasar black hole mass scaling relationships from
 reverberation mapped samples}.
\newblock {\em \mnras} {\bf 2015}, {\em 451},~1290--1298.
\newblock {\url{https://doi.org/10.1093/mnras/stv767}}.

\bibitem[{Savi{\'c}} {et~al.}(2018){Savi{\'c}}, {Goosmann}, {Popovi{\'c}},
 {Marin}, and {Afanasiev}]{savicetal18}
{Savi{\'c}}, D.; {Goosmann}, R.; {Popovi{\'c}}, L.{\v{C}}.; {Marin}, F.;
 {Afanasiev}, V.L.
\newblock {AGN black hole mass estimates using polarization in broad emission
 lines}.
\newblock {\em \aap} {\bf 2018}, {\em 614},~A120.
\newblock {\url{https://doi.org/10.1051/0004-6361/201732220}}.

\bibitem[{Afanasiev} {et~al.}(2019){Afanasiev}, {Popovi{\'c}}, and
 {Shapovalova}]{afanasievetal19}
{Afanasiev}, V.L.; {Popovi{\'c}}, L.{\v{C}}.; {Shapovalova}, A.I.
\newblock {Spectropolarimetry of Seyfert 1 galaxies with equatorial scattering:
 black hole masses and broad-line region characteristics}.
\newblock {\em \mnras} {\bf 2019}, {\em 482},~4985--4999.
\newblock {\url{https://doi.org/10.1093/mnras/sty2995}}.

\bibitem[{Savi{\'c}} {et~al.}(2020){Savi{\'c}}, {Popovi{\'c}},
 {Shablovinskaya}, and {Afanasiev}]{savicetal20}
{Savi{\'c}}, {\DJ}.; {Popovi{\'c}}, L.{\v{C}}.; {Shablovinskaya}, E.;
 {Afanasiev}, V.L.
\newblock {Estimating supermassive black hole masses in active galactic nuclei
 using polarization of broad Mg II, H {\ensuremath{\alpha}}, and H
 {\ensuremath{\beta}} lines}.
\newblock {\em \mnras} {\bf 2020}, {\em 497},~3047--3054.
\newblock {\url{https://doi.org/10.1093/mnras/staa2039}}.

\bibitem[{Marziani} {et~al.}(2019){Marziani}, {del Olmo},
 {Mart{\'\i}nez-Carballo}, {Mart{\'\i}nez-Aldama}, {Stirpe}, {Negrete},
 {Dultzin}, {D'Onofrio}, {Bon}, and {Bon}]{marzianietal19}
{Marziani}, P.; {del Olmo}, A.; {Mart{\'\i}nez-Carballo}, M.A.;
 {Mart{\'\i}nez-Aldama}, M.L.; {Stirpe}, G.M.; {Negrete}, C.A.; {Dultzin}, D.;
 {D'Onofrio}, M.; {Bon}, E.; {Bon}, N.
\newblock {Black hole mass estimates in quasars. A comparative analysis of
 high- and low-ionization lines}.
\newblock {\em \aap} {\bf 2019}, {\em 627},~A88.
\newblock {\url{https://doi.org/10.1051/0004-6361/201935265}}.

\bibitem[{D'Onofrio} {et~al.}(2024){D'Onofrio}, {Marziani}, {Chiosi}, and
 {Negrete}]{donofrioetal24}
{D'Onofrio}, M.; {Marziani}, P.; {Chiosi}, C.; {Negrete}, C.A.
\newblock {The Correlation Luminosity-Velocity Dispersion of Galaxies and
 Active Galactic Nuclei}.
\newblock {\em Universe} {\bf 2024}, {\em 10},~254.
\newblock {\url{https://doi.org/10.3390/universe10060254}}.

\bibitem[{Wolf} {et~al.}(2024){Wolf}, {Lai}, {Onken}, {Amrutha}, {Bian},
 {Hon}, {Tisserand}, and {Webster}]{wolfetal24}
{Wolf}, C.; {Lai}, S.; {Onken}, C.A.; {Amrutha}, N.; {Bian}, F.; {Hon}, W.J.;
 {Tisserand}, P.; {Webster}, R.L.
\newblock {The accretion of a solar mass per day by a 17-billion solar mass
 black hole}.
\newblock {\em Nat. Astron.} {\bf 2024}, {\em 8},~520--529.
\newblock {\url{https://doi.org/10.1038/s41550-024-02195-x}}.

\bibitem[{Bosman} {et~al.}(2024){Bosman}, {{\'A}lvarez-M{\'a}rquez},
 {Colina}, {Walter}, {Alonso-Herrero}, {Ward}, {{\~A}-stlin}, {Greve},
 {Wright}, {Bik}, {Boogaard}, {Caputi}, {Costantin}, {Eckart},
 {Garc{\'\i}a-Mar{\'\i}n}, {Gillman}, {Hjorth}, {Iani}, {Ilbert}, {Jermann},
 {Labiano}, {Langeroodi}, {Pei{\ss}ker}, {Rinaldi}, {Topinka}, {van der Werf},
 {G{\"u}del}, {Henning}, {Lagage}, {Ray}, {van Dishoeck}, and
 {Vandenbussche}]{bosmanetal24}
{Bosman}, S.E.I.; {{\'A}lvarez-M{\'a}rquez}, J.; {Colina}, L.; {Walter}, F.;
 {Alonso-Herrero}, A.; {Ward}, M.J.; {{\~A}-stlin}, G.; {Greve}, T.R.;
 {Wright}, G.; {Bik}, A.; et~al.
\newblock {A mature quasar at cosmic dawn revealed by JWST rest-frame infrared
 spectroscopy}.
\newblock {\em Nat. Astron.} {\bf 2024}, {\em 8},~1054--1065.
\newblock {\url{https://doi.org/10.1038/s41550-024-02273-0}}.

\bibitem[{Bogd{\'a}n} {et~al.}(2024){Bogd{\'a}n}, {Goulding}, {Natarajan},
 {Kov{\'a}cs}, {Tremblay}, {Chadayammuri}, {Volonteri}, {Kraft}, {Forman},
 {Jones}, {Churazov}, and {Zhuravleva}]{bogdanetal24}
{Bogd{\'a}n}, {\'A}.; {Goulding}, A.D.; {Natarajan}, P.; {Kov{\'a}cs}, O.E.;
 {Tremblay}, G.R.; {Chadayammuri}, U.; {Volonteri}, M.; {Kraft}, R.P.;
 {Forman}, W.R.; {Jones}, C.; et~al.
\newblock {Evidence for heavy-seed origin of early supermassive black holes
 from a z {\ensuremath{\approx}} 10 X-ray quasar}.
\newblock {\em Nat. Astron.} {\bf 2024}, {\em 8},~126--133.
\newblock {\url{https://doi.org/10.1038/s41550-023-02111-9}}.

\bibitem[{Melia}(2024{\natexlab{a}})]{melia24}
{Melia}, F.
\newblock {The cosmic timeline implied by the JWST reionization crisis}.
\newblock {\em \aap} {\bf 2024}, {\em 689},~A10.
\newblock {\url{https://doi.org/10.1051/0004-6361/202450835}}.

\bibitem[{Melia}(2024{\natexlab{b}})]{melia24a}
{Melia}, F.
\newblock {Strong observational support for the Rh=ct timeline in the early
 universe}.
\newblock {\em Phys. Dark Universe} {\bf 2024}, {\em 46},~101587.
\newblock {\url{https://doi.org/10.1016/j.dark.2024.101587}}.

\bibitem[{Hu{\v{s}}ko} {et~al.}(2024){Hu{\v{s}}ko}, {Lacey}, {Roper},
 {Schaye}, {Briggs}, and {Schaller}]{huskoetal24}
{Hu{\v{s}}ko}, F.; {Lacey}, C.G.; {Roper}, W.J.; {Schaye}, J.; {Briggs}, J.M.;
 {Schaller}, M.
\newblock {The effects of super-Eddington accretion and feedback on the growth
 of early supermassive black holes and galaxies}.
\newblock {\em arXiv} {\bf 2024}, arXiv:2410.09450.

\bibitem[{Jiang} {et~al.}(2014){Jiang}, {Stone}, and {Davis}]{jiangetal14}
{Jiang}, Y.F.; {Stone}, J.M.; {Davis}, S.W.
\newblock {A Global Three-dimensional Radiation Magneto-hydrodynamic Simulation
 of Super-Eddington Accretion Disks}.
\newblock {\em \apj} {\bf 2014}, {\em 796},~106.
\newblock {\url{https://doi.org/10.1088/0004-637X/796/2/106}}.

\bibitem[{S{\k{a}}dowski} {et~al.}(2015){S{\k{a}}dowski}, {Narayan},
 {Tchekhovskoy}, {Abarca}, {Zhu}, and {McKinney}]{sadowskietal15}
{S{\k{a}}dowski}, A.; {Narayan}, R.; {Tchekhovskoy}, A.; {Abarca}, D.; {Zhu},
 Y.; {McKinney}, J.C.
\newblock {Global simulations of axisymmetric radiative black hole accretion
 discs in general relativity with a mean-field magnetic dynamo}.
\newblock {\em \mnras} {\bf 2015}, {\em 447},~49--71.
\newblock {\url{https://doi.org/10.1093/mnras/stu2387}}.

\bibitem[{Huang} {et~al.}(2019){Huang}, {Hu}, {Zhao}, {Zhang}, {Lu}, {Wang},
 {Zhang}, {Du}, {Li}, {Bai}, {Ho}, {Bian}, {Yuan}, and {Wang}]{huangetal19}
{Huang}, Y.K.; {Hu}, C.; {Zhao}, Y.L.; {Zhang}, Z.X.; {Lu}, K.X.; {Wang}, K.;
 {Zhang}, Y.; {Du}, P.; {Li}, Y.R.; {Bai}, J.M.; et~al.
\newblock {Reverberation Mapping of the Narrow-line Seyfert 1 Galaxy I Zwicky
 1: Black Hole Mass}.
\newblock {\em \apj} {\bf 2019}, {\em 876},~102.
\newblock {\url{https://doi.org/10.3847/1538-4357/ab16ef}}.

\bibitem[{Marinello} {et~al.}(2020){Marinello}, {Rodr{\'\i}guez-Ardila},
 {Marziani}, {Sigut}, and {Pradhan}]{marinelloetal20b}
{Marinello}, M.; {Rodr{\'\i}guez-Ardila}, A.; {Marziani}, P.; {Sigut}, A.;
 {Pradhan}, A.
\newblock {Panchromatic Properties of the Extreme Fe ii Emitter PHL 1092}.
\newblock {\em \mnras} {\bf 2020}, \emph{494}, 4187–4202.
\newblock {\url{https://doi.org/10.1093/mnras/staa934}}.

\bibitem[{Li} {et~al.}(2024){Li}, {Hu}, {Yao}, {Chen}, {Bai}, {Yang}, {Du},
 {Fang}, {Fu}, {Liu}, {Peng}, {Songsheng}, {Wang}, {Xiao}, {Zhai}, {Winkler},
 {Bai}, {Ho}, {Petrov}, {Aceituno}, {Wang}, and {SARM
 Collaboration}]{lietal24}
{Li}, Y.R.; {Hu}, C.; {Yao}, Z.H.; {Chen}, Y.J.; {Bai}, H.R.; {Yang}, S.; {Du},
 P.; {Fang}, F.N.; {Fu}, Y.X.; {Liu}, J.R.; et~al.
\newblock {Spectroastrometry and Reverberation Mapping of Active Galactic
 Nuclei. I. The H{\ensuremath{\beta}} Broad-line Region Structure and Black
 Hole Masses of Five Quasars}.
\newblock {\em \apj} {\bf 2024}, {\em 974},~86.
\newblock {\url{https://doi.org/10.3847/1538-4357/ad6906}}.

\bibitem[Mart{\'\i}nez-Aldama {et~al.}(2017)Mart{\'\i}nez-Aldama, Del~Olmo,
 Marziani, Negrete, Dultzin, and
 Mart{\'\i}nez-Carballo]{martinez-aldamaetal17}
Mart{\'\i}nez-Aldama, M.L.; Del~Olmo, A.; Marziani, P.; Negrete, C.A.; Dultzin,
 D.; Mart{\'\i}nez-Carballo, M.A.
\newblock HE0359-3959: An Extremely Radiating Quasar.
\newblock {\em Front. Astron. Space Sci.} {\bf 2017}, {\em
 4},~29.
\newblock {\url{https://doi.org/10.3389/fspas.2017.00029}}.

\bibitem[{Mushano} {et~al.}(2024){Mushano}, {Ogawa}, {Ohsuga}, {Yajima}, and
 {Omukai}]{mushanoetal24}
{Mushano}, T.; {Ogawa}, T.; {Ohsuga}, K.; {Yajima}, H.; {Omukai}, K.
\newblock {Impact of Ly$\alpha$ radiation force on super-Eddington accretion
 onto a massive black hole}.
\newblock {\em arXiv} {\bf 2024}, arXiv:2410.04378.
\newblock {\url{https://doi.org/10.48550/arXiv.2410.04378}}.

\bibitem[{Milosavljevi{\'c}} {et~al.}(2009){Milosavljevi{\'c}}, {Couch}, and
 {Bromm}]{milosavlijevicetal09}
{Milosavljevi{\'c}}, M.; {Couch}, S.M.; {Bromm}, V.
\newblock {Accretion Onto Intermediate-Mass Black Holes in Dense Protogalactic
 Clouds}.
\newblock {\em \apjl} {\bf 2009}, {\em 696},~L146--L149.
\newblock {\url{https://doi.org/10.1088/0004-637X/696/2/L146}}.

\bibitem[{D'Onofrio} and {Burigana}(2009)]{donofrioburigana09}
{D'Onofrio}, M.; {Burigana}, C.
\newblock {\em {Questions of Modern Cosmology: Galileo's Legacy}}; Springer: Berlin/Heidelberg, Germany, 2009.
\newblock {\url{https://doi.org/10.1007/978-3-642-00792-7}}.

\bibitem[{Zamfir} {et~al.}(2010){Zamfir}, {Sulentic}, {Marziani}, and
 {Dultzin}]{zamfiretal10}
{Zamfir}, S.; {Sulentic}, J.W.; {Marziani}, P.; {Dultzin}, D.
\newblock {Detailed characterization of H{$\beta$} emission line profile in
 low-z SDSS quasars}.
\newblock {\em \mnras} {\bf 2010}, {\em 403},~1759.
\newblock {\url{https://doi.org/10.1111/j.1365-2966.2009.16236.x}}.

\bibitem[{Trakhtenbrot} and {Netzer}(2012)]{trakhtenbrotnetzer12}
{Trakhtenbrot}, B.; {Netzer}, H.
\newblock {Black hole growth to z = 2 - I. Improved virial methods for
 measuring M$_{BH}$ and L/L$_{Edd}$}.
\newblock {\em \mnras} {\bf 2012}, {\em 427},~3081--3102.
\newblock {\url{https://doi.org/10.1111/j.1365-2966.2012.22056.x}}.

\bibitem[{Mapelli}(2020)]{mapelli20}
{Mapelli}, M.
\newblock {Binary black hole mergers: Formation and populations}.
\newblock {\em Front. Astron. Space Sci.} {\bf 2020}, {\em
 7},~38. 
\newblock {\url{https://doi.org/10.3389/fspas.2020.00038}}.

\bibitem[{Greene} {et~al.}(2020){Greene}, {Strader}, and {Ho}]{greeneetal20}
{Greene}, J.E.; {Strader}, J.; {Ho}, L.C.
\newblock {Intermediate-Mass Black Holes}.
\newblock {\em \araa} {\bf 2020}, {\em 58},~257--312.
\newblock {\url{https://doi.org/10.1146/annurev-astro-032620-021835}}.

\bibitem[{Davis} {et~al.}(2024){Davis}, {Graham}, {Soria}, {Jin},
 {Karachentsev}, {Karachentseva}, and {D'Onghia}]{davisetal24}
{Davis}, B.L.; {Graham}, A.W.; {Soria}, R.; {Jin}, Z.; {Karachentsev}, I.D.;
 {Karachentseva}, V.E.; {D'Onghia}, E.
\newblock {Identification of Intermediate-mass Black Hole Candidates among a
 Sample of Sd Galaxies}.
\newblock {\em \apj} {\bf 2024}, {\em 971},~123.
\newblock {\url{https://doi.org/10.3847/1538-4357/ad55eb}}.

\bibitem[{Ebisuzaki} {et~al.}(2001){Ebisuzaki}, {Makino}, {Tsuru}, {Funato},
 {Portegies Zwart}, {Hut}, {McMillan}, {Matsushita}, {Matsumoto}, and
 {Kawabe}]{ebisuzakietal01}
{Ebisuzaki}, T.; {Makino}, J.; {Tsuru}, T.G.; {Funato}, Y.; {Portegies Zwart},
 S.; {Hut}, P.; {McMillan}, S.; {Matsushita}, S.; {Matsumoto}, H.; {Kawabe},
 R.
\newblock {Missing Link Found? The ``Runaway'' Path to Supermassive Black
 Holes}.
\newblock {\em \apjl} {\bf 2001}, {\em 562},~L19--L22.
\newblock {\url{https://doi.org/10.1086/338118}}.

\bibitem[{Mapelli}(2016)]{mapelli16}
{Mapelli}, M.
\newblock {Massive black hole binaries from runaway collisions: The impact of
 metallicity}.
\newblock {\em \mnras} {\bf 2016}, {\em 459},~3432--3446.
\newblock {\url{https://doi.org/10.1093/mnras/stw869}}.

\bibitem[{Portegies Zwart} and {McMillan}(2002)]{portegiesetal02}
{Portegies Zwart}, S.F.; {McMillan}, S.L.W.
\newblock {The Runaway Growth of Intermediate-Mass Black Holes in Dense Star
 Clusters}.
\newblock {\em \apj} {\bf 2002}, {\em 576},~899--907.
\newblock {\url{https://doi.org/10.1086/341798}}.

\bibitem[{Portegies Zwart} {et~al.}(2004){Portegies Zwart}, {Baumgardt},
 {Hut}, {Makino}, and {McMillan}]{portegiesetal04}
{Portegies Zwart}, S.F.; {Baumgardt}, H.; {Hut}, P.; {Makino}, J.; {McMillan},
 S.L.W.
\newblock {Formation of massive black holes through runaway collisions in dense
 young star clusters}.
\newblock {\em \nat} {\bf 2004}, {\em 428},~724--726.
\newblock {\url{https://doi.org/10.1038/nature02448}}.

\bibitem[{Reines} {et~al.}(2013){Reines}, {Greene}, and
 {Geha}]{reinesetal13}
{Reines}, A.E.; {Greene}, J.E.; {Geha}, M.
\newblock {Dwarf Galaxies with Optical Signatures of Active Massive Black
 Holes}.
\newblock {\em \apj} {\bf 2013}, {\em 775},~116.
\newblock {\url{https://doi.org/10.1088/0004-637X/775/2/116}}.

\bibitem[{Chilingarian} {et~al.}(2018){Chilingarian}, {Katkov},
 {Zolotukhin}, {Grishin}, {Beletsky}, {Boutsia}, and
 {Osip}]{chilingarianetal18}
{Chilingarian}, I.V.; {Katkov}, I.Y.; {Zolotukhin}, I.Y.; {Grishin}, K.A.;
 {Beletsky}, Y.; {Boutsia}, K.; {Osip}, D.J.
\newblock {A Population of Bona Fide Intermediate-mass Black Holes Identified
 as Low-luminosity Active Galactic Nuclei}.
\newblock {\em \apj} {\bf 2018}, {\em 863},~1.
\newblock {\url{https://doi.org/10.3847/1538-4357/aad184}}.

\bibitem[{Yue} {et~al.}(2024){Yue}, {Eilers}, {Simcoe}, {Mackenzie},
 {Matthee}, {Kashino}, {Bordoloi}, {Lilly}, and {Naidu}]{yueetal24}
{Yue}, M.; {Eilers}, A.C.; {Simcoe}, R.A.; {Mackenzie}, R.; {Matthee}, J.;
 {Kashino}, D.; {Bordoloi}, R.; {Lilly}, S.J.; {Naidu}, R.P.
\newblock {EIGER. V. Characterizing the Host Galaxies of Luminous Quasars at z
 {\ensuremath{\gtrsim}} 6}.
\newblock {\em \apj} {\bf 2024}, {\em 966},~176.
\newblock {\url{https://doi.org/10.3847/1538-4357/ad3914}}.

\bibitem[{Stone} {et~al.}(2024){Stone}, {Lyu}, {Rieke}, {Alberts}, and
 {Hainline}]{stoneetal24}
{Stone}, M.A.; {Lyu}, J.; {Rieke}, G.H.; {Alberts}, S.; {Hainline}, K.N.
\newblock {Undermassive Host Galaxies of Five z {\ensuremath{\sim}} 6 Luminous
 Quasars Detected with JWST}.
\newblock {\em \apj} {\bf 2024}, {\em 964},~90.
\newblock {\url{https://doi.org/10.3847/1538-4357/ad2a57}}.

\bibitem[{Kokorev} {et~al.}(2023){Kokorev}, {Fujimoto}, {Labbe}, {Greene},
 {Bezanson}, {Dayal}, {Nelson}, {Atek}, {Brammer}, {Caputi}, {Chemerynska},
 {Cutler}, {Feldmann}, {Fudamoto}, {Furtak}, {Goulding}, {de Graaff}, {Leja},
 {Marchesini}, {Miller}, {Nanayakkara}, {Oesch}, {Pan}, {Price}, {Setton},
 {Smit}, {Stefanon}, {Wang}, {Weaver}, {Whitaker}, {Williams}, and
 {Zitrin}]{kokorevetal23}
{Kokorev}, V.; {Fujimoto}, S.; {Labbe}, I.; {Greene}, J.E.; {Bezanson}, R.;
 {Dayal}, P.; {Nelson}, E.J.; {Atek}, H.; {Brammer}, G.; {Caputi}, K.I.;
 et~al.
\newblock {UNCOVER: A NIRSpec Identification of a Broad-line AGN at z = 8.50}.
\newblock {\em \apjl} {\bf 2023}, {\em 957},~L7.
\newblock {\url{https://doi.org/10.3847/2041-8213/ad037a}}.

\bibitem[{Mazzucchelli} {et~al.}(2023){Mazzucchelli}, {Bischetti},
 {D'Odorico}, {Feruglio}, {Schindler}, {Onoue}, {Ba{\~n}ados}, {Becker},
 {Bian}, {Carniani}, {Decarli}, {Eilers}, {Farina}, {Gallerani}, {Lai},
 {Meyer}, {Rojas-Ruiz}, {Satyavolu}, {Venemans}, {Wang}, {Yang}, and
 {Zhu}]{mazzucchellietal23}
{Mazzucchelli}, C.; {Bischetti}, M.; {D'Odorico}, V.; {Feruglio}, C.;
 {Schindler}, J.T.; {Onoue}, M.; {Ba{\~n}ados}, E.; {Becker}, G.D.; {Bian},
 F.; {Carniani}, S.; et~al.
\newblock {XQR-30: Black hole masses and accretion rates of 42 z
 {\ensuremath{\gtrsim}} 6 quasars}.
\newblock {\em \aap} {\bf 2023}, {\em 676},~A71.
\newblock {\url{https://doi.org/10.1051/0004-6361/202346317}}.

\bibitem[{{\"U}bler} {et~al.}(2023){{\"U}bler}, {Maiolino}, {Curtis-Lake},
 {P{\'e}rez-Gonz{\'a}lez}, {Curti}, {Perna}, {Arribas}, {Charlot}, {Marshall},
 {D'Eugenio}, {Scholtz}, {Bunker}, {Carniani}, {Ferruit}, {Jakobsen}, {Rix},
 {Rodr{\'\i}guez Del Pino}, {Willott}, {Boeker}, {Cresci}, {Jones}, {Kumari},
 and {Rawle}]{ubleretal23}
{{\"U}bler}, H.; {Maiolino}, R.; {Curtis-Lake}, E.; {P{\'e}rez-Gonz{\'a}lez},
 P.G.; {Curti}, M.; {Perna}, M.; {Arribas}, S.; {Charlot}, S.; {Marshall},
 M.A.; {D'Eugenio}, F.; et~al.
\newblock {GA-NIFS: A massive black hole in a low-metallicity AGN at z
 {\ensuremath{\sim}} 5.55 revealed by JWST/NIRSpec IFS}.
\newblock {\em \aap} {\bf 2023}, {\em 677},~A145.
\newblock {\url{https://doi.org/10.1051/0004-6361/202346137}}.

\bibitem[{Narayan} {et~al.}(2003){Narayan}, {Igumenshchev}, and
 {Abramowicz}]{narayanetal03}
{Narayan}, R.; {Igumenshchev}, I.V.; {Abramowicz}, M.A.
\newblock {Magnetically Arrested Disk: An Energetically Efficient Accretion
 Flow}.
\newblock {\em \pasj} {\bf 2003}, {\em 55},~L69--L72.
\newblock {\url{https://doi.org/10.1093/pasj/55.6.L69}}.

\bibitem[{Blandford} and {Znajek}(1977)]{blandfordznajek77}
{Blandford}, R.D.; {Znajek}, R.L.
\newblock {Electromagnetic extraction of energy from Kerr black holes}.
\newblock {\em \mnras} {\bf 1977}, {\em 179},~433--456.

\bibitem[{McKinney} {et~al.}(2015){McKinney}, {Dai}, and
 {Avara}]{mckinneyetal15}
{McKinney}, J.C.; {Dai}, L.; {Avara}, M.J.
\newblock {Efficiency of super-Eddington magnetically-arrested accretion}.
\newblock {\em \mnras} {\bf 2015}, {\em 454},~L6--L10.
\newblock {\url{https://doi.org/10.1093/mnrasl/slv115}}.

\bibitem[{Ricarte} {et~al.}(2023){Ricarte}, {Narayan}, and
 {Curd}]{ricarteetal23}
{Ricarte}, A.; {Narayan}, R.; {Curd}, B.
\newblock {Recipes for Jet Feedback and Spin Evolution of Black Holes with
 Strongly Magnetized Super-Eddington Accretion Disks}.
\newblock {\em \apjl} {\bf 2023}, {\em 954},~L22.
\newblock {\url{https://doi.org/10.3847/2041-8213/aceda5}}.

\bibitem[{Lowell} {et~al.}(2024){Lowell}, {Jacquemin-Ide}, {Tchekhovskoy},
 and {Duncan}]{lowelletal24}
{Lowell}, B.; {Jacquemin-Ide}, J.; {Tchekhovskoy}, A.; {Duncan}, A.
\newblock {Rapid Black Hole Spin-down by Thick Magnetically Arrested Disks}.
\newblock {\em \apj} {\bf 2024}, {\em 960},~82.
\newblock {\url{https://doi.org/10.3847/1538-4357/ad09af}}.

\bibitem[{S{\k{a}}dowski} and {Narayan}(2015)]{sadowskinarayan15}
{S{\k{a}}dowski}, A.; {Narayan}, R.
\newblock {Powerful radiative jets in supercritical accretion discs around
 non-spinning black holes}.
\newblock {\em \mnras} {\bf 2015}, {\em 453},~3213--3221.
\newblock {\url{https://doi.org/10.1093/mnras/stv1802}}.

\bibitem[{Yang} {et~al.}(2023){Yang}, {Yuan}, {Kwan}, and {Dai}]{yangetal23}
{Yang}, H.; {Yuan}, F.; {Kwan}, T.; {Dai}, L.
\newblock {The properties of wind and jet from a super-Eddington accretion flow
 around a supermassive black hole}.
\newblock {\em \mnras} {\bf 2023}, {\em 523},~208--220.
\newblock {\url{https://doi.org/10.1093/mnras/stad1444}}.

\bibitem[{Czerny} {et~al.}(2023){Czerny}, {Sniegowska}, {Janiuk}, and
 {You}]{czernyetal23}
{Czerny}, B.; {Sniegowska}, M.; {Janiuk}, A.; {You}, B.
\newblock {Accretion processes onto black holes: Theoretical problems,
 observational constraints}.
\newblock {\em arXiv} {\bf 2023}, arXiv:2312.02911.
\newblock {\url{https://doi.org/10.48550/arXiv.2312.02911}}.

\bibitem[{Paliya} {et~al.}(2024){Paliya}, {Stalin}, {Dom{\'\i}nguez}, and
 {Saikia}]{paliyaetal24}
{Paliya}, V.S.; {Stalin}, C.S.; {Dom{\'\i}nguez}, A.; {Saikia}, D.J.
\newblock {Narrow-line Seyfert 1 galaxies in Sloan Digital Sky Survey: A new
 optical spectroscopic catalogue}.
\newblock {\em \mnras} {\bf 2024}, {\em 527},~7055--7069.
\newblock {\url{https://doi.org/10.1093/mnras/stad3650}}.

\bibitem[{Foschini} {et~al.}(2024){Foschini}, {Dalla Barba}, {Tornikoski},
 {Andernach}, {Marziani}, {Marscher}, {Jorstad}, {J{\"a}rvel{\"a}},
 {Ant{\'o}n}, and {Dalla Bont{\`a}}]{foschinietal24}
{Foschini}, L.; {Dalla Barba}, B.; {Tornikoski}, M.; {Andernach}, H.;
 {Marziani}, P.; {Marscher}, A.P.; {Jorstad}, S.G.; {J{\"a}rvel{\"a}}, E.;
 {Ant{\'o}n}, S.; {Dalla Bont{\`a}}, E.
\newblock {The Power of Relativistic Jets: A Comparative Study}.
\newblock {\em Universe} {\bf 2024}, {\em 10},~156.
\newblock {\url{https://doi.org/10.3390/universe10040156}}.

\bibitem[{Lewin} {et~al.}(1997){Lewin}, {van Paradijs}, and {van den
 Heuvel}]{lewinetal97}
{Lewin}, W.H.G.; {van Paradijs}, J.; {van den Heuvel}, E.P.J.
\newblock {\em {X-Ray Binaries}}; Cambridge University Press: Cambridge, UK, 1997.

\bibitem[Remillard and McClintock(2006)]{remillard2006}
Remillard, R.A.; McClintock, J.E.
\newblock X-ray properties of black-hole binaries.
\newblock {\em Annu. Rev. Astron. Astrophys.} {\bf 2006}, {\em
 44},~49--92.

\bibitem[{Fender} {et~al.}(2004){Fender}, {Belloni}, and
 {Gallo}]{fenderetal04}
{Fender}, R.P.; {Belloni}, T.M.; {Gallo}, E.
\newblock {Towards a unified model for black hole X-ray binary jets}.
\newblock {\em \mnras} {\bf 2004}, {\em 355},~1105--1118.
\newblock {\url{https://doi.org/10.1111/j.1365-2966.2004.08384.x}}.

\bibitem[{Homan} and {Belloni}(2005)]{homanbelloni05}
{Homan}, J.; {Belloni}, T.
\newblock {The Evolution of Black Hole States}.
\newblock {\em \apss} {\bf 2005}, {\em 300},~107--117.
\newblock {\url{https://doi.org/10.1007/s10509-005-1197-4}}.

\bibitem[{Belloni}(2010)]{belloni10}
{Belloni}, T.M.
\newblock {States and Transitions in Black Hole Binaries}. 
In {\em The Jet Paradigm}; {Belloni}, T., Ed.; Lecture Notes in Physics; Springer: Berlin/Heidelberg, Germany, 2010; 
Volume 794, p.~53.
\newblock {\url{https://doi.org/10.1007/978-3-540-76937-8_3}}.

\bibitem[{Koljonen} {et~al.}(2018){Koljonen}, {Maccarone}, {McCollough},
 {Gurwell}, {Trushkin}, {Pooley}, {Piano}, and {Tavani}]{koljonenetal18}
{Koljonen}, K.I.I.; {Maccarone}, T.; {McCollough}, M.L.; {Gurwell}, M.;
 {Trushkin}, S.A.; {Pooley}, G.G.; {Piano}, G.; {Tavani}, M.
\newblock {The hypersoft state of Cygnus X-3. A key to jet quenching in X-ray
 binaries?}
\newblock {\em \aap} {\bf 2018}, {\em 612},~A27.
\newblock {\url{https://doi.org/10.1051/0004-6361/201732284}}.

\bibitem[{Done} {et~al.}(2007){Done}, {Gierli{\'n}ski}, and
 {Kubota}]{doneetal07}
{Done}, C.; {Gierli{\'n}ski}, M.; {Kubota}, A.
\newblock {Modelling the behaviour of accretion flows in X-ray binaries.
 Everything you always wanted to know about accretion but were afraid to ask}.
\newblock {\em Astron. Astroph. Rev.} {\bf 2007}, {\em 15},~1--66.
\newblock {\url{https://doi.org/10.1007/s00159-007-0006-1}}.

\bibitem[{Gladstone} {et~al.}(2009){Gladstone}, {Roberts}, and
 {Done}]{gladstoneetal09}
{Gladstone}, J.C.; {Roberts}, T.P.; {Done}, C.
\newblock {The ultraluminous state}.
\newblock {\em \mnras} {\bf 2009}, {\em 397},~1836--1851.
\newblock {\url{https://doi.org/10.1111/j.1365-2966.2009.15123.x}}.

\bibitem[{Middleton} {et~al.}(2013){Middleton}, {Miller-Jones}, {Markoff},
 {Fender}, {Henze}, {Hurley-Walker}, {Scaife}, {Roberts}, {Walton},
 {Carpenter}, {Macquart}, {Bower}, {Gurwell}, {Pietsch}, {Haberl}, {Harris},
 {Daniel}, {Miah}, {Done}, {Morgan}, {Dickinson}, {Charles}, {Burwitz}, {Della
 Valle}, {Freyberg}, {Greiner}, {Hernanz}, {Hartmann}, {Hatzidimitriou},
 {Riffeser}, {Sala}, {Seitz}, {Reig}, {Rau}, {Orio}, {Titterington}, and
 {Grainge}]{middletonetal13}
{Middleton}, M.J.; {Miller-Jones}, J.C.A.; {Markoff}, S.; {Fender}, R.;
 {Henze}, M.; {Hurley-Walker}, N.; {Scaife}, A.M.M.; {Roberts}, T.P.;
 {Walton}, D.; {Carpenter}, J.; et~al.
\newblock {Bright radio emission from an ultraluminous stellar-mass microquasar
 in M 31}.
\newblock {\em \nat} {\bf 2013}, {\em 493},~187--190.
\newblock {\url{https://doi.org/10.1038/nature11697}}.

\bibitem[{Bachetti} {et~al.}(2014){Bachetti}, {Harrison}, {Walton},
 {Grefenstette}, {Chakrabarty}, {F{\"u}rst}, {Barret}, {Beloborodov}, {Boggs},
 {Christensen}, {Craig}, {Fabian}, {Hailey}, {Hornschemeier}, {Kaspi},
 {Kulkarni}, {Maccarone}, {Miller}, {Rana}, {Stern}, {Tendulkar}, {Tomsick},
 {Webb}, and {Zhang}]{bachettietal14}
{Bachetti}, M.; {Harrison}, F.A.; {Walton}, D.J.; {Grefenstette}, B.W.;
 {Chakrabarty}, D.; {F{\"u}rst}, F.; {Barret}, D.; {Beloborodov}, A.; {Boggs},
 S.E.; {Christensen}, F.E.; et~al.
\newblock {An ultraluminous X-ray source powered by an accreting neutron star}.
\newblock {\em \nat} {\bf 2014}, {\em 514},~202--204.
\newblock {\url{https://doi.org/10.1038/nature13791}}.

\bibitem[{Kaaret} {et~al.}(2017){Kaaret}, {Feng}, and
 {Roberts}]{kaaretetal17}
{Kaaret}, P.; {Feng}, H.; {Roberts}, T.P.
\newblock {Ultraluminous X-Ray Sources}.
\newblock {\em \araa} {\bf 2017}, {\em 55},~303--341.
\newblock {\url{https://doi.org/10.1146/annurev-astro-091916-055259}}.

\bibitem[{MacKenzie} {et~al.}(2023){MacKenzie}, {Roberts}, and
 {Walton}]{mckenzieetal23}
{MacKenzie}, A.D.A.; {Roberts}, T.P.; {Walton}, D.J.
\newblock {The hyperluminous X‑ray source population}.
\newblock {\em Astron. Nachrichten} {\bf 2023}, {\em 344},~e20230028.
\newblock {\url{https://doi.org/10.1002/asna.20230028}}.

\bibitem[{Makishima} {et~al.}(1986){Makishima}, {Maejima}, {Mitsuda},
 {Bradt}, {Remillard}, {Tuohy}, {Hoshi}, and {Nakagawa}]{makishimaetal86}
{Makishima}, K.; {Maejima}, Y.; {Mitsuda}, K.; {Bradt}, H.V.; {Remillard},
 R.A.; {Tuohy}, I.R.; {Hoshi}, R.; {Nakagawa}, M.
\newblock {Simultaneous X-Ray and Optical Observations of GX 339-4 in an X-Ray
 High State}.
\newblock {\em \apj} {\bf 1986}, {\em 308},~635.
\newblock {\url{https://doi.org/10.1086/164534}}.

\bibitem[{Dunn} {et~al.}(2010){Dunn}, {Fender}, {K{\"o}rding}, {Belloni},
 and {Cabanac}]{dunnetal10}
{Dunn}, R.J.H.; {Fender}, R.P.; {K{\"o}rding}, E.G.; {Belloni}, T.; {Cabanac},
 C.
\newblock {A global spectral study of black hole X-ray binaries}.
\newblock {\em \mnras} {\bf 2010}, {\em 403},~61--82.
\newblock {\url{https://doi.org/10.1111/j.1365-2966.2010.16114.x}}.

\bibitem[{Corbel} {et~al.}(2000){Corbel}, {Fender}, {Tzioumis}, {Nowak},
 {McIntyre}, {Durouchoux}, and {Sood}]{corbeletal00}
{Corbel}, S.; {Fender}, R.P.; {Tzioumis}, A.K.; {Nowak}, M.; {McIntyre}, V.;
 {Durouchoux}, P.; {Sood}, R.
\newblock {Coupling of the X-ray and radio emission in the black hole candidate
 and compact jet source GX 339-4}.
\newblock {\em \aap} {\bf 2000}, {\em 359},~251--268.
\newblock {\url{https://doi.org/10.48550/arXiv.astro-ph/0003460}}.

\bibitem[{Homan} {et~al.}(2005){Homan}, {Buxton}, {Markoff}, {Bailyn},
 {Nespoli}, and {Belloni}]{homanetal05}
{Homan}, J.; {Buxton}, M.; {Markoff}, S.; {Bailyn}, C.D.; {Nespoli}, E.;
 {Belloni}, T.
\newblock {Multiwavelength Observations of the 2002 Outburst of GX 339-4: Two
 Patterns of X-Ray-Optical/Near-Infrared Behavior}.
\newblock {\em \apj} {\bf 2005}, {\em 624},~295--306.
\newblock {\url{https://doi.org/10.1086/428722}}.

\bibitem[{Fender} {et~al.}(2009){Fender}, {Homan}, and
 {Belloni}]{fenderetal09}
{Fender}, R.P.; {Homan}, J.; {Belloni}, T.M.
\newblock {Jets from black hole X-ray binaries: Testing, refining and extending
 empirical models for the coupling to X-rays}.
\newblock {\em \mnras} {\bf 2009}, {\em 396},~1370--1382.
\newblock {\url{https://doi.org/10.1111/j.1365-2966.2009.14841.x}}.

\bibitem[{Belloni} {et~al.}(2005){Belloni}, {Homan}, {Casella}, {van der
 Klis}, {Nespoli}, {Lewin}, {Miller}, and {M{\'e}ndez}]{bellonietal05}
{Belloni}, T.; {Homan}, J.; {Casella}, P.; {van der Klis}, M.; {Nespoli}, E.;
 {Lewin}, W.H.G.; {Miller}, J.M.; {M{\'e}ndez}, M.
\newblock {The evolution of the timing properties of the black-hole transient
 GX 339-4 during its 2002/2003 outburst}.
\newblock {\em \aap} {\bf 2005}, {\em 440},~207--222.
\newblock {\url{https://doi.org/10.1051/0004-6361:20042457}}.

\bibitem[{Motta} {et~al.}(2011){Motta}, {Mu{\~n}oz-Darias}, {Casella},
 {Belloni}, and {Homan}]{mottaetal11}
{Motta}, S.; {Mu{\~n}oz-Darias}, T.; {Casella}, P.; {Belloni}, T.; {Homan}, J.
\newblock {Low-frequency oscillations in black holes: A spectral-timing
 approach to the case of GX 339-4}.
\newblock {\em \mnras} {\bf 2011}, {\em 418},~2292--2307.
\newblock {\url{https://doi.org/10.1111/j.1365-2966.2011.19566.x}}.

\bibitem[{Dihingia} {et~al.}(2023){Dihingia}, {Mizuno}, and
 {Sharma}]{dihingiaetal23}
{Dihingia}, I.K.; {Mizuno}, Y.; {Sharma}, P.
\newblock {High-soft to Low-hard State Transition in Black Hole X-Ray Binaries
 with GRMHD Simulations}.
\newblock {\em \apj} {\bf 2023}, {\em 958},~105.
\newblock {\url{https://doi.org/10.3847/1538-4357/ad0049}}.

\bibitem[{Corbel} and {Fender}(2002)]{corbelfender02}
{Corbel}, S.; {Fender}, R.P.
\newblock {Near-Infrared Synchrotron Emission from the Compact Jet of GX
 339-4}.
\newblock {\em \apjl} {\bf 2002}, {\em 573},~L35--L39.
\newblock {\url{https://doi.org/10.1086/341870}}.

\bibitem[{Corbel} {et~al.}(2003){Corbel}, {Nowak}, {Fender}, {Tzioumis}, and
 {Markoff}]{corbeletal03}
{Corbel}, S.; {Nowak}, M.A.; {Fender}, R.P.; {Tzioumis}, A.K.; {Markoff}, S.
\newblock {Radio/X-ray correlation in the low/hard state of GX 339-4}.
\newblock {\em \aap} {\bf 2003}, {\em 400},~1007--1012.
\newblock {\url{https://doi.org/10.1051/0004-6361:20030090}}.

\bibitem[{Gallo} {et~al.}(2003){Gallo}, {Fender}, and {Pooley}]{galloetal03}
{Gallo}, E.; {Fender}, R.P.; {Pooley}, G.G.
\newblock {A universal radio-X-ray correlation in low/hard state black hole
 binaries}.
\newblock {\em \mnras} {\bf 2003}, {\em 344},~60--72.
\newblock {\url{https://doi.org/10.1046/j.1365-8711.2003.06791.x}}.

\bibitem[Esin {et~al.}(1997)Esin, McClintock, and Narayan]{esin1997}
Esin, A.A.; McClintock, J.E.; Narayan, R.
\newblock Models of black hole X-ray binaries: Spectral energy distributions of
 LHSs and VHSs.
\newblock {\em Astrophys. J.} {\bf 1997}, {\em 489},~865--889.

\bibitem[Maccarone(2003)]{maccarone2003}
Maccarone, T.J.
\newblock On the evolution of the compact jet and the nature of X-ray states in
 black hole binaries.
\newblock {\em Astron. Astrophys.} {\bf 2003}, {\em 409},~697--706.

\bibitem[Done {et~al.}(2007)Done, Gierli{\'n}ski, and Kubota]{done2007}
Done, C.; Gierli{\'n}ski, M.; Kubota, A.
\newblock Modelling the behaviour of accretion flows in X-ray binaries: A
 review.
\newblock {\em Astron. Astrophys. Rev.} {\bf 2007}, {\em 15},~1--66.

\bibitem[{Siemiginowska} {et~al.}(1996){Siemiginowska}, {Czerny}, and
 {Kostyunin}]{siemiginowskaetal96}
{Siemiginowska}, A.; {Czerny}, B.; {Kostyunin}, V.
\newblock {Evolution of an Accretion Disk in an Active Galactic Nucleus}.
\newblock {\em \apj} {\bf 1996}, {\em 458},~491.
\newblock {\url{https://doi.org/10.1086/176831}}.

\bibitem[{Zamanov} and {Marziani}(2002)]{zamanovmarziani02}
{Zamanov}, R.; {Marziani}, P.
\newblock {Searching for the Physical Drivers of Eigenvector 1: From Quasars to
 Nanoquasars}.
\newblock {\em ApJ} {\bf 2002}, {\em 571},~L77--L80.
\newblock {\url{https://doi.org/10.1086/341367}}.

\bibitem[{Bon} {et~al.}(2018){Bon}, {Bon}, and {Marziani}]{bonetal18}
{Bon}, N.; {Bon}, E.; {Marziani}, P.
\newblock {AGN Broad Line Region variability in the context of Eigenvector 1:
 case of NGC 5548}.
\newblock {\em Front. Astron. Space Sci.} {\bf 2018}, {\em
 5},~3.
\newblock {\url{https://doi.org/10.3389/fspas.2018.00003}}.

\bibitem[{Panda} and {{\'S}niegowska}(2024)]{pandasniegowska24}
{Panda}, S.; {{\'S}niegowska}, M.
\newblock {Changing-look Active Galactic Nuclei. I. Tracking the Transition on
 the Main Sequence of Quasars}.
\newblock {\em \apjs} {\bf 2024}, {\em 272},~13.
\newblock {\url{https://doi.org/10.3847/1538-4365/ad344f}}.

\bibitem[{Komossa} and {Greiner}(1999)]{komossagreiner99}
{Komossa}, S.; {Greiner}, J.
\newblock {Discovery of a giant and luminous X-ray outburst from the optically
 inactive galaxy pair RX J1242.6-1119}.
\newblock {\em \aap} {\bf 1999}, {\em 349},~L45--L48.
\newblock {\url{https://doi.org/10.48550/arXiv.astro-ph/9908216}}.

\bibitem[{Sniegowska} {et~al.}(2020){Sniegowska}, {Czerny}, {Bon}, and
 {Bon}]{sniegowskaetal20}
{Sniegowska}, M.; {Czerny}, B.; {Bon}, E.; {Bon}, N.
\newblock {Possible mechanism for multiple changing-look phenomena in active
 galactic nuclei}.
\newblock {\em \aap} {\bf 2020}, {\em 641},~A167.
\newblock {\url{https://doi.org/10.1051/0004-6361/202038575}}.

\bibitem[{Dai} {et~al.}(2018){Dai}, {McKinney}, {Roth}, {Ramirez-Ruiz}, and
 {Miller}]{daietal18}
{Dai}, L.; {McKinney}, J.C.; {Roth}, N.; {Ramirez-Ruiz}, E.; {Miller}, M.C.
\newblock {A Unified Model for Tidal Disruption Events}.
\newblock {\em \apjl} {\bf 2018}, {\em 859},~L20.
\newblock {\url{https://doi.org/10.3847/2041-8213/aab429}}.

\bibitem[{Petrushevska} {et~al.}(2023){Petrushevska}, {Leloudas},
 {Ili{\'c}}, {Bronikowski}, {Charalampopoulos}, {Jaisawal}, {Paraskeva},
 {Pursiainen}, {Raki{\'c}}, {Schulze}, {Taggart}, {Wedderkopp}, {Anderson},
 {de Boer}, {Chambers}, {Chen}, {Damljanovi{\'c}}, {Fraser}, {Gao}, {Gomboc},
 {Gromadzki}, {Ihanec}, {Maguire}, {Mar{\v{c}}un}, {M{\"u}ller-Bravo},
 {Nicholl}, {Onori}, {Reynolds}, {Smartt}, {Sollerman}, {Smith}, {Wevers}, and
 {Wyrzykowski}]{petrushevskaetal23}
{Petrushevska}, T.; {Leloudas}, G.; {Ili{\'c}}, D.; {Bronikowski}, M.;
 {Charalampopoulos}, P.; {Jaisawal}, G.K.; {Paraskeva}, E.; {Pursiainen}, M.;
 {Raki{\'c}}, N.; {Schulze}, S.; et~al.
\newblock {The rise and fall of the iron-strong nuclear transient PS16dtm}.
\newblock {\em \aap} {\bf 2023}, {\em 669},~A140.
\newblock {\url{https://doi.org/10.1051/0004-6361/202244623}}.

\bibitem[{Ili{\'c}} {et~al.}(2020){Ili{\'c}}, {Oknyansky}, {Popovi{\'c}},
 {Tsygankov}, {Belinski}, {Tatarnikov}, {Dodin}, {Shatsky}, {Ikonnikova},
 {Raki{\'c}}, {Kova{\v{c}}evi{\'c}}, {Mar{\v{c}}eta-Mandi{\'c}}, {Burlak},
 {Mishin}, {Metlova}, {Potanin}, and {Zheltoukhov}]{ilicetal20}
{Ili{\'c}}, D.; {Oknyansky}, V.; {Popovi{\'c}}, L.{\v{C}}.; {Tsygankov}, S.S.;
 {Belinski}, A.A.; {Tatarnikov}, A.M.; {Dodin}, A.V.; {Shatsky}, N.I.;
 {Ikonnikova}, N.P.; {Raki{\'c}}, N.; et~al.
\newblock {A flare in the optical spotted in the changing-look Seyfert NGC
 3516}.
\newblock {\em \aap} {\bf 2020}, {\em 638},~A13.
\newblock {\url{https://doi.org/10.1051/0004-6361/202037532}}.

\bibitem[{Oknyansky} {et~al.}(2020){Oknyansky}, {Tsygankov}, {Dodin},
 {Tatarnikov}, {Belinski}, {Ikonnikova}, {Burlak}, {Fedoteva}, {Shatsky},
 {Mishin}, {Zheltoukhov}, and {Potanin}]{oknyanskyetal20}
{Oknyansky}, V.L.; {Tsygankov}, S.S.; {Dodin}, A.V.; {Tatarnikov}, A.M.;
 {Belinski}, A.A.; {Ikonnikova}, N.P.; {Burlak}, M.A.; {Fedoteva}, A.A.;
 {Shatsky}, N.I.; {Mishin}, E.O.; et~al.
\newblock {Changing Look AGN NGC 3516 brightens again after long being in a
 very low state}.
\newblock {\em Astron. Telegr.} {\bf 2020}, {\em 13691},~1.

\bibitem[{Oknyansky} {et~al.}(2021){Oknyansky}, {Tsygankov}, {Lipunov},
 {Gorbovskoy}, and {Tyurina}]{oknyanskyetal21}
{Oknyansky}, V.L.; {Tsygankov}, S.S.; {Lipunov}, V.M.; {Gorbovskoy}, E.S.;
 {Tyurina}, N.V.
\newblock {Discovery of new changing look in NGC 1566}.
\newblock In \emph{Proceedings of the Nuclear Activity in Galaxies Across Cosmic Time}; 
 {Povi{\'c}}, M., {Marziani}, P., {Masegosa}, J., {Netzer}, H., {Negu}, S.H., {Tessema}, S.B., Eds.; 
 IAU Symposium; 
 {Cambridge University Press: Cambridge, UK}, 2021; Volume 356, pp. 127--131.
\newblock {\url{https://doi.org/10.1017/S1743921320002720}}.

\bibitem[{Oknyansky}(2022)]{oknyansky22}
{Oknyansky}, V.
\newblock {Changing looks of the nucleus of the Seyfert galaxy NGC 1566
 compared with other changing-look AGNs}.
\newblock {\em Astron. Nachrichten} {\bf 2022}, {\em 343},~e210080.
\newblock {\url{https://doi.org/10.1002/asna.20210080}}.

\bibitem[{Xu} {et~al.}(2024){Xu}, {Komossa}, {Grupe}, {Wang}, {Xin}, {Han},
 {Wei}, {Bai}, {Bon}, {Cangemi}, {Cordier}, {Dennefeld}, {Gallo},
 {Kollatschny}, {Kong}, {Ochmann}, {Qiu}, and {Schartel}]{xuetal24}
{Xu}, D.W.; {Komossa}, S.; {Grupe}, D.; {Wang}, J.; {Xin}, L.P.; {Han}, X.H.;
 {Wei}, J.Y.; {Bai}, J.Y.; {Bon}, E.; {Cangemi}, F.; et~al.
\newblock {Changing-Look Narrow-Line Seyfert 1 Galaxies, their Detection with
 SVOM, and the Case of NGC 1566}.
\newblock {\em Universe} {\bf 2024}, {\em 10},~61.
\newblock {\url{https://doi.org/10.3390/universe10020061}}.

\bibitem[{Ulrich} {et~al.}(1997){Ulrich}, {Maraschi}, and
 {Urry}]{ulrichetal97}
{Ulrich}, M.H.; {Maraschi}, L.; {Urry}, C.M.
\newblock {Variability of Active Galactic Nuclei}.
\newblock {\em \araa} {\bf 1997}, {\em 35},~445--502.
\newblock {\url{https://doi.org/10.1146/annurev.astro.35.1.445}}.

\bibitem[{Drake} {et~al.}(2009){Drake}, {Djorgovski}, {Mahabal}, {Beshore},
 {Larson}, {Graham}, {Williams}, {Christensen}, {Catelan}, {Boattini},
 {Gibbs}, {Hill}, and {Kowalski}]{drakeetal09}
{Drake}, A.J.; {Djorgovski}, S.G.; {Mahabal}, A.; {Beshore}, E.; {Larson}, S.;
 {Graham}, M.J.; {Williams}, R.; {Christensen}, E.; {Catelan}, M.; {Boattini},
 A.; et~al.
\newblock {First Results from the Catalina Real-Time Transient Survey}.
\newblock {\em \apj} {\bf 2009}, {\em 696},~870--884.
\newblock {\url{https://doi.org/10.1088/0004-637X/696/1/870}}.

\bibitem[{Graham} {et~al.}(2019){Graham}, {Kulkarni}, {Bellm}, {Adams},
 {Barbarino}, {Blagorodnova}, {Bodewits}, {Bolin}, {Brady}, {Cenko}, {Chang},
 {Coughlin}, {De}, {Eadie}, {Farnham}, {Feindt}, {Franckowiak}, {Fremling},
 {Gezari}, {Ghosh}, {Goldstein}, {Golkhou}, {Goobar}, {Ho}, {Huppenkothen},
 {Ivezi{\'c}}, {Jones}, {Juric}, {Kaplan}, {Kasliwal}, {Kelley}, {Kupfer},
 {Lee}, {Lin}, {Lunnan}, {Mahabal}, {Miller}, {Ngeow}, {Nugent}, {Ofek},
 {Prince}, {Rauch}, {van Roestel}, {Schulze}, {Singer}, {Sollerman}, {Taddia},
 {Yan}, {Ye}, {Yu}, {Barlow}, {Bauer}, {Beck}, {Belicki}, {Biswas}, {Brinnel},
 {Brooke}, {Bue}, {Bulla}, {Burruss}, {Connolly}, {Cromer}, {Cunningham},
 {Dekany}, {Delacroix}, {Desai}, {Duev}, {Feeney}, {Flynn}, {Frederick},
 {Gal-Yam}, {Giomi}, {Groom}, {Hacopians}, {Hale}, {Helou}, {Henning},
 {Hover}, {Hillenbrand}, {Howell}, {Hung}, {Imel}, {Ip}, {Jackson}, {Kaspi},
 {Kaye}, {Kowalski}, {Kramer}, {Kuhn}, {Landry}, {Laher}, {Mao}, {Masci},
 {Monkewitz}, {Murphy}, {Nordin}, {Patterson}, {Penprase}, {Porter},
 {Rebbapragada}, {Reiley}, {Riddle}, {Rigault}, {Rodriguez}, {Rusholme}, {van
 Santen}, {Shupe}, {Smith}, {Soumagnac}, {Stein}, {Surace}, {Szkody}, {Terek},
 {Van Sistine}, {van Velzen}, {Vestrand}, {Walters}, {Ward}, {Zhang}, and
 {Zolkower}]{grahametal19}
{Graham}, M.J.; {Kulkarni}, S.R.; {Bellm}, E.C.; {Adams}, S.M.; {Barbarino},
 C.; {Blagorodnova}, N.; {Bodewits}, D.; {Bolin}, B.; {Brady}, P.R.; {Cenko},
 S.B.; et~al.
\newblock {The Zwicky Transient Facility: Science Objectives}.
\newblock {\em \pasp} {\bf 2019}, {\em 131},~078001.
\newblock {\url{https://doi.org/10.1088/1538-3873/ab006c}}.

\bibitem[{Masci} {et~al.}(2019){Masci}, {Laher}, {Rusholme}, {Shupe},
 {Groom}, {Surace}, {Jackson}, {Monkewitz}, {Beck}, {Flynn}, {Terek},
 {Landry}, {Hacopians}, {Desai}, {Howell}, {Brooke}, {Imel}, {Wachter}, {Ye},
 {Lin}, {Cenko}, {Cunningham}, {Rebbapragada}, {Bue}, {Miller}, {Mahabal},
 {Bellm}, {Patterson}, {Juri{\'c}}, {Golkhou}, {Ofek}, {Walters}, {Graham},
 {Kasliwal}, {Dekany}, {Kupfer}, {Burdge}, {Cannella}, {Barlow}, {Van
 Sistine}, {Giomi}, {Fremling}, {Blagorodnova}, {Levitan}, {Riddle}, {Smith},
 {Helou}, {Prince}, and {Kulkarni}]{mascietal19}
{Masci}, F.J.; {Laher}, R.R.; {Rusholme}, B.; {Shupe}, D.L.; {Groom}, S.;
 {Surace}, J.; {Jackson}, E.; {Monkewitz}, S.; {Beck}, R.; {Flynn}, D.;
 et~al.
\newblock {The Zwicky Transient Facility: Data Processing, Products, and
 Archive}.
\newblock {\em \pasp} {\bf 2019}, {\em 131},~018003.
\newblock {\url{https://doi.org/10.1088/1538-3873/aae8ac}}.

\bibitem[{Bellm} {et~al.}(2019){Bellm}, {Kulkarni}, {Graham}, {Dekany},
 {Smith}, {Riddle}, {Masci}, {Helou}, {Prince}, {Adams}, {Barbarino},
 {Barlow}, {Bauer}, {Beck}, {Belicki}, {Biswas}, {Blagorodnova}, {Bodewits},
 {Bolin}, {Brinnel}, {Brooke}, {Bue}, {Bulla}, {Burruss}, {Cenko}, {Chang},
 {Connolly}, {Coughlin}, {Cromer}, {Cunningham}, {De}, {Delacroix}, {Desai},
 {Duev}, {Eadie}, {Farnham}, {Feeney}, {Feindt}, {Flynn}, {Franckowiak},
 {Frederick}, {Fremling}, {Gal-Yam}, {Gezari}, {Giomi}, {Goldstein},
 {Golkhou}, {Goobar}, {Groom}, {Hacopians}, {Hale}, {Henning}, {Ho}, {Hover},
 {Howell}, {Hung}, {Huppenkothen}, {Imel}, {Ip}, {Ivezi{\'c}}, {Jackson},
 {Jones}, {Juric}, {Kasliwal}, {Kaspi}, {Kaye}, {Kelley}, {Kowalski},
 {Kramer}, {Kupfer}, {Landry}, {Laher}, {Lee}, {Lin}, {Lin}, {Lunnan},
 {Giomi}, {Mahabal}, {Mao}, {Miller}, {Monkewitz}, {Murphy}, {Ngeow},
 {Nordin}, {Nugent}, {Ofek}, {Patterson}, {Penprase}, {Porter}, {Rauch},
 {Rebbapragada}, {Reiley}, {Rigault}, {Rodriguez}, {van Roestel}, {Rusholme},
 {van Santen}, {Schulze}, {Shupe}, {Singer}, {Soumagnac}, {Stein}, {Surace},
 {Sollerman}, {Szkody}, {Taddia}, {Terek}, {Van Sistine}, {van Velzen},
 {Vestrand}, {Walters}, {Ward}, {Ye}, {Yu}, {Yan}, and
 {Zolkower}]{bellmetal19}
{Bellm}, E.C.; {Kulkarni}, S.R.; {Graham}, M.J.; {Dekany}, R.; {Smith}, R.M.;
 {Riddle}, R.; {Masci}, F.J.; {Helou}, G.; {Prince}, T.A.; {Adams}, S.M.;
 et~al.
\newblock {The Zwicky Transient Facility: System Overview, Performance, and
 First Results}.
\newblock {\em \pasp} {\bf 2019}, {\em 131},~018002.
\newblock {\url{https://doi.org/10.1088/1538-3873/aaecbe}}.

\bibitem[{Chambers} {et~al.}(2016){Chambers}, {Magnier}, {Metcalfe},
 {Flewelling}, {Huber}, {Waters}, {Denneau}, {Draper}, {Farrow}, {Finkbeiner},
 {Holmberg}, {Koppenhoefer}, {Price}, {Rest}, {Saglia}, {Schlafly}, {Smartt},
 {Sweeney}, {Wainscoat}, {Burgett}, {Chastel}, {Grav}, {Heasley}, {Hodapp},
 {Jedicke}, {Kaiser}, {Kudritzki}, {Luppino}, {Lupton}, {Monet}, {Morgan},
 {Onaka}, {Shiao}, {Stubbs}, {Tonry}, {White}, {Ba{\~n}ados}, {Bell},
 {Bender}, {Bernard}, {Boegner}, {Boffi}, {Botticella}, {Calamida},
 {Casertano}, {Chen}, {Chen}, {Cole}, {Deacon}, {Frenk}, {Fitzsimmons},
 {Gezari}, {Gibbs}, {Goessl}, {Goggia}, {Gourgue}, {Goldman}, {Grant},
 {Grebel}, {Hambly}, {Hasinger}, {Heavens}, {Heckman}, {Henderson}, {Henning},
 {Holman}, {Hopp}, {Ip}, {Isani}, {Jackson}, {Keyes}, {Koekemoer}, {Kotak},
 {Le}, {Liska}, {Long}, {Lucey}, {Liu}, {Martin}, {Masci}, {McLean}, {Mindel},
 {Misra}, {Morganson}, {Murphy}, {Obaika}, {Narayan}, {Nieto-Santisteban},
 {Norberg}, {Peacock}, {Pier}, {Postman}, {Primak}, {Rae}, {Rai}, {Riess},
 {Riffeser}, {Rix}, {R{\"o}ser}, {Russel}, {Rutz}, {Schilbach}, {Schultz},
 {Scolnic}, {Strolger}, {Szalay}, {Seitz}, {Small}, {Smith}, {Soderblom},
 {Taylor}, {Thomson}, {Taylor}, {Thakar}, {Thiel}, {Thilker}, {Unger},
 {Urata}, {Valenti}, {Wagner}, {Walder}, {Walter}, {Watters}, {Werner},
 {Wood-Vasey}, and {Wyse}]{chambersetal16}
{Chambers}, K.C.; {Magnier}, E.A.; {Metcalfe}, N.; {Flewelling}, H.A.; {Huber},
 M.E.; {Waters}, C.Z.; {Denneau}, L.; {Draper}, P.W.; {Farrow}, D.;
 {Finkbeiner}, D.P.; et~al.
\newblock {The Pan-STARRS1 Surveys}.
\newblock {\em arXiv} {\bf 2016}, arXiv:1612.05560.
\newblock {\url{https://doi.org/10.48550/arXiv.1612.05560}}.

\bibitem[{Kochanek} {et~al.}(2017){Kochanek}, {Shappee}, {Stanek},
 {Holoien}, {Thompson}, {Prieto}, {Dong}, {Shields}, {Will}, {Britt},
 {Perzanowski}, and {Pojma{\'n}ski}]{kochaneketal17}
{Kochanek}, C.S.; {Shappee}, B.J.; {Stanek}, K.Z.; {Holoien}, T.W.S.;
 {Thompson}, T.A.; {Prieto}, J.L.; {Dong}, S.; {Shields}, J.V.; {Will}, D.;
 {Britt}, C.; et~al.
\newblock {The All-Sky Automated Survey for Supernovae (ASAS-SN) Light Curve
 Server v1.0}.
\newblock {\em \pasp} {\bf 2017}, {\em 129},~104502.
\newblock {\url{https://doi.org/10.1088/1538-3873/aa80d9}}.

\bibitem[{Ivezi{\'c}} {et~al.}(2019){Ivezi{\'c}}, {Kahn}, {Tyson}, {Abel},
 {Acosta}, {Allsman}, {Alonso}, {AlSayyad}, {Anderson}, {Andrew}, {Angel},
 {Angeli}, {Ansari}, {Antilogus}, {Araujo}, {Armstrong}, {Arndt}, {Astier},
 {Aubourg}, {Auza}, {Axelrod}, {Bard}, {Barr}, {Barrau}, {Bartlett}, {Bauer},
 {Bauman}, {Baumont}, {Bechtol}, {Bechtol}, {Becker}, {Becla}, {Beldica},
 {Bellavia}, {Bianco}, {Biswas}, {Blanc}, {Blazek}, {Blandford}, {Bloom},
 {Bogart}, {Bond}, {Booth}, {Borgland}, {Borne}, {Bosch}, {Boutigny},
 {Brackett}, {Bradshaw}, {Brandt}, {Brown}, {Bullock}, {Burchat}, {Burke},
 {Cagnoli}, {Calabrese}, {Callahan}, {Callen}, {Carlin}, {Carlson},
 {Chandrasekharan}, {Charles-Emerson}, {Chesley}, {Cheu}, {Chiang}, {Chiang},
 {Chirino}, {Chow}, {Ciardi}, {Claver}, {Cohen-Tanugi}, {Cockrum}, {Coles},
 {Connolly}, {Cook}, {Cooray}, {Covey}, {Cribbs}, {Cui}, {Cutri}, {Daly},
 {Daniel}, {Daruich}, {Daubard}, {Daues}, {Dawson}, {Delgado}, {Dellapenna},
 {de Peyster}, {de Val-Borro}, {Digel}, {Doherty}, {Dubois},
 {Dubois-Felsmann}, {Durech}, {Economou}, {Eifler}, {Eracleous}, {Emmons},
 {Fausti Neto}, {Ferguson}, {Figueroa}, {Fisher-Levine}, {Focke}, {Foss},
 {Frank}, {Freemon}, {Gangler}, {Gawiser}, {Geary}, {Gee}, {Geha}, {Gessner},
 {Gibson}, {Gilmore}, {Glanzman}, {Glick}, {Goldina}, {Goldstein}, {Goodenow},
 {Graham}, {Gressler}, {Gris}, {Guy}, {Guyonnet}, {Haller}, {Harris},
 {Hascall}, {Haupt}, {Hernandez}, {Herrmann}, {Hileman}, {Hoblitt}, {Hodgson},
 {Hogan}, {Howard}, {Huang}, {Huffer}, {Ingraham}, {Innes}, {Jacoby}, {Jain},
 {Jammes}, {Jee}, {Jenness}, {Jernigan}, {Jevremovi{\'c}}, {Johns}, {Johnson},
 {Johnson}, {Jones}, {Juramy-Gilles}, {Juri{\'c}}, {Kalirai}, {Kallivayalil},
 {Kalmbach}, {Kantor}, {Karst}, {Kasliwal}, {Kelly}, {Kessler}, {Kinnison},
 {Kirkby}, {Knox}, {Kotov}, {Krabbendam}, {Krughoff}, {Kub{\'a}nek},
 {Kuczewski}, {Kulkarni}, {Ku}, {Kurita}, {Lage}, {Lambert}, {Lange},
 {Langton}, {Le Guillou}, {Levine}, {Liang}, {Lim}, {Lintott}, {Long},
 {Lopez}, {Lotz}, {Lupton}, {Lust}, {MacArthur}, {Mahabal}, {Mandelbaum},
 {Markiewicz}, {Marsh}, {Marshall}, {Marshall}, {May}, {McKercher}, {McQueen},
 {Meyers}, {Migliore}, {Miller}, {Mills}, {Miraval}, {Moeyens}, {Moolekamp},
 {Monet}, {Moniez}, {Monkewitz}, {Montgomery}, {Morrison}, {Mueller},
 {Muller}, {Mu{\~n}oz Arancibia}, {Neill}, {Newbry}, {Nief}, {Nomerotski},
 {Nordby}, {O'Connor}, {Oliver}, {Olivier}, {Olsen}, {O'Mullane}, {Ortiz},
 {Osier}, {Owen}, {Pain}, {Palecek}, {Parejko}, {Parsons}, {Pease},
 {Peterson}, {Peterson}, {Petravick}, {Libby Petrick}, {Petry},
 {Pierfederici}, {Pietrowicz}, {Pike}, {Pinto}, {Plante}, {Plate}, {Plutchak},
 {Price}, {Prouza}, {Radeka}, {Rajagopal}, {Rasmussen}, {Regnault}, {Reil},
 {Reiss}, {Reuter}, {Ridgway}, {Riot}, {Ritz}, {Robinson}, {Roby}, {Roodman},
 {Rosing}, {Roucelle}, {Rumore}, {Russo}, {Saha}, {Sassolas}, {Schalk},
 {Schellart}, {Schindler}, {Schmidt}, {Schneider}, {Schneider}, {Schoening},
 {Schumacher}, {Schwamb}, {Sebag}, {Selvy}, {Sembroski}, {Seppala}, {Serio},
 {Serrano}, {Shaw}, {Shipsey}, {Sick}, {Silvestri}, {Slater}, {Smith},
 {Smith}, {Sobhani}, {Soldahl}, {Storrie-Lombardi}, {Stover}, {Strauss},
 {Street}, {Stubbs}, {Sullivan}, {Sweeney}, {Swinbank}, {Szalay}, {Takacs},
 {Tether}, {Thaler}, {Thayer}, {Thomas}, {Thornton}, {Thukral}, {Tice},
 {Trilling}, {Turri}, {Van Berg}, {Vanden Berk}, {Vetter}, {Virieux},
 {Vucina}, {Wahl}, {Walkowicz}, {Walsh}, {Walter}, {Wang}, {Wang}, {Warner},
 {Wiecha}, {Willman}, {Winters}, {Wittman}, {Wolff}, {Wood-Vasey}, {Wu},
 {Xin}, {Yoachim}, and {Zhan}]{izevicetal19}
{Ivezi{\'c}}, {\v{Z}}.; {Kahn}, S.M.; {Tyson}, J.A.; {Abel}, B.; {Acosta}, E.;
 {Allsman}, R.; {Alonso}, D.; {AlSayyad}, Y.; {Anderson}, S.F.; {Andrew}, J.;
 et~al.
\newblock {LSST: From Science Drivers to Reference Design and Anticipated Data
 Products}.
\newblock {\em \apj} {\bf 2019}, {\em 873},~111.
\newblock {\url{https://doi.org/10.3847/1538-4357/ab042c}}.

\end{thebibliography}
\end{document}